\documentclass[lettersize,journal]{IEEEtran}
\usepackage{enumitem}
\usepackage{amsfonts}
\usepackage{enumitem,kantlipsum}
\usepackage{textcomp}
\usepackage{stfloats}
\usepackage{url}
\usepackage{caption}
\usepackage{graphics}
\usepackage{tabularx}
\usepackage[dvipsnames,table]{xcolor}
\usepackage{verbatim}
\usepackage{pifont}
\usepackage{float}
\usepackage{amsmath}
\usepackage{cancel}
\usepackage{libertinust1math}
\usepackage{pgfplots}
\usepackage{amsmath}
\usepackage{forest}
\usepackage{graphicx}
\usepackage{cite}
\usepackage{etoolbox}
\usepackage{graphicx}
\usepackage{float}
\usepackage{lscape}
\usepackage{longtable}
\usepackage{threeparttable}
\usepackage{ltxtable}
\usepackage{hhline}
\usepackage{caption}
\usepackage{wasysym}
\usepackage{subcaption}
\usepackage{algorithm} 
\usepackage{color} 
\usetikzlibrary{patterns}
\usepackage{colortbl}
\usepackage{algpseudocode}
\usepackage{verbatim}
\usepackage{lipsum}
\usetikzlibrary{automata, arrows.meta, positioning}
\usetikzlibrary{decorations.pathreplacing}
\usepackage{amsthm}
\usepackage{booktabs}
\usepackage{amsmath}
\usepackage{algorithm}
\usepackage{inputenc}
\usepackage{adjustbox}
\usepackage{url} 
\usepackage{multirow}
\usepackage{color,soul}
\usepackage{lipsum}
\usepackage{mathtools}
\usepackage{cuted}
\usepackage{xcolor}
\usepackage{wasysym}
\usepackage{pgfplotstable}
\usepackage[normalem]{ulem}
\usetikzlibrary{arrows.meta}
\usepackage{smartdiagram}
\usepackage{tikz}
\usetikzlibrary{mindmap, shapes, shadows}
\usepackage{xcolor}

\usepackage{enumitem}
\usepackage{amsfonts}
\usepackage{enumitem,kantlipsum}
\usepackage{textcomp}
\usepackage{stfloats}
\usepackage{url}
\usepackage{caption}
\usepackage{graphics}
\usepackage{tabularx}
\usepackage[dvipsnames,table]{xcolor}
\usepackage{verbatim}
\usepackage{pifont}
\usepackage{float}
\usepackage{amsmath}
\usepackage{cancel}
\usepackage{libertinust1math}
\usepackage{pgfplots}
\usepackage{amsmath}
\usepackage{forest}
\usepackage{graphicx}
\usepackage{cite}
\usepackage{etoolbox}
\usepackage{graphicx}
\usepackage{float}
\usepackage{lscape}
\usepackage{longtable}
\usepackage{ltxtable}
\usepackage{hhline}
\usepackage{caption}
\usepackage{wasysym}
\usepackage{subcaption}
\usepackage{algorithm} 
\usepackage{color} 
\usepackage{colortbl}
\usepackage{algpseudocode}
\usepackage{verbatim}
\usepackage{lipsum}
\usetikzlibrary{automata, arrows.meta, positioning}
\usetikzlibrary{decorations.pathreplacing}
\usepackage{amsthm}
\usepackage{booktabs}
\usepackage{amsmath}
\usepackage{algorithm}
\usepackage{inputenc}
\usepackage{adjustbox}
\usepackage{url} 
\usepackage{multirow}
\usepackage{color,soul}
\usepackage{lipsum}
\usepackage{mathtools}
\usepackage{cuted}
\usepackage{xcolor}
\usepackage{wasysym}
\usepackage{tikz}
\usepackage{adjustbox}
\usetikzlibrary{shapes.geometric, arrows}

\usepackage{url}    
\usepackage{xurl}   

\usepackage[acronym]{glossaries}
\newacronym{AES}{AES}{Advanced Encryption Standard}
\newacronym{CC}{CC}{Cloud Computing}
\newacronym{RSA}{RSA}{Rivest-Shamir-Adleman}
\newacronym{ECC}{ECC}{Elliptic Curve Cryptography}
\newacronym{DHE}{DHE}{Diffie-Hellman Exchange}
\newacronym{CICD}{CI/CD}{Continuous Integration and Continuous Deployment}
\newacronym{RPKI}{RPKI}{Resource Public Key Infrastructure}
\newacronym{PQC}{PQC}{Post-Quantum Cryptography}
\newacronym{PQ}{PQ}{Post-Quantum}
\newacronym{KEM/ENC}{KEM/ENC}{Key Encapsulation Mechanism/Encryption}
\newacronym{SIG}{SIG}{Signature}
\newacronym{CFI}{CFI}{Control-Flow Integrity}
\newacronym{SCA}{SCA}{Side-Channel Attack}
\newacronym{BIKE}{BIKE}{Bit Flipping Key Encapsulation}
\newacronym{HQC}{HQC}{Hamming Quasi-Cyclic}
\newacronym{OS}{OS}{Operating System}
\newacronym{VMs}{VMs}{Virtual Machines}
\newacronym{FA}{FA}{Fault Injection Attacks}
\newacronym{TPM}{TPM}{Trusted Platform Module}
\newacronym{TLS}{TLS}{Transport Layer Security}
\newacronym{VM}{VM}{Virtual Machine}
\newacronym{AEAD}{AEAD}{Authenticated Encryption with Associated Data}
\newacronym{EoP}{EoP}{Elevation of Privilege}
\newacronym{ASLR}{ASLR}{Address Space Layout Randomization}
\newacronym{ROP}{ROP}{Return-Oriented Programming}
\newacronym{JOP}{JOP}{Jump-Oriented Programming}
\newacronym{DEP}{DEP}{Data Execution Prevention}
\newacronym{JIT}{JIT}{Just-in-Time}
\newacronym{ACL}{ACL}{Access Control List}
\newacronym{vTPM}{vTPM}{virtual Trusted Platform Module}
\newacronym{TEE}{TEE}{Trusted Execution Environment}
\newacronym{DoS}{DoS}{Denial-of-Service}
\newacronym{API}{API}{Application Programming Interface}
\newacronym{IaaS}{IaaS}{Infrastructure as a Service}
\newacronym{PaaS}{PaaS}{Platform as a Service}
\newacronym{SaaS}{SaaS}{Software as a Service}
\newacronym{TA}{TA}{Timing Attacks}
\newacronym{HNDL}{HNDL}{Harvest-Now, Decrypt-Later}
\newacronym{SPA}{SPA}{Simple Power Analysis}
\newacronym{APA}{APA}{Advanced Power Analysis}
\newacronym{EM}{EM}{Electromagnetic Attacks}
\newacronym{FIPS}{FIPS}{Federal Information Processing Standards}
\newacronym{TMP}{TMP}{Template Attacks}
\newacronym{CB}{CB}{Cold-Boot Attacks}
\newacronym{CA}{CA}{Certificate Authority}
\newacronym{QAML}{QAML}{Quantum-Accelerated Machine Learning}

\newacronym{GCP}{GCP}{Google Cloud Platform}
\newacronym{AWS}{AWS}{Amazon Web Services}
\newacronym{CDN}{CDN}{Content Delivery Network}
\newacronym{VNF}{VNF}{Virtual Network Function}
\newacronym{SDN}{SDN}{Software-Defined Networking}
\newacronym{KMS}{KMS}{Key Management Service} 
\newacronym{CSP}{CSP}{Cloud Service Provider}
\newacronym{OID}{OID}{Object Identifier}
\newacronym{QKD}{QKD}{Quantum Key Distribution}
\newacronym{QC}{QC}{Quantum Computing}
\newacronym{FHE}{FHE}{Fully Homomorphic Encryption}
\newacronym{HSM}{HSM}{Hardware Security Module}
\newacronym{NTT}{NTT}{Number Theoretic Transform}
\newacronym{STRIDE}{STRIDE}{Spoofing, Tampering, Repudiation, Information Disclosure, Denial of Service, Elevation of Privilege}
\newacronym{MITM}{MITM}{Man in the Middle}
\newacronym{NIST}{NIST}{National Institute of Standards and Technology}
\newacronym{PKI}{PKI}{Public Key Infrastructure}

\newacronym{VPC}{VPC}{Virtual Private Cloud}
\newacronym{CBOM}{CBOM}{Cryptographic Bill of Materials}
\newacronym{KBOM}{KBOM}{Key Bill of Materials}
\newacronym{MTTR}{MTTR}{Mean Time To Repair}
\newacronym{SSH}{SSH}{Secure Shell}
\newacronym{ML}{ML}{Machine Learning}
\newacronym{SAML}{SAML}{Security Assertion Markup Language}
\newacronym{NFV}{NFV}{Network Function Virtualization}
\newacronym{IIoT}{IIoT}{Industrial Internet of Things}

\usepackage{pgfplotstable}
\usepackage[normalem]{ulem}
\usetikzlibrary{arrows.meta}
\usepackage{smartdiagram}
\usetikzlibrary{positioning, fit, calc, shapes, arrows,trees}
\usepackage{hhline}
\usetikzlibrary{positioning, fit, calc, shapes, arrows,trees}
\usepackage{hhline}
\definecolor{columbiablue}{rgb}{0.61, 0.87, 1.0}
\DeclareRobustCommand{\hlpink}[1]{{\sethlcolor{pink}{#1}}}
\newcommand*{\Scale}[2][4]{\scalebox{#1}{$#2$}}%
\newcommand*{\Resize}[2]{\resizebox{#1}{!}{$#2$}}%

\tikzset{%
 thick arrow/.style={
 -{Triangle[angle=120:1pt 1]},
 line width=0.8cm, 
 draw=blue!20
 },
 arrow label/.style={
 text=black,
 align=center
 },
 set mark/.style={
 insert path={
 node [midway, arrow label, node contents=#1]
 }
 }
}
\newcommand\deleted{\bgroup\markoverwith{\textcolor{red}{\rule[0.5ex]{2pt}{0.4pt}}}\ULon}

\usetikzlibrary{pgfplots.groupplots}

\newcommand\doublecheck{\textcolor{black}{\checkmark\kern-0em\checkmark}}
\newcommand\semidoublecheck{\textcolor{black}{\checkmark\kern-0em\bcancel{\checkmark}}}

\definecolor{lemon}{rgb}{1.0, 1.0, 0.13}

\newcommand\low{\cellcolor{green!60}L}
\newcommand\med{\cellcolor{lemon!80}M}
\newcommand\high{\cellcolor{red!80}H}

\usepackage[breaklinks=true,hidelinks]{hyperref}

\begin{document}
\definecolor{BulletsColor}{rgb}{0, 0, 0.9}
\newlist{myBullets}{itemize}{1}

\setlist[myBullets]{
 label={\textbullet},
 leftmargin=*,
 topsep=0ex,
 partopsep=0ex,
 parsep=0ex,
 itemsep=0ex,
}

\newlist{myBullets1}{itemize}{1}

\setlist[myBullets1]{
 label={\textbullet},
 leftmargin=*,
 topsep=0ex,
 partopsep=0ex,
 parsep=0ex,
 itemsep=0ex,
}
\title{Future-Proofing Cloud Security Against Quantum Attacks: Risk, Transition, and Mitigation Strategies
}

\author{Yaser Baseri, Abdelhakim Hafid, Arash Habibi Lashkari
 
\IEEEcompsocitemizethanks{\IEEEcompsocthanksitem Yaser Baseri and Abdelhakim Hafid are with University of Montreal, Montreal,  Canada.
Emails: yaser.baseri@umontreal.ca; ahafid@umontreal.ca. Arash Habibi Lashkari is with the School of Information Technology, York University, Toronto,  Canada. Email: ahabibil@yorku.ca. \protect} 
}

\maketitle
\begin{abstract}
Quantum Computing (QC) threatens the cryptographic foundations of Cloud 
Computing (CC), exposing distributed infrastructures to novel attack vectors. 
This survey provides comprehensive analysis of quantum-safe cloud security, 
examining vulnerabilities, transition strategies, and layer-specific 
countermeasures across nine architectural layers (application, data, runtime, 
middleware, OS, virtualization, server, storage, networking). We employ 
STRIDE-based risk assessment aligned with NIST SP 800-30 to evaluate quantum 
threats through three transition phases: pre-transition (classical cryptography 
vulnerabilities), hybrid (migration risks), and post-transition (PQC 
implementation weaknesses including side-channel attacks). Our security 
framework integrates hybrid cryptographic strategies (algorithmic combiners, 
dual/composite certificates, protocol-level migration), cryptographic agility, 
and risk-prioritized mitigation tailored to cloud environments. We benchmark 
NIST-standardized PQC algorithms   for 
performance and deployment suitability, assess side-channel and implementation 
vulnerabilities, and analyze quantum-safe strategies from leading CSPs (AWS, 
Azure, GCP). The survey delivers layer-specific threat taxonomies, 
likelihood-impact risk matrices, and CSP-informed deployment roadmaps for cloud 
architects, policymakers, and researchers. We identify six critical research 
directions: standardization and interoperability, hardware acceleration and 
performance optimization, AI-enhanced security and threat mitigation, 
integration with emerging cloud technologies, systemic preparedness and 
workforce development, and migration frameworks with crypto-agility.

\end{abstract}


\section{Introduction}\label{sec:intro}
\gls{QC} has evolved from theory into a technological force threatening 
contemporary cryptographic systems. Enabled by quantum mechanical phenomena 
such as superposition and entanglement, \gls{QC} offers exponential speedups 
for solving problems considered intractable for classical computers~\cite{9800933}. 
Critically, advances in quantum hardware threaten the foundational pillars of 
modern security, including RSA, DHE, and ECC, by exploiting their underlying 
mathematical hardness via Shor's~\cite{shor1999polynomial} and 
Grover's~\cite{grover1996fast} algorithms.

The implications for \gls{CC} security are profound. Cloud infrastructures 
process vast quantities of sensitive data, from personal information and 
financial records to critical infrastructure controls, across extensively 
virtualized, containerized, and distributed microservices architectures. As 
many deployments still rely on classical cryptography, organizations face 
growing exposure to quantum-capable adversaries. Cryptographic failures may 
cascade throughout the entire \gls{CC} ecosystem, threatening public safety, 
disrupting essential services, and undermining trust in digital infrastructures 
at national and global levels~\cite{mosca2018cybersecurity, gidney2021factor, 
NISTPQC, CISAPQC}.

Transitioning to quantum-safe algorithms is necessary but insufficient to 
eliminate quantum-era risks. Even with \gls{PQC} deployed, adversaries may 
exploit migration-induced weaknesses, protocol downgrade paths, and 
implementation-level vulnerabilities, including \gls{SCA}, fault attacks, 
and quantum-amplified \gls{DoS}. Moreover, the significantly larger 
cryptographic artifacts introduced by \gls{PQC} stress protocol parsing 
and memory management, elevating the risk of memory-safety vulnerabilities 
and attack-surface exploitation (Table~\ref{table:Post-Transition}). 
Rapid advances in quantum hardware further underscore the urgency of 
proactive, defense-in-depth security strategies. In response, major 
\gls{CC} providers (\gls{AWS}, \gls{GCP}, Azure) have initiated hybrid 
and post-quantum deployments~\cite{aws2025quantum,google2023quantum,azure2023quantum}.

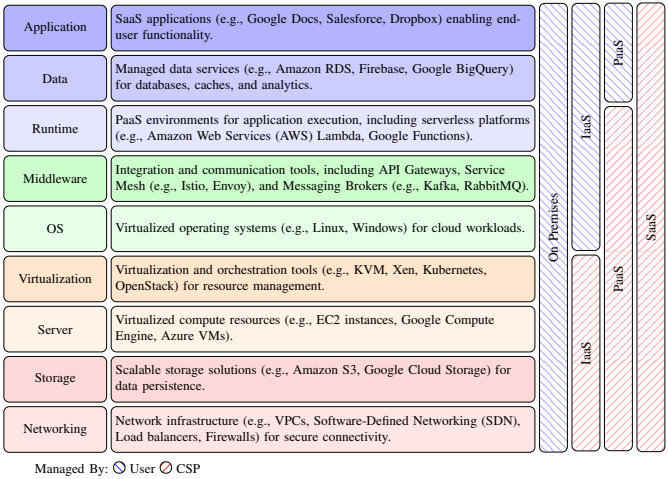
\begin{figure}[!htbp]
\centering
{
 \resizebox{\linewidth}{!}{
\begin{tikzpicture}[
node distance=0pt,
every node/.style={
rectangle, draw=black, rounded corners, minimum height=1.2cm, align=center, font = {\large}
},
cloudlayer/.style={minimum width=2.7cm},
description/.style={ minimum width=10.9cm, text width=10.9cm, minimum height=1.2cm,align=left},
cloudmodel/.style={minimum width=0.5cm,text width=0.5cm},
managedbyuser/.style={pattern=north west lines, pattern color=blue!80},
managedbyprovider/.style={pattern=north east lines, pattern color=red!80, font = {\large}}
]

\node[cloudlayer,fill=blue!30] (application) {Application};
\node[right=3pt of application, description, fill=blue!30] (appdesc) {SaaS applications (e.g., Google Docs, Salesforce, Dropbox) enabling end-user functionality.};

\node[cloudlayer, below=3pt of application, fill=blue!20] (data) {Data};
\node[right=3pt of data, description,, fill=blue!20] (datadesc) {Managed data services (e.g., Amazon RDS, Firebase, Google BigQuery) for databases, caches, and analytics.};

\node[cloudlayer, below=3pt of data, fill=blue!10] (runtime) {Runtime};
\node[right=3pt of runtime, description, fill=blue!10] (runtimedesc) {PaaS environments for application execution, including serverless platforms (e.g., \glsentryshort{AWS} Lambda, Google Functions).};

\node[cloudlayer, below=3pt of runtime, fill=green!20] (middleware) {Middleware};
\node[right=3pt of middleware, description, fill=green!20] (middlewaredesc) {Integration and communication tools, including \glsentryshort{API} Gateways, Service Mesh (e.g., Istio, Envoy), and Messaging Brokers (e.g., Kafka, RabbitMQ).};

\node[cloudlayer, below=3pt of middleware, fill=green!10] (os) {OS};
\node[right=3pt of os, description, fill=green!10] (osdesc) {Virtualized operating systems (e.g., Linux, Windows) for cloud workloads.};

\node[cloudlayer, below=3pt of os, fill=orange!20] (virtualization) {Virtualization};
\node[right=3pt of virtualization, description, fill=orange!20] (virtualizationdesc) {Virtualization and orchestration tools (e.g., KVM, Xen, Kubernetes, OpenStack) for resource management.};

\node[cloudlayer, below=3pt of virtualization, fill=orange!10] (server) {Server};
\node[right=3pt of server, description, fill=orange!10] (serverdesc) {Virtualized compute resources (e.g., EC2 instances, Google Compute Engine, Azure VMs).};

\node[cloudlayer, below=3pt of server, fill=red!15] (storage) {Storage};
\node[right=3pt of storage, description, fill=red!15] (storagedesc) {Scalable storage solutions (e.g., Amazon S3, Google Cloud Storage) for data persistence.};

\node[cloudlayer, below=3pt of storage, fill=red!10] (network) {Networking};
\node[right=3pt of network, description, fill=red!10] (networkdesc) {Network infrastructure (e.g., \glsentrylong{VPC}s, \glsentrylong{SDN}, Load balancers, Firewalls) for secure connectivity.};

\node[cloudmodel,right=3pt of appdesc, align=center,minimum height=11.8cm, yshift=-5.25cm,
pattern=north west lines, pattern color=blue!30] (onpremises) {\rotatebox{90}{On Premises}};

\node[cloudmodel,right=3pt of onpremises, align=center,minimum height=6.5cm, yshift=2.65cm,
pattern=north west lines, pattern color=blue!30] (iaas_user) {\rotatebox{90}{IaaS}};

\node[cloudmodel,below=3pt of iaas_user, align=center,minimum height=5.15cm,
pattern=north east lines, pattern color=red!30] (iaas_provider) {\rotatebox{90}{IaaS}};

\node[cloudmodel,right=3pt of iaas_provider, align=center,minimum height=2.6cm, yshift=7.9cm,
pattern=north west lines, pattern color=blue!30] (paas_user) {\rotatebox{90}{PaaS}};

\node[,cloudmodel,below=3pt of paas_user, align=center,minimum height=9.05cm,
pattern=north east lines, pattern color=red!30] (paas_provider) {\rotatebox{90}{PaaS}};

\node[cloudmodel,right=3pt of paas_provider, align=center,minimum height=11.8cm, yshift=1.35cm,
pattern=north east lines, pattern color=red!30] (saas_user) {\rotatebox{90}{SaaS}};

\node[below=-5pt of network, align=left, xshift=2.7cm, draw=none] (legend) {
{Managed By:}
\tikz\draw[managedbyuser] (0,0) rectangle (0.3,0.3); {User} 
\tikz\draw[managedbyprovider] (0,0) rectangle (0.3,0.3); {Cloud Service Provider (CSP)}
};

\end{tikzpicture}}}

\caption{Layered Architecture of a \gls{CC} Stack, Highlighting Components Requiring Quantum-Resilient Security Evaluation}
\label{fig:stack}

\end{figure}

As illustrated in Figure~\ref{fig:stack}, \gls{CC} is inherently multilayered, 
spanning application services to networking, with each layer presenting distinct 
security challenges. These layers underpin \gls{IaaS}, \gls{PaaS}, and \gls{SaaS} 
models, each defining distinct user and provider responsibility 
boundaries~\cite{parast2022cloud}. This complexity necessitates systematic, 
layer-specific risk analysis, as quantum vulnerabilities manifest differently 
across service models.

To address these challenges, we systematically investigate \gls{QC} 
impacts on \gls{CC} security across nine architectural layers  before and after 
quantum-safe transitions. We employ structured risk assessment incorporating STRIDE 
(Spoofing, Tampering, Repudiation, Information Disclosure, \gls{DoS}, and 
\gls{EoP})~\cite{khan2017stride, khalil2023threat}, aligned with NIST SP 800-30 
and additional security frameworks, to evaluate quantum-induced attack vectors 
across three transition phases: pre-transition, hybrid, and post-transition.
To this end, our primary contributions are:

\begin{itemize}[leftmargin=*, noitemsep, topsep=2pt]
 \item \textbf{Comprehensive Quantum Threat Analysis:} Systematic assessment 
 of quantum-induced risks across all nine \gls{CC} layers, addressing 
 pre-transition classical vulnerabilities and post-transition challenges in 
 quantum-resistant systems.
 
 \item \textbf{STRIDE-Based Threat Modeling:} Structured application of STRIDE 
 to categorize quantum-enabled threats, mapping their impact across \gls{CC} 
 layers and service models for precise vulnerability assessment.
 
 \item \textbf{Structured Risk Assessment Framework:} Quantum-specific 
 methodology integrating STRIDE with NIST SP 800-30 guidelines to evaluate 
 threat likelihood and impact across three transition phases.
 
 \item \textbf{Multi-Layered Mitigation Strategies:} Targeted defense mechanisms 
 for each \gls{CC} layer, addressing specific vulnerabilities and attack vectors 
 throughout the cloud stack.
 
 \item \textbf{Hybrid Cryptographic Transition Framework:} Backward-compatible 
 hybrid strategy supporting secure \gls{PQC} migration at algorithmic, 
 certificate, and protocol levels.
 
 \item \textbf{PQC Performance Evaluation:} Empirical benchmarking of 
 NIST-standardized \gls{PQC} algorithms, analyzing computational overhead, 
 latency, and cloud deployment suitability.
 
 \item \textbf{Platform-Specific CSP Analysis:} Comprehensive synthesis of 
 quantum-safe initiatives from major \glspl{CSP} (\gls{AWS}, \gls{GCP}, Azure), 
 comparing implementation strategies and identifying deployment patterns.
\end{itemize}

\subsection{Related Work}
\label{sec:related-works}

\begin{table}[!h]
\scriptsize
\renewcommand{\arraystretch}{1.2}
\small
\caption{{Comparison of Our Survey With Prior Work on Quantum-Safe \gls{CC} Security}}
\vspace{-0.3cm}
\label{tab:survey}
\begin{center}
\resizebox{\linewidth}{!}{
\begin{tabular}{|c|
>{\centering\arraybackslash}b{0.45cm}|
>{\centering\arraybackslash}b{0.45cm}|
>{\centering\arraybackslash}b{0.45cm}|
>{\centering\arraybackslash}b{0.45cm}|
>{\centering\arraybackslash}b{0.45cm}|
>{\centering\arraybackslash}b{0.45cm}|
>{\centering\arraybackslash}b{0.45cm}|
>{\centering\arraybackslash}b{0.45cm}|
>{\centering\arraybackslash}b{0.45cm}|
>{\centering\arraybackslash}b{0.45cm}|}
\hline
\rotatebox{90}{\textbf{Study}} &
\rotatebox{90}{\textbf{Risk Assessment}} &
\rotatebox{90}{\textbf{STRIDE-Based Threat Modeling\ \ }} &
\rotatebox{90}{\textbf{Vulnerability Analysis}} &
\rotatebox{90}{\textbf{CSP Security/Posture Analysis}} &
\rotatebox{90}{\textbf{Mitigation Strategy (PQC/QKD)}} &
\rotatebox{90}{\textbf{NIST-Std. \gls{PQC} Adoption}} &
\rotatebox{90}{\textbf{Performance \& Scalability}} &
\rotatebox{90}{\textbf{Post-Migration Strategy}} &
\rotatebox{90}{\textbf{Hybrid Cloud Architecture}} &
\rotatebox{90}{\textbf{Actionable Migration Roadmap}} \\ \hline

\cite{kaiiali2019cloud} &
$\times$ & 
$\times$ & 
\checkmark & 
$\times$ & 
$\checkmark_{QC}$ & 
$\times$ & 
$\times$ & 
$\times$ & 
$\times$ & 
$\times$ \\ 
\hline

\cite{leymann2020quantum} &
$\times$ &  
$\times$ &  
$\times$ &  
$\times$ &  
$\times$ &  
$\times$ &  
$\times$ &  
$\times$ &  
$\times$ &  
$\times$ \\  
\hline

\cite{mangla2023mitigating} &
$\times$ &  
$\times$ &  
\checkmark &  
$\times$ &  
$\checkmark_{QC}$ &  
$\times$ &  
$\times$ &  
$\times$ &  
$\times$ &  
$\times$ \\  
\hline

\cite{yang2023survey} &
$\times$ &  
$\times$ &  
\checkmark &  
$\times$ &  
$\checkmark_{QC}$ &  
$\times$ &  
$\times$ &  
$\times$ &  
$\times$ &  
$\times$ \\  
\hline

\cite{nguyen2024quantum} &
$\times$ &  
$\times$ &  
\checkmark &  
$\times$ &  
$\checkmark_{QC}$ &  
$\times$ &  
$\times$ &  
$\times$ &  
$\times$ &  
$\times$ \\  
\hline

\cite{coupel2025security} &
$\times$ &  
$\times$ &  
\checkmark &  
$\times$ &  
$\times$ &  
$\times$ &  
$\times$ &  
$\times$ &  
$\times$ &  
$\times$ \\  
\hline

\textbf{Our Work} &
\checkmark & 
\checkmark & 
\checkmark & 
\checkmark & 
$\checkmark_{\mathrm{PQC}}$ & 
\checkmark & 
\checkmark & 
\checkmark & 
\checkmark & 
\checkmark \\ 
\hline
\end{tabular}}
\end{center}
\begin{tablenotes}[para,flushleft] \scriptsize \checkmark\;=\;addressed;\quad 
$\times$\;=\;not addressed / not a primary focus;\quad 
$\checkmark_{\mathrm{PQC}}$\;=\;addressed using \emph{NIST-standard PQC} (e.g., ML\text{-}KEM, ML\text{-}DSA); \quad 
$\checkmark_{\mathrm{QC}}$\;=\;addressed using \emph{quantum cryptography} (e.g., QKD).\;\quad 
\emph{Hybrid Cloud Architecture} refers to hybrid \emph{cryptographic} deployment for PQC/QKD transition (not hybrid quantum--classical compute).
\end{tablenotes}
\end{table}

Prior literature on quantum-era cloud security clusters into three strands:
\emph{(i) vulnerability-centric studies} that catalog weaknesses without formal, 
NIST-aligned risk modeling or \gls{CSP}-specific guidance~\cite{kaiiali2019cloud, 
mangla2023mitigating, yang2023survey};
\emph{(ii) quantum/architecture works} outlining integration patterns but offering 
neither evidence of NIST-standard \gls{PQC} adoption nor hybrid cryptographic 
migration orchestration~\cite{leymann2020quantum, nguyen2024quantum}; and
\emph{(iii) attack-taxonomy surveys} mapping threats yet lacking implementation-level 
roadmaps and post-migration hardening~\cite{coupel2025security}.
Across these strands, we observe consistent gaps in (a) STRIDE-based, NIST-aligned 
cloud risk assessment; (b) \gls{CSP} security/posture analysis for \gls{AWS}, Azure, 
and \gls{GCP}; (c) governance-aware hybrid transition planning; and (d) post-migration 
assurance.

Our work addresses these gaps by presenting a technically rigorous, multi-dimensional 
framework. First, we operationalize a NIST-aligned risk model using STRIDE and 
likelihood–impact evaluation to prioritize quantum-era threats across the cloud stack. 
Second, we conduct a granular \gls{CSP} security/posture analysis of \gls{AWS}, Azure, 
and \gls{GCP}. Third, we propose a layered mitigation strategy spanning pre-quantum 
exposure, hybrid cryptographic deployment, and post-quantum hardening, covering 
defense-in-depth, crypto-agility, and fault/intrusion tolerance. Fourth, we provide 
actionable guidance and deployment blueprints for \glspl{CSP}, auditors, and enterprise 
users. Finally, we introduce a hybrid cryptographic transition orchestration model that 
balances robustness with compatibility and operational continuity. This contrasts with 
prior works that remain conceptual or fragmented, as summarized in Table~\ref{tab:survey}.

 \begin{table*}[!htbp]
\centering
\caption{List of Abbreviations}

\label{tab:abbreviation}
\resizebox{\linewidth}{!}{%
\begin{tabular}{|l|p{0.45\textwidth}|l|p{0.45\textwidth}|l|p{0.45\textwidth}|}
\hline
\textbf{Abbr.} & \textbf{Description} & \textbf{Abbr.} & \textbf{Description} & \textbf{Abbr.} & \textbf{Description} \\
\hline
\glsentryshort{ACL} & \glsentrylong{ACL} & \glsentryshort{GCP} & \glsentrylong{GCP} & \glsentryshort{PQC} & \glsentrylong{PQC} \\ 
\glsentryshort{AEAD} & \glsentrylong{AEAD} & \glsentryshort{HNDL} & \glsentrylong{HNDL} & \glsentryshort{QAML} & \glsentrylong{QAML} \\ 
\glsentryshort{AES} & \glsentrylong{AES} & \glsentryshort{HSM} & \glsentrylong{HSM} & \glsentryshort{QC} & \glsentrylong{QC} \\ 
\glsentryshort{APA} & \glsentrylong{APA} & \glsentryshort{IaaS} & \glsentrylong{IaaS} & \glsentryshort{ROP} & \glsentrylong{ROP} \\ 
\glsentryshort{API} & \glsentrylong{API} & \glsentryshort{IIoT} & \glsentrylong{IIoT} & \glsentryshort{RPKI} & \glsentrylong{RPKI} \\ 
\glsentryshort{ASLR} & \glsentrylong{ASLR} & \glsentryshort{JIT} & \glsentrylong{JIT} & \glsentryshort{SaaS} & \glsentrylong{SaaS} \\ 
\glsentryshort{AWS} & \glsentrylong{AWS} & \glsentryshort{JOP} & \glsentrylong{JOP} & \glsentryshort{SAML} & \glsentrylong{SAML} \\ 
\glsentryshort{CA} & \glsentrylong{CA} & \glsentryshort{KBOM} & \glsentrylong{KBOM} & \glsentryshort{SCA} & \glsentrylong{SCA} \\ 
\glsentryshort{CB} & \glsentrylong{CB} & \glsentryshort{KEM/ENC} & \glsentrylong{KEM/ENC} & \glsentryshort{SDN} & \glsentrylong{SDN} \\ 
\glsentryshort{CBOM} & \glsentrylong{CBOM} & \glsentryshort{KMS} & \glsentrylong{KMS} & \glsentryshort{SPA} & \glsentrylong{SPA} \\ 
\glsentryshort{CC} & \glsentrylong{CC} & \glsentryshort{MITM} & \glsentrylong{MITM} & \glsentryshort{SSH} & \glsentrylong{SSH} \\ 
\glsentryshort{CFI} & \glsentrylong{CFI} & \glsentryshort{ML} & \glsentrylong{ML} & \glsentryshort{STRIDE} & Spoofing, Tampering, Repudiation, Info. Disclosure, \\ 
\glsentryshort{CICD} & Continuous Integration and Continuous Deployment & \glsentryshort{MTTR} & \glsentrylong{MTTR} & &   \glsentrylong{DoS},  \glsentrylong{EoP} \\ 
\glsentryshort{CSP} & \glsentrylong{CSP} & \glsentryshort{NFV} & \glsentrylong{NFV} & \glsentryshort{TA} & \glsentrylong{TA} \\ 
\glsentryshort{DEP} & \glsentrylong{DEP} & \glsentryshort{NIST} & \glsentrylong{NIST} & \glsentryshort{TEE} & \glsentrylong{TEE} \\ 
\glsentryshort{DHE} & \glsentrylong{DHE} & \glsentryshort{NTT} & \glsentrylong{NTT} & \glsentryshort{TLS} & \glsentrylong{TLS} \\ 
\glsentryshort{DoS} & \glsentrylong{DoS} & \glsentryshort{OID} & \glsentrylong{OID} & \glsentryshort{TMP} & \glsentrylong{TMP} \\ 
\glsentryshort{ECC} & \glsentrylong{ECC} & \glsentryshort{OS} & \glsentrylong{OS} & \glsentryshort{TPM} & \glsentrylong{TPM} \\ 
\glsentryshort{EM} & \glsentrylong{EM} & \glsentryshort{PaaS} & \glsentrylong{PaaS} & \glsentryshort{VM} & \glsentrylong{VM} \\ 
\glsentryshort{EoP} & \glsentrylong{EoP} & \glsentryshort{PKI} & \glsentrylong{PKI} & \glsentryshort{vTPM} & \glsentrylong{vTPM} \\ 
\glsentryshort{FA} & \glsentrylong{FA} & \glsentryshort{PQ} & \glsentrylong{PQ} & & \\ 
\hline
\end{tabular}}

\end{table*}

\subsection{Survey Methodology and Literature Sourcing}
\label{methodology} 
\begin{table}[!htbp]
\centering
\caption{Representative Search Phrases and Result Volumes}
\small
\label{tab:search-phrases-results}
\resizebox{\linewidth}{!}{%
\begin{tabular}{|p{9cm}|l|}
\hline
\textbf{Search Phrase} & \textbf{Results Found} \\
\hline
"cloud security" AND ("quantum" OR "post-quantum") AND ("cryptography" OR "encryption") & $\approx$ 5,970 \\
\hline
("PQC" OR "post-quantum cryptography") AND ("cloud computing" OR "cloud services") AND ("migration" OR "transition") & $\approx$ 1,930\\
\hline
"quantum attacks" AND "cloud infrastructure" AND ("risk" OR "vulnerability" OR "STRIDE") & $\approx$ 345 \\
\hline
("quantum-safe" OR "quantum-resilient") AND "cloud" AND ("mitigation" OR "framework") & $\approx$ 3,680 \\
\hline
"NIST {PQC}" AND ("standardization" OR ("Kyber" OR "Dilithium" OR "SPHINCS+" OR "HQC")) & $\approx$ 3,240 \\
\hline
"hybrid cryptography" AND ("cloud migration" OR "quantum-safe transition") & $\approx$ 35 \\
\hline
"post-quantum cryptography" AND ("{AWS}" OR "Google Cloud" OR "Microsoft Azure") AND ("roadmap" OR "deployment" OR "migration") & $\approx$ 850 \\
\hline
\end{tabular}}
\end{table}

We adopted a structured multi-stage review methodology spanning \gls{PQC}, layered 
cloud architectures, threat modeling, and \gls{CSP} quantum-safe migration strategies. 
Studies were included if they addressed: (a) quantum vulnerabilities across 
transition stages; (b) formal threat modeling frameworks (STRIDE, NIST SP 800-30); 
(c) hybrid cryptographic migration; (d) NIST \gls{PQC} candidate evaluation; or (e) 
\gls{CSP} quantum readiness initiatives. Studies lacking cloud specificity or formal 
modeling were excluded. 
We searched IEEE Xplore, ACM Digital Library, SpringerLink, Scopus, arXiv, 
PQCrypto/CRYPTO/EUROCRYPT proceedings, top journals (ACM Computing Surveys, 
IEEE Communications Surveys \& Tutorials, IEEE TIFS, Journal of Cryptology), 
and \gls{CSP} technical documentation. Search queries targeted \gls{PQC} adoption, 
cloud-layer security, quantum threats, and migration strategies, yielding 
~15,000 initial records. After duplicate removal and screening, 200+ studies 
were mapped to our nine-layer architecture (Figure~\ref{fig:stack}), 
classified by transition phase relevance (Sections~\ref{sec:pre},~\ref{sec:post}), 
aligned with STRIDE and NIST SP 800-30, and analyzed for \gls{CSP} quantum-safe 
deployment strategies.

\subsection{Organization}
\begin{figure}[!h]
 \centering
 \begin{adjustbox}{width=\linewidth}
 \begin{tikzpicture}[node distance=1cm]


\definecolor{color0}{RGB}{230, 50, 50} 
\definecolor{color1}{RGB}{255, 102, 102} 
\definecolor{color2}{RGB}{255, 153, 102} 
\definecolor{color3}{RGB}{255, 204, 102} 
\definecolor{color4}{RGB}{255, 255, 102} 
\definecolor{color5}{RGB}{178, 255, 102} 
\definecolor{color6}{RGB}{102, 255, 178} 
\definecolor{color7}{RGB}{102, 204, 255} 
\definecolor{color8}{RGB}{102, 153, 255} 
\definecolor{color9}{RGB}{178, 102, 255} 

\large
 \tikzstyle{title} = [rectangle, draw=black, thick, fill=gray!35, text width=5cm, text centered, minimum height=2cm]
 \tikzstyle{section} = [rectangle, draw=black, thick, fill=Orange!35, text width=5cm, text centered, minimum height=1.2cm]
 \tikzstyle{subsection} = [rectangle, draw=black, thick, fill=orange!35, text width=11cm, text centered, minimum height=0.8cm]
 \tikzstyle{arrow} = [thick,->,>=stealth]

 \node (title) [title] {Cloud Security in the Quantum Era: Risks, Transition Strategies, and Mitigations};

 \node (intro) [section, right of=title,fill=color9!55, xshift=5cm,yshift=13.4cm] {Section~\ref{sec:intro}: Introduction};
 \node (threats) [section, below of=intro,fill=color8!55, yshift=-2cm] {Section~\ref{sec:risk-analysis-approach}: Quantum-Safe Transition Risk Assessment Approach};
 \node (crypto) [section, below of=threats,fill=color7!55,yshift=-1.5cm] {Section~\ref{sec:risk}: Cryptographic Standards and Quantum Cyber Impact};
 \node (pretrans) [section, below of=crypto,fill=color6!55,yshift=-2.5cm] {Section~\ref{sec:pre}: Pre-Transition Quantum Threat Landscape};
 \node (posttrans) [section, below of=pretrans,fill=color5!55,yshift=-3.4cm] {Section~\ref{sec:post}: Post-Transition Quantum Security Measures};
 \node (hybrid) [section, below of=posttrans,fill=color4!55,yshift=-2.9cm] {Section~\ref{sec:hybrid-cloud-security}: Hybrid Security Approaches in Cloud Environments};
 \node (performance) [section, below of=hybrid,fill=color3!55,yshift=-1.9cm] {Section~\ref{sec:performance_evaluation}: Performance Evaluation of \gls{PQC}};
 \node (providers) [section, below of=performance,fill=color2!55,yshift=-1.4cm] {Section~\ref{sec:pqc_cloud_providers}: \gls{PQC} in Major \glspl{CSP}};
 \node (future) [section, below of=providers,fill=color1!55,yshift=-3.4cm] {Section~\ref{sec:future_directions}: Future Research Directions};
 \node (conclusion) [section, below of=future,fill=color0!55,yshift=-2.8cm] {Section~\ref{sec:conclusion}: Conclusion};

 \node (intro2) [subsection, right of=intro, xshift=8cm,yshift=0cm,fill=color9!35] {\ref{sec:intro}(B): Methodology and Literature Sourcing};
\node (intro1) [subsection, above of=intro2,fill=color9!35] 
{\ref{sec:intro}(A): Related Work};
\node (intro3) [subsection, below of=intro2,fill=color9!35] {\ref{sec:intro}(C):  Organization};

 \node (threats1) [subsection, right of=threats,xshift=8cm,yshift=0.5cm,fill=color8!35] {\ref{sec:risk-analysis-approach}(A): Risk Scoping and Preparation};
 \node (threats2) [subsection, below of=threats1,fill=color8!35] {\ref{sec:risk-analysis-approach}(B): Conducting Risk Assessment};


 \node (crypto1) [subsection, right of=crypto,xshift=8cm,yshift=0.5cm, fill=color7!35] {\ref{sec:risk}(A): Classical Cryptographic Risks};
  \node (crypto2) [subsection, below of=crypto1,fill=color7!35] {\ref{sec:risk}(B): Quantum-Safe Cryptographic Risks};

 \node (pretrans1) [subsection, right of=pretrans, xshift=8cm,yshift=1.5cm, fill=color6!35] {\ref{sec:pre}(A): Pre-Transition Vulnerabilities};
 \node (pretrans2) [subsection, below of=pretrans1, fill=color6!35] {\ref{sec:pre}(B): Pre-Transition Attack Vectors};
 \node (pretrans3) [subsection, below of=pretrans2, fill=color6!35] {\ref{sec:pre}(C): Threat Landscape in \gls{CC} Layers};
 \node (pretrans4) [subsection, below of=pretrans3, fill=color6!35] {\ref{sec:pre}(D): Risk Assessment Across Layers};

 \node (posttrans1) [subsection, right of=posttrans, xshift=8cm, yshift=1.5cm, fill=color5!35] {\ref{sec:post}(A): Post-Transition Vulnerabilities};
 \node (posttrans2) [subsection, below of=posttrans1, fill=color5!35] {\ref{sec:post}(B): Post-Transition Attack Vectors};
 \node (posttrans3) [subsection, below of=posttrans2, fill=color5!35] {\ref{sec:post}(C): Threat Landscape in \gls{CC} Layers};
 \node (posttrans4) [subsection, below of=posttrans3, fill=color5!35] {\ref{sec:post}(D): Risk Assessment Across Layers};

 \node (hybrid2) [subsection, right of=hybrid,xshift=8cm,fill=color4!35] {\ref{sec:hybrid-cloud-security}(B): Hybrid Certificate Strategies in Cloud PKI};
 \node (hybrid1) [subsection, above of=hybrid2,fill=color4!35] {\ref{sec:hybrid-cloud-security}(A): Hybrid Algorithmic Strategies for Cloud Workloads};

 \node (hybrid3) [subsection, below of=hybrid2,fill=color4!35] {\ref{sec:hybrid-cloud-security}(C): Hybrid Protocol Strategies in Cloud Services};

 \node (performance1) [subsection, right of=performance,xshift=8cm,yshift=0.5cm,fill=color3!35] {\ref{sec:performance_evaluation}(A): Benchmarking Methodology};
 \node (performance2) [subsection, below of=performance1,fill=color3!35] {\ref{sec:performance_evaluation}(B): Key Observations and Cloud Deployment Suitability};
 
 \node (providers1) [subsection, right of=providers,xshift=8cm,yshift=0.5cm,fill=color2!35] {\ref{sec:pqc_cloud_providers}(A): Comparative Analysis of Implementation Strategies};
 \node (providers2) [subsection, below of=providers1,fill=color2!35] {\ref{sec:pqc_cloud_providers}(B): Synthesis: Current State of \gls{PQC} in \gls{CC}};

\node (future4) [subsection, right of=future,xshift=8cm,yshift=-0.5cm,fill=color1!35] {\ref{sec:future_directions}(D): Integration with Emerging Cloud Technologies};
\node (future5) [subsection, below of=future4,fill=color1!35] {\ref{sec:future_directions}(E): Workforce Development and Systemic Preparedness};
\node (future6) [subsection, below of=future5,fill=color1!35] {\ref{sec:future_directions}(F): Migration Frameworks and Crypto-Agility};

\node (future3) [subsection, above of=future4,fill=color1!35] {\ref{sec:future_directions}(C): AI-Enhanced Security and Advanced Threat Mitigation};
\node (future2) [subsection, above of=future3,fill=color1!35] {\ref{sec:future_directions}(B): Hardware Acceleration and Performance Optimization};
\node (future1) [subsection, above of=future2,fill=color1!35] {\ref{sec:future_directions}(A): Standardization and Interoperability};


 \draw [arrow] (title) |- (intro);
 \draw [arrow] (title) |- (threats);
 \draw [arrow] (title) |- (crypto);
 \draw [arrow] (title) |- (pretrans);
 \draw [arrow] (title) -- (posttrans);
 \draw [arrow] (title) |- (hybrid);
 \draw [arrow] (title) |- (performance);
 \draw [arrow] (title) |- (providers);
 \draw [arrow] (title) |- (future);
 \draw [arrow] (title) |- (conclusion);
 \draw [arrow] (intro) |- (intro1);
 \draw [arrow] (intro.east) -- (intro2);
 \draw [arrow] (intro) |- (intro3);

 \draw [arrow] (threats.east) |- (threats1);
 \draw [arrow] (threats.east) |- (threats2);

 \draw [arrow] (crypto.east) |- (crypto1);
 \draw [arrow] (crypto.east) |- (crypto2);

 \draw [arrow] (pretrans) |- (pretrans1);
 \draw [arrow] (pretrans.east) |- (pretrans2);
 \draw [arrow] (pretrans.east) |- (pretrans3);
 \draw [arrow] (pretrans) |- (pretrans4);

 \draw [arrow] (posttrans) |- (posttrans1);
 \draw [arrow] (posttrans.east) |- (posttrans2);
 \draw [arrow] (posttrans.east) |- (posttrans3);
 \draw [arrow] (posttrans) |- (posttrans4);

 \draw [arrow] (hybrid) |- (hybrid1);
 \draw [arrow] (hybrid) -- (hybrid2);
 \draw [arrow] (hybrid) |- (hybrid3);
 \draw [arrow] (performance.east) |- (performance1);
 \draw [arrow] (performance.east) |- (performance2);
 \draw [arrow] (providers.east) |- (providers1);
 \draw [arrow] (providers.east) |- (providers2);
 \draw [arrow] (future) |- (future1);
 \draw [arrow] (future) |- (future2);
 \draw [arrow] (future.east) |- (future3);
 \draw [arrow] (future.east) |- (future4);
 \draw [arrow] (future) |- (future5);
 \draw [arrow] (future) |- (future6);
 \end{tikzpicture}
 \end{adjustbox}
 \caption{Organizational Structure of This Survey}
 \vspace{-10pt}
 \label{fig:survey_structure}
\end{figure}
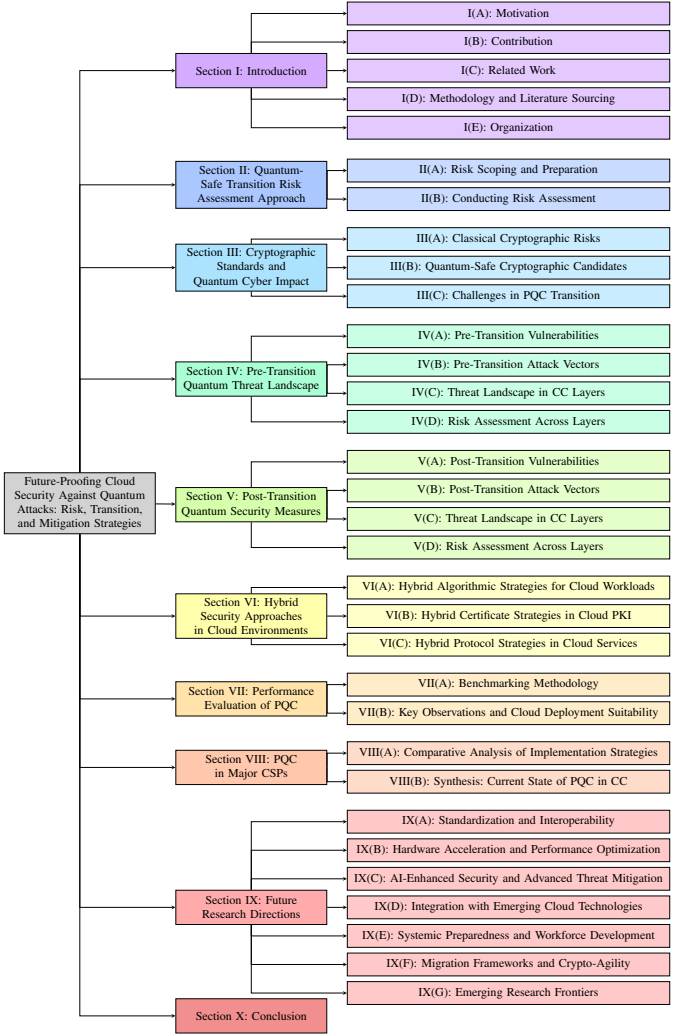

Figure~\ref{fig:survey_structure} illustrates the comprehensive structure of this survey, which systematically addresses the quantum-safe transition across eleven interconnected sections. 
Section~\ref{sec:risk-analysis-approach} 
presents structured risk assessment methodology to evaluate quantum-specific threats 
during cloud migration. Section~\ref{sec:risk} examines classical cryptographic 
vulnerabilities exposed by \gls{QC} and emerging \gls{PQC} risks. 
Sections~\ref{sec:pre} and~\ref{sec:post} analyze threat landscapes across cloud 
layers, contrasting pre-transition vulnerabilities with quantum security postures. 
Section~\ref{sec:hybrid-cloud-security} explores hybrid deployment strategies enabling 
phased, backward-compatible quantum-safe adoption. Section~\ref{sec:performance_evaluation} 
benchmarks NIST-standardized PQC algorithms, evaluating computational efficiency and 
cloud deployment suitability. Section~\ref{sec:pqc_cloud_providers} presents comparative 
analysis of quantum-resistant implementations by major cloud providers, examining their 
technical approaches and capabilities. Section~\ref{sec:future_directions} identifies 
research priorities in standardization, scalability, and emerging technology integration. 
Section~\ref{sec:conclusion} synthesizes key insights and provides actionable 
recommendations for strengthening cloud infrastructure resilience against quantum-era 
threats.
\vspace{-0.2cm}
\section{Quantum-Safe Transition Risk Assessment Approach}\label{sec:risk-analysis-approach}

\begin{figure*}[!ht]
 \begin{center} 
\resizebox{0.8\linewidth}{!}{%
\begin{tikzpicture}[
 title/.style={minimum height=1cm,minimum width=4.1cm,font = {\large}},
 body/.style={draw,top color=white, bottom color=blue!20, rounded corners,minimum width=3.8cm,,minimum height=1cm,font = {\footnotesize}},
 typetag/.style={rectangle, draw=black!100, anchor=west}]
 \node (d0) [draw,top color=white, bottom color=blue!20, rounded corners,minimum height=1cm,minimum width=1.27\textwidth,font = {\large}] at (0,0) {{NIST} Cybersecurity Framework};
 
\node (d3) [title,below=of d0.center] {Detect};
 \node (d31) [body,below=of d3.west, typetag, xshift=2mm,yshift=-0.9cm,minimum height=2.1cm] {\begin{minipage}[c]{3.4cm}\centering
Anomalies and Events 
\end{minipage}};
 \node (d32) [body,below=of d31.west, typetag,yshift=-1.4cm,minimum height=2.1cm] {\begin{minipage}[c]{3.4cm}\centering
Security Continuous Monitoring
\end{minipage}};
 \node (d33) [body,below=of d32.west, typetag,yshift=-1.35cm,minimum height=2.1cm] {\begin{minipage}[c]{3.4cm}\centering
Detection Processes
\end{minipage}};
\node [top color=blue!40, bottom color=blue!40, rounded corners,minimum height=1cm,draw=black!100,fill opacity=0.1, fit={ (d3) (d31) (d32) (d33)}] {};
 
\node (d2) [title, left of=d3,xshift=-0.2\textwidth] {Protect};
 \node (d21) [body,below=of d2.west, typetag, yshift=-0.35cm, xshift=2mm] {\begin{minipage}[c]{3.4cm}\centering
Identity Management and Access Control
\end{minipage}};
 \node (d22) [body,below=1.17 of d21.west, typetag] {\begin{minipage}[c]{3.4cm}\centering
Awareness and Training
\end{minipage}};
 \node (d23) [body,below=1.17 of d22.west, typetag] {\begin{minipage}[c]{3.4cm}\centering
Data Security 
\end{minipage}};
\node (d24) [body,below=1.17 of d23.west, typetag] {\begin{minipage}[c]{3.4cm}\centering
Info. Protection Processes and Procedures
\end{minipage}};
\node (d25) [body,below=1.17 of d24.west, typetag] {\begin{minipage}[c]{3.4cm}\centering
Maintenance
\end{minipage}};
\node (d26) [body,below=1.17 of d25.west, typetag]{\begin{minipage}[c]{3.4cm}\centering
Protective Technology
\end{minipage}};
\node [top color=blue!40, bottom color=blue!40, rounded corners,minimum height=1cm,draw=black!100,fill opacity=0.1, fit={ (d2) (d21) (d22) (d23) (d24) (d25) (d26)}] {};

\node (d1) [title,left of=d2,xshift=-0.2\textwidth] {Identify};
 \node (d11) [body,below=of d1.west, typetag, yshift=-0.35cm, xshift=2mm] {\begin{minipage}[c]{3.4cm}\centering
Asset Management
\end{minipage}};
 \node (d12) [body,below=1.17 of d11.west, typetag] {\begin{minipage}[c]{3.4cm}\centering
Business Environment
\end{minipage}};
 \node (d13) [body,below=1.17 of d12.west, typetag] {\begin{minipage}[c]{3.4cm}\centering
Governance
\end{minipage}};
\node (d14) [body,below=1.17 of d13.west, typetag] {\begin{minipage}[c]{3.4cm}\centering
Risk Assessment
\end{minipage}};
\node (d15) [body,below=1.17 of d14.west, typetag] {\begin{minipage}[c]{3.4cm}\centering
Risk Management Strategy
\end{minipage}};
\node (d16) [body,below=1.17 of d15.west, typetag]{\begin{minipage}[c]{3.4cm}\centering Supply Chain Risk Management
\end{minipage}};
\node [top color=blue!40, bottom color=blue!40, rounded corners,minimum height=1cm,draw=black!100,fill opacity=0.1, fit={ (d1) (d11) (d12) (d13) (d14) (d15) (d16)}] {}; 
 
\node (d4) [title, right of=d3,xshift=0.2\textwidth] {Respond};
 \node (d41) [body,below=of d4.west, typetag, yshift=-0.6cm, xshift=2mm,minimum height=1.55cm] {\begin{minipage}[c]{3.4cm}\centering
Response Planning
\end{minipage}};
 \node (d42) [body,below=of d41.west, typetag, yshift=-0.8cm,minimum height=1.55cm] {\begin{minipage}[c]{3.4cm}\centering
Communications 
\end{minipage}};
 \node (d43) [body,below=of d42.west, typetag, yshift=-0.75cm,minimum height=1.55cm] {\begin{minipage}[c]{3.4cm}\centering
Analysis
\end{minipage}};
\node (d44) [body,below=of d43.west, typetag, yshift=-0.8cm,minimum height=1.55cm] {\begin{minipage}[c]{3.4cm}\centering
Mitigation Improvements 
\end{minipage}};
\node [top color=blue!40, bottom color=blue!40, rounded corners,minimum height=1cm,draw=black!100,fill opacity=0.1, fit={ (d4) (d41) (d42) (d43) (d44)}] {};

\node (d5) [title, right of=d4,xshift=0.2\textwidth] {Recover};
 \node (d51) [body,below=of d5.west, typetag, xshift=2mm,yshift=-0.9cm,minimum height=2.1cm] {\begin{minipage}[c]{3.4cm}\centering
Recovery Planning
\end{minipage}};
 \node (d52) [body,below=of d51.west, typetag,yshift=-1.4cm,minimum height=2.1cm] {\begin{minipage}[c]{3.4cm}\centering
Improvements 
\end{minipage}};
 \node (d53) [body,below=of d52.west, typetag,yshift=-1.35cm,minimum height=2.1cm] {\begin{minipage}[c]{3.4cm}\centering
Communications
\end{minipage}};
\node (d1to5)[top color=blue!40, bottom color=blue!40, rounded corners,minimum height=1cm,draw=black!100,fill opacity=0.1, fit={ (d5) (d51) (d52) (d53)}] {};

 \node (steps) [below=of d1to5.south, yshift= 0cm, xshift=-13.5cm]{
 \begin{tikzpicture}
 
 \fill[blue!35] (0,0) -- (7.5,0) -- (8,0.5) -- (7.5,1) -- (0,1) -- cycle;
 \node at (3.2,0.8) {\text{Identify Threat Sources and Events}};

 \fill[red!35] (0,-1.2) -- (8.5,-1.2) -- (9,-0.7) -- (8.5,-0.2) -- (0,-0.2) -- cycle;
 \node at (4.4,-0.4) {\text{Identify Vulnerabilities and Predisposing Conditions}};

 \fill[green!35] (0,-2.4) -- (9.5,-2.4) -- (10,-1.9) -- (9.5,-1.4) -- (0,-1.4) -- cycle;
 \node at (3.2,-1.6) {\text{Determine Likelihood of Occurrence}};

 \fill[orange!35] (0,-3.6) -- (10.5,-3.6) -- (11,-3.1) -- (10.5,-2.6) -- (0,-2.6) -- cycle;
 \node at (2.9,-2.8) {\text{Determine Magnitude of Impact}};

 \fill[blue!35] (0,-4.8) -- (11.5,-4.8) -- (12,-4.3) -- (11.5,-3.8) -- (0,-3.8) -- cycle;
 \node at (1.3,-4) {\text{Assess Risk}};

 \node [below=-1cm, align=center] at (6, -6) {\large{Risk Assessment Process Phases According to \gls{NIST}}};
 \end{tikzpicture}};

 \node (risk) [below=of d1to5.south, yshift=0.5cm, xshift=-3cm] {\renewcommand{\arraystretch}{1.8}{
\subfloat{}{\begin{math}
 \qquad \raisebox{-1\normalbaselineskip}{Likelihood$\ \left\{\rule{0pt}{3\normalbaselineskip}\right.$}
\hspace{-0.15cm}
\begin{tabular}{p{0.1cm}}\\
\multicolumn{1}{l}{Low} \\ 
\multicolumn{1}{l}{Medium} \\
\multicolumn{1}{l}{High}
\end{tabular}
\end{math}
}
\hspace{-0.3cm}
\begin{math}
 \overbrace{\
\begin{tabular}{p{1.5cm}p{1.5cm}p{1.5cm}}
Low & Medium & High \\ \hline
\multicolumn{1}{|l|}{\cellcolor{green!60}Low} & \multicolumn{1}{l|}{\cellcolor{green!60}Low} & \multicolumn{1}{l|}{\cellcolor{lemon!80}Medium} \\ \hline
\multicolumn{1}{|l|}{\cellcolor{green!60}Low} & \multicolumn{1}{l|}{\cellcolor{lemon!80}Medium} & \multicolumn{1}{l|}{\cellcolor{red!80}High} \\ \hline
\multicolumn{1}{|l|}{\cellcolor{lemon!80}Medium} & \multicolumn{1}{l|}{\cellcolor{red!80}High} & \multicolumn{1}{l|}{\cellcolor{red!80}High} \\ \hline
\end{tabular}
}^{\mbox{Impact}}
\end{math}
 }};

 \node (risk-evaluation) [below=of risk,xshift=2cm,yshift=1cm]{\large{Qualitative Risk Assessment Matrix}};

\draw [thick arrow] (d14.west) -- ++(-1,0) |- ($ (steps.west) + (2.7,0) $);
\draw [thick arrow] ($ (steps.east) + (-2.5,-2) $)-| (risk-evaluation.south);
 \end{tikzpicture}
}
\caption{Quantum-Safe Transition Risk Assessment Approach~\cite{baseri2025blockchain}}
\label{fig:risk-assesment}
\vspace{-0.5cm}
\end{center}
\end{figure*}
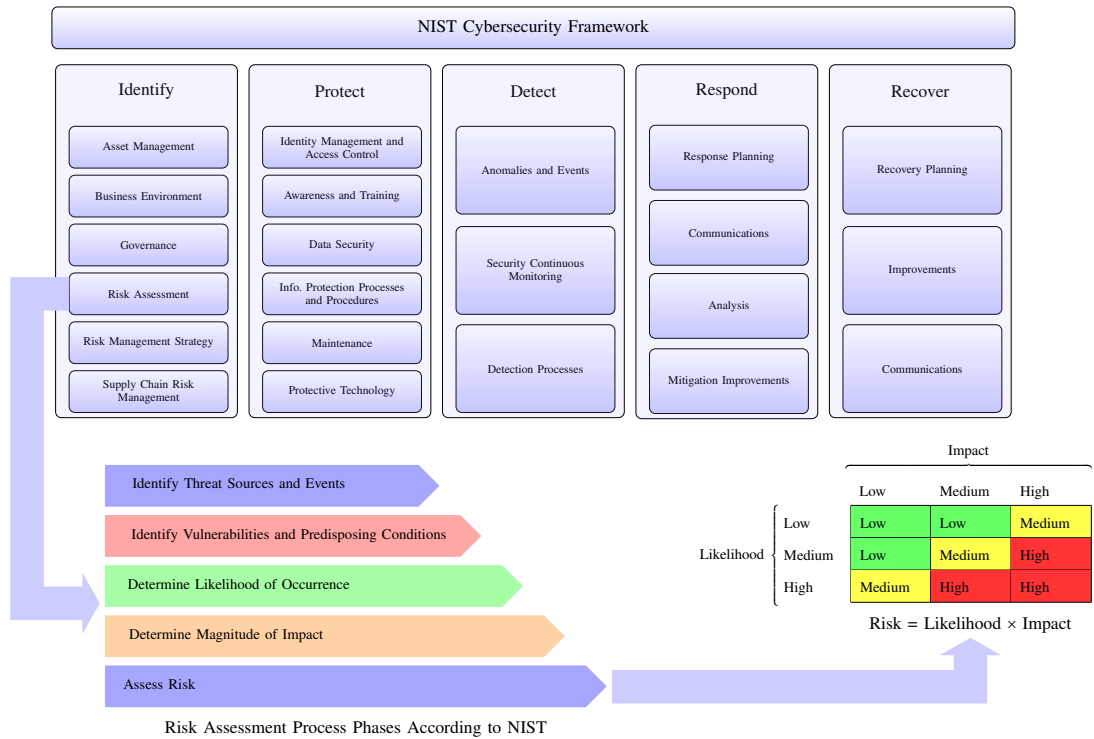 
We adopt a structured risk assessment methodology aligned with NIST SP 800-30~\cite{NIST-SP800-30} 
and the NIST Cybersecurity Framework~\cite{NIST-CSF2024} to evaluate quantum-safe transitions 
in \gls{CC}. Our four-phase approach (preparation, conduct, communication, maintenance) 
identifies quantum-specific threats, analyzes vulnerabilities, and quantifies risk through 
likelihood-impact assessments~\cite{BASERI2025104272,baseri2025blockchain}. As illustrated in 
Figure~\ref{fig:risk-assesment}, this methodology enables systematic risk prioritization across 
cloud migration stages.

\subsection{Risk Scoping and Preparation}\label{subsec:risk_scoping}

We establish assessment scope through purpose, boundaries, assumptions, and threat models 
for systematic risk identification across transition stages. Our threat model assumes 
adversaries possess quantum resources to execute Shor's and Grover's algorithms, while 
accounting for current PQC limitations in efficiency, interoperability, and implementation 
readiness. We employ STRIDE-based threat categorization~\cite{van2021descriptive, jelacic2017stride} 
with discrete likelihood-impact criteria (Low, Medium, High), enabling risk-prioritized 
mitigation strategies across cloud components.

\subsection{Conduct Assessment}

Risk assessment comprises five tasks: (1) identifying threat sources and events, 
(2) identifying vulnerabilities, (3) determining likelihood, (4) determining impact, 
and (5) assessing risk. Figure~\ref{fig:risk-assesment} summarizes this process.

\subsubsection{Identify Threat Sources and Events} 

We identify quantum-specific threats across three transition phases: (a) \textit{pre-transition} 
threats target classical cryptography (e.g., Shor's and Grover's algorithms breaking RSA, 
ECC, symmetric encryption); (b) \textit{transition} threats exploit hybrid implementation 
weaknesses (e.g., downgrade attacks, protocol misconfigurations, PQC side-channel 
vulnerabilities); (c) \textit{post-transition} threats target quantum-resistant systems 
via implementation flaws and advanced vectors (e.g., fault injection, quantum-enhanced 
side-channel attacks). These adversaries exploit vulnerabilities across cloud layers, 
compromising cryptographic foundations and critical services.

\subsubsection{Identify Vulnerabilities and Predisposing Conditions} 

We identify quantum-exploitable vulnerabilities across cloud components: classical 
encryption susceptibility requiring \gls{PQC} adoption with key management, efficiency, and 
interoperability challenges; side-channel risks in multi-tenant \glspl{VM}/containers; \gls{PQC} 
computational overhead impacting scalability; security gaps during hybrid transition; 
and evolving quantum-safe regulatory requirements. Addressing these vulnerabilities 
is critical for quantum-resistant cloud security.

\begin{table}
\centering
\caption{Evaluation Criteria for Likelihood Levels}

\label{table:likelihood}
\scriptsize
\resizebox{\linewidth}{!}{%
\begin{tabular}{|l|l|p{0.76\linewidth}|}
\hline
\multirow{43}{*}{{Likelihood}} 
& \cellcolor{red!80}\multirow{15}{*}{{High}} 
& \begin{myBullets}
\item \textbf{Critical Vulnerabilities:} Significant cryptographic or architectural weaknesses exist, making quantum-based exploitation straightforward (e.g., susceptible to Grover's algorithm with insufficient key sizes).
\item \textbf{Substantial Exposure \& Capability:} System or network is publicly accessible or widely distributed, enabling well-resourced quantum adversaries to launch targeted attacks.
\item \textbf{Evolving Threat Actor Intent:} High motivation from nation-state or financially driven adversaries possessing relevant quantum resources and expertise.
\item \textbf{Minimal Mitigations \& Monitoring:} Existing countermeasures (e.g., side-channel protections, post-quantum patches) are inadequate or not yet implemented.
\vspace{-7pt}
\end{myBullets} \\ \cline{2-3} 
& \cellcolor{lemon!80}\multirow{14}{*}{{Medium}}
& \begin{myBullets}
\item \textbf{Partial Vulnerabilities:} Known weaknesses or legacy components, but partial post-quantum cryptographic measures (e.g., hybrid PKI) limit easy exploitation.
\item \textbf{Moderate Exposure:} Some access control and segmentation exist; quantum adversaries may still locate attack paths through specific interfaces or less-secured nodes.
\item \textbf{Limited Advanced Resources:} Adversaries possess quantum knowledge but rely on partial or developing quantum computational power; exploitation is feasible with effort.
\item \textbf{Active Monitoring \& Patching:} Ongoing patches or threat intelligence reduce the likelihood, but do not fully eliminate risk. 
\vspace{-7pt} \end{myBullets} \\ \cline{2-3}
& \cellcolor{green!60}\multirow{15}{*}{{Low}}
& \begin{myBullets}\vspace{-2pt}
\item \textbf{Robust Quantum-Secure Readiness:} Core systems are updated with quantum-resistant algorithms (e.g., lattice-based KEMs/ENCs, large AES keys) and effectively hardened.
\item \textbf{Minimal Network Footprint:} Highly restricted interfaces or air-gapped segments require advanced or specialized quantum attacks that are impractical at present.
\item \textbf{Strong Security Culture:} Regular penetration testing, code audits, and side-channel defense mechanisms (fault attack countermeasures, noise generation) are well-established.
\item \textbf{Threat Actors Deterred:} The complexity, cost, or resource intensity of a quantum attack is prohibitive, rendering exploitation unlikely in the near to mid-term.
\item \textbf{Beyond Practical Quantum Availability:} Attacks that rely on advanced methods, such as \gls{QAML}, lie significantly beyond the foreseeable availability of practical, scalable quantum hardware.
  \end{myBullets}
\\ \hline
\end{tabular}
}
\end{table}
\subsubsection{Determine Likelihood of Occurrence} 

We evaluate exploitation likelihood using qualitative assessment adapted from NIST SP 800-30 
Appendix G~\cite{NIST-SP800-30}. Likelihood is categorized as \textit{Low (L)}, 
\textit{Medium (M)}, or \textit{High (H)} based on exploitability complexity, quantum 
computational requirements, adversary capabilities, cloud component exposure, and 
countermeasure effectiveness. Detailed criteria are in Table~\ref{table:likelihood}.

\begin{table}[!htbp]
\centering
\caption{Evaluation Criteria for Impact Levels}
\label{table:impact}
\scriptsize
\resizebox{\linewidth}{!}{%
\begin{tabular}{|l|l|p{0.715\linewidth}|}
\hline
\multirow{42}{*}{{Impact}} 
& \cellcolor{red!80}\multirow{15}{*}{{High}} 
& \begin{myBullets}
\item \textbf{Severe Cryptographic Breach:} Rapid decryption or compromise of critical data (e.g., PKI certificates, stored encryption keys) undermines system integrity.
\item \textbf{Extended Operational Downtime:} Quantum-enabled DoS or hypervisor exploit disrupts vital services (e.g., multi-\gls{VM} outages) for a prolonged period.
\item \textbf{Major Reputational \& Financial Damage:} Regulatory non-compliance, lawsuits, or extended brand erosion could follow large-scale disclosures or interruptions.
\item \textbf{Widespread Ecosystem Effects:} Breach extends beyond organizational boundaries (e.g., third-party services or supply chains), resulting in national-level impact.
\vspace{-7pt} \end{myBullets} \\ \cline{2-3}

& \cellcolor{lemon!80}\multirow{15}{*}{{Medium}} 
& \begin{myBullets}\vspace{-0.1cm}
\item \textbf{Partial Data Compromise:} Core encryption may remain intact, but peripheral or secondary data (e.g., logs, non-critical credentials) is exposed or altered.
\item \textbf{Moderate Disruption:} Services experience slowdowns or brief outages; backup and recovery procedures manage to limit extended downtime.
\item \textbf{Regulatory Concerns:} Incidents trigger short-term investigations or require notifications; fines or legal actions remain a possibility if mitigations fail.
\item \textbf{Scoped Impact to Specific Layers:} Issues contained at the middleware or application level, without cascading into the entire cloud infrastructure.
 \vspace{-7pt}  \end{myBullets} \\ \cline{2-3}

& \cellcolor{green!60}\multirow{15}{*}{{Low}} 
& \begin{myBullets}
\item \textbf{Minimal Disclosure or Tampering:} Attempts at quantum exploitation yield negligible data leaks or ephemeral system disruptions quickly resolved by standard controls.
\item \textbf{Limited Operational Effect:} System recovers rapidly from any intrusion, with no long-term effects on data integrity or user trust.
\item \textbf{Low Financial \& Reputational Impact:} Minimal brand exposure, minor incident handling costs, and no regulatory violations.
\item \textbf{Isolated Local Event:} Event remains contained to a single VM, test environment, or network segment, with no broader risk propagation. 
\end{myBullets} \\ \hline
\end{tabular}
}
\end{table}

\subsubsection{Determine Magnitude of Impact} 
Using criteria from NIST SP 800-30 Appendix H~\cite{NIST-SP800-30}, we assess quantum 
threat severity across \textit{Low (L)}, \textit{Medium (M)}, and \textit{High (H)} 
impact levels, evaluating cryptographic compromise, operational disruption, data breaches, 
system stability degradation, and reputation damage. Detailed criteria are in 
Table~\ref{table:impact}.

\subsubsection{Assess Risk}  
Risk is assessed by combining likelihood and impact using a qualitative matrix 
(Figure~\ref{fig:risk-matrix}) mapping combinations to risk levels. High-risk scenarios 
demand immediate mitigation; medium-risk scenarios require planned responses; low-risk 
scenarios may be monitored or accepted based on risk tolerance. This systematic assessment 
prioritizes quantum risks, enabling informed quantum-safe transition decisions.

\begin{figure}[!h]
\renewcommand{\arraystretch}{1.5}
 \scriptsize
  \centering
\renewcommand{\arraystretch}{1.5}
 \scriptsize
\centering{
\subfloat{}{\begin{math}
 \raisebox{-1\normalbaselineskip}{Likelihood$\ \left\{\rule{0pt}{2.7\normalbaselineskip}\right.$}
\hspace{-0.15cm}
\begin{tabular}{p{0.1cm}}\\
\multicolumn{1}{l}{Low} \\ 
\multicolumn{1}{l}{Medium} \\
\multicolumn{1}{l}{High}
\end{tabular}
\end{math}
}
\begin{math}
 \overbrace{\
\begin{tabular}{p{0.7cm}p{.7cm}p{0.7cm}}
Low & Medium & High \\ \hline
\multicolumn{1}{|l|}{\cellcolor{green!60}Low} & \multicolumn{1}{l|}{\cellcolor{green!60}Low} & \multicolumn{1}{l|}{\cellcolor{lemon!80}Medium} \\ \hline
\multicolumn{1}{|l|}{\cellcolor{green!60}Low} & \multicolumn{1}{l|}{\cellcolor{lemon!80}Medium} & \multicolumn{1}{l|}{\cellcolor{red!80}High} \\ \hline
\multicolumn{1}{|l|}{\cellcolor{lemon!80}Medium} & \multicolumn{1}{l|}{\cellcolor{red!80}High} & \multicolumn{1}{l|}{\cellcolor{red!80}High} \\ \hline
\end{tabular}
}^{\mbox{Impact}}
\end{math}
 }

  \caption{Qualitative Risk Assessment based on Likelihood and Impact Levels}
 \label{fig:risk-matrix}
\end{figure}

\section{Cryptographic Standards and \gls{QC}: Cyber Impact and Risk Assessment}
\label{sec:risk}

This section evaluates \gls{QC} impacts on cryptographic standards in cloud 
infrastructures, building on the quantum risk assessment framework 
from~\cite{baseri2025evaluation}. Classical cryptographic vulnerabilities to Shor's 
and Grover's algorithms, as well as PQC residual risks from \glspl{SCA}, 
manifest distinctly in multi-tenant, virtualized cloud environments. These evaluations 
establish the likelihood-impact metrics that underpin systematic threat modeling across 
cloud layers and transition phases (Sections~\ref{sec:pre}–\ref{sec:post}).

\subsection{Classic Cryptographic Standards and \gls{QC}: Assessing Cyber Risks}
\label{sec:classical-crypto-risks}

Classical cryptographic primitives (symmetric encryption, asymmetric cryptography, and 
hash functions) face critical quantum threats. Shor's 
algorithm~\cite{shor1999polynomial,shor1994algorithms} can efficiently break asymmetric 
cryptosystems including RSA and ECC, while Grover's algorithm~\cite{grover1996fast} 
accelerates brute-force attacks on symmetric\break schemes and hash functions. The risk is 
further exacerbated by \gls{HNDL}  
attacks~\cite{barenkamp2022steal}, where adversaries store encrypted data today to 
decrypt it once quantum computers become viable. This subsection evaluates these 
vulnerabilities to provide foundation for quantum-safe transition.

\begin{table*}[!hbpt]

\caption{Classic Cryptographic Standards and \gls{QC}: Assessing Cyber Risks}

\small
\renewcommand{\arraystretch}{1.1}
\label{tab:Pre-Migration-Alg}
\resizebox{\textwidth}{!}{%
\begin{tabular}
{|l|l|l|l|l|l|p{0.22\linewidth}|p{0.5\linewidth}|l|l|l|p{0.27\linewidth}|}
\hline

\multirow{2}{*}{\textbf{Crypto Type}} & \multirow{2}{*}{\textbf{Algorithms}} & \multirow{2}{*}{\textbf{Variants}} & \multirow{2}{*}{\textbf{Key Length (bits)}} & \multicolumn{2}{l|}{\textbf{Strengths (bits)}}&\multirow{2}{*}{\textbf{Vulnerabilities}} &\multirow{2}{*}{\textbf{Quantum Threats (STRIDE)}}& \multirow{2}{*}{\textbf{L}} & \multirow{2}{*}{\textbf{I}} & \multirow{2}{*}{\textbf{R}} & \multirow{2}{*}{\textbf{Possible QC-resistant Solutions}} \\ \cline{5-6}
 & & & & \multicolumn{1}{l|}{\textbf{Classic}} & \textbf{Quantum} & & & & & & \\ \hline
\multirow{9}{*}{Asymmetric} & \multirow{3}{*}{ECC~\cite{RFC8813, RFC7748}} & ECC 256 & 256 & \multicolumn{1}{l|}{128} & 0 & \multirow{8}{*}{{\begin{minipage}{\linewidth}
Broken by Shor's Algorithm~\cite{shor1999polynomial,shor1994algorithms}.
\end{minipage}}} & &\med &\high&\high& \multirow{8}{*}{{\begin{minipage}{\linewidth}Algorithms presented in Table~\ref{tab:post-alg}.\end{minipage}}}\\ \cline{3-6}\cline{9-11}
 & & ECC 384 & 384 & \multicolumn{1}{l|}{256} & 0 & & \multirow{5}{*}{{\begin{minipage}{\linewidth}
 \vspace{-8pt}
{For digital signature:}
\begin{myBullets}
\item {\textbf{Spoofing:} Shor's Algorithm allows forging of digital signatures.}
\item {\textbf{Tampering:} Integrity checks can be bypassed due to signature forgery.}
\item {\textbf{Repudiation:} Valid signatures can be forged, denying the origin of the message.} \end{myBullets}
{For KEM/ENC:}
\begin{myBullets}
\item {\textbf{Info. Disclosure:} KEM/ENC algorithms can be broken, revealing encrypted data.}
 \end{myBullets}
\end{minipage}}}\vspace{0.1cm}&\med& \high &\high& \\ \cline{3-6}\cline{9-11}
 & & ECC 521 & 521 & \multicolumn{1}{l|}{256} & 0 & & &\med & \high&\high& \\ \cline{2-6}\cline{9-11}
 & \multirow{2}{*}{FFDHE~\cite{rfc7919}} & DHE2048 & 2048 & \multicolumn{1}{l|}{112} & 0 & &&\med & \high&\high &\\ \cline{3-6}\cline{9-11}
 & & DHE3072 & 3072 & \multicolumn{1}{l|}{128} & 0 & & &\med& \high& \high&\\ \cline{2-6}\cline{9-11}
 & \multirow{3}{*}{RSA~\cite{moriarty2016pkcs}} & RSA 1024 & 1024 & \multicolumn{1}{l|}{80} & 0 & 
& & \med& \high& \high & \\ \cline{3-6}\cline{9-11}
 & & RSA 2048 & 2048 & \multicolumn{1}{l|}{112} & 0 & & &\med& \high &\high& \\ \cline{3-6}\cline{9-11}
 & & RSA 3072 & 3072 & \multicolumn{1}{l|}{128} & 0 & & &\med& \high &\high& \\ \hline
\multirow{4}{*}{Symmetric} & \multirow{3}{*}{AES~\cite{FIPS197}} & AES 128 & 128 & \multicolumn{1}{l|}{128} & 64 & \multirow{3}{*}{{\begin{minipage}{\linewidth}
Weakened by Grover's Algorithm~\cite{grover1996fast}.
\end{minipage}}}
& \multirow{3}{*}{{\begin{minipage}{\linewidth}
\begin{myBullets}
\vspace{0.1cm}
\item {\textbf{Info. Disclosure:} Grover's algorithm reduces the effective key length, making brute-force attacks feasible.}
 \vspace{3pt} \end{myBullets}
\end{minipage}}}&\med& \med & \med & \multirow{3}{*}{{\begin{minipage}{\linewidth}Larger key sizes are needed.\end{minipage}}} \\
\cline{3-6}\cline{9-11}
 & & AES 192 & 192 & \multicolumn{1}{l|}{192} & 96 & & &\med & \med &\med & \\ \cline{3-6}\cline{9-11}
 & & AES 256 & 256 & \multicolumn{1}{l|}{256} & 128 & & &\med& \low&\low& \\ 
 \cline{2-12}
 & \multirow{3}{*}{SHA2~\cite{eastlake2011us}} & SHA 256 & - & \multicolumn{1}{l|}{128} & {85} & \multirow{6}{*}{{\begin{minipage}{\linewidth}
Weakened by Brassard et al.'s Algorithm~\cite{brassard1997quantum}.
\end{minipage}}} & \multirow{6}{*}{{\begin{minipage}{\linewidth}
\begin{myBullets}
\item {\textbf{Spoofing:} Fake hash values can be created.}
\item {\textbf{Tampering:} Data integrity can be compromised by finding collisions.}
 \vspace{3pt} \end{myBullets}
\end{minipage}}}&\med& \med &\med & \multirow{6}{*}{{\begin{minipage}{\linewidth}Larger hash values are needed.\end{minipage}}} \\ \cline{3-6}\cline{9-11} 
 & & SHA 384 & - & \multicolumn{1}{l|}{192} & 128 & & &\med & \low&\low& \\ \cline{3-6}\cline{9-11}
 & & SHA 512 & - & \multicolumn{1}{l|}{256} & 170 & & & \med& \low&\low& \\ \cline{2-6}\cline{9-11}
 & \multirow{3}{*}{SHA3~\cite{eastlake2011us}} & SHA3 256 & - & \multicolumn{1}{l|}{128} & {85} && &\med& \med &\med & \\ \cline{3-6}\cline{9-11} 
 & & SHA3 384 & - & \multicolumn{1}{l|}{192} & 128 & & &\med & \low&\low & \\ \cline{3-6}\cline{9-11}
 & & SHA3 512 & - & \multicolumn{1}{l|}{256} & 170 & & &\med& \low&\low & \\ \hline
\end{tabular}%
}
\begin{tablenotes}[flushleft] \scriptsize \item[\dag] Likelihood evaluation assumes an adjustable 15-year horizon for the emergence of cryptographically relevant quantum computers (see Figure~\ref{fig:chart2}).\\
\end{tablenotes}\vspace{-0.3cm}
\end{table*}

\subsubsection{Quantifying Quantum Threats}

\begin{figure}[!h]
  \centering
 \fbox{\includegraphics[trim=0.2cm 0.1cm 0.2cm 0.1cm, clip=true, width=0.9\linewidth, height=0.54\linewidth]{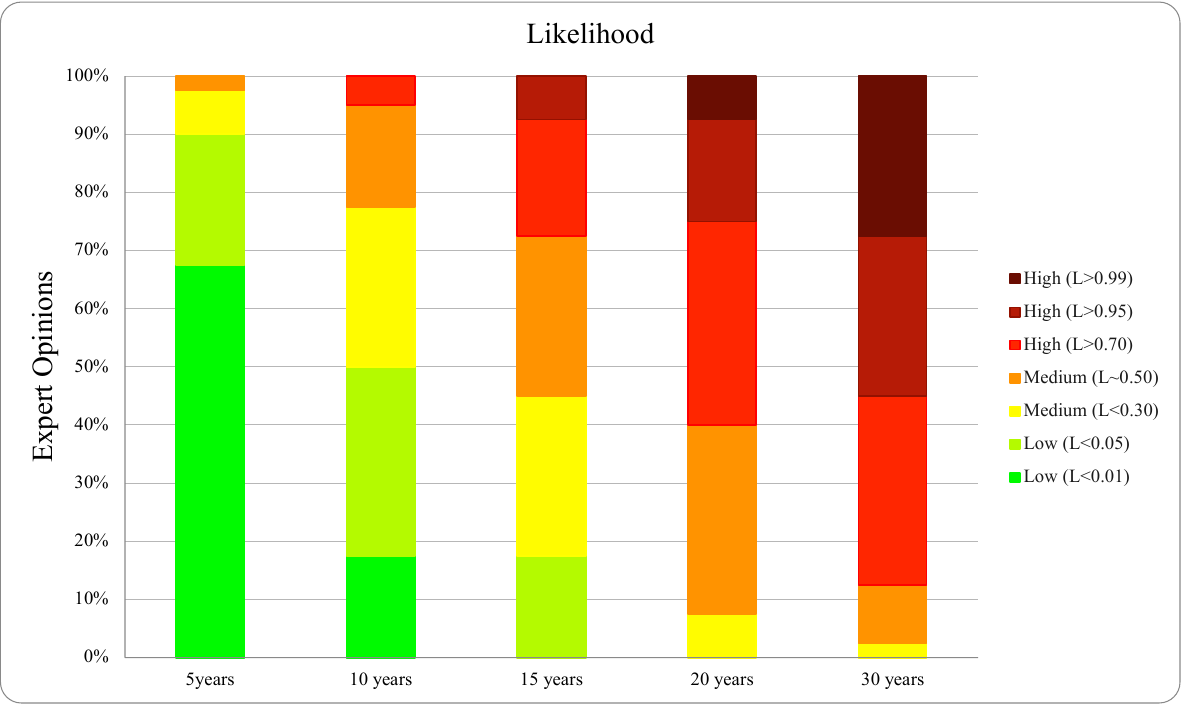}}
 
  \captionof{figure}{Cumulative Expert Opinions Related to Quantum Threat to Classic Cryptography}
  \label{fig:chart1}
\end{figure}

To understand the risks posed by \gls{QC}, we analyze quantum computer emergence 
timelines spanning 5 to 30 years, based on cumulative expert likelihood 
assessments~\cite{mosca2024quantum}. Figure~\ref{fig:chart1} summarizes expert opinions 
on when quantum computers capable of breaking classical cryptography will emerge. We 
define quantum threat as the probability that a quantum computer can break RSA-2048 
encryption within 24 hours. These probability assessments can be extended to evaluate 
the likelihood of breaking other cryptographic algorithms based on their quantum security 
levels, as illustrated in Figure~\ref{fig:classic-impact}.

\begin{figure}[!h]
 \centering
\fbox{\includegraphics[trim=0.2cm 0.1cm 0.1cm 0.6cm, clip=true, width=0.9\linewidth, height=0.54\linewidth]{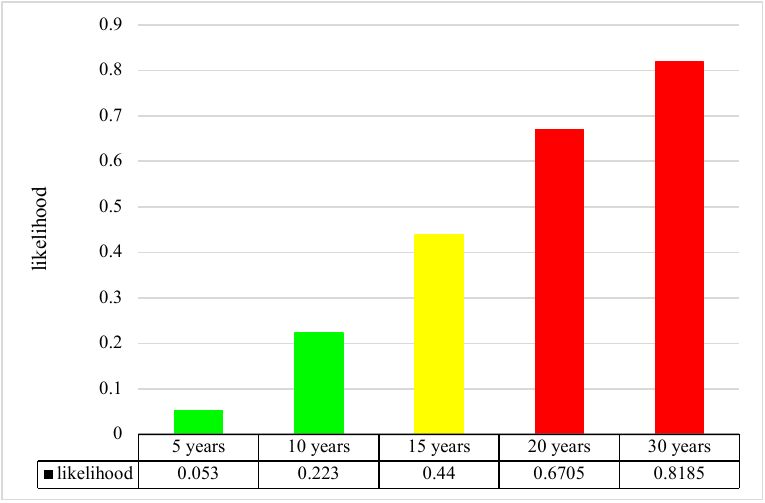}}
 
  \captionof{figure}{Expected Likelihood of Quantum Threat to Classic Cryptography Over a 30-Year Horizon.}
 \label{fig:chart2}
\end{figure}
 \subsubsection{Expected Likelihood of Quantum Threat}

We assess the expected likelihood that a quantum threat to classical cryptosystems occurs within the
5-, 10-, 15-, 20-, and 30-year horizons by aggregating expert responses from~\cite{mosca2024quantum}.
For each horizon $j$, we group the reported probabilities into bins $B_i$ with representative values
$l_i\in[0,1]$ (e.g., bin midpoints) and empirical frequencies $p_{j,i}$ (fraction of responses in $B_i$,
so $\sum_i p_{j,i}=1$). The expected likelihood is the weighted average \(\mathbb{E}[L_j] \;=\; \sum_i l_i\, p_{j,i}.\)
As shown in Figure~\ref{fig:chart2}, we categorize $\mathbb{E}[L_j]$ as low (10~years), medium (15~years),
and high (20+~years). For this analysis, we adopt a \emph{medium} likelihood at the 15-year horizon.

\subsubsection{Quantum Impact Assessment}

\begin{figure}[!h]
 \centering
\fbox{\includegraphics[trim=0.1cm 0.1cm 0.1cm 0.8cm, clip=true, width=0.9\linewidth, height=0.54\linewidth]{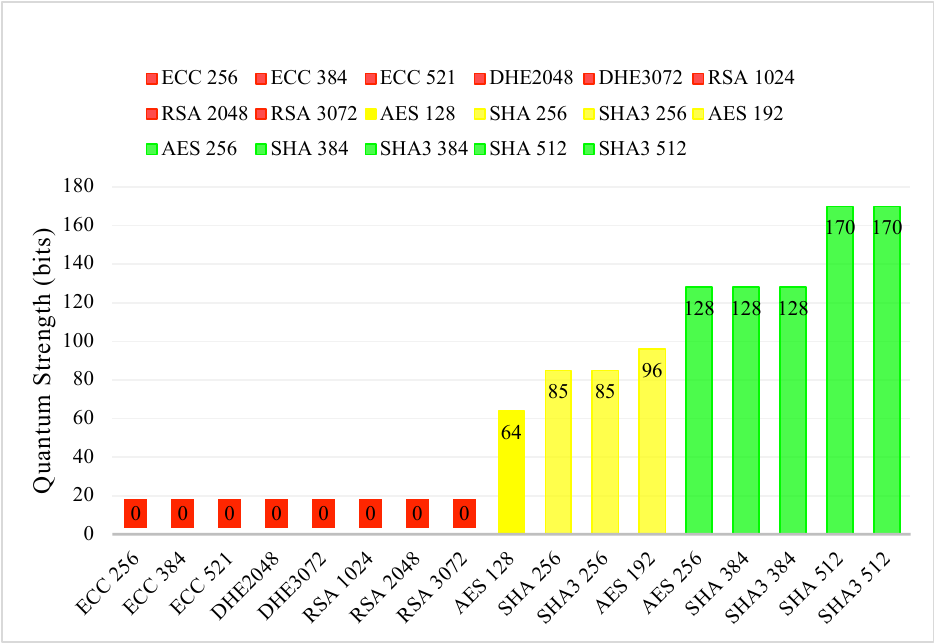}}
 
  \captionof{figure}{Expected Impact of Quantum Threat for Classic Cryptography}
  \label{fig:classic-impact}
\end{figure}

To assess the risk, we evaluate the impact of quantum threats on different classical cryptographic algorithms. This impact is determined by each algorithm's quantum security strength, as illustrated in Figure~\ref{fig:classic-impact}. An impact is considered high if the algorithm's quantum strength is less than 64 bits, medium if it falls between 64 and 127 bits, and low if it is 128 bits or greater. 

\subsubsection{Risk Evaluation}

Risk assessment combines quantum threat likelihood and potential impact using a 
qualitative risk matrix (Figure~\ref{fig:risk-matrix}), which underpins the systematic 
\gls{CC} stack threat modeling presented in Section~\ref{sec:pre}. Our evaluation 
adopts an adjustable 15-year time horizon with medium likelihood 
(Figure~\ref{fig:chart2}) and applies the algorithm-specific impact levels established 
in Figure~\ref{fig:classic-impact}. Table~\ref{tab:Pre-Migration-Alg} synthesizes the 
pre-transition security posture of classic cryptographic standards by presenting 
classical and quantum security strengths, identifying specific vulnerabilities, and 
mapping emerging quantum threats to the STRIDE threat modeling framework.

\subsection{Quantum-Safe Cryptographic Standards: Addressing \gls{QC} and Assessing 
Residual Risk}\label{sec:post-alg}

As established in the preceding subsection, symmetric and hash-based algorithms can 
be strengthened against quantum attacks through increased key sizes, whereas public-key 
cryptosystems face fundamental vulnerabilities requiring complete replacement. This 
subsection examines \gls{NIST}-standardized quantum-safe solutions addressing this critical 
transition.

\begin{table*}[!htbp]

\scriptsize
\caption{NIST-Standardized Quantum-Resistant Cryptographic Algorithms: 
Cyber Impact and Risk Assessment}

\label{tab:post-alg}
\resizebox{\textwidth}{!}{%
\begin{tabular}{|l|p{0.13\linewidth}|p{0.11\linewidth}|p{0.12\linewidth}|p{0.48\linewidth}|p{0.40\linewidth}|l|l|l|}
\hline
\multirow{1}{*}{\textbf{Algorithms}} & \multirow{1}{*}{\textbf{Description}} & \multirow{1}{*}{\textbf{FIPS Compliance}} & \multirow{1}{*}{\textbf{Attacks $\dag$}} & \multirow{1}{*}{\textbf{Possible Countermeasures}} & \multirow{1}{*}{\textbf{STRIDE Threats}} & \multirow{1}{*}{\textbf{L}} & \multirow{1}{*}{\textbf{I}} & \multirow{1}{*}{\textbf{R}} \\ \hline

\multirow{10}{*}{Kyber~\cite{bos2018crystals}} & \multirow{10}{*}{\begin{minipage}{\linewidth}Lattice-based \gls{KEM/ENC} using M-LWE over cyclotomic rings\end{minipage}}&\multirow{10}{*}{\begin{minipage}{\linewidth}FIPS 203~\cite{nistfips203}\end{minipage}}&  \begin{minipage}{\linewidth}\glsentryshort{FA}~\cite{ravi2019number,oder2018practical,ravi2020drop}
\end{minipage}  & 
{\begin{minipage}{\linewidth}
\begin{myBullets}\vspace{2pt}
\item Mask decryption by splitting secret key~\cite{oder2018practical,ravi2020drop}
\item Check secret/error components for trivial weaknesses~\cite{ravi2019number}
\vspace{1pt}\end{myBullets}
\end{minipage}}
 & 
{\begin{minipage}{\linewidth}
\begin{myBullets}\vspace{2pt}
\item \textbf{Info. Disclosure:} message and key recovery~\cite{ravi2019number,oder2018practical,ravi2020drop}
\vspace{1pt}\end{myBullets}
\end{minipage}}
 &\med &\med &\med \\ \cline{4-9} 

& & & \begin{minipage}{\linewidth}\glsentryshort{SPA}~\cite{hamburg2021chosen}
\end{minipage} & 
{\begin{minipage}{\linewidth}
\begin{myBullets}\vspace{2pt}
\item Mask inputs~\cite{hamburg2021chosen}
\item Randomize NTT operation order or insert dummy operations~\cite{hamburg2021chosen}
\vspace{1pt}\end{myBullets}
\end{minipage}}
 & 
{\begin{minipage}{\linewidth}
\begin{myBullets}\vspace{2pt}
\item \textbf{Info. Disclosure:} key recovery~\cite{hamburg2021chosen}
\vspace{1pt}\end{myBullets}
\end{minipage}}
 &\med & \med&\med \\ \cline{4-9} 

& & & \begin{minipage}{\linewidth}\glsentryshort{APA}~\cite{pessl2019more,kamucheka2021power,dubrova2022breaking} 
\end{minipage} & 
{\begin{minipage}{\linewidth}
\begin{myBullets}\vspace{2pt}
\item Mask NTT operations~\cite{pessl2019more}
\item No known countermeasures~\cite{dubrova2022breaking}
\vspace{1pt}\end{myBullets}
\end{minipage}}
 & 
{\begin{minipage}{\linewidth}
\begin{myBullets}\vspace{2pt}
\item \textbf{Info. Disclosure:} symmetric key recovery~\cite{pessl2019more}
\vspace{1pt}\end{myBullets}
\end{minipage}}
 &\high & \med&\high\\ \cline{4-9} 

& & &\glsentryshort{EM}~\cite{ravi2020generic,xu2021magnifying,ravi2020drop} & 
{\begin{minipage}{\linewidth}
\begin{myBullets}\vspace{2pt}
\item Mask ECC decryption/decapsulation and FO transform~\cite{ravi2020generic,oder2018practical,xu2021magnifying}
\item Discard low-entropy ciphertexts~\cite{xu2021magnifying,dwivedi2024novel}
\item Split secret into random shares and randomize decryption~\cite{xu2021magnifying}
\vspace{1pt}\end{myBullets}
\end{minipage}}
 & 
{\begin{minipage}{\linewidth}
\begin{myBullets}\vspace{2pt}
\item \textbf{Info. Disclosure:} full key extraction~\cite{xu2021magnifying,ravi2020generic} or message bits~\cite{ravi2020drop}
\vspace{1pt}\end{myBullets}
\end{minipage}}
 &\low &\med & \low\\ \cline{4-9} 

& & &\glsentryshort{TMP}~\cite{ravi2021exploiting} & 
{\begin{minipage}{\linewidth}
\begin{myBullets}\vspace{2pt}
\item No known countermeasures~\cite{ravi2021exploiting}
\vspace{1pt}\end{myBullets}
\end{minipage}}
 & 
{\begin{minipage}{\linewidth}
\begin{myBullets}\vspace{2pt}
\item \textbf{Info. Disclosure:} message recovery~\cite{ravi2021exploiting}
\vspace{1pt}\end{myBullets}
\end{minipage}}
&\med & \med&\med\\ \cline{4-9} 

& & &\glsentryshort{CB}~\cite{albrecht2018cold} & 
{\begin{minipage}{\linewidth}
\begin{myBullets}\vspace{2pt}
\item Store secret in time domain instead of frequency domain~\cite{albrecht2018cold}
\vspace{1pt}\end{myBullets}
\end{minipage}}
 & 
{\begin{minipage}{\linewidth}
\begin{myBullets}\vspace{2pt}
\item \textbf{Info. Disclosure:} secret key recovery~\cite{albrecht2018cold}
\vspace{1pt}\end{myBullets}
\end{minipage}}
 & \low & \med & \low \\ \hline

\multirow{10}{*}{Dilithium~\cite{ducas2018crystals}} & \multirow{10}{*}{\begin{minipage}{\linewidth}Lattice-based signature using Fiat-Shamir with Aborts and rejection sampling\end{minipage}} &\multirow{10}{*}{\begin{minipage}{\linewidth}FIPS 204~\cite{nistfips204}\end{minipage}}& \glsentryshort{FA}~\cite{ravi2019number,bruinderink2018differential} & 
{\begin{minipage}{\linewidth}
\begin{myBullets}\vspace{2pt}
\item Check secret/error components for trivial weaknesses~\cite{ravi2019number}
\item Double computation, verification-after-sign, additional randomness~\cite{bruinderink2018differential}
\vspace{1pt}\end{myBullets}
\end{minipage}}
 & 
{\begin{minipage}{\linewidth}
\begin{myBullets}\vspace{2pt}
\item \textbf{Spoofing, Tampering:} key recovery~\cite{ravi2019number,bruinderink2018differential}
\item \textbf{Tampering, Repudiation:} signature forgery~\cite{bruinderink2018differential}
\item \textbf{EoP:} privilege escalation via forged signatures~\cite{bruinderink2018differential}
\vspace{1pt}\end{myBullets}
\end{minipage}}
&\med & \med&\med \\ \cline{4-9} 

& & &\glsentryshort{APA}~\cite{migliore2019masking,marzougui2022profiling} & 
{\begin{minipage}{\linewidth}
\begin{myBullets}\vspace{2pt}
\item Linear secret sharing masking~\cite{migliore2019masking}
\item Boolean and arithmetic masking via variable splitting/sharing~\cite{marzougui2022profiling}
\vspace{1pt}\end{myBullets}
\end{minipage}}
 & 
{\begin{minipage}{\linewidth}
\begin{myBullets}\vspace{2pt}
\item \textbf{Spoofing, Tampering, Repudiation:} signature forgery~\cite{marzougui2022profiling}
\item \textbf{EoP:} privilege escalation via forged signatures~\cite{marzougui2022profiling}
\item \textbf{Spoofing, Tampering, Repudiation, EoP:} secret variable disclosure~\cite{migliore2019masking}
\vspace{1pt}\end{myBullets}
\end{minipage}}
 &\med & \med&\med \\ \cline{4-9} 

& & &\glsentryshort{EM}~\cite{ravi2019exploiting,singh2024end} & 
{\begin{minipage}{\linewidth}
\begin{myBullets}\vspace{2pt}
\item Re-order operations and embed vulnerable addition deep within signing~\cite{ravi2019exploiting}
\item Bit-slicing design for NTT with spatial intra-instruction redundancy~\cite{singh2024end}
\vspace{1pt}\end{myBullets}
\end{minipage}}
 & 
{\begin{minipage}{\linewidth}
\begin{myBullets}\vspace{2pt}
\item \textbf{Spoofing, Tampering:} secret key information disclosure~\cite{singh2024end}
\item \textbf{Tampering, Repudiation, EoP:} signature forgery~\cite{ravi2019exploiting}
\vspace{1pt}\end{myBullets}
\end{minipage}}
 &\low & \med&\low\\ \cline{4-9}

& & &\glsentryshort{TMP}~\cite{cryptoeprint:2023/050} & 
{\begin{minipage}{\linewidth}
\begin{myBullets}\vspace{2pt}
\item Shuffling and secret sharing~\cite{cryptoeprint:2023/050}
\vspace{1pt}\end{myBullets}
\end{minipage}}
 & 
{\begin{minipage}{\linewidth}
\begin{myBullets}\vspace{2pt}
\item \textbf{Spoofing, Tampering:} signer's secret key disclosure~\cite{cryptoeprint:2023/050}
\item \textbf{Repudiation, EoP:} signature forgery via revealed key~\cite{cryptoeprint:2023/050}
\vspace{1pt}\end{myBullets}
\end{minipage}}
&\med & \med&\med\\ \hline 

\multirow{3}{*}{SPHINCS+~\cite{bernstein2019sphincs+}} & \multirow{3}{*}{\begin{minipage}{\linewidth}Stateless hash-based signature relying on collision resistance\end{minipage}} &\multirow{3}{*}{\begin{minipage}{\linewidth}FIPS 205~\cite{nistfips205}\end{minipage}}& \glsentryshort{FA}~\cite{castelnovi2018grafting,genet2018practical} & 
{\begin{minipage}{\linewidth}
\begin{myBullets}\vspace{2pt}
\item Redundant signature computation and public index derivation~\cite{castelnovi2018grafting}
\item Tree integrity checks via cross-layer linkage and subtree recomputation~\cite{castelnovi2018grafting,genet2018practical}
\item Enhanced hash functions and one-time signature caching~\cite{genet2018practical}
\item Fault detection via instruction duplication~\cite{genet2018practical}
\vspace{1pt}\end{myBullets}
\end{minipage}}
 & 
\begin{minipage}{\linewidth}
\begin{myBullets}\vspace{2pt}
\item \textbf{Spoofing, Tampering:} secret key recovery~\cite{castelnovi2018grafting} or universal forgery~\cite{genet2018practical}
\item \textbf{Tampering, Repudiation:} message signature forgery~\cite{castelnovi2018grafting} or universal forgery via voltage glitch~\cite{genet2018practical}
\vspace{1pt}\end{myBullets}
\end{minipage}
&\med & \med&\med \\ \cline{4-9} 

& & & \begin{minipage}{\linewidth}\glsentryshort{APA}~\cite{kannwischer2018differential} 
\end{minipage} & 
{\begin{minipage}{\linewidth}
\begin{myBullets}\vspace{2pt}
\item Hide Mix procedure order~\cite{kannwischer2018differential}
\vspace{1pt}\end{myBullets}
\end{minipage}}
 & 
{\begin{minipage}{\linewidth}
\begin{myBullets}\vspace{2pt}
\item \textbf{Spoofing, Tampering:} secret key recovery~\cite{kannwischer2018differential}
\item \textbf{Tampering, Repudiation, EoP:} arbitrary message signature~\cite{kannwischer2018differential}
\vspace{1pt}\end{myBullets}
\end{minipage}}
 &\med & \med&\med \\ \hline

\multirow{8}{*}{Falcon~\cite{fouque2018falcon}} & \multirow{8}{*}{{\begin{minipage}{\linewidth}Lattice-based signature using NTRU shortest vector problem\end{minipage}}} & \multirow{8}{*}{\begin{minipage}{\linewidth}Pending FIPS 206~\cite{NISTIR8214C}\end{minipage}}&\glsentryshort{FA}~\cite{mccarthy2019bearz} & 
{\begin{minipage}{\linewidth}
\begin{myBullets}\vspace{2pt}
\item Double signature computation~\cite{mccarthy2019bearz}
\item Immediate verification after signing~\cite{mccarthy2019bearz}
\item Zero-check sampled vector~\cite{mccarthy2019bearz}
\vspace{1pt}\end{myBullets}
\end{minipage}}
 & 
{\begin{minipage}{\linewidth}
\begin{myBullets}\vspace{2pt}
\item \textbf{Spoofing, Tampering:} private key retrieval~\cite{mccarthy2019bearz}
\item \textbf{Repudiation, EoP:} signature forgery via retrieved key~\cite{mccarthy2019bearz}
\vspace{1pt}\end{myBullets}
\end{minipage}}
&\med & \med&\med \\ \cline{4-9} 

& & &\glsentryshort{TA}~\cite{mccarthy2019bearz} & 
{\begin{minipage}{\linewidth}
\begin{myBullets}\vspace{2pt}
\item Blind-Vector algorithm with Fisher-Yates shuffling~\cite{mccarthy2019bearz}
\item Sample discard with random cache reads~\cite{mccarthy2019bearz}
\vspace{1pt}\end{myBullets}
\end{minipage}}
 & 
{\begin{minipage}{\linewidth}
\begin{myBullets}\vspace{2pt}
\item \textbf{Spoofing, Tampering:} private key retrieval~\cite{mccarthy2019bearz}
\item \textbf{Repudiation, EoP:} signature forgery via retrieved key~\cite{mccarthy2019bearz}
\vspace{1pt}\end{myBullets}
\end{minipage}}
&\med & \med&\med \\ \cline{4-9} 

& & & \begin{minipage}{\linewidth}\glsentryshort{SPA}~\cite{guerreau2022hidden}
\end{minipage}  & 
{\begin{minipage}{\linewidth}
\begin{myBullets}\vspace{2pt}
\item Lower Hamming weight gap~\cite{guerreau2022hidden}
\vspace{1pt}\end{myBullets}
\end{minipage}}
 & 
{\begin{minipage}{\linewidth}
\begin{myBullets}\vspace{2pt}
\item \textbf{Spoofing, Tampering, Repudiation, EoP:} complete secret key recovery~\cite{guerreau2022hidden}
\vspace{1pt}\end{myBullets}
\end{minipage}}
&\med & \med&\med\\ \cline{4-9} 

& & & \begin{minipage}{\linewidth}\glsentryshort{EM}~\cite{karabulut2021falcon}
\end{minipage} & 
{\begin{minipage}{\linewidth}
\begin{myBullets}\vspace{2pt}
\item Constant power consumption hiding~\cite{karabulut2021falcon}
\item Masking via randomizing intermediate values~\cite{karabulut2021falcon}
\vspace{1pt}\end{myBullets}
\end{minipage}}
 & 
{\begin{minipage}{\linewidth}
\begin{myBullets}\vspace{2pt}
\item \textbf{Spoofing, Tampering:} secret signing key extraction~\cite{karabulut2021falcon}
\item \textbf{Tampering, Repudiation, EoP:} arbitrary message signature forgery~\cite{karabulut2021falcon}
\vspace{1pt}\end{myBullets}
\end{minipage}}
 & \low & \med& \low \\ \hline

\multirow{8}{*}{HQC~\cite{melchor2018hamming}} &
\multirow{8}{*}{\begin{minipage}{\linewidth}Code-based \gls{KEM/ENC} using Hamming metric decoding\end{minipage}} &
\multirow{8}{*}{\begin{minipage}{\linewidth}Pending FIPS 207~\cite{NISTIR8545}\end{minipage}} &  \begin{minipage}{\linewidth}\glsentryshort{FA}~\cite{cayrel2020message,xagawa2021fault}
\end{minipage}  &
\begin{minipage}{\linewidth}
\begin{myBullets}\vspace{2pt}
\item Constant-time error handling~\cite{xagawa2021fault}
\item Instruction duplication and random delays~\cite{xagawa2021fault}
\item No known countermeasures~\cite{cayrel2020message}
\vspace{1pt}\end{myBullets}
\end{minipage} &
\begin{minipage}{\linewidth}
\begin{myBullets}\vspace{2pt}
\item \textbf{Info. Disclosure:} message recovery and key leakage~\cite{cayrel2020message,xagawa2021fault}
\vspace{1pt}\end{myBullets}
\end{minipage} & \med & \med & \med \\ \cline{4-9}

& & &  \begin{minipage}{\linewidth}\glsentryshort{TA}~\cite{guo2020key,guo2022don,wafo2020practicable}
\end{minipage}  &
\begin{minipage}{\linewidth}
\begin{myBullets}\vspace{2pt}
\item Constant-time decoding, field operations, and RNG~\cite{wafo2020practicable,guo2022don}
\item No known countermeasures~\cite{guo2020key}
\vspace{1pt}\end{myBullets}
\end{minipage} &
\begin{minipage}{\linewidth}
\begin{myBullets}\vspace{2pt}
\item \textbf{Info. Disclosure:} key recovery via timing side-channels~\cite{guo2020key,guo2022don}
\vspace{1pt}\end{myBullets}
\end{minipage} & \med & \med & \med \\ \cline{4-9}

& & &  \begin{minipage}{\linewidth}\glsentryshort{SPA}~\cite{schamberger2020power}
\end{minipage}  &
\begin{minipage}{\linewidth}
\begin{myBullets}\vspace{2pt}
\item No known countermeasures~\cite{schamberger2020power}
\vspace{1pt}\end{myBullets}
\end{minipage} &
\begin{minipage}{\linewidth}
\begin{myBullets}\vspace{2pt}
\item \textbf{Info. Disclosure:} partial key recovery from Hamming weight leakage~\cite{schamberger2020power}
\vspace{1pt}\end{myBullets}
\end{minipage} & \high & \med & \high \\ \cline{4-9}

& & & \begin{minipage}{\linewidth}\glsentryshort{EM}~\cite{goy2022new,paiva2025tu} 
\end{minipage} 
&
\begin{minipage}{\linewidth}
\begin{myBullets}\vspace{2pt}
\item Linear secret sharing to mask sensitive values~\cite{goy2022new}
\item No known countermeasures~\cite{paiva2025tu}
\vspace{1pt}\end{myBullets}
\end{minipage} &
\begin{minipage}{\linewidth}
\begin{myBullets}\vspace{2pt}
\item \textbf{Info. Disclosure:} full key recovery from EM leakage~\cite{goy2022new}
\vspace{1pt}\end{myBullets}
\end{minipage} & \med & \med & \med \\ \hline

\end{tabular}%
}
\begin{tablenotes}[para,flushleft]
\scriptsize 
 \item[\dag] Attack abbreviations: \glsentryshort{FA}=\glsentrylong{FA}, \glsentryshort{SPA}=\glsentrylong{SPA}, \glsentryshort{APA}=\glsentrylong{APA}, \glsentryshort{EM}=\glsentrylong{EM}, \glsentryshort{TMP}=\glsentrylong{TMP}, \glsentryshort{CB}=\glsentrylong{CB}, \glsentryshort{TA}=\glsentrylong{TA}.
 \item[\ddag] We perform risk evaluation with the presumption of considering the countermeasures mentioned in the table.
\end{tablenotes}
\end{table*}
\begin{enumerate}[topsep=0ex, itemsep=0ex, wide, font=\itshape, labelwidth=!, 
labelindent=0pt, label*=3.2.\arabic*]
\item \textit{Transition to Quantum-Safe Cryptographic Algorithms.}
\gls{NIST}'s standardization process evaluated PQC algorithms for KEM/ENC and digital 
signatures. By August 2024, \gls{NIST} released three standards~\cite{NIST2024}: FIPS 
203~\cite{nistfips203} (ML-KEM/CRYSTALS-Kyber for encryption), FIPS 
204~\cite{nistfips204} (ML-DSA/CRYSTALS-Dilithium for signatures), and FIPS 
205~\cite{nistfips205} (SLH-DSA/SPHINCS+ as backup signature method). Additionally, 
\gls{NIST} selected HQC (code-based KEM/ENC) for standardization, with draft FIPS expected 
in 2026~\cite{NISTIR8545}. Table~\ref{tab:post-alg} summarizes algorithm descriptions, 
FIPS compliance status, identified attacks, countermeasures, and STRIDE-mapped threats.

\item \textit{Challenges Beyond Quantum-Resistant Algorithms.}
While \gls{NIST} \gls{PQC} algorithms resist quantum attacks, they remain vulnerable to 
\glspl{SCA} exploiting implementation-level information leakage. \glspl{SCA} 
leverage power consumption, electromagnetic radiation, and timing variations to extract 
cryptographic secrets. These attacks comprise passive techniques (electromagnetic 
analysis, template attacks, cold-boot attacks, power analysis, timing analysis) and 
active techniques (fault injection). Numerous \glspl{SCA} against \gls{NIST} \gls{PQC} candidates have 
been documented~\cite{ravi2019number,oder2018practical,hamburg2021chosen,pessl2019more,bruinderink2018differential,migliore2019masking,castelnovi2018grafting,mccarthy2019bearz,cayrel2020message,guo2020key}, 
with ongoing research revealing additional vulnerabilities. Continuous assessment of 
implementation security and countermeasure effectiveness remains critical for 
\gls{PQC} deployment.

\item \textit{Quantum Attack Vectors for Quantum-Safe Cryptography.}
Quantum adversaries may exploit \glspl{SCA} vulnerabilities in \gls{PQC} implementations, 
compromising \gls{NIST}-standardized algorithms despite their mathematical quantum resistance. 
We assess risks for each attack vector documented in Table~\ref{tab:post-alg} through 
comprehensive qualitative risk assessment integrating likelihood evaluation and impact 
analysis (Figure~\ref{fig:risk-matrix}).

\item \textit{Likelihood and Impact Evaluation.}
Likelihood assessment examines three factors: exploitability characteristics 
(physical/network/internet access requirements), countermeasure availability and 
effectiveness (detailed in Table~\ref{tab:post-alg}), and NIST SP 800-30 Appendix G 
criteria~\cite{NIST-SP800-30} adapted for quantum threat scenarios 
(Table~\ref{table:likelihood}). We categorize likelihood as: (a) \textit{High} - 
known remote exploit with no effective countermeasures; (b) \textit{Medium} - known 
exploit requiring physical access without adequate countermeasures; (c) \textit{Low} - 
no known exploit, or exploit mitigated by effective countermeasures, or attack requires 
elevated privileges. 
Impact assessment applies NIST SP 800-30 Appendix H criteria~\cite{NIST-SP800-30} with 
quantum-specific adaptations (Table~\ref{table:impact}). Quantum-enabled threats can 
compromise data confidentiality and integrity, disrupt service availability, damage 
organizational reputation, or trigger regulatory violations. Given these potential 
consequences, quantum threats warrant minimum \textit{medium} baseline impact.
Risk evaluation synthesizes likelihood and impact assessments using the qualitative 
matrix in Figure~\ref{fig:risk-matrix}. Table~\ref{tab:post-alg} presents comprehensive 
algorithm-specific analysis including vulnerability characterization, attack vector 
documentation, countermeasure strategies, and quantified risk levels essential for 
robust PQC transition planning.

\end{enumerate}

\section{Pre-Transition Quantum Threat Landscape in \gls{CC}}
\label{sec:pre}

\gls{CC} infrastructures are structured across nine architectural layers  (applications, data, runtime, middleware, OS, virtualization, server, storage, and networks) as illustrated in Figure~\ref{fig:stack}. This section assesses the 
quantum threat landscape during the pre-transition phase, when cloud environments  still rely on vulnerable classical cryptographic standards. Building on vulnerabilities  discussed in Section~\ref{sec:risk}, we systematically analyze how quantum-enabled  adversaries exploit each layer's cryptographic and architectural dependencies, exposing  systemic risks. Delayed deployment of quantum-resistant primitives further amplifies  exposure to attacks such as key recovery, ciphertext harvesting for future decryption, 
and signature forgeries. To guide early-stage defense planning, we emphasize the need  for tailored threat modeling and layered mitigation strategies aligned with each  architectural layer.
\subsection{Pre-Transition Vulnerabilities}

Classical cryptographic primitives in \gls{CC} architectures exhibit fundamental vulnerabilities to quantum computational attacks. Unlike patchable implementation flaws, these vulnerabilities stem from mathematical limitations of established cryptographic schemes and their pervasive role as security foundations across cloud architectural layers. This subsection categorizes domains that quantum adversaries may exploit, forming the basis for subsequent attack vector analysis.

\begin{enumerate}[topsep=0ex, itemsep=0ex, wide, font=\itshape, labelwidth=!, labelindent=0pt, label*=4.1.\arabic*]
\item \textit{Cryptographic Vulnerabilities.}
Quantum computers can decrypt standard encryption algorithms (RSA, DHE, ECC), leading to cryptographic breaches and unauthorized access to critical information. Shor's and Grover's algorithms break weak encryption algorithms, both symmetric and asymmetric, exposing sensitive data stored or transmitted in the cloud~\cite{lu2019attacking,shen2020lightweight}. Quantum computers can also forge signatures, enabling impersonation of legitimate users or entities and unauthorized access to high-security systems~\cite{compastie2020virtualization}.

\item \textit{Identity and Access Management Vulnerabilities.}
Quantum computers exploit digital signatures for identity theft, allowing impersonation of legitimate users~\cite{esparza2019understanding}. Quantum algorithms bypass weak authentication mechanisms, enabling unauthorized access to cloud resources~\cite{madhu2025multi, baseri2024cybersecurity}. Quantum computers can predict or forge session tokens, allowing session hijacking and unauthorized access to cloud services~\cite{chawla2023roadmap}.

\item \textit{Data Integrity and Confidentiality Vulnerabilities.}
Quantum computers alter encrypted data storage, potentially manipulating sensitive records through data tampering~\cite{jangjou2022comprehensive}. Quantum algorithms decrypt sensitive data stored in the cloud, leading to data exfiltration and data breaches. Attackers collect encrypted data now for later quantum decryption, compromising long-term confidentiality through \gls{HNDL} attacks~\cite{barenkamp2022steal}.

\item \textit{Network and Communication Vulnerabilities.}
Quantum computers decrypt communication channels, enabling interception and manipulation of data in transit via \gls{MITM} attacks. Quantum adversaries force weaker cryptographic protocols, exposing encrypted communications to decryption and tampering through protocol downgrade attacks~\cite{BASERI2024103883,compastie2020virtualization}. Quantum-forged \gls{RPKI} credentials facilitate BGP route hijacking, maliciously rerouting or intercepting network traffic~\cite{barenkamp2022steal}.

\item \textit{Virtualization and Runtime Vulnerabilities.}
Quantum computers decrypt \glspl{VM} during migration, exposing sensitive data. Quantum-assisted \glspl{SCA} extract cryptographic keys from hypervisors, compromising hosted VMs. Quantum-enabled adversaries intercept and decrypt inter-\gls{VM} communication~\cite{saeed2020cross}.

\item \textit{Storage Vulnerabilities.}
Quantum algorithms decrypt data stored in cloud storage, leading to encrypted data breaches. Attackers exploit deduplication techniques~\cite{garcia2024mapping,prajapati2022review} for quantum-assisted storage deduplication attacks compromising data integrity. Quantum algorithms forge file hashes, tampering with integrity verification mechanisms through quantum-optimized integrity poisoning.

\item \textit{Middleware and Application Vulnerabilities.}
Quantum computers accelerate cryptographic attacks, recovering \gls{API} credentials~\cite{mazzocca2025survey} and cryptographic keys by breaking signatures and weakening symmetric encryption. This enables quantum-enhanced \gls{API} abuse through forged authentication and impersonation, leading to unauthorized access, privilege escalation, and data exfiltration.

\item \textit{Firmware and Hardware Vulnerabilities.}
Quantum-forged signatures bypass firmware validation, enabling backdoored 
updates that compromise system integrity~\cite{10.1145/3432893}. 
Quantum-assisted malware exploits weakened cryptographic primitives to embed 
persistent rootkits surviving reboots and updates~\cite{ThreatsProtection}. 
Quantum-enhanced \glspl{SCA} extract cryptographic keys from \glspl{HSM} and 
\glspl{TPM}  via electromagnetic emissions, power fluctuations, or timing analysis, 
leading to unauthorized decryption and system compromise~\cite{crypto4aHSM}.

\item \textit{\gls{DoS} and Resource Exhaustion.}
Quantum algorithms optimize resource exhaustion attacks, leading to quantum-optimized DoS attacks. Quantum-assisted DoS attacks overload hypervisors, leading to hypervisor overload~\cite{baseri2025evaluation,BASERI2024103883}.

\item \textit{Supply Chain and Software Integrity Vulnerabilities.}
Quantum-forged signatures compromise software dependencies through quantum-assisted software supply chain attacks. Quantum computers compromise cryptographic channels securing \gls{SDN} controllers, leading to quantum-forged \gls{SDN} controller exploits~\cite{abdou2018comparative}.
\end{enumerate}

\subsection{Pre-Transition Attack Vectors}

\gls{QC} introduces significant risks to classical cryptographic systems in \gls{CC} environments. Quantum computers, leveraging Shor's and Grover's algorithms, can break widely used encryption schemes, exposing vulnerabilities across \gls{CC} infrastructure layers. This subsection categorizes primary attack vectors introduced by quantum capabilities, focusing on their impact on classical systems before transition to quantum-safe cryptography.
\begin{enumerate}[topsep=0ex, itemsep=0ex, wide, font=\itshape, labelwidth=!, labelindent=0pt, label*=4.2.\arabic*]
\item \textit{Cryptographic Attacks.} Grover's algorithm reduces symmetric key encryption security (e.g., AES) by square-rooting the key space, enabling significantly faster brute-force attacks~\cite{grover1996fast}. Shor's algorithm factors large integers and computes discrete logarithms, breaking RSA and ECC-based encryption, enabling communication decryption, digital signature forgery, and entity impersonation~\cite{lu2019attacking}. Quantum algorithms accelerate hash function collision discovery, undermining digital signature and authentication integrity~\cite{baseri2025evaluation}.

\item \textit{Data and Storage Exploitation.} \gls{HNDL} attacks enable attackers to store encrypted data today for future quantum decryption, threatening long-term confidentiality~\cite{barenkamp2022steal}. Quantum algorithms decrypt data-at-rest, exposing sensitive information in databases, backups, and cloud storage systems~\cite{jangjou2022comprehensive}. Quantum-enhanced attacks modify encrypted data without detection, compromising data integrity and trust~\cite{baseri2025evaluation, BASERI2024103883,li2024secure}.

\item \textit{Network and Communication Vulnerabilities.} Quantum algorithms break encryption used in TLS and IPsec, enabling eavesdropping or manipulation of network traffic~\cite{BASERI2024103883}. Quantum attacks break WPA2/WPA3 encryption, allowing unauthorized Wi-Fi access and sensitive data interception~\cite{lounis2020attacks}. Quantum-forged \gls{RPKI} credentials enable BGP route hijacking, rerouting or blackholing traffic for eavesdropping or \gls{DoS} attacks~\cite{compastie2020virtualization}.

\item \textit{Virtualization and Cloud Infrastructure.} Quantum algorithms decrypt \gls{VM} data during migration, exposing sensitive information and compromising \gls{VM} integrity~\cite{compastie2020virtualization}. Quantum attacks bypass hypervisor security, granting elevated privileges and control over multiple VMs. Quantum-enhanced attacks intercept and decrypt inter-\gls{VM} communication, leading to data leakage and tampering~\cite{gruss2019page}.

\item \textit{Software and Firmware Exploitation.} Quantum-forged digital signatures inject malicious code into software updates, compromising system integrity~\cite{hoffman2020web}. Quantum algorithms break firmware encryption, enabling malicious firmware injection into IoT devices, servers, and cloud infrastructure~\cite{appiah2025secure,ThreatsProtection}. Quantum-enhanced malware evades detection and persists across reboots, enabling long-term system compromise.

\item \textit{Authentication and Access Control.} Grover's algorithm accelerates password cracking, compromising weak credentials stored in cloud applications~\cite{baseri2025evaluation, BASERI2024103883}. Quantum algorithms predict or forge session tokens, allowing hijacking of authenticated sessions and privilege escalation~\cite{grassi2020digital}. Quantum-assisted brute-force attacks recover \gls{API} keys, enabling unauthorized access to cloud services and middleware~\cite{madhu2025multi}.

\item \textit{\gls{DoS} and Resource Exhaustion.} Quantum algorithms identify optimal traffic patterns for maximum resource exhaustion, leading to service degradation or downtime~\cite{tabrizchi2020survey}. Quantum-enhanced hash collisions craft duplicate POST requests, bypassing rate limits and overloading servers~\cite{baseri2024cybersecurity}.

\item \textit{Side-Channel and Physical Attacks.} Quantum-assisted techniques extract cryptographic keys from hardware via timing, power, or electromagnetic analysis~\cite{chowdhury2021physical}. Quantum algorithms exploit \glspl{HSM} and firmware backdoors, enabling persistent system access~\cite{ThreatsProtection}.

\item \textit{AI and Machine Learning Exploitation.} Quantum-optimized adversarial examples bypass AI-powered security models, leading to misclassification and system compromise~\cite{west2023towards,lu2020quantum}. Quantum-assisted techniques generate poisoned datasets, manipulating AI/ML models stored in the cloud.

\item \textit{Supply Chain and Software Updates.} Attackers use quantum-forged signatures to inject malicious updates into software repositories, compromising entire systems~\cite{akter2024integrated}. Quantum-enhanced attacks compromise middleware components, enabling unauthorized access and data manipulation~\cite{hoffman2020web}.
\end{enumerate}
These attack vectors highlight the urgent need for transitioning to quantum-resistant cryptographic standards and implementing robust security measures across all layers of \gls{CC} infrastructure~\cite{reece2023systemic}.

\subsection{Pre-Transition Quantum Threat Landscape in Each \gls{CC} Layer}

The transition to \gls{QC} introduces systemic security challenges across the entire \gls{CC} stack. Quantum algorithms, specifically Shor's and Grover's~\cite{shor1999polynomial,grover1996fast}, jeopardize classical cryptographic foundations, undermining encryption, authentication, and data integrity protocols. This subsection taxonomizes layer-specific vulnerabilities through STRIDE threat modeling aligned with NIST SP~800-30 risk quantification~\cite{NIST-SP800-30}, mapping attack vectors, exploitation pathways, and quantum-resistant countermeasures. Table~\ref{table:Pre-Transition} consolidates pre-transition threats across nine architectural layers, categorizing vulnerabilities by likelihood (quantum timeline feasibility), impact (breach severity), and risk (composite assessment).
\begin{enumerate}[topsep=0ex, itemsep=0ex, wide, font=\itshape, labelwidth=!, labelindent=0pt, label*=4.3.\arabic*]
\item \textit{Application Layer.} The application layer delivers end-user functionality through \gls{SaaS} platforms relying on RSA/ECC-based authentication (\gls{API} keys, OAuth tokens) and AES encryption for data-at-rest and data-in-transit. Shor's algorithm compromises public-key authentication, enabling credential forgery and session hijacking, while Grover's algorithm reduces password hashing security margins (Argon2, PBKDF2), necessitating parameter hardening~\cite{merhav2019universal,bernstein2017postquantum}. Quantum-assisted traffic analysis amplifies \gls{MITM} and replay attack risks, while accelerated pattern discovery may enhance injection-based exploits~\cite{bhargavan2016downgrade}. Mitigation strategies include NIST \gls{PQC} standards (ML-KEM, ML-DSA/SLH-DSA, Table~\ref{tab:post-alg})~\cite{nistfips203,nistfips204,nistfips205}, FIDO2/WebAuthn with \gls{PQC}, and KEMTLS~\cite{schwabe2020post}.

\item \textit{Data Layer.} Managed data services (Amazon RDS, Firebase, BigQuery) provide databases, caches, and analytics for cloud workloads. These services store encrypted datasets vulnerable to Shor's algorithm (key protection compromise) and Grover's algorithm (AES-128 requiring migration to AES-256)~\cite{mosca2018cybersecurity}. \gls{HNDL} attacks enable long-term exposure of archived backups~\cite{barenkamp2022steal}, while quantum-assisted analysis exacerbates AI/ML data poisoning risks. Storage side channels (metadata inference, access-pattern leakage) further elevate reconstruction risks~\cite{gruss2019page}. Defensive measures include lattice-based encryption (ML-KEM), quantum-resistant hashing (SHA-3), distributed backup sharding, and \gls{CBOM}/\gls{KBOM} inventories~\cite{nist2024transition}.

\item \textit{Runtime Layer.} \gls{PaaS} platforms (\gls{AWS} Lambda, Google Functions) provide execution environments for applications, managing in-memory cryptographic operations susceptible to side-channel leakage (cache-timing, power analysis) potentially amplified by advanced machine-learning techniques~\cite{refml25,refml31}. \gls{VM} migration exposes encrypted runtime states to interception, while forged code-signing credentials enable malicious software injection. Protocol downgrade attacks may force weaker TLS/SSH configurations~\cite{bhargavan2016downgrade}. Protection mechanisms include \gls{TEE}-based isolation\break (SGX, TDX, SEV), constant-time implementations, and\break quantum-resistant \gls{VM} migration protocols~\cite{COPPOLINO2025104457}.

\item \textit{Middleware Layer.} Integration and communication tools (\gls{API} gateways, service meshes like Istio and Envoy, message brokers like Kafka and RabbitMQ) coordinate inter-service communication using RSA/ECC-based authentication. Grover's algorithm weakens symmetric and hash-based \gls{PKI} components~\cite{bernstein2017postquantum}, while forged signatures enable command-and-control channel abuse~\cite{bates2012detecting}. Cross-\gls{VM} cache-based side channels threaten cryptographic material on shared hardware~\cite{saeed2020cross}. Resilience strategies include stateful hash-based signatures (SPHINCS+), hybrid TLS, and zero-trust \gls{API} authentication using \gls{PQC}.

\item \textit{\gls{OS} Layer.} Virtualized operating systems (Linux, Windows) for cloud workloads rely on kernel-level cryptographic modules (\glspl{HSM}, \glspl{TPM}) for secure boot, code signing, and integrity verification. Quantum attacks target firmware signature validation and symmetric key security, while cross-\gls{VM} side channels enable credential leakage~\cite{compastie2020virtualization,gruss2019page}. Hardening approaches include quantum-resistant \glspl{HSM}/\glspl{TPM}, ML-DSA/SLH-DSA-based secure boot, and hardware-backed memory encryption~\cite{nistfips204,nistfips205}.

\item \textit{Virtualization Layer.} Hypervisors and orchestration tools (KVM, Xen, Kubernetes, OpenStack) manage multi-tenant workloads with RSA/ECC-protected \gls{VM} images and control-plane authentication. Shor's algorithm threatens encrypted migration channels, while forged orchestration credentials enable hypervisor compromise~\cite{9889331}. Cross-\gls{VM} side channels amplify risk propagation~\cite{NARAYANA20216465}. Security measures include hardware-enforced \gls{VM} isolation (AMD SEV-SNP, Intel TDX), quantum-safe migration protocols, and \gls{PQC}-based hypervisor authentication.

\item \textit{Server Layer.} Virtualized compute resources (EC2 instances, Google Compute Engine, Azure VMs) perform firmware-level cryptographic validation vulnerable to Shor-enabled key compromise and forged UEFI signatures~\cite{lu2019attacking}. Supply-chain attacks may inject malicious dependencies, while physical side channels (EM, power analysis) expose CPU-level secrets~\cite{10.1145/3432893}. Safeguards include \gls{PQC}-based firmware signing, hardware roots of trust (\gls{TPM}~2.0), and SBOM auditing~\cite{nist2024transition}.

\item \textit{Storage Layer.} Scalable storage solutions (Amazon S3, Google Cloud Storage) for data persistence encrypt data-at-rest using AES and manage access with RSA/ECC-based credentials. Grover's algorithm reduces symmetric security margins, while \gls{HNDL} attacks enable deferred decryption of archived data~\cite{barenkamp2022steal}. Metadata correlation and deduplication attacks expose sensitive usage patterns~\cite{jia2024machine}. Protection strategies include lattice-based encryption, SHA-3-based deduplication, re-encryption of legacy backups, and \gls{PQC}-compatible searchable encryption.

\item \textit{Network Layer.} Infrastructure (\glspl{VPC}, \gls{SDN}, load balancers, firewalls) for secure connectivity relies on TLS/IPsec and \gls{RPKI}-protected routing. Shor's algorithm compromises asymmetric key exchange, while Grover's reduces symmetric tunnel protection margins~\cite{abdou2018comparative}. \gls{HNDL} attacks capture encrypted traffic for future decryption, and control-plane compromise enables malicious flow manipulation~\cite{lounis2020attacks}. Defensive strategies include hybrid and \gls{PQC}-based TLS, quantum-safe \gls{RPKI}, zero-trust \gls{SDN} validation, and \gls{HSM}-backed key protection.
\end{enumerate}

Table~\ref{table:Pre-Transition} summarizes these threats by STRIDE category, likelihood, impact, and composite risk. Cross-layer dependencies—credential compromise enabling lateral movement, hypervisor breaches exposing all guest \glspl{VM}—amplify systemic risk~\cite{mosca2018cybersecurity}, necessitating defense-in-depth strategies combining \gls{PQC} migration, cryptographic agility, and zero-trust architectures.

\begin{table*}[!htbp]
\caption{Pre-Transition Quantum Threat Landscape in Each \gls{CC} Layer}
\large
\renewcommand{\arraystretch}{1.1}
\resizebox{\linewidth}{!}{%
\begin{tabular}{|p{0.13\linewidth}|p{0.25\linewidth}|p{0.35\linewidth}|p{0.48\linewidth}|p{0.4\linewidth}|p{0.65\linewidth}|l|l|l|}
\hline 
\textbf{Layer} & \textbf{Vulnerability} & \textbf{Exploitation Method} & \textbf{Attack Vector} & \textbf{Countermeasures} & \textbf{STRIDE Threats}& {\textbf{L}} & {\textbf{I}} & {\textbf{R}}\\ \hline

\multirow{30}{2cm}{Application}
& \begin{minipage}{\linewidth} Quantum-Enhanced \gls{API} Abuse
\end{minipage} & \begin{minipage}{\linewidth} Credential Theft, Token Forgery~\cite{madhu2025multi,esparza2019understanding}
\end{minipage} & \begin{minipage}{\linewidth} Quantum-enhanced cryptographic attacks recover \gls{API} keys or exploit authentication weaknesses to bypass security restrictions
\end{minipage} & \begin{minipage}{\linewidth} Adopt quantum-secre \gls{API} authentication mechanisms (e.g., FIDO2/WebAuthn with \gls{PQC}, mutual authentication) \end{minipage} 
& \begin{minipage}{\linewidth}
\begin{myBullets}\vspace{2pt}

\item \textbf{Spoofing:} Forged \gls{API} credentials impersonate legitimate users.
\item \textbf{Tampering:} Authentication tokens are modified.
\item \textbf{EoP:} \gls{API} abuse leads to unauthorized access.
 \vspace{3pt} \end{myBullets}
\end{minipage} &\med&\high&\high\\ \cline{2-9}

& \begin{minipage}{\linewidth}Weak Encryption Algorithms
\end{minipage} & \begin{minipage}{\linewidth} Brute Force Attack~\cite{merhav2019universal}
\end{minipage} & \begin{minipage}{\linewidth} Shor’s Algorithm could break RSA/ECC-based encryption protecting application-layer transactions
\end{minipage} & \begin{minipage}{\linewidth} Transition to NIST-approved \gls{PQC} encryption as mentioned in Table~\ref{tab:post-alg}
\end{minipage} 
& \begin{minipage}{\linewidth}
\begin{myBullets}\vspace{2pt}

\item \textbf{Spoofing:} Broken encryption enables identity forgery.
\item \textbf{Tampering:}	 data is decrypted and modified and its integrity is compromised.
\item \textbf{Repudiation:}	 cryptographic proofs is forged making actions non-repudiable.
\item \textbf{Info. Disclosure:} Sensitive application data is decrypted.
 \vspace{3pt} \end{myBullets}
\end{minipage} &\med&\high&\high \\ \cline{2-9}

& \begin{minipage}{\linewidth} Quantum-Accelerated Credential Stuffing
\end{minipage} & \begin{minipage}{\linewidth} Password Hash Exploitation~\cite{baseri2025evaluation,BASERI2024103883}
\end{minipage} & \begin{minipage}{\linewidth} \vspace{0.1cm} Grover’s Algorithm significantly reduces password cracking time, breaking weak credentials stored in cloud applications \vspace{0.1cm}
\end{minipage} & \begin{minipage}{\linewidth} Use quantum-safe password hashing algorithms (e.g., Argon2, BLAKE3) and deploy multi-factor authentication (MFA)
\end{minipage} & \begin{minipage}{\linewidth}
\begin{myBullets}\vspace{2pt}

\item \textbf{Spoofing:} Compromised credentials enable user impersonation.
 \vspace{3pt} \end{myBullets}
\end{minipage} &\low&\high&\med \\ \cline{2-9}

& \begin{minipage}{\linewidth}Insecure Software Updates
\end{minipage} & \begin{minipage}{\linewidth} Quantum Forged Code Signing
\end{minipage} & \begin{minipage}{\linewidth} Exploiting quantum-broken digital signatures to inject malicious software updates
\end{minipage} & \begin{minipage}{\linewidth} Use \gls{PQC}  signature schemes mentioned in Table~\ref{tab:post-alg}
\end{minipage} & \begin{minipage}{\linewidth}
\begin{myBullets}\vspace{2pt}

\item \textbf{Tampering:} Software updates are modified.
\item \textbf{Spoofing:} Forged digital signatures appear legitimate.
 \vspace{3pt} \end{myBullets}
\end{minipage} &\med&\med&\med \\ \cline{2-9}


& \begin{minipage}{\linewidth} Quantum-Accelerated Injection Attacks
\end{minipage} & \begin{minipage}{\linewidth} Quantum-Guided SQL Injection
\end{minipage} & \begin{minipage}{\linewidth} Quantum-assisted techniques optimize SQL injection payloads, bypassing input validation
\end{minipage} & \begin{minipage}{\linewidth} Implement quantum-resistant input validation, deploy AI-based anomaly detection for SQL queries
\end{minipage} & \begin{minipage}{\linewidth}
\begin{myBullets}\vspace{2pt}

\item \textbf{Tampering:} Database queries are manipulated.
\item \textbf{Info. Disclosure:}Sensitive SQL query results are exfiltrated.
\item \textbf{EoP:} Injection exploits escalate access.
 \vspace{3pt} \end{myBullets}
\end{minipage}&\low&\med&\low\\ \cline{2-9}

& \begin{minipage}{\linewidth} Adversarial AI Attacks~\cite{west2023towards,lu2020quantum} on Cloud AI Services\vspace{0.1cm}
\end{minipage} & \begin{minipage}{\linewidth} Quantum Optimization for Adversarial Examples
\end{minipage} & \begin{minipage}{\linewidth} \gls{QC} generates adversarial inputs that bypass AI-powered security models
\end{minipage} & \begin{minipage}{\linewidth} Integrate adversarial AI defense mechanisms trained against quantum-enhanced attacks
\end{minipage} & \begin{minipage}{\linewidth}
\begin{myBullets}\vspace{2pt}

\item \textbf{Tampering:} AI models are manipulated.
\item \textbf{Info. Disclosure:} Private AI training data is extracted.
 \vspace{3pt} \end{myBullets}
 \end{minipage}&\low&\med&\low\\ \cline{2-9}
& \begin{minipage}{\linewidth} Weak Authentication Mechanisms
\end{minipage} & \begin{minipage}{\linewidth} \vspace{3pt}
Vulnerabilities such Hash Collisions, Assertion Repudiation, and Impersonation or Masquerade attacks~\cite{grassi2020digital,8939382}
\end{minipage} & \begin{minipage}{\linewidth} Quantum algorithms forge authentication signatures or break hash functions, enabling impersonation
\end{minipage} & \begin{minipage}{\linewidth} Deploy quantum-resistant authentication protocols (e.g., \gls{PQC}-based OAuth, quantum-safe HSMs)
\end{minipage} & \begin{minipage}{\linewidth}
\begin{myBullets}\vspace{2pt}

\item \textbf{Spoofing:} Authentication credentials are forged.
\item \textbf{Repudiation:} Malicious actions are denied.
\item \textbf{EoP:} Authentication bypass enables unauthorized access.
 \vspace{3pt} \end{myBullets}
\end{minipage} &\med&\high&\high \\ \cline{2-9}

& \begin{minipage}{\linewidth} Insufficient Data Protection
\end{minipage} & \begin{minipage}{\linewidth} Application Software Exploitation, Cryptanalysis Attacks
\end{minipage} & \begin{minipage}{\linewidth}  attacks on application-layer encryption (TLS termination), cryptographic breaches
\end{minipage} & \begin{minipage}{\linewidth} Implement post-quantum  TLS (e.g., KEMTLS~\cite{schwabe2020post}), encrypt all data-at-rest using \gls{PQC} standards \vspace{0.1cm}
\end{minipage} & \begin{minipage}{\linewidth}
\begin{myBullets}\vspace{2pt}

\item \textbf{Info. Disclosure:} Sensitive data is decrypted.
\item \textbf{Tampering:} Encrypted data integrity is compromised.
 \vspace{3pt} \end{myBullets}
\end{minipage}&\med&\high&\high \\ \cline{2-9}

& \begin{minipage}{\linewidth} Vulnerable Network Interfaces
\end{minipage} & \begin{minipage}{\linewidth} POST Request Collision~\cite{smith2019postcollision} Enhanced by \gls{QC}
\end{minipage} & \begin{minipage}{\linewidth} Exploiting quantum-accelerated hash collisions to craft duplicate POST requests, bypassing rate limits
\end{minipage} & \begin{minipage}{\linewidth} Implement quantum-resistant rate-limiting, deploy quantum-secure hash functions like SHA-3~\cite{nist2024transition}
\end{minipage} & \begin{minipage}{\linewidth}
\begin{myBullets}\vspace{2pt}

\item \textbf{DoS:} Services are overloaded via hash collisions.
 \vspace{3pt} \end{myBullets}
\end{minipage}&\med&\med&\med\\ \cline{2-9}
& \begin{minipage}{\linewidth} Session Hijacking via Quantum Attacks
\end{minipage} & \begin{minipage}{\linewidth} Quantum Token Forgery
\end{minipage} & \begin{minipage}{\linewidth} \gls{QC} enables the prediction or forging of session tokens used in authentication
\end{minipage} & \begin{minipage}{\linewidth} Deploy quantum-secure session tokens and use \gls{PQC}-based TLS~\cite{post-quantum-tls}
\end{minipage} & \begin{minipage}{\linewidth}
\begin{myBullets}\vspace{2pt}

\item \textbf{Spoofing:} Session tokens are forged.
\item \textbf{EoP:} Authenticated sessions are hijacked.
\item \textbf{Info. Disclosure:} Session data is exposed.

 \vspace{3pt} \end{myBullets}
\end{minipage} &\med&\high&\high\\ \hline

\multirow{35}{*}{Data}

& \begin{minipage}{\linewidth}Susceptible encryption under quantum attacks
\end{minipage} & \begin{minipage}{\linewidth} Brute Force Attack~\cite{gort2022relational}
\end{minipage} & \begin{minipage}{\linewidth} Quantum algorithms  reduce the effective security of symmetric encryption keys
\end{minipage} & \begin{minipage}{\linewidth} 
 Implement quantum-resistant encryption standards~\cite{nist2024transition}; and  promote best practices in cryptographic implementations.
\end{minipage} & \begin{minipage}{\linewidth}
\begin{myBullets}\vspace{2pt} 
\item \textbf{Info. Disclosure:} Encrypted data can be decrypted, exposing sensitive information.
\item \textbf{Tampering:} Attackers can modify decrypted data, affecting its integrity.
 \vspace{3pt} \end{myBullets}
\end{minipage} & \low & \high & \med \\ \cline{2-9} 

& \begin{minipage}{\linewidth} Encrypted Data Storage Tampering
\end{minipage} & \begin{minipage}{\linewidth} Data Modification Attack~\cite{jangjou2022comprehensive}
\end{minipage} & \begin{minipage}{\linewidth} Quantum attackers can breach and modify encrypted data in storage systems
\end{minipage} & \begin{minipage}{\linewidth} 
Regular integrity verification, and quantum-safe cryptographic hashes and \gls{PQC} encryption mechanisms.
\end{minipage} & \begin{minipage}{\linewidth}
\begin{myBullets}\vspace{2pt} 
\item \textbf{Tampering:} Attackers modify encrypted storage, leading to unauthorized data manipulation.
\item \textbf{Info. Disclosure:} If attackers decrypt data, sensitive information may be leaked.
 \vspace{3pt} \end{myBullets}
\end{minipage} & \med & \high & \high \\ \cline{2-9} 

& \begin{minipage}{\linewidth} Malware Injection via Quantum-Forged Code Signing
\end{minipage} & \begin{minipage}{\linewidth} Malware Injection Attacks
\end{minipage} & \begin{minipage}{\linewidth} Code injection via quantum-broken signatures
\end{minipage} & \begin{minipage}{\linewidth} 
Upgrade to quantum-resistant digital signatures; and strengthen software delivery authentication.
\end{minipage} & \begin{minipage}{\linewidth}
\begin{myBullets}\vspace{2pt} 
\item \textbf{Spoofing:} Attackers use quantum-forged signatures to impersonate legitimate software publishers.
\item \textbf{Tampering:} Malicious software updates are injected, altering system functionality.
 \vspace{3pt} \end{myBullets}
\end{minipage} & \med & \med & \med \\ \cline{2-9} 




& \begin{minipage}{\linewidth} Vulnerable classical encryption  to quantum cryptanalysis
\end{minipage} & \begin{minipage}{\linewidth} Cryptanalysis Attacks~\cite{brassard1997quantum,grover1996fast,shor1994algorithms}
\end{minipage} & \begin{minipage}{\linewidth} Quantum decryption via Shor's, Grover's, and Brassard et al.'s algorithms
\end{minipage} & \begin{minipage}{\linewidth} Migrate to NIST-approved \gls{PQC} schemes across all encryption layers; and update encryption protocols accordingly.
\end{minipage} & \begin{minipage}{\linewidth}
\begin{myBullets}\vspace{2pt} 
\item \textbf{Info. Disclosure:} Cryptographic keys and encrypted data are exposed after decryption.
\item \textbf{Tampering:} Attackers can modify decrypted data, affecting data integrity.
 \vspace{3pt} \end{myBullets}
\end{minipage} & \med & \high & \high \\ \cline{2-9} 

& \begin{minipage}{\linewidth} Quantum-enhanced collision attacks on hash functions
\end{minipage} & \begin{minipage}{\linewidth} Hash-based DoS~\cite{masdari2016survey}
\end{minipage} & \begin{minipage}{\linewidth} POST request collision using \gls{QC}
\end{minipage} & \begin{minipage}{\linewidth} Rate limiting, quantum resistant hash functions, and advanced intrusion detection.
\end{minipage} & \begin{minipage}{\linewidth}
\begin{myBullets}\vspace{2pt} 
\item \textbf{Spoofing:} Attackers exploit hash collisions to impersonate users in authentication systems.
\item \textbf{Tampering:} Data integrity is compromised by forging hash collisions.
\item \textbf{ \gls{DoS}:} Overloaded hash processing results in service degradation or downtime.
 \vspace{3pt} \end{myBullets}
\end{minipage} & \med & \med & \med \\ \cline{2-9} 

& \begin{minipage}{\linewidth} Weak Cryptographic Access Controls
\end{minipage} & \begin{minipage}{\linewidth} Unauthorized Access and Data Theft~\cite{Threatinsiderdef}
\end{minipage} & \begin{minipage}{\linewidth} Quantum attacks bypass access policies
\end{minipage} & \begin{minipage}{\linewidth} Enforce multi-factor authentication with \gls{PQC}; and  establish strict zero-trust access control policies using quantum-secure methods.
\end{minipage} & \begin{minipage}{\linewidth}
\begin{myBullets}\vspace{2pt} 
\item \textbf{EoP:} Attackers gain unauthorized access by breaking cryptographic access control mechanisms.
\item \textbf{Spoofing:} Stolen credentials allow attackers to impersonate users.
\item \textbf{Info. Disclosure:} Sensitive data may be accessed once authentication is bypassed.
 \vspace{3pt} \end{myBullets}
\end{minipage} & \med & \high & \high \\ \cline{2-9} 

& \begin{minipage}{\linewidth} Quantum-Accelerated Data Poisoning
\end{minipage} & \begin{minipage}{\linewidth} Data Integrity Attacks~\cite{li2024secure}
\end{minipage} & \begin{minipage}{\linewidth} Quantum-optimized adversarial techniques generate poisoned datasets that manipulate AI/ML models stored in the cloud.
\end{minipage} & \begin{minipage}{\linewidth} Use quantum-resistant anomaly detection, adversarial training, and quantum-secure ML models.
\end{minipage} & \begin{minipage}{\linewidth}
\begin{myBullets}\vspace{2pt} 
\item \textbf{Tampering:} AI models are manipulated by injecting adversarial samples.
\item \textbf{Info. Disclosure:} Poisoned models can leak sensitive training data.
 \vspace{3pt} \end{myBullets}
\end{minipage} & \low & \med & \low \\ \cline{2-9}


& \begin{minipage}{\linewidth} Quantum-Enhanced Storage \glspl{SCA}
\end{minipage} & \begin{minipage}{\linewidth} Side-Channel Analysis~\cite{chowdhury2021physical}
\end{minipage} & \begin{minipage}{\linewidth} Quantum-assisted side-channel techniques extract encryption keys or access control policies from cloud storage systems by analyzing metadata leakage or storage access patterns.
\end{minipage} & \begin{minipage}{\linewidth} Deploy quantum-secure homomorphic encryption; prevent side-channel analysis through randomized memory access patterns and secure enclave processing.
\end{minipage} & \begin{minipage}{\linewidth}
\begin{myBullets}\vspace{2pt} 
\item \textbf{Info. Disclosure:} Storage encryption keys are extracted via side-channel techniques.
\item \textbf{Tampering:} Access policies and metadata are modified to manipulate storage permissions.
 \vspace{3pt} \end{myBullets}
\end{minipage} & \med & \high & \high \\ \cline{2-9}

& \begin{minipage}{\linewidth} Quantum-Assisted Data Reconstruction from Encrypted Cloud Backups
\end{minipage} & \begin{minipage}{\linewidth} Backup Attacks
\end{minipage} & \begin{minipage}{\linewidth} Quantum decryption techniques leveraging Shor’s or Grover’s algorithms can decrypt backup data decryption of cloud-stored encrypted backups, reconstructing sensitive files without directly breaking encryption keys. \vspace{0.1cm}
\end{minipage} & \begin{minipage}{\linewidth} Use lattice-based encryption for cloud backups, shard and distribute backup data across different nodes.
\end{minipage} & \begin{minipage}{\linewidth}
\begin{myBullets}\vspace{2pt} 
\item \textbf{Info. Disclosure:} Sensitive backup data is reconstructed using quantum pattern matching.
\item \textbf{Tampering:} Reconstructed data can be modified, affecting integrity.
 \vspace{3pt} \end{myBullets}
\end{minipage} & \med & \high & \high \\ \cline{2-9}
& \begin{minipage}{\linewidth} \vspace{0.1cm} Quantum-Enhanced Ransomware \vspace{0.1cm} 
\end{minipage} & \begin{minipage}{\linewidth} Ransomware~\cite{craciun2019trends} 
\end{minipage} & \begin{minipage}{\linewidth} Data encrypted with \gls{PQC} algorithms, preventing recovery 
\end{minipage} & \begin{minipage}{\linewidth} Backup Systems, Anti-Ransomware Detection 
\end{minipage} & \begin{minipage}{\linewidth}
 \begin{myBullets}\vspace{2pt}
 \item \textbf{DoS:} Data becomes permanently inaccessible.
 \vspace{3pt} \end{myBullets}
\end{minipage} & \med & \high & \high \\ 
\hline

\multirow{24}{*}{Runtime}

& \begin{minipage}{\linewidth} Susceptibility to quantum decryption
\end{minipage} & \begin{minipage}{\linewidth} {\gls{VM} Mobility}~\cite{compastie2020virtualization,huang2012security}
\end{minipage} & \begin{minipage}{\linewidth} 
Encrypted VMs may be exposed to decryption during host-to-host transfers
\end{minipage} & \begin{minipage}{\linewidth} 
Adopt NIST-approved \gls{PQC} (e.g., ML-KEM); and apply hybrid cryptography  during the transition phase.
\end{minipage} & \begin{minipage}{\linewidth}
\begin{myBullets}\vspace{2pt} 
\item \textbf{Spoofing:} Decrypted \gls{VM} contents enable identity impersonation.
\item \textbf{Info. Disclosure:} Attackers gain access to \gls{VM} data.
 \vspace{3pt} \end{myBullets}
\end{minipage} & \med & \high & \high \\ \cline{2-9}

& \begin{minipage}{\linewidth} Weak encryption methods
\end{minipage} & \begin{minipage}{\linewidth} Brute Force Attack~\cite{tirado2018new}
\end{minipage} & \begin{minipage}{\linewidth} Quantum-accelerated brute-force attacks exploit Grover’s algorithm to guess encryption keys
\end{minipage} & \begin{minipage}{\linewidth} Early adoption of \gls{PQC} algorithms; Regular security updates
\end{minipage} & \begin{minipage}{\linewidth}
\begin{myBullets}\vspace{2pt} 
\item \textbf{Spoofing:} Attackers use brute-forced keys to forge identities.
\item \textbf{Tampering:} Broken encryption enables unauthorized modification.
\item \textbf{Info. Disclosure:} Sensitive runtime data is exposed.
 \vspace{3pt} \end{myBullets}
\end{minipage} & \med & \high & \high \\ \cline{2-9} 

& \begin{minipage}{\linewidth} Quantum-Accelerated \glspl{SCA}
\end{minipage} & \begin{minipage}{\linewidth} Side-Channel Analysis~\cite{chowdhury2021physical}
\end{minipage} & \begin{minipage}{\linewidth} Quantum-assisted techniques extract cryptographic keys from runtime memory via cache timing, power, or electromagnetic analysis.
\end{minipage} & \begin{minipage}{\linewidth} Implement quantum-resistant cryptographic protocols; use secure enclaves; deploy memory obfuscation techniques.
\end{minipage} & \begin{minipage}{\linewidth}
\begin{myBullets}\vspace{2pt} 
\item \textbf{Info. Disclosure:} Sensitive runtime cryptographic operations are exposed.
\item \textbf{Tampering:} Encryption keys are extracted for later use.
 \vspace{3pt} \end{myBullets}
\end{minipage} & \low & \med & \low \\ \cline{2-9} 

& \begin{minipage}{\linewidth} Quantum-Assisted Cryptographic Backdoor Exploitation
\end{minipage} & \begin{minipage}{\linewidth} Cryptanalysis Attacks
\end{minipage} & \begin{minipage}{\linewidth} Exploiting weak or hidden cryptographic backdoors at runtime using quantum speedups
\end{minipage} & \begin{minipage}{\linewidth} Transition to verified \gls{PQC} algorithms; strengthen runtime cryptographic library audits
\end{minipage} & \begin{minipage}{\linewidth}
\begin{myBullets}\vspace{2pt} 
\item \textbf{Spoofing:} Quantum exploits enable bypassing authentication.
\item \textbf{Info. Disclosure:} Runtime secrets (e.g., \gls{API} keys, TLS certs) are extracted.
 \vspace{3pt} \end{myBullets}
\end{minipage} & \med & \high & \high \\  \cline{2-9} 

&\begin{minipage}{\linewidth} Quantum-Enabled Runtime Code Injection
\end{minipage} & \begin{minipage}{\linewidth} Memory Corruption Exploitation~\cite{cloosters2020teerex}
\end{minipage} & \begin{minipage}{\linewidth} Quantum-optimized fuzzing locates vulnerabilities, injecting malicious runtime code
\end{minipage} & \begin{minipage}{\linewidth} Employ runtime integrity verification; use AI-based anomaly detection
\end{minipage} & \begin{minipage}{\linewidth}
\begin{myBullets}\vspace{2pt} 
\item \textbf{Tampering:} Code execution is altered with injected payloads.
\item \textbf{EoP:} Malicious code runs with elevated permissions.
 \vspace{3pt} \end{myBullets}
\end{minipage} & \med & \high & \high \\  \cline{2-9}

& \begin{minipage}{\linewidth} Manipulated or forged  software update 
\end{minipage} & \begin{minipage}{\linewidth} Application Software Exploitation~\cite{compastie2020virtualization,hoffman2020web}
\end{minipage} & \begin{minipage}{\linewidth} Code injection or manipulation with forged signature during software update
\end{minipage} & \begin{minipage}{\linewidth} Deployment of quantum-safe digital signature schemes; Use of secure and authenticated update channels
\end{minipage} & \begin{minipage}{\linewidth}
\begin{myBullets}\vspace{2pt} 
\item \textbf{Tampering:} Altered software updates introduce vulnerabilities.
\item \textbf{Spoofing:} Quantum-forged signatures enable unauthorized updates.
 \vspace{3pt} \end{myBullets}
\end{minipage} & \med & \med & \med \\ \cline{2-9}
& \begin{minipage}{\linewidth} Inadequate encryption
\end{minipage} & \begin{minipage}{\linewidth}Application Software Exploitation~\cite{compastie2020virtualization,hoffman2020web}
\end{minipage} & \begin{minipage}{\linewidth} Quantum-enabled encryption downgrade or key compromise during runtime execution, exposing sensitive in-memory data
\end{minipage} & \begin{minipage}{\linewidth} Transition to quantum-resistant encryption protocols; Use of post-quantum TLS (Kyber-TLS) and encrypted inter-\gls{VM} communication
\end{minipage} &\begin{minipage}{\linewidth}
\begin{myBullets}\vspace{2pt} 
\item \textbf{Info. Disclosure:} Leaked runtime encryption keys expose sensitive data.
\item \textbf{Tampering:} Attackers modify decrypted runtime memory, affecting execution integrity.
 \vspace{3pt} \end{myBullets}
\end{minipage} & \med & \high & \high \\  
\cline{2-9} 

& \begin{minipage}{\linewidth} Quantum-Assisted Cryptographic Downgrade Attacks
\end{minipage} & \begin{minipage}{\linewidth} Protocol Downgrade Attacks~\cite{bhargavan2016downgrade}
\end{minipage} & \begin{minipage}{\linewidth} Quantum attackers force runtime environments to negotiate weaker TLS, SSH, or VPN encryption
\end{minipage} & \begin{minipage}{\linewidth} Enforce post-quantum TLS (Kyber-TLS), detect downgrade attempts, enable strict encryption policies
\end{minipage} & \begin{minipage}{\linewidth}
\begin{myBullets}\vspace{2pt} 
\item \textbf{Tampering:} Encryption negotiation is manipulated to use weak cryptographic protocols.
\item \textbf{Info. Disclosure:} Runtime communication is exposed due to weaker cryptography.
 \vspace{3pt} \end{myBullets}
\end{minipage} & \med & \high & \high \\ 
\cline{2-9}
& \begin{minipage}{\linewidth}Weak cryptosystems
\end{minipage} & \begin{minipage}{\linewidth} Cryptanalysis Attacks~\cite{brassard1997quantum,grover1996fast,shor1994algorithms}
\end{minipage} & \begin{minipage}{\linewidth} Data exfiltration using quantum algorithms to break encryption
\end{minipage} & \begin{minipage}{\linewidth} Upgrading to quantum-resistant algorithms; Conducting regular security audits
\end{minipage} & \begin{minipage}{\linewidth}
\begin{myBullets}\vspace{2pt} 
\item \textbf{Info. Disclosure:} Cryptographic failure exposes sensitive data.
\item \textbf{Tampering:} Attackers modify decrypted runtime data.
 \vspace{3pt} \end{myBullets}
\end{minipage} & \med & \high & \high \\ \cline{2-9}

& \begin{minipage}{\linewidth}Compromised update integrity
\end{minipage} & \begin{minipage}{\linewidth} Application Software Exploitation~\cite{compastie2020virtualization,hoffman2020web}
\end{minipage} & \begin{minipage}{\linewidth} Forged updates granting elevated access
\end{minipage} & \begin{minipage}{\linewidth} Secure bootstrapping and update processes; Multi-factor authentication systems
\end{minipage} & \begin{minipage}{\linewidth}
\begin{myBullets}\vspace{2pt} 
\item \textbf{EoP:} Unauthorized update enables root access.
\item \textbf{Tampering:} Software integrity is compromised.
 \vspace{3pt} \end{myBullets}
\end{minipage} & \med & \high & \high \\ \cline{2-9}

& \begin{minipage}{\linewidth}Inter-\gls{VM} Communication  
\end{minipage} & \begin{minipage}{\linewidth} Quantum-Assisted MITM~\cite{compastie2020virtualization} 
\end{minipage} & \begin{minipage}{\linewidth} Quantum adversaries exploit runtime communication between \glspl{VM}, intercepting traffic via quantum-decrypting MITM techniques. 
\end{minipage} & \begin{minipage}{\linewidth} Implement quantum secure communication channels; Enforce strict isolation policies for VM-to-\gls{VM} interactions; Deploy real-time anomaly detection for unusual network patterns. 
\end{minipage} & \begin{minipage}{\linewidth}
 \begin{myBullets}\vspace{2pt} 
 \item \textbf{Info. Disclosure:} Quantum MITM attacks decrypt and expose VM-to-\gls{VM} communication.
 \item \textbf{Tampering:} Adversaries manipulate live network traffic between running VMs.
 \item \textbf{Spoofing:} Attackers impersonate trusted \gls{VM} entities to intercept sensitive data.
 \vspace{3pt} \end{myBullets}
\end{minipage} & \med & \high & \high \\ \hline

\end{tabular}%
}\label{table:Pre-Transition}
\begin{tablenotes}[flushleft] \scriptsize \item[\dag] To evaluate the likelihood, we consider 15 years as the timeline of emerging quantum computers.
\end{tablenotes} \end{table*}

\begin{table*}[!htbp]
\ContinuedFloat 
\caption{Pre-Transition Quantum Threat Landscape in Each \gls{CC} Layer (Continued)} 
\large
\renewcommand{\arraystretch}{1.1}
\resizebox{\linewidth}{!}{%
\begin{tabular}{|p{0.13\linewidth}|p{0.25\linewidth}|p{0.35\linewidth}|p{0.48\linewidth}|p{0.4\linewidth}|p{0.65\linewidth}|l|l|l|}
\hline 
\textbf{Layer} & \textbf{Vulnerability} & \textbf{Exploitation Method} & \textbf{Attack Vector} & \textbf{Countermeasures} & \textbf{STRIDE Threats}& {\textbf{L}} & {\textbf{I}} & {\textbf{R}}\\ \hline

\multirow{20}{*}{Middleware} 

& \begin{minipage}{\linewidth} Cryptographic Key Cracking 
\end{minipage} & \begin{minipage}{\linewidth} Brute Force Attack~\cite{bates2012detecting} 
\end{minipage} & \begin{minipage}{\linewidth} Quantum algorithms like Grover's accelerating \gls{PKI} and \gls{API} key cracking 
\end{minipage} & \begin{minipage}{\linewidth} Adoption of quantum-resistant \glspl{PKI} 
\end{minipage} & \begin{minipage}{\linewidth}
 \begin{myBullets}\vspace{2pt} 
 \item \textbf{Spoofing:} Attackers forge cryptographic keys to impersonate valid users.
 \item \textbf{Info. Disclosure:} Cracked keys expose sensitive middleware authentication mechanisms.
 \vspace{3pt} \end{myBullets}
\end{minipage} & \low & \high & \med \\ \cline{2-9} 

& \begin{minipage}{\linewidth} Code Injection 
\end{minipage} & \begin{minipage}{\linewidth} C\&C Channel Exploitation~\cite{compastie2020virtualization} 
\end{minipage} & \begin{minipage}{\linewidth} MITM attacks with forged signatures 
\end{minipage} & \begin{minipage}{\linewidth} Usage of stateful hash-based signatures for code and communications 
\end{minipage} & \begin{minipage}{\linewidth}
 \begin{myBullets}\vspace{2pt} 
 \item \textbf{Tampering:} Malicious code is injected into middleware components.
 \item \textbf{Spoofing:} Forged signatures make malicious components appear legitimate.
 \vspace{3pt} \end{myBullets}
\end{minipage} & \med & \high & \high \\ \cline{2-9} 

& \begin{minipage}{\linewidth} Non-repudiation Failure 
\end{minipage} & \begin{minipage}{\linewidth} C\&C Channel Exploitation~\cite{compastie2020virtualization} 
\end{minipage} & \begin{minipage}{\linewidth} Forging signatures on C\&C channels 
\end{minipage} & \begin{minipage}{\linewidth} Immutable logging mechanisms with quantum-resistant signatures 
\end{minipage} & \begin{minipage}{\linewidth}
 \begin{myBullets}\vspace{2pt} 
 \item \textbf{Repudiation:} Attackers forge transaction records or logs.
 \item \textbf{Spoofing:} Fake signatures allow unauthorized access.
 \vspace{3pt} \end{myBullets}
\end{minipage} & \med & \med & \med \\ \cline{2-9} 

& \begin{minipage}{\linewidth} Quantum-Assisted \gls{API} Key Leakage 
\end{minipage} & \begin{minipage}{\linewidth} Credential Theft, Brute Force~\cite{esparza2019understanding} 
\end{minipage} & \begin{minipage}{\linewidth} \gls{QC} accelerates brute-force attacks on middleware \gls{API} keys 
\end{minipage} & \begin{minipage}{\linewidth} Deploy \gls{PQC}-based \gls{API} authentication, enforce rate-limiting, zero-trust policies 
\end{minipage} & \begin{minipage}{\linewidth}
 \begin{myBullets}\vspace{2pt} 
 \item \textbf{Spoofing:} Stolen \gls{API} keys allow adversaries to impersonate services.
 \item \textbf{Info. Disclosure:} Compromised keys expose confidential middleware interactions.
 \vspace{3pt} \end{myBullets}
\end{minipage} & \med & \high & \high \\ \cline{2-9} 

& \begin{minipage}{\linewidth} Quantum-Accelerated Middleware Protocol Downgrade Attacks 
\end{minipage} & \begin{minipage}{\linewidth} Protocol Downgrade Attacks~\cite{bhargavan2016downgrade} 
\end{minipage} & \begin{minipage}{\linewidth} Quantum adversaries force weaker cryptographic protocols (e.g., TLS 1.2 instead of Kyber-TLS)~\cite{compastie2020virtualization} 
\end{minipage} & \begin{minipage}{\linewidth} Strict enforcement of post-quantum  TLS (e.g., KEMTLS~\cite{schwabe2020post}), detect downgrade attempts 
\end{minipage} & \begin{minipage}{\linewidth}
 \begin{myBullets}\vspace{2pt} 
 \item \textbf{Tampering:} Attackers manipulate protocol negotiation.
 \item \textbf{Info. Disclosure:} Downgraded protocols expose encrypted middleware communications.
 \vspace{3pt} \end{myBullets}
\end{minipage} & \med & \high & \high \\ \cline{2-9} 

& \begin{minipage}{\linewidth} Quantum-Enhanced Middleware Service Injection 
\end{minipage} & \begin{minipage}{\linewidth} \glspl{SCA} \& Code Injection~\cite{owasp_code_injection,hoffman2020web} 
\end{minipage} & \begin{minipage}{\linewidth} Quantum-forged signatures allow rogue middleware services to be injected 
\end{minipage} & \begin{minipage}{\linewidth} Deploy \gls{PQC}  signatures, secure middleware runtime integrity 
\end{minipage} & \begin{minipage}{\linewidth}
 \begin{myBullets}\vspace{2pt} 
 \item \textbf{Tampering:} Unauthorized services modify middleware configurations.
 \item \textbf{EoP:} Malicious services escalate access beyond intended permissions.
 \vspace{3pt} \end{myBullets}
\end{minipage} & \med & \high & \high \\ \cline{2-9} 

& \begin{minipage}{\linewidth} Quantum-Accelerated Middleware Configuration Tampering 
\end{minipage} & \begin{minipage}{\linewidth} Access Control Manipulation~\cite{tabrizchi2020survey} 
\end{minipage} & \begin{minipage}{\linewidth} Quantum-aided cryptanalysis modifies security configurations in middleware 
\end{minipage} & \begin{minipage}{\linewidth} Enforce post-quantum cryptographic controls for access configurations 
\end{minipage} & \begin{minipage}{\linewidth}
 \begin{myBullets}\vspace{2pt} 
 \item \textbf{Tampering:} Middleware security policies are modified.
 \item \textbf{Info. Disclosure:} Misconfigured access controls expose sensitive middleware operations.
 \vspace{3pt} \end{myBullets}
\end{minipage} & \med & \high & \high \\ \hline 

\multirow{40}{*}{OS} 

& \begin{minipage}{\linewidth} Quantum Algorithm Exploits 
\end{minipage} & \begin{minipage}{\linewidth} Injecting Virus~\cite{ThreatsProtection} 
\end{minipage} & \begin{minipage}{\linewidth} Code injection with forged signature during software update 
\end{minipage} & \begin{minipage}{\linewidth} Implementation of Quantum-Secure Signature Schemes, Transition to Quantum-Resistant Certificates 
\end{minipage} & \begin{minipage}{\linewidth}
 \begin{myBullets}\vspace{2pt}
 \item \textbf{Spoofing:} Attackers inject quantum-forged signatures to impersonate valid software.
 \item \textbf{Tampering:} Injected malware alters \gls{OS} functionality.
 \vspace{3pt} \end{myBullets}
\end{minipage} & \med & \high & \high \\ \cline{2-9} 

& \begin{minipage}{\linewidth} Cryptographic Breakdown
\end{minipage} & \begin{minipage}{\linewidth} Injecting Insecure Code~\cite{owasp_code_injection,hoffman2020web} 
\end{minipage} & \begin{minipage}{\linewidth} Code injection with forged signature during software update 
\end{minipage} & \begin{minipage}{\linewidth} Adoption of \gls{PQC} Libraries, Secure Software Development Lifecycle (SDLC) Practices 
\end{minipage} & \begin{minipage}{\linewidth}
 \begin{myBullets}\vspace{2pt}
 \item \textbf{Spoofing:} Attackers forge cryptographic keys to bypass authentication.
 \item \textbf{Tampering:} Malicious code modifies \gls{OS} kernel or execution flow.
 \vspace{3pt} \end{myBullets}
\end{minipage} & \med & \high & \high \\ \cline{2-9} 

& \begin{minipage}{\linewidth} Symmetric Key Exposure 
\end{minipage} & \begin{minipage}{\linewidth} Brute Force Attack~\cite{tirado2018new} 
\end{minipage} & \begin{minipage}{\linewidth} Exploitation of symmetric key cryptography via Grover's algorithm 
\end{minipage} & \begin{minipage}{\linewidth} Transition to Larger Symmetric Key Sizes, Quantum Key Distribution (QKD) Integration 
\end{minipage} & \begin{minipage}{\linewidth}
 \begin{myBullets}\vspace{2pt} 
 \item \textbf{Spoofing:} Stolen symmetric keys enable attacker impersonation.
 \item \textbf{Info. Disclosure:} Decryption of OS-protected data exposes credentials and files.
 \vspace{3pt} \end{myBullets}
\end{minipage} 
& \low 
& \high 
& \med 
\\ \cline{2-9} 

&\begin{minipage}{\linewidth} Kernel Integrity Compromise
\end{minipage} & \begin{minipage}{\linewidth} \gls{OS} Kernel Exploitation~\cite{compastie2020virtualization} 
\end{minipage} & \begin{minipage}{\linewidth} Exploitation through breaking hardware security modules (HSM) using quantum algorithms 
\end{minipage} & \begin{minipage}{\linewidth} Secure Kernel Development Techniques, Quantum-Resistant HSMs 
\end{minipage} & \begin{minipage}{\linewidth}
 \begin{myBullets}\vspace{2pt} 
 \item \textbf{Tampering:} Quantum-aided attacks modify kernel security mechanisms.
 \item \textbf{EoP:} Kernel takeover enables full \gls{OS} compromise.
 \vspace{3pt} \end{myBullets}
\end{minipage} 
& \med 
& \high 
& \high 
\\ \cline{2-9} 

& \begin{minipage}{\linewidth} Malware Insertion 
\end{minipage} & \begin{minipage}{\linewidth} Injecting Virus~\cite{ThreatsProtection} 
\end{minipage} & \begin{minipage}{\linewidth} Code injection with forged signature during software update 
\end{minipage} & \begin{minipage}{\linewidth} Advanced Endpoint Detection and Response (EDR) Tools, Regular Security Updates 
\end{minipage} & \begin{minipage}{\linewidth}
 \begin{myBullets}\vspace{2pt} 
 \item \textbf{Tampering:} Malware alters \gls{OS} files, configurations, or binaries.
 \item \textbf{DoS:} Persistent malware infections disrupt \gls{OS} functionality.
 \vspace{3pt} \end{myBullets}
\end{minipage} 
& \med 
& \high 
& \high 
\\ \cline{2-9} 
& \begin{minipage}{\linewidth} Non-repudiation Failure 
\end{minipage} & \begin{minipage}{\linewidth} Injecting Virus~\cite{ThreatsProtection}, Injecting Insecure Code~\cite{owasp_code_injection,hoffman2020web} 
\end{minipage} & \begin{minipage}{\linewidth} Code injection with forged signature during software update 
\end{minipage} & \begin{minipage}{\linewidth} \vspace{0.1cm} Implement quantum-resistant cryptographic solutions for signature verification; Enhance audit and logging mechanisms; Strengthen tamper-proof logging using quantum-resistant hashing schemes. 
 \vspace{0.1cm}\end{minipage} & \begin{minipage}{\linewidth}
 \begin{myBullets}\vspace{2pt}
 \item \textbf{Repudiation:} Quantum attacks on digital signatures allow attackers to deny actions.
 \item \textbf{Tampering:} Modified logs hide unauthorized activities and manipulate forensic evidence.
 \vspace{3pt} \end{myBullets}
\end{minipage} & \med & \med & \med \\ \cline{2-9} 

& \begin{minipage}{\linewidth} Quantum Cryptanalytic Attacks 
\end{minipage} & \begin{minipage}{\linewidth} \gls{OS} Kernel Exploitation~\cite{compastie2020virtualization} 
\end{minipage} & \begin{minipage}{\linewidth} Exploiting HSMs via Shor’s, Grover’s, and Brassard et al.'s algorithms 
\end{minipage} & \begin{minipage}{\linewidth} \vspace{0.1cm} Enforce quantum-resistant cryptographic hardware; Secure boot enforcement; Employ \glspl{TPM}  to mitigate key exposure risks. \vspace{0.1cm}
\end{minipage} &\begin{minipage}{\linewidth}
 \begin{myBullets}\vspace{2pt}
 \item \textbf{Info. Disclosure:} \gls{HSM} compromise exposes sensitive \gls{OS} cryptographic keys.
 \vspace{3pt} \end{myBullets}
\end{minipage} 
& \med 
& \high 
& \high 
\\ \cline{2-9} 

& \begin{minipage}{\linewidth} Inter-\gls{VM} Communication MITM 
\end{minipage} & \begin{minipage}{\linewidth} Quantum-Assisted MITM~\cite{compastie2020virtualization} 
\end{minipage} & \begin{minipage}{\linewidth} Quantum-enabled adversaries intercept \gls{VM} traffic 
\end{minipage} & \begin{minipage}{\linewidth} Transition to Quantum-Secure Channels, Anomaly Detection 
\end{minipage} &\begin{minipage}{\linewidth}
 \begin{myBullets}\vspace{2pt}
 \item \textbf{Info. Disclosure:} MITM attacks expose \gls{OS} network data.
 \item \textbf{Tampering:} Attacker manipulates encrypted \gls{VM} communication.
 \vspace{3pt} \end{myBullets}
\end{minipage} & \med & \high & \high \\ \cline{2-9} 
& \begin{minipage}{\linewidth} Malicious Code Injection 
\end{minipage} & \begin{minipage}{\linewidth} Injecting Insecure Code~\cite{owasp_code_injection,hoffman2020web}, Keylogger~\cite{commomtypmalware}, Trojan Horse~\cite{commomtypmalware}, Fileless Malware~\cite{commomtypmalware}, Worm~\cite{rawal2023malware}, Virus~\cite{ThreatsProtection}, Insecure Code~\cite{owasp_code_injection,hoffman2020web}, Trap Door~\cite{cox2025fifty,10964397} 
\end{minipage} & \begin{minipage}{\linewidth} Code injection with forged signature during software update 
\end{minipage} & \begin{minipage}{\linewidth} Quantum-resistant cryptographic algorithms for code signing; Secure Software Update Mechanisms; Incorporate behavior-based malware detection to counter quantum-assisted stealth attacks. 
\end{minipage} & \begin{minipage}{\linewidth}
 \begin{myBullets}\vspace{2pt}
 \item \textbf{Info. Disclosure:} Malicious code insertion enables unauthorized surveillance.
 \item \textbf{Tampering:} Attackers modify system files and processes for persistent compromise.
 \vspace{3pt} \end{myBullets}
\end{minipage} & \med & \high & \high \\ \cline{2-9} 

& \begin{minipage}{\linewidth} Cross-\gls{VM} \glspl{SCA}
\end{minipage} & \begin{minipage}{\linewidth} \glspl{SCA}~\cite{gruss2019page} 
\end{minipage} & \begin{minipage}{\linewidth} Quantum-assisted exfiltration of cryptographic keys 
\end{minipage} & \begin{minipage}{\linewidth} Strengthen \gls{VM} isolation,  \gls{SCA} protections 
\end{minipage} & \begin{minipage}{\linewidth}
 \begin{myBullets}\vspace{2pt} 
 \item \textbf{Info. Disclosure:} Cryptographic keys are extracted via side-channel analysis.
 \item \textbf{EoP:} Stolen keys allow higher access levels.
 \vspace{3pt} \end{myBullets}
\end{minipage} 
& \med 
& \high 
& \high 
\\ \cline{2-9} 

& \begin{minipage}{\linewidth} Trap Door Injection 
\end{minipage} & \begin{minipage}{\linewidth} Malware Exploitation~\cite{cox2025fifty,10964397} 
\end{minipage} & \begin{minipage}{\linewidth} Malicious code with quantum-enhanced trap doors is injected into OS-level software, allowing persistent access, privilege escalation, and stealthy security evasion. \gls{QC} accelerates backdoor discovery, cryptographic forgery, and evasion techniques. 
\end{minipage} & \begin{minipage}{\linewidth} Enhanced Software Supply Chain Security; Continuous \gls{OS} Integrity Monitoring; Deployment of \gls{PQC}  signature Validation for Software Updates; Secure Boot Mechanisms with Quantum-Resistant Cryptography. 
\end{minipage} & \begin{minipage}{\linewidth}
 \begin{myBullets}\vspace{2pt} 
 \item \textbf{Spoofing:} trap doors are used to maintain unauthorized access while appearing as legitimate users or processes, bypassing authentication.
 \item \textbf{Tampering:} Security measures and \gls{OS} integrity mechanisms are bypassed via quantum-forged signatures, making modifications undetectable.
 \item \textbf{EoP:} Malware exploits quantum-broken cryptographic protections to escalate access rights, enabling control over the OS.
 \vspace{3pt} \end{myBullets}
\end{minipage} & \med & \high & \high \\ \cline{2-9} 

& \begin{minipage}{\linewidth} Kernel Integrity Compromise 
\end{minipage} & \begin{minipage}{\linewidth} \gls{OS} Kernel Exploitation~\cite{compastie2020virtualization} 
\end{minipage} & \begin{minipage}{\linewidth} Exploiting hardware security modules (HSM) using quantum algorithms 
\end{minipage} & \begin{minipage}{\linewidth} Secure Kernel Development Techniques, Quantum-Resistant \glspl{HSM}, and \glspl{TPM}  to prevent unauthorized kernel access. 
\end{minipage} & \begin{minipage}{\linewidth}
 \begin{myBullets}\vspace{2pt}
 \item \textbf{Tampering:} Quantum-aided attacks modify kernel security mechanisms.
 \item \textbf{EoP:} Kernel takeover enables full \gls{OS} compromise.
 \vspace{3pt} \end{myBullets}
\end{minipage} & \med & \high & \high \\ \hline 

\multirow{16}{*}{Virtualization} 
& \begin{minipage}{\linewidth} Cryptographic Key Vulnerability 
\end{minipage} & \begin{minipage}{\linewidth} Brute Force Attack~\cite{grover1996fast} 
\end{minipage} & \begin{minipage}{\linewidth} Brute-force cryptographic keys through Grover's algorithm
\end{minipage} & \begin{minipage}{\linewidth} Upgrade to cryptographic algorithms with larger key sizes and early adoption of \gls{PQC} standards 
\end{minipage} & \begin{minipage}{\linewidth}
 \begin{myBullets}\vspace{2pt} 
 \item \textbf{Spoofing:} Attackers forge \gls{VM} identities using cracked keys. 
 \item \textbf{Info. Disclosure:} Decryption of stored or transmitted \gls{VM} data.
 \vspace{3pt} \end{myBullets}
\end{minipage} 
& \med 
& \high 
& \high 
\\ \cline{2-9} 

& \begin{minipage}{\linewidth} Weak Encryption Algorithms 
\end{minipage} & \begin{minipage}{\linewidth} \gls{VM} Mobility~\cite{compastie2020virtualization,huang2012security} 
\end{minipage} & \begin{minipage}{\linewidth} Quantum algorithms like Shor's and Grover's could decrypt VMs during migration or brute-force cryptographic keys 
\end{minipage} & \begin{minipage}{\linewidth} Transition to  \gls{NIST}-approved \gls{PQC} standards; strengthen key management practices 
\end{minipage} & \begin{minipage}{\linewidth} \begin{myBullets}\vspace{2pt}
 \item \textbf{Spoofing:} Attackers forge identities by breaking encryption.
 \item \textbf{Info. Disclosure:} Decryption exposes VM-sensitive data.
 \vspace{3pt} \end{myBullets} \end{minipage}
& \med 
& \high 
& \high 
\\ \cline{2-9} 

& \begin{minipage}{\linewidth} Hypervisor Vulnerabilities 
\end{minipage} & \begin{minipage}{\linewidth} Side-channel cache attacks~\cite{irazoqui2015s}, Cryptanalysis~\cite{brassard1997quantum,grover1996fast,shor1994algorithms} 
\end{minipage} & \begin{minipage}{\linewidth} Data exfiltration via quantum algorithms 
\end{minipage} & \begin{minipage}{\linewidth} Hardware-enforced isolation, intrusion detection systems, and quantum-resistant algorithms for sensitive operations 
\end{minipage} & \begin{minipage}{\linewidth} \begin{myBullets}\vspace{2pt}
 \item \textbf{Info. Disclosure:} Attackers extract sensitive hypervisor data.
 \item \textbf{Tampering:} Hypervisor configurations are modified to weaken security.
 \item \textbf{EoP:} A compromised hypervisor enables full control over guest VMs.
 \vspace{3pt} \end{myBullets}
\end{minipage} & \med 
& \high 
& \high 
\\ \cline{2-9}

& \begin{minipage}{\linewidth} Cross-\gls{VM} Attacks 
\end{minipage} & \begin{minipage}{\linewidth} Side-channel Cache Attacks~\cite{NARAYANA20216465,anwar2017cross,saeed2020cross}, Cryptanalysis~\cite{brassard1997quantum,grover1996fast,shor1994algorithms} 
\end{minipage} & \begin{minipage}{\linewidth} Data exfiltration via quantum algorithms 
\end{minipage} & \begin{minipage}{\linewidth} \gls{VM} segregation, strict access control, and continuous monitoring to detect \glspl{SCA} 
\end{minipage} & \begin{minipage}{\linewidth} \begin{myBullets}\vspace{2pt}
 \item \textbf{Info. Disclosure:} Cryptographic keys or \gls{VM} data are leaked via \glspl{SCA}.
 \item \textbf{Tampering:} Side-channel exploitation allows attackers to modify or inject malicious data.
 \item \textbf{EoP:} Attackers escalate privileges by accessing another VM’s resources.
 \vspace{3pt} \end{myBullets} 
\end{minipage} & \med 
& \high 
& \high 
\\ \cline{2-9}

& \begin{minipage}{\linewidth} \gls{VM} Migration Risks 
\end{minipage} & \begin{minipage}{\linewidth} \gls{VM} Migration~\cite{9889331} 
\end{minipage} & \begin{minipage}{\linewidth} Data exfiltration during \gls{VM} migration 
\end{minipage} & \begin{minipage}{\linewidth} End-to-end encryption, secure migration protocols, and security audits for migration processes 
\end{minipage} & \begin{minipage}{\linewidth} \begin{myBullets}\vspace{2pt}
 \item \textbf{Info. Disclosure:} Migration data is exposed if encryption is broken. 
 \item \textbf{Tampering:} Attackers inject malicious data during migration.
 \vspace{3pt} \end{myBullets} 
\end{minipage} & \med 
& \high 
& \high 
\\ \cline{2-9} 

& \begin{minipage}{\linewidth} Hypervisor Overload 
\end{minipage} & \begin{minipage}{\linewidth} Hypervisor Exploitation~\cite{compastie2020virtualization} 
\end{minipage} & \begin{minipage}{\linewidth} Quantum DoS attacks via hypervisor exploitation 
\end{minipage} & \begin{minipage}{\linewidth} Redundant, distributed architectures and quantum-aware rate limiting 
\end{minipage} & \begin{minipage}{\linewidth} \begin{myBullets}\vspace{2pt}
 \item \textbf{DoS:} The hypervisor becomes overwhelmed, impacting \gls{VM} availability.
 \vspace{3pt} \end{myBullets} 
\end{minipage} & \med 
& \med 
& \med 

\\ \cline{2-9}


& \begin{minipage}{\linewidth} \gls{VM} Identity Forgery 
\end{minipage} & \begin{minipage}{\linewidth} \gls{VM} Mobility~\cite{compastie2020virtualization,huang2012security} 
\end{minipage} & \begin{minipage}{\linewidth} Decryption of an encrypted \gls{VM} during migration 
\end{minipage} & \begin{minipage}{\linewidth} Quantum-resistant cryptographic protocols for secure \gls{VM} migration 
\end{minipage} & \begin{minipage}{\linewidth}
 \begin{myBullets}\vspace{2pt} 
 \item \textbf{Spoofing:} Attackers forge \gls{VM} identities, allowing unauthorized execution of rogue workloads.
 \item \textbf{Info. Disclosure:} If encryption is broken, attackers gain access to \gls{VM} data during migration.
 \vspace{3pt} \end{myBullets}
\end{minipage} 
& \med 
& \high 
& \high 
 \\ \cline{2-9}
& \begin{minipage}{\linewidth} Hypervisor Integrity Compromise 
\end{minipage} & \begin{minipage}{\linewidth} Hypervisor Exploitation~\cite{compastie2020virtualization} 
\end{minipage} & \begin{minipage}{\linewidth} Direct attacks on the hypervisor 
\end{minipage} & \begin{minipage}{\linewidth} Verified boot, attestation mechanisms 
\end{minipage} & \begin{minipage}{\linewidth} \begin{myBullets}\vspace{2pt}
 \item \textbf{Info. Disclosure:} A compromised hypervisor exposes all guest \gls{VM} data.
 \item \textbf{Tampering:} Attackers modify the hypervisor to introduce backdoors.
 \item \textbf{EoP:} Exploiting hypervisor flaws allows attackers to control guest VMs.
 \vspace{3pt} \end{myBullets}
 
\end{minipage} &
\med 
& \high 
& \high 
\\ \cline{2-9} 

& \begin{minipage}{\linewidth} Inter-\gls{VM} Communication Eavesdropping 
\end{minipage} & \begin{minipage}{\linewidth} Cross-\gls{VM} Side-channel Cache Attacks~\cite{NARAYANA20216465,10.1145/3715001,anwar2017cross,saeed2020cross} 

\end{minipage} & \begin{minipage}{\linewidth} Cryptanalysis with quantum algorithms 
\end{minipage} & \begin{minipage}{\linewidth} Quantum-safe encrypted channels 
\end{minipage} & \begin{minipage}{\linewidth}
 \begin{myBullets}\vspace{2pt} 
 \item \textbf{Info. Disclosure:} Quantum-assisted cryptanalysis extracts confidential inter-\gls{VM} data.
 \item \textbf{Tampering:} Attackers manipulate intercepted data, injecting malicious payloads into middleware.
 \vspace{3pt} \end{myBullets}
\end{minipage} 
& \med 
& \high 
& \high 
\\ \cline{2-9} 

& \begin{minipage}{\linewidth} Unauthorized Access via Hypervisor Exploitation 
\end{minipage} & \begin{minipage}{\linewidth} Hypervisor Exploitation~\cite{compastie2020virtualization} 
\end{minipage} & \begin{minipage}{\linewidth} Gaining higher privileges via hypervisor vulnerabilities 
\end{minipage} & \begin{minipage}{\linewidth} Multifactor authentication with \gls{PQC} 
\end{minipage} & \begin{minipage}{\linewidth}
 \begin{myBullets}\vspace{2pt} 
 \item \textbf{Info. Disclosure:} A compromised hypervisor exposes sensitive \gls{VM} data.
 \item \textbf{EoP:} Attackers escalate privileges and gain unauthorized control over middleware services.
 \vspace{3pt} \end{myBullets}
\end{minipage}
& \med 
& \high 
& \high 
\\ \hline

\end{tabular}%
}
\label{table:Pre-Transition}
 \end{table*}

\begin{table*}[!htbp]
\ContinuedFloat 
\caption{Pre-Transition Quantum Threat Landscape in Each \gls{CC} Layer (Continued)} 
\large
\renewcommand{\arraystretch}{1.1}
\resizebox{\linewidth}{!}{%
\begin{tabular}{|p{0.13\linewidth}|p{0.25\linewidth}|p{0.3\linewidth}|p{0.48\linewidth}|p{0.47\linewidth}|p{0.65\linewidth}|l|l|l|}
\hline 
\textbf{Layer} & \textbf{Vulnerability} & \textbf{Exploitation Method} & \textbf{Attack Vector} & \textbf{Countermeasures} & \textbf{STRIDE Threats}& {\textbf{L}} & {\textbf{I}} & {\textbf{R}}\\ \hline

\multirow{30}{*}{Server} 

& \begin{minipage}{\linewidth} Quantum Key-Breaking 
\end{minipage} & \begin{minipage}{\linewidth} Brute Force Attack~\cite{lu2019attacking} 
\end{minipage} & \begin{minipage}{\linewidth} Exploiting quantum algorithms to break cryptographic keys 
\end{minipage} & \begin{minipage}{\linewidth} Implement quantum-resistant cryptographic algorithms 
\end{minipage} & \begin{minipage}{\linewidth} 
 \begin{myBullets}\vspace{2pt} 
 \item \textbf{Spoofing:} Attackers forge authentication credentials. 
 \item \textbf{Info. Disclosure:} Stolen keys enable decryption of server communications. 
 \vspace{3pt} \end{myBullets} 
\end{minipage} 
& \med 
& \high 
& \high 
\\ \cline{2-9} 

& \begin{minipage}{\linewidth} Firmware Integrity Compromise 
\end{minipage} & \begin{minipage}{\linewidth} Firmware Exploitation~\cite{10.1145/3432893} 
\end{minipage} & \begin{minipage}{\linewidth} Forged firmware updates with malicious code 
\end{minipage} & \begin{minipage}{\linewidth} Secure boot, firmware signing, and verification 
\end{minipage} & \begin{minipage}{\linewidth} 
 \begin{myBullets}\vspace{2pt} 
 \item \textbf{Tampering:} Malicious firmware alters system behavior. 
 \item \textbf{EoP:} Attackers inject firmware-level backdoors. 
 \vspace{3pt} \end{myBullets} 
\end{minipage} 
& \med 
& \high 
& \high 
\\ \cline{2-9} 

& \begin{minipage}{\linewidth} Quantum-Assisted Software Supply Chain Attacks 
\end{minipage} & \begin{minipage}{\linewidth} Supply Chain Attacks~\cite{akter2024integrated} 
\end{minipage} & \begin{minipage}{\linewidth} Quantum-forged signatures compromise package repositories and inject malicious software dependencies 
\end{minipage} & \begin{minipage}{\linewidth} Implement \gls{PQC}  signature verification; Enforce SBOM (Software Bill of Materials) auditing 
\end{minipage} & \begin{minipage}{\linewidth} 
 \begin{myBullets}\vspace{2pt} 
 \item \textbf{Tampering:} Attackers inject compromised dependencies into software updates. 
 \item \textbf{Spoofing:} Quantum-forged signatures make malicious software appear legitimate. 
 \vspace{3pt} \end{myBullets} 
\end{minipage} 
& \med 
& \high 
& \high 
\\ \cline{2-9} 

& \begin{minipage}{\linewidth}Traceability of Changes 
\end{minipage} & \begin{minipage}{\linewidth} Firmware Exploitation~\cite{10.1145/3432893} 
\end{minipage} & \begin{minipage}{\linewidth} Unauthorized firmware updates to erase security logs 
\end{minipage} & \begin{minipage}{\linewidth} Immutable logs and event tracking systems 
\end{minipage} & \begin{minipage}{\linewidth} 
 \begin{myBullets}\vspace{2pt} 
 \item \textbf{Repudiation:} Attackers erase evidence of tampering. 
 \item \textbf{Tampering:} Attackers modify security logs to mislead forensic analysis. 
 \vspace{3pt} \end{myBullets} 
\end{minipage} 
& \med 
& \med 
& \med 
\\ \cline{2-9} 

& \begin{minipage}{\linewidth} Encrypted Data Breach 
\end{minipage} & \begin{minipage}{\linewidth} Cryptanalysis Attacks~\cite{brassard1997quantum,grover1996fast,shor1994algorithms} 
\end{minipage} & \begin{minipage}{\linewidth} Advanced cryptanalysis leveraging quantum algorithms 
\end{minipage} & \begin{minipage}{\linewidth} Enhance encryption with quantum-safe solutions 
\end{minipage} & \begin{minipage}{\linewidth} 
 \begin{myBullets}\vspace{2pt} 
 \item \textbf{Info. Disclosure:} Decryption exposes sensitive server data. 
 \vspace{3pt} \end{myBullets} 
\end{minipage} 
& \med 
& \high 
& \high 
\\ \cline{2-9} 

& \begin{minipage}{\linewidth} Unauthorized Access Control 
\end{minipage} & \begin{minipage}{\linewidth} Firmware Exploitation~\cite{10.1145/3432893} 
\end{minipage} & \begin{minipage}{\linewidth} Injecting malicious firmware to gain higher privileges 
\end{minipage} & \begin{minipage}{\linewidth} Access control mechanisms, multi-factor authentication 
\end{minipage} & \begin{minipage}{\linewidth} 
 \begin{myBullets}\vspace{2pt} 
 \item \textbf{EoP:} Attackers escalate access via firmware backdoors. 
 \item \textbf{Tampering:} Attackers modify access control settings to disable security policies. 
 \vspace{3pt} \end{myBullets} 
\end{minipage} 
& \med 
& \high 
& \high 
\\ \cline{2-9} 

& \begin{minipage}{\linewidth} Quantum-Accelerated \glspl{SCA} 
\end{minipage} & \begin{minipage}{\linewidth} Side-Channel Analysis~\cite{chowdhury2021physical}
\end{minipage} & \begin{minipage}{\linewidth} Quantum-enhanced analysis extracts cryptographic keys from CPUs 
\end{minipage} & \begin{minipage}{\linewidth} Hardware-level mitigations, noise injection, side-channel-resistant cryptography 
\end{minipage} & \begin{minipage}{\linewidth} 
 \begin{myBullets}\vspace{2pt} 
 \item \textbf{Info. Disclosure:} Attackers extract cryptographic keys from power or cache analysis. 
 \vspace{3pt} \end{myBullets} 
\end{minipage} 
& \med 
& \high 
& \high \\ \cline{2-9}

& Quantum-Optimized DoS on Server Resources 
& DoS Attacks 
& Quantum-assisted attacks overload CPU, network bandwidth, and storage I/O 
& Adaptive rate limiting, quantum-aware server resource allocation 
& \begin{minipage}{\linewidth} 
 \begin{myBullets}\vspace{2pt} 
 \item \textbf{DoS:} Servers become unresponsive due to quantum-optimized resource exhaustion. 
 \vspace{3pt} \end{myBullets} 
\end{minipage} 
& \med 
& \med 
& \med 
\\ \cline{2-9}
& \begin{minipage}{\linewidth} Quantum-Persistent Rootkits in Server Firmware 
\end{minipage} 
& \begin{minipage}{\linewidth} Firmware Exploitation~\cite{10.1145/3432893} 
\end{minipage} 
& \begin{minipage}{\linewidth} Quantum-assisted malware embeds rootkits within server firmware, persisting across reboots and firmware updates. 
\end{minipage} 
& \begin{minipage}{\linewidth} Implement firmware integrity monitoring, secure boot mechanisms, and hardware-assisted malware detection. 
\end{minipage} 
& \begin{minipage}{\linewidth} 
 \begin{myBullets}\vspace{2pt} 
 \item \textbf{Tampering:} Rootkits modify firmware, evading detection even after system reboots. 
 \item \textbf{EoP:} Attackers gain persistent low-level access to the server’s hardware. 
 \vspace{3pt} \end{myBullets} 
\end{minipage} 
& \med 
& \high 
& \high 
\\ \cline{2-9} 

& \begin{minipage}{\linewidth} Quantum-Accelerated Insider Privilege Escalation 
\end{minipage} 
& \begin{minipage}{\linewidth} Insider Threats~\cite{Threatinsiderdef} 
\end{minipage} 
& \begin{minipage}{\linewidth} Malicious insiders leverage quantum-assisted brute-force to escalate privileges beyond authorized access. 
\end{minipage} 
& \begin{minipage}{\linewidth} Enforce strict multi-factor authentication (MFA), use post-quantum authentication, and implement real-time privilege monitoring. 
\end{minipage} 
& \begin{minipage}{\linewidth} 
 \begin{myBullets}\vspace{2pt} 
 \item \textbf{EoP:} Attackers bypass authentication controls using quantum-accelerated attacks. 
 \item \textbf{Tampering:} Insiders modify access control configurations to retain unauthorized privileges. 
 \vspace{3pt} \end{myBullets} 
\end{minipage} 
& \med 
& \high 
& \high 
\\ \hline

\multirow{35}{*}{Storage} 
& \begin{minipage}{\linewidth} Symmetric Cryptographic Key Vulnerability
\end{minipage}
& \begin{minipage}{\linewidth} Brute Force Attack~\cite{shen2020lightweight}
\end{minipage}
& \begin{minipage}{\linewidth} Grover's algorithm used to brute-force an 
n-bit symmetric key
\end{minipage}
& \begin{minipage}{\linewidth} Enhanced symmetric encryption with longer key sizes, cryptographic agility, and hybrid encryption schemes
\end{minipage}
& \begin{minipage}{\linewidth}
\begin{myBullets}\vspace{2pt}
\item \textbf{Spoofing:} Attackers forge authentication credentials after brute-forcing keys.
\item \textbf{Info. Disclosure:} Decryption of encrypted storage data.\vspace{0.1cm}
 \vspace{3pt} \end{myBullets}
\end{minipage}
& \med
& \high
& \high
\\ \cline{2-9}

& \begin{minipage}{\linewidth} Vulnerabilities in Classic Cryptosystems \end{minipage} & \begin{minipage}{\linewidth} Cryptanalysis Attacks~\cite{brassard1997quantum,grover1996fast,shor1994algorithms} \end{minipage} & \begin{minipage}{\linewidth} Data exfiltration via breaking classic cryptosystems using quantum algorithms \end{minipage} & \begin{minipage}{\linewidth}\vspace{2pt} Early adoption of quantum-resistant encryption methods (e.g., lattice-based, hash-based, and code-based cryptography), Regular security audits \vspace{2pt}\end{minipage} & \begin{minipage}{\linewidth} \begin{myBullets}\vspace{2pt} \item \textbf{Info. Disclosure:} Decryption of stored and in-transit sensitive data. \item \textbf{Tampering:} Attackers modify decrypted data undetected.\vspace{3pt} \end{myBullets} \end{minipage} & \med & \high & \high 
\\ \cline{2-9} 
& \begin{minipage}{\linewidth} Susceptibility to Quantum-Complex Encryption \end{minipage} & \begin{minipage}{\linewidth} Ransomware~\cite{craciun2019trends,11177145} \end{minipage} & \begin{minipage}{\linewidth} Quantum-enhanced ransomware encrypting storage using quantum-generated complex cryptographic methods \end{minipage} & \begin{minipage}{\linewidth} \vspace{0.1cm}Intrusion detection systems, secure data backups, quantum-resilient backup strategies, ransomware-specific defenses \vspace{0.1cm}\end{minipage} & \begin{minipage}{\linewidth} \begin{myBullets}\vspace{2pt} \item \textbf{DoS:} Attackers encrypt cloud storage with complex quantum-encrypted ransomware, blocking data access. \vspace{3pt} \end{myBullets} \vspace{0.1cm}\end{minipage} & \low & \high &\med
 
\\ \cline{2-9} 
& \begin{minipage}{\linewidth} Forgery of Digital Signatures \end{minipage} & \begin{minipage}{\linewidth} Data Theft~\cite{Threatinsiderdef} \end{minipage} & \begin{minipage}{\linewidth} Code injection with forged signatures during updates or data modification in storage \end{minipage} & \begin{minipage}{\linewidth} Transitional digital signature standards (e.g., \gls{PQC}-based digital signatures), Multi-factor authentication for update processes \end{minipage} & 
\begin{minipage}{\linewidth} \begin{myBullets}\vspace{2pt} 
\item \textbf{EoP:} Attackers use quantum-forged signatures to escalate permissions. 
\item \textbf{Tampering:} Attackers modify storage data undetected. \vspace{3pt} \end{myBullets} \end{minipage} & \med 
& \high 
& \high 
\\ \cline{2-9} 
& \begin{minipage}{\linewidth} Quantum-Assisted Storage Deduplication Attacks \end{minipage} & \begin{minipage}{\linewidth} Storage Optimization Weaknesses \end{minipage} & \begin{minipage}{\linewidth} Exploiting deduplication techniques~\cite{garcia2024mapping,prajapati2022review} to overwrite legitimate storage data \end{minipage} & \begin{minipage}{\linewidth} Secure hash-based deduplication with quantum-resistant integrity verification \end{minipage} & \begin{minipage}{\linewidth} \begin{myBullets}\vspace{2pt} \item \textbf{Tampering:} Attackers exploit deduplication to replace legitimate files. \item \textbf{DoS:} Overwritten data can make critical files unavailable.\vspace{3pt} \end{myBullets} \end{minipage} & \low & \med & \low 
\\ \cline{2-9} 
& \begin{minipage}{\linewidth} Quantum-Powered Searchable Encryption Attacks \end{minipage} & \begin{minipage}{\linewidth} Query Optimization Weaknesses \end{minipage} & \begin{minipage}{\linewidth} Quantum-aided attacks decrypt and analyze search patterns on encrypted cloud storage \end{minipage} & \begin{minipage}{\linewidth} Post-quantum secure searchable encryption protocols \end{minipage} & \begin{minipage}{\linewidth} \begin{myBullets}\vspace{2pt} \item \textbf{Info. Disclosure:} Attackers infer sensitive data by analyzing search queries. \vspace{3pt} \end{myBullets} \end{minipage} & \low&\high&\med
\\ \cline{2-9}
& \begin{minipage}{\linewidth} Quantum-Assisted Metadata Correlation Attacks \end{minipage} 
& \begin{minipage}{\linewidth} Storage Metadata Analysis~\cite{jia2024machine} \end{minipage} 
& \begin{minipage}{\linewidth} Quantum algorithms accelerate metadata analysis to infer sensitive storage patterns without decrypting the actual data \end{minipage} 
& \begin{minipage}{\linewidth} Metadata anonymization, post-quantum searchable encryption, and obfuscation techniques \end{minipage} 
& \begin{minipage}{\linewidth} 
 \begin{myBullets}\vspace{2pt} 
 \item \textbf{Info. Disclosure:} Attackers infer sensitive storage metadata such as file access logs, timestamps, and access frequency. 
 \item \textbf{Tampering:} Attackers modify storage metadata to manipulate forensic analysis. 
 \vspace{3pt} \end{myBullets} 
\end{minipage} 
& \med &\med&\med
\\ \cline{2-9}


& \begin{minipage}{\linewidth} Quantum-Accelerated ‘\gls{HNDL}’ Exploit~\cite{barenkamp2022steal} \end{minipage} 
& \begin{minipage}{\linewidth} Long-Term Encrypted Data Harvesting \end{minipage} 
& \begin{minipage}{\linewidth} Attackers steal encrypted cloud storage data today and decrypt it in the future when quantum computers can break RSA/ECC encryption  (\gls{HNDL})\end{minipage} 
& \begin{minipage}{\linewidth} Transition to quantum-resistant encryption now, regularly re-encrypt older backups using \gls{PQC} algorithms \end{minipage} 
& \begin{minipage}{\linewidth} 
 \begin{myBullets}\vspace{2pt} 
 \item \textbf{Info. Disclosure:} Attackers decrypt stolen storage data once \gls{QC} becomes powerful enough. 
 \item \textbf{Tampering:} Encrypted backups remain valid attack surfaces for future exploitation. 
 \vspace{3pt} \end{myBullets} 
\end{minipage} 
& \high & \high & \high 
\\ \cline{2-9} 
& \begin{minipage}{\linewidth} Quantum-Accelerated Analysis of Encrypted Backups \end{minipage} 
& \begin{minipage}{\linewidth} Quantum Cryptanalysis on Stored Data \end{minipage} 
& \begin{minipage}{\linewidth} Attackers leverage quantum brute-force methods to decrypt older cloud backups using Grover’s algorithm \end{minipage} 
& \begin{minipage}{\linewidth} Re-encrypt backups using \gls{PQC} algorithms, avoid key reuse across backup generations, use quantum-safe encryption for at-rest data \end{minipage} 
& \begin{minipage}{\linewidth} 
 \begin{myBullets}\vspace{2pt} 
 \item \textbf{Info. Disclosure:} Decryption of stored backups exposes sensitive historical data. 
 \item \textbf{Tampering:} Attackers modify backup archives after decrypting them. 
 \vspace{3pt} \end{myBullets} 
\end{minipage} 
& \med 
& \high 
& \high 
\\ \cline{2-9} 
& \begin{minipage}{\linewidth} Quantum-Optimized Integrity Poisoning on Shared Storage \end{minipage} 
& \begin{minipage}{\linewidth} Hash Collision Exploits~\cite{baseri2025evaluation,BASERI2024103883} \end{minipage} 
& \begin{minipage}{\linewidth} Attackers use Grover’s algorithm to forge file hashes and tamper with integrity verification mechanisms in cloud storage \end{minipage} 
& \begin{minipage}{\linewidth} Implement quantum-secure cryptographic hash functions (e.g., SHA-3), enforce integrity checks with cryptographic proofs \end{minipage} 
& \begin{minipage}{\linewidth} 
 \begin{myBullets}\vspace{2pt} 
 \item \textbf{Tampering:} Attackers modify cloud storage files while maintaining valid hash values. 
 \item \textbf{Info. Disclosure:} Altered files lead to unauthorized data manipulation. 
 \vspace{3pt} \end{myBullets} 
\end{minipage} 
& \med 
& \high 
& \high 
\\ \cline{2-9} 
& \begin{minipage}{\linewidth} Hybrid Cloud Data Orchestration Exploits with Quantum Assistance \end{minipage} 
& \begin{minipage}{\linewidth} Multi-Cloud Data Sync Vulnerabilities~\cite{reece2023systemic} \end{minipage} 
& \begin{minipage}{\linewidth} Attackers intercept data during hybrid-cloud storage syncs and decrypt in-transit storage data using Shor’s algorithm \end{minipage} 
& \begin{minipage}{\linewidth} Enforce end-to-end \gls{PQC} encryption for cross-cloud storage replication, implement zero-trust security models for data synchronization \end{minipage} 
& \begin{minipage}{\linewidth} 
 \begin{myBullets}\vspace{2pt} 
 \item \textbf{Info. Disclosure:} Intercepted storage sync data can be decrypted by quantum adversaries. 
 \item \textbf{Tampering:} Attackers modify replicated storage data between \glspl{CSP}. 
 \vspace{3pt} \end{myBullets} 
\end{minipage} 
& \med 
& \high 
& \high 
\\ \hline 

\multirow{12}{*}{Networking}
& Vulnerable Symmetric Encryption
& Brute Force Attack~\cite{idhom2020network}
& Grover's algorithm to brute-force a symmetric key
& Use larger key lengths (e.g. 256-bit), adopt hybrid or \gls{PQC} encryption
& \begin{minipage}{\linewidth}
\begin{myBullets}\vspace{2pt} 
\item \textbf{Spoofing}: Broken keys let attackers impersonate entities
\item \textbf{Info. Disclosure}: Decrypting traffic in-flight
 \vspace{3pt} \end{myBullets} 
\end{minipage}
& \med & \high & \high \\ \cline{2-9}

& {Insecure Wi-Fi Protocols}
& Wi-Fi Wardriving~\cite{lounis2020attacks}
& Shor’s/Grover’s to break WPA2/WPA3 keys
& Upgrade to WPA3 with quantum resistance; robust passphrases
& \begin{minipage}{\linewidth}
\begin{myBullets}\vspace{2pt} 
\item \textbf{Tampering}: Inject or modify Wi-Fi traffic
\item \textbf{Info. Disclosure}: Read frames, capturing user data
\item \textbf{Spoofing}: Impersonate AP/clients
 \vspace{3pt} \end{myBullets} 
\end{minipage}
& \med & \high & \high\\
\cline{2-9}

& Physical AP Vulnerabilities
& Physical AP Attacks~\cite{chakraborty2017minimum,lounis2020attacks}
& Quantum-aided cryptanalysis + physical tampering
& Regular AP security reviews; patch firmware
& \begin{minipage}{\linewidth}
\begin{myBullets}\vspace{2pt} 
\item \textbf{Tampering}: Direct hardware/firmware compromise
\item \textbf{Info. Disclosure}: Sniff traffic or exfil data
\item \textbf{EoP}: Seizing AP control for extended network access
 \vspace{3pt} \end{myBullets} 
\end{minipage}
& \med & \high & \high \\
\cline{2-9}

& {Susceptible Cryptographic Protocols}
& Cryptanalysis Attacks~\cite{zhang2018novel}
& Breaking TLS/IPsec with Shor/Grover
& Quantum-safe TLS (TLS 1.3 \gls{PQC} drafts); frequent cryptographic reviews
& \begin{minipage}{\linewidth}
\begin{myBullets}\vspace{2pt} 
\item \textbf{Tampering}: Modify packets once the cipher is broken
\item \textbf{Info. Disclosure}: Eavesdrop on network sessions
 \vspace{3pt} \end{myBullets} 
\end{minipage}
& \med & \high & \high \\ \cline{2-9}
& {Signature Forgery Risk}
& MITM~\cite{chakraborty2017minimum}
& Forging digital signatures with quantum algorithms
& \gls{PQC}-based signatures; vigilant certificate management
& \begin{minipage}{\linewidth}
\begin{myBullets}\vspace{2pt} 
\item \textbf{Tampering}: Replace/alter messages unnoticed
\item \textbf{Spoofing}: Impersonate signers for code or data injection
 \vspace{3pt} \end{myBullets} 
\end{minipage}
& \med & \high & \high \\
\cline{2-9}

& {Lack of Non-Repudiation Controls}
& Message Repudiation Attack
& Forged signatures enable the user to deny authorship of a message
& Quantum-safe non-repudiation solutions
& \begin{minipage}{\linewidth}
\begin{myBullets}\vspace{2pt} 
\item \textbf{Repudiation}: Attackers or malicious users disclaim or rewrite message origin
 \vspace{3pt} \end{myBullets} 
\end{minipage}
& \med & \high & \high\\
\cline{2-9}
& {Encryption Weaknesses}
& Packet Sniffing (Encrypted)~\cite{shibly2020threat}
& Data exfiltration via cipher decryption
& \gls{PQC} encryption; ongoing protocol updates
& \begin{minipage}{\linewidth}
\begin{myBullets}\vspace{2pt} 
\item \textbf{Info. Disclosure}: Attackers decrypt captured packets
\item \textbf{Spoofing}: Possibly re-encrypt traffic to pose as genuine nodes
 \vspace{3pt} \end{myBullets} 
\end{minipage}
& \med & \high & \high \\
\cline{2-9}
& {Vulnerable Data Transmission}
& \gls{MITM}~\cite{compastie2020virtualization,10725789}
& Intercepting data channels with quantum-aided decryption
& Secure quantum-safe protocols; frequent security testing
& \begin{minipage}{\linewidth}
\begin{myBullets}\vspace{2pt} 
\item \textbf{Info. Disclosure}: Eavesdrop logs traffic
\item \textbf{Tampering}: Inject or alter data in flight
 \vspace{3pt} \end{myBullets} 
\end{minipage}
& \med & \high & \high \\ \hline


\multirow{30}{*}{Networking}
& {Weak Cryptanalysis Defenses}
& Cryptanalysis Attacks~\cite{brassard1997quantum,grover1996fast,shor1994algorithms}
& Data theft via advanced quantum cryptanalysis
& Migrate to \gls{PQC} algorithms; frequent audits
& \begin{minipage}{\linewidth}
\begin{myBullets}\vspace{2pt} 
\item \textbf{Info. Disclosure}: Decrypt network streams
\item \textbf{Spoofing}: Potential forging of session keys/credentials
 \vspace{3pt} \end{myBullets} 
\end{minipage}
& \med & \high & \high \\ \cline{2-9}
& {Compromised Update Mechanisms}
& System Sabotage~\cite{Threatinsiderdef}
& Forged signatures enable the injection of malicious code into software updates
& Authenticated update channels; \gls{PQC}-based code signing
& \begin{minipage}{\linewidth}
\begin{myBullets}\vspace{2pt} 
\item \textbf{EoP}: Malicious updates yield admin-level access
\item \textbf{Tampering}: Replacing legitimate updates with harmful ones
 \vspace{3pt} \end{myBullets} 
\end{minipage}
& \med & \high & \high \\ \cline{2-9}
& {Vulnerable \gls{VM} Communication} 
& Inter-\gls{VM} Communication Introspection~\cite{compastie2020virtualization}
& MITM using quantum decryption~\cite{compastie2020virtualization,10725789}
& Quantum-secure channels; regular pentesting
& \begin{minipage}{\linewidth}
\begin{myBullets}\vspace{2pt}
\item \textbf{Info. Disclosure}:\quad Eavesdropping on sensitive inter-\gls{VM} messages.
\item \textbf{Tampering}:\quad Inject or modify data in transit among VMs.
 \vspace{3pt} \end{myBullets}
\end{minipage}& \med & \high & \high \\ \cline{2-9}

& \begin{minipage}{\linewidth} BGP Route Hijacking \end{minipage}
& \begin{minipage}{\linewidth} Quantum-Forged \gls{RPKI}  Credentials \end{minipage}
& \begin{minipage}{\linewidth} 
Attackers use quantum capabilities to break or forge \gls{RPKI}  route certificates, 
announcing fraudulent BGP routes that reroute or blackhole traffic. 
\end{minipage}
& \begin{minipage}{\linewidth} \vspace{2pt} 
Adopt post-quantum \gls{RPKI}  signatures; 
monitor route announcements for anomalies; 
deploy BGPsec or route-filtering best practices.
\end{minipage}
& \begin{minipage}{\linewidth}
\begin{myBullets}\vspace{2pt} 
\item \textbf{Tampering}: Maliciously inject incorrect routes.
\item \textbf{Info. Disclosure}: Divert traffic for eavesdropping.
\item \textbf{DoS}: Hijacking can blackhole or disrupt critical paths.
 \vspace{3pt} \end{myBullets} 
\end{minipage}
& \med 
& \high 
& \high 
\\

\cline{2-9}
& \begin{minipage}{\linewidth} Quantum Side-Channel on Network Devices \end{minipage}
& \begin{minipage}{\linewidth} Advanced Side-Channel Analysis (e.g., QAML) \end{minipage}
& \begin{minipage}{\linewidth}
Attackers apply quantum-accelerated machine learning to glean cryptographic keys or credentials 
from routers, load balancers, or switches via timing/EM/power side-channels.
\end{minipage}
& \begin{minipage}{\linewidth}\vspace{0.1cm}
Deploy tamper-resistant \glspl{HSM}, \gls{SCA}-resistant firmware, \gls{PQC}-based cryptography, shielding, and operation randomization to 
mitigate \glsentryshort{EM} and power analysis\vspace{0.1cm}
\end{minipage}
& \begin{minipage}{\linewidth}
\begin{myBullets}\vspace{2pt} 
\item \textbf{Info. Disclosure}: Extract keys from network appliances.
\item \textbf{EoP}: Stolen keys let attackers impersonate or control devices.
 \vspace{3pt} \end{myBullets} 
\end{minipage}
& \low 
& \high 
& \med 
\\

\cline{2-9}
& \begin{minipage}{\linewidth} Quantum-Optimized DDoS/Route-Flooding \end{minipage}
& \begin{minipage}{\linewidth} Combinatorial Optimization \end{minipage}
& \begin{minipage}{\linewidth}
Attackers leverage quantum algorithms (QAOA, annealing) to identify 
optimal traffic patterns or paths, intensifying resource exhaustion or route flooding.
\end{minipage}
& \begin{minipage}{\linewidth}\vspace{0.1cm}
Adaptive DDoS protection (AI-based anomaly detection), 
\gls{SDN}-based dynamic rerouting, 
rate-limiting specifically tested against quantum-optimized patterns.\vspace{0.1cm}
\end{minipage}
& \begin{minipage}{\linewidth}
\begin{myBullets}\vspace{2pt} 
\item \textbf{DoS}: Flooded routes or links degrade or disrupt network availability.
 \vspace{3pt} \end{myBullets} 
\end{minipage}
& \med 
& \med 
& \med 
\\

\cline{2-9}
& \begin{minipage}{\linewidth} Quantum-Forged \gls{SDN} Controller Exploits \end{minipage}
& \begin{minipage}{\linewidth} Control Plane Compromise~\cite{abdou2018comparative} \end{minipage}
& \begin{minipage}{\linewidth}
Attackers break or forge the cryptographic channels securing \gls{SDN} controllers, injecting malicious flow rules that compromise the entire network fabric.

\end{minipage}
& \begin{minipage}{\linewidth}\vspace{0.1cm}
\gls{PQC} or ephemeral encryption for controller-plane communications; 
zero-trust validation of \gls{SDN} rules; 
integrity checks on flow-table updates.\vspace{0.1cm}
\end{minipage}
& \begin{minipage}{\linewidth}
\begin{myBullets}\vspace{2pt} 
\item \textbf{Tampering}: Inject flow rules to route traffic improperly.
\item \textbf{EoP}: Gains control of entire network via \gls{SDN}.
 \vspace{3pt} \end{myBullets} 
\end{minipage}
& \med 
& \high 
& \high
\\

\cline{2-9}
& \begin{minipage}{\linewidth} Quantum-Late Decryption of Network Tunnels \end{minipage}
& \begin{minipage}{\linewidth} \gls{HNDL}~\cite{barenkamp2022steal} \end{minipage}
& \begin{minipage}{\linewidth}
Attackers capture large volumes of encrypted traffic (VPN, TLS) now, 
planning to decrypt it in 10-15+ years once quantum hardware matures.
\end{minipage}
& \begin{minipage}{\linewidth}\vspace{0.1cm}
Immediate switch to \gls{PQC}/hybrid ciphers for all high-value communications; 
implement forward secrecy and ephemeral key exchanges.\vspace{0.1cm}
\end{minipage}
& \begin{minipage}{\linewidth}
\begin{myBullets}\vspace{2pt} 
\item \textbf{Info. Disclosure}: Future quantum decryption reveals historical traffic, credentials, or secrets.
 \vspace{3pt} \end{myBullets} 
\end{minipage}
& \med
& \high
& \high
\\ \hline
 \end{tabular}%
}
\label{table:Pre-Transition-Cont}
\vspace{-1cm}
 \end{table*}

\subsection{Risk Assessment of Pre-Transition Quantum Threats Across \gls{CC} Layers}
\label{sec:risk_assessment_pretransition}

This section provides a comprehensive risk assessment of the {pre-transition} quantum threat landscape across
different \gls{CC} layers, using the likelihood and impact criteria defined in
Tables~\ref{table:likelihood} and~\ref{table:impact}. All likelihood determinations assume a 15-year timeline
for the emergence of large-scale \gls{QC} capable of executing Shor's or Grover's algorithms to break classical
encryption.
The likelihood (L) of each pre-transition quantum threat is evaluated based on three key factors: (1) Intrinsic exploitability, which assesses whether known classical vulnerabilities (e.g., weak keys, legacy algorithms) can be accelerated by quantum techniques; (2) Timeline for large-scale \gls{QC}, determining whether the exploit strictly depends on Shor’s or Grover’s algorithms at scale (expected in 15+ years) or on more advanced capabilities such as \gls{QAML}; and (3) Existing mitigations, which include partial or hybrid \gls{PQC} deployment, robust side-channel defenses, and strong network segmentation. Based on these considerations, we categorize likelihood (L) as follows: (a) \textit{High (H)}, 
for threats with critical vulnerabilities (e.g., short RSA keys, unpatched software) and minimal 
quantum-safe protections, coupled with broad exposure or nation-state adversaries investing in 
quantum R\&D, implying straightforward exploitation once large-scale \gls{QC} emerges; 
(b) \textit{Medium (M)}, for threats partially mitigated by hybrid or longer-key cryptographic 
approaches (e.g., \gls{HNDL} scenarios), where adversaries rely on partial quantum power realistic 
in 15+ years—classical techniques remain effective, but quantum feasibility heightens concern; 
and (c) \textit{Low (L)}, for threats where systems have robust quantum readiness (PQC-based TLS, 
large symmetric keys, side-channel noise injection) or require advanced \gls{QAML} capabilities 
beyond the 15-year horizon, making exploits improbable under foreseeable constraints or deterred 
by high cost—proactive upgrades to AES-256 and SHA-3 significantly reduce feasibility, hence 
Grover-dependent threats (e.g., symmetric key weakening, hash pre-image attacks) are rated Low.



The {impact} ({I}) of each quantum threat is determined by three key factors: (1) Scope of data compromise, assessing whether highly sensitive information such as \gls{VM} images, PKI certificates, or financial records is exposed when cryptographic defenses fail; (2) Operational and reputational damage, evaluating the potential consequences, including extended downtime, lawsuits, regulatory non-compliance, or brand erosion; and (3) Cascade effect, considering whether the breach extends beyond a single \gls{VM} or application, such as a hypervisor exploit in a multi-tenant architecture. Based on these factors, we define {Impact (I)} as follows: (a) \textit{High (H)}, for attacks undermining critical cryptographic primitives, enabling complete data decryption or causing major reputational harm and persistent outages that may spread across core infrastructure layers; (b) \textit{Medium (M)}, for attacks that remain contained to a specific layer, leading to partial data leaks or localized slowdowns while core systems might remain intact; and (c) \textit{Low (L)}, for threats resulting in minimal disclosure or ephemeral disruptions quickly resolved by standard controls and confined to small user bases.
The overall risk is determined based on likelihood and impact levels using the risk matrix in Figure~\ref{fig:risk-matrix}. The comprehensive analysis for each vulnerability is detailed in Table~\ref{table:Pre-Transition}.

\section{Post-Transition Quantum Threat Landscape in \gls{CC}}
\label{sec:post}

The adoption of quantum-safe cryptography will significantly reshape the threat 
landscape across all layers of \gls{CC} as illustrated in Figure~\ref{fig:stack}. While \gls{PQC} and other quantum-secure algorithms address direct cryptographic vulnerabilities, their deployment introduces new integration risks, performance trade-offs, and residual attack vectors across interdependent cloud layers. This section examines evolving cyber risks during the post-transition phase, where \gls{PQC} mechanisms are operational but not yet fully mature or uniformly deployed. We assess how implementation flaws, hybrid-mode inconsistencies, and supply chain dependencies may compromise intended security guarantees. Each subsection analyzes layer-specific concerns and highlights areas requiring proactive hardening and continuous adaptation.
\subsection{Post-Transition Vulnerabilities}

The transition to \gls{PQC} mitigates direct quantum threats but introduces 
new vulnerabilities across cryptographic design, implementation, infrastructure, 
and adversarial exploitation. This subsection categorizes key vulnerabilities 
that emerge in the post-transition phase, forming the basis for subsequent 
attack vector analysis as detailed in Table~\ref{table:Post-Transition}.

\begin{enumerate}[topsep=0ex, itemsep=0ex, wide, font=\itshape, labelwidth=!, labelindent=0pt, label*=5.1.\arabic*]

\item \textit{Cryptographic Weaknesses.} 
\gls{PQC} schemes may exhibit structural weaknesses susceptible to classical and hybrid cryptanalysis~\cite{lounis2020attacks,NIST2024,baseri2025blockchain}. \glspl{SCA} (timing, power, electromagnetic analysis per Table~\ref{tab:post-alg}), cold boot attacks, and key migration failures pose significant risks~\cite{villanueva2020cold}. Quantum-assisted cryptanalysis and fault injection attacks pose additional threats~\cite{lounis2020attacks}. Key management risks (weak selection, entropy deficiencies, protocol downgrade) compromise security~\cite{xu2021magnifying,dwivedi2024novel,strenzke2024legacy,bhargavan2016downgrade}.

\item \textit{Implementation Vulnerabilities.} 
\gls{PQC} computational complexity increases memory corruption~\cite{cloosters2020teerex}, buffer overflow~\cite{akter2024integrated,bruinderink2018differential}, and fault injection~\cite{xagawa2021fault,kreuzer2020fault} risks. Optimized implementations may introduce side-channel leaks via cache-based attacks or data-dependent timing~\cite{kim2020cache,brumley2015cache, guo2020key,guo2022don,wafo2020practicable}. Inadequate \gls{API} security, parameter misconfigurations, and incomplete implementations create vulnerabilities~\cite{mousavi2023detecting}. Inappropriate logging hinders forensics~\cite{otmani2010cryptanalysis}.

\item \textit{Infrastructure and Protocol Risks.} 
Large \gls{PQC} key and ciphertext sizes affect storage, processing, and network performance, causing fragmentation, latency, and DoS risks~\cite{compastie2020virtualization, ficco2015stealthy}. \gls{PQC}-TLS, IPSec, and other protocols may suffer misconfigurations and key recovery threats~\cite{zhang2018novel, lounis2020attacks}. Poorly designed hybrid cryptosystems introduce inconsistencies and expanded attack surfaces~\cite{mansoor2025securing}. Storage systems face performance degradation and backup failures~\cite{joshi2018standards}.

\item \textit{Hardware and Supply Chain Threats.} 
Malicious implants and compromised hardware undermine \gls{PQC} security~\cite{10.1145/3634737.3657016}. \glspl{TEE} remain vulnerable to \glspl{SCA}~\cite{schellenberg2018remote}. Firmware exploitation and inadequate integrity checks expose systems to persistent threats~\cite{10.1145/3432893}. Micro-architectural vulnerabilities (Rowhammer, Spectre/Meltdown) complicate hardware security~\cite{shen2021micro, lou2021survey}.

\item \textit{Quantum-Enabled Adversary Capabilities.} 
Quantum-enhanced attacks include \gls{HNDL} strategies targeting long-term encrypted data~\cite{barenkamp2022steal}, quantum-assisted AI enhancing social engineering (deepfakes, phishing), quantum-powered malware and adversarial AI~\cite{arif2024future}, and quantum-assisted brute-force attacks on symmetric crypto and hash functions~\cite{baseri2025blockchain}.

\end{enumerate}
\subsection{Post-Transition Attack Vectors}

As \gls{PQC} transitions into deployment, its security landscape is shaped by emerging attack vectors exploiting \gls{PQC} implementations across \gls{CC} architecture levels. We categorize these attacks based on Table~\ref{table:Post-Transition}.

\begin{enumerate}[topsep=0ex, itemsep=0ex, wide, font=\itshape, labelwidth=!, labelindent=0pt, label*=5.2.\arabic*]

\item \textit{\acrlongpl{SCA}.} \gls{PQC} algorithms are vulnerable to \glspl{SCA} exploiting power consumption, electromagnetic radiation, cache states, and timing behavior~\cite{roy2022self, blomer2014tampering}. Microarchitectural exploits (cache attacks, branch prediction, row hammer-based fault injection) demonstrated against classical cryptosystems pose similar \gls{PQC} threats~\cite{shen2021micro,lou2021survey}. Optical and acoustic emissions could reveal secret keys~\cite{roy2022self}. Constant-time implementations and randomized masking mitigate these risks~\cite{schellenberg2018remote}.

\item \textit{Code Injection Attacks.} Buffer overflows, \gls{ROP}, and \gls{JOP} enable malicious payload injection into \gls{PQC} processes, causing privilege escalation or code execution~\cite{compastie2020virtualization, hoffman2020web, poulios2015ropinjector}. \gls{PQC} implementation complexity, especially in hybrid stacks, expands attack surfaces~\cite{mishra2025modern}. \gls{ASLR}, \gls{CFI}, and stack canaries are critical hardening measures~\cite{kim2022threat}.

\item \textit{File System Exploitation.} Attackers manipulate cryptographic key storage via race conditions or concurrency bottlenecks, inducing unauthorized file modifications~\cite{joshi2018standards}. Wiper malware targeting \gls{PQC} key stores disrupts cryptographic operations, necessitating secure enclaves and \glspl{TPM}~\cite{compastie2020virtualization}.

\item \textit{\gls{OS} and Kernel Attacks.} Kernel vulnerabilities (privilege escalation, memory corruption, side-channel leaks in process scheduling) compromise \gls{PQC} applications~\cite{compastie2020virtualization, hoffman2020web}. Race conditions in cryptographic libraries may enable predictable key derivation~\cite{masdari2016survey}. Kernel integrity monitoring and sandboxing are required~\cite{mishra2025modern}.

\item \textit{Firmware Attacks.} Firmware represents a major \gls{PQC} attack surface~\cite{10.1145/3432893,10.1145/3634737.3657016}. Supply-chain attacks, malicious OTA updates, and embedded backdoors threaten key management~\cite{294541}. Hardware-based validation (secure boot) is necessary~\cite{ThreatsProtection}.

\item \textit{Hypervisor Exploitation.} Virtualized \gls{PQC} deployments face \gls{VM} hopping, guest jumping, and hypervisor-based \gls{DoS} risks~\cite{compastie2020virtualization, almutairy2019taxonomy, huang2012security}. Misconfigured hypervisors enable \gls{VM} escape and \gls{PQC} exploits with escalated privileges~\cite{compastie2020virtualization}. Strict access controls and hardware-backed virtualization (Intel VT-x, AMD-V) reduce threats~\cite{mishra2025modern}.

\item \textit{\gls{VM} Migration Exploitation.} Live migration of VMs running \gls{PQC} workloads enables cryptographic secret interception if encryption is incomplete or misconfigured~\cite{9889331}. Attackers exploit memory snapshot leaks or migration traffic injection to manipulate cryptographic state~\cite{9889331}. Post-quantum TLS (OQS-OpenSSL~\cite{post-quantum-tls, open-quantum-safe-tls}, KEMTLS~\cite{schwabe2020post}) during migration ensures confidentiality.

\item \textit{\gls{TEE} Exploits.} \glspl{SCA} on Intel SGX, AMD SEV, and ARM TrustZone enable \gls{PQC} key extraction from secure enclaves~\cite{schellenberg2018remote}. Secure enclave rollback vulnerabilities enable key reuse attacks, necessitating quantum-safe \gls{TEE} designs with stronger attestation~\cite{tasso2021resistance}.

\item \textit{Resource Exhaustion in \gls{CC}.} \gls{PQC} computational overhead enables \gls{DoS} scenarios where attackers overwhelm cloud resources via repeated \gls{PQC} invocations~\cite{masdari2016survey}. Poorly optimized implementations exacerbate risks, requiring adaptive rate-limiting and hardware acceleration~\cite{jafarigiv2020scalable}.

\item \textit{Storage Exploitation and Data Remanence.} Large \gls{PQC} key sizes create storage/retrieval challenges, leading to out-of-bounds reads/writes exploitable for data extraction~\cite{joshi2018standards}. Cold boot attacks targeting \gls{PQC} memory remnants require volatile-only key storage~\cite{polanco2019cold}.

\item \textit{Logging and Audit Manipulation.} Attackers tamper with logs to erase \gls{PQC} exploit forensic evidence~\cite{compastie2020virtualization}. Log poisoning (injecting false cryptographic errors) misleads incident response~\cite{commomtypmalware}. Cryptographically verifiable logging (blockchain-based audit trails) enhances traceability~\cite{joshi2018standards}.

\item \textit{Network Exploitation in \gls{PQC} Protocols.} Network-based cryptanalysis of \gls{PQC} handshakes (PQC-TLS, PQC-VPN, PQC-IPSec) reveals downgrade attack vectors~\cite{zhang2018novel,bhargavan2016downgrade}. Fragmentation-based attacks on oversized key exchanges cause packet reassembly vulnerabilities~\cite{essay89509}. Strict protocol enforcement and quantum-resistant KEMs are crucial~\cite{NIST2024}.

\item \textit{Cryptanalysis.} Emerging techniques (lattice-reduction, quantum-assisted brute-force, mathematical backdoors) may compromise \gls{PQC}~\cite{zhang2018novel, otmani2010cryptanalysis}. Continuous algorithm evaluation and cryptographic agility remain necessary~\cite{marchesi2025survey}.
\end{enumerate}
Addressing quantum-security requires holistic protection across hardware, firmware, hypervisors, \gls{OS}, storage, and networking. These threats reinforce \emph{security-by-design} principles, necessitating the integration of quantum-resistant mitigations throughout the entire cloud computing stack.

\subsection{Post-Transition Quantum Threat Landscape in Each \gls{CC} Layer}

The transition to \gls{PQC} introduces new vulnerabilities across all \gls{CC} layers due to increased computational complexity, larger cryptographic payloads, emerging side-channel threats, insecure cryptographic implementations, resource exhaustion, and an expanded attack surface driven by evolving cryptanalytic techniques and system-level weaknesses. Mitigating these risks requires robust security mechanisms. Table~\ref{table:Post-Transition} details quantum-era threats and corresponding countermeasures for each layer.

\begin{enumerate}[topsep=0ex, itemsep=0ex, wide, font=\itshape, labelwidth=!, labelindent=0pt, label*=5.3.\arabic*]
\item \textit{Application Layer.} Vulnerable to multiple security threats including buffer overflow attacks, \glspl{SCA}, \gls{VM} exploitation, oversized cryptographic payloads, and command injection~\cite{compastie2020virtualization, hoffman2020web, masdari2016survey, kim2022threat}. Buffer overflow attacks occur when applications overwrite memory boundaries, allowing attackers to execute arbitrary code through injecting malicious payloads, often leading to privilege escalation~\cite{hoffman2020web}. Effective countermeasures include stack canaries, \gls{ASLR}, and \gls{CFI}~\cite{wang2016sigdrop}. \glspl{SCA} enable adversaries to infer cryptographic keys or sensitive data through power analysis, electromagnetic leakage, and timing variations~\cite{roy2022self, schellenberg2018remote, sayakkara2019survey, accikkapi2019side, genkin2017acoustic, le2019algebraic, breier2020countermeasure}. These threats can be mitigated using constant-time algorithms, hardware shielding, and cryptographic implementations resistant to differential power analysis~\cite{tasso2021resistance}. \gls{VM} exploitation occurs when attackers leverage vulnerabilities in \gls{VM}-based applications to escalate privileges or access unauthorized resources~\cite{compastie2020virtualization}, mitigated using hardened hypervisors, secure \gls{API} access controls, and proper \gls{VM} isolation~\cite{huang2012security}. Oversized cryptographic payloads can overload system resources, causing \gls{DoS}~\cite{ficco2015stealthy, masdari2016survey}, which can be mitigated by enforcing rate limiting and payload size restrictions.

\item \textit{Data Layer.} Faces security risks such as data remanence, integrity violations, cryptanalysis, and \gls{PQC} key migration risks~\cite{zhang2018novel, otmani2010cryptanalysis, lounis2020attacks}. Although \gls{PQC} mitigates decryption attacks, attackers can still exploit \glspl{SCA}  to leak sensitive data. Vendors and implementers must ensure secure hardware/software configurations. Side-channel vulnerabilities such as timing attacks and fault injection can extract cryptographic keys~\cite{blomer2014tampering,roy2022self}. Mitigation strategies include constant-time algorithms, noise injection, and tamper-resistant hardware. Data remanence occurs when residual information remains on storage devices even after deletion, leading to unintended data exposure~\cite{Threatinsiderdef}. Secure erasure techniques such as cryptographic wiping and physical destruction help prevent this risk~\cite{blomer2014tampering}. Integrity attacks involve unauthorized data modifications or the injection of corrupted data, which compromises system reliability~\cite{compastie2020virtualization}. Countermeasures include cryptographic hash functions, Merkle tree-based verification, and tamper-proof logging~\cite{otmani2010cryptanalysis}. The increasing risk posed by quantum algorithms to traditional encryption methods necessitates the transition to \gls{PQC}-resistant mechanisms~\cite{zhang2018novel}. Hybrid cryptographic approaches combining classical and quantum-safe encryption ensure long-term security~\cite{lounis2020attacks}.

\item \textit{Runtime Layer.} Susceptible to memory corruption attacks, \gls{JIT} compilation vulnerabilities, and insecure logging mechanisms~\cite{poulios2015ropinjector, blomer2014tampering}. Attackers may exploit out-of-bounds memory access or cryptographic overhead to compromise application runtimes. Runtime vulnerabilities such as buffer overflows, \gls{ROP}/\gls{JOP} attacks~\cite{poulios2015ropinjector,wang2016sigdrop}, and side-channel leaks can be mitigated by implementing stack canaries, \gls{CFI}, and memory-safe languages. Memory corruption exploits, such as \gls{ROP} and \gls{JOP}, manipulate memory execution flows to gain control over system processes~\cite{wang2016sigdrop, poulios2015ropinjector}. Secure execution environments utilizing \gls{CFI}, \gls{DEP}, and \gls{ASLR} mitigate these threats~\cite{kim2022threat}. Emerging quantum-based fuzzing techniques accelerate fault injection in runtime environments, leading to unpredictable execution behavior~\cite{blomer2014tampering, accikkapi2019side, schellenberg2018remote, genkin2017acoustic, le2019algebraic, breier2020countermeasure}, which can be countered through quantum-safe runtime validation and execution integrity monitoring~\cite{tasso2021resistance}. Insecure logging allows attackers to manipulate logs, either to erase traces of their activity or inject misleading records~\cite{Threatinsiderdef}. Ensuring secure logging through append-only records and cryptographic timestamping can mitigate these vulnerabilities.

\item \textit{Middleware Layer.} Exposed to micro-architectural vulnerabilities, hypervisor misconfiguration, and \gls{VM} hopping attacks~\cite{10.1145/3715001,compastie2020virtualization, NARAYANA20216465, anwar2017cross, saeed2020cross}. Middleware, which bridges the \gls{OS} and application layers, is vulnerable to buffer overflows and hypervisor misconfigurations. Continuous hypervisor security audits, resource monitoring, and advanced intrusion detection are essential to prevent tampering or \gls{DoS} attacks. Hypervisor exploitation and \gls{VM} hopping are key threats~\cite{compastie2020virtualization,almutairy2019taxonomy,huang2012security}, with countermeasures including strict \gls{VM} isolation and secure hypervisor configurations. Micro-architectural vulnerabilities, such as Rowhammer and transient execution leaks (e.g., Spectre and Meltdown), compromise data integrity and allow attackers to read sensitive information across process boundaries~\cite{shen2021micro, lou2021survey}. Cache partitioning, speculative execution barriers, and memory fencing are recommended countermeasures~\cite{gruss2019page, esfahani2021enhanced}. \gls{VM} hopping occurs when adversaries exploit resource misconfigurations within the middleware, depriving legitimate \gls{VM}s of resources~\cite{almutairy2019taxonomy, huang2012security}. Dynamic resource allocation and robust \gls{ACL} effectively mitigate this risk~\cite{9889331}.

\item \textit{Virtualization Layer.} Subject to \gls{VM} image tampering, hyperjacking, and \gls{vTPM} exploitation~\cite{compastie2020virtualization, mishra2025modern, sun2025survey}. Shared hardware resources (e.g., cache, memory) facilitate potent \glspl{SCA}. \gls{VM} escape, hyperjacking, and micro-architectural vulnerabilities are key threats~\cite{compastie2020virtualization,shen2021micro}. Layered micro-architectural defenses, periodic patching, and noise generation can mitigate cross-\gls{VM} breaches and \gls{DoS} attacks caused by \gls{PQC} overhead. \gls{VM} image tampering occurs when attackers manipulate virtual machine snapshots to introduce persistent malware or backdoors~\cite{rawal2023malware, commomtypmalware}. To prevent this, cryptographic integrity verification mechanisms should be used to ensure \gls{VM} image authenticity before deployment~\cite{muller2020retrofitting}. Hyperjacking is an advanced attack in which adversaries replace or compromise the hypervisor itself, granting them complete control over all hosted \gls{VM}s~\cite{compastie2020virtualization}. Countermeasures such as Secure Boot, runtime hypervisor monitoring, and \glspl{TEE} help mitigate this threat~\cite{sun2025survey}.

\item \textit{Server Layer.} Firmware vulnerabilities, \glspl{SCA}, and outdated boot processes remain pressing concerns. Server vendors must adopt side-channel-resistant designs~\cite{schellenberg2018remote,sayakkara2019survey,accikkapi2019side,genkin2017acoustic,le2019algebraic,breier2020countermeasure}, secure firmware updates, and rigorous code attestation to mitigate firmware exploitation~\cite{10.1145/3432893}, hardware supply chain attacks, and \gls{TEE} vulnerabilities.

\item \textit{Storage Layer.} Buffer overflows and oversized cryptographic payloads can fuel large-scale wiper or ransomware attacks, compromising entire storage subsystems. \gls{PQC}-induced storage risks include key size strain and fault injection\break attacks~\cite{craciun2019trends,marchesi2025survey}. Implementers must deploy cryptographic agility, data redundancy, and continuous integrity checks to secure storage systems.

\item \textit{Network Layer.} Susceptible to oversized cryptographic payload risks, adaptive encryption protocol exploits, and traffic analysis attacks~\cite{jafarigiv2020scalable, muller2020retrofitting, essay89509}. Fragmentation of oversized \gls{PQC} packets and cryptanalysis tactics threaten data in transit. Table~\ref{table:Post-Transition-Networking} identifies \gls{PQC} cryptanalysis, \glspl{SCA}, and misconfiguration exploits as key network vulnerabilities~\cite{lounis2020attacks,blomer2014tampering,accikkapi2019side,schellenberg2018remote,genkin2017acoustic,le2019algebraic,breier2020countermeasure,tasso2021resistance}. Employing agile encryption protocols, robust network management, and thorough packet inspection ensures secure communication~\cite{marchesi2025survey,ISACAQuantumThreat}. Traffic analysis attacks allow adversaries to infer user behavior or data patterns even when encryption is in place~\cite{sayakkara2019survey}. Padding, traffic obfuscation, and morphing techniques serve as effective countermeasures against such risks.
\end{enumerate}

Although \gls{PQC} algorithms mitigate large-scale cryptanalytic breaks, post-transition systems remain exposed to implementation flaws, micro-architectural leakage, side-channel exploitation, migration-induced supply-chain and deployment risks, and \gls{PQC}-amplified resource exhaustion. A defense-in-depth posture prioritizing cryptographic agility~\cite{marchesi2025survey}, memory safety, and continuous monitoring is essential for quantum-era resilience. This taxonomy across the nine \gls{CC} layers provides a structured foundation for aligning mitigation strategies with the evolving post-quantum threat landscape (see Table~\ref{table:Post-Transition}).

\begin{table*}[!htbp]
\caption{Post-Transition Quantum Threat Landscape in Each \gls{CC} Layer}\label{table:Post-Transition}

\large
\renewcommand{\arraystretch}{1.1}
\resizebox{\linewidth}{!}{%
\begin{tabular}{|p{0.13\linewidth}|p{0.25\linewidth}|p{0.32\linewidth}|p{0.45\linewidth}|p{0.47\linewidth}|p{0.72\linewidth}|l|l|l|}
\hline 
\textbf{Layer} & \textbf{Vulnerability} & \textbf{Exploitation Method} & \textbf{Attack Vector} & \textbf{Countermeasures} & \textbf{STRIDE Threat} & {\textbf{L}} & {\textbf{I}} & {\textbf{R}}\\ \hline 

\multirow{30}{*}{\begin{minipage}{\linewidth} Application \end{minipage}} 
& \multirow{4}{*}{\begin{minipage}{\linewidth} Buffer Overflow Vulnerabilities \end{minipage}} 
& \begin{minipage}{\linewidth} Application Software Exploitation~\cite{compastie2020virtualization,hoffman2020web} \end{minipage} 
& \begin{minipage}{\linewidth} Code injection exploiting memory buffer limits in \gls{PQC} cryptographic operations. \end{minipage} 
& \begin{minipage}{\linewidth} Use memory-safe programming languages, enforce strict buffer allocation policies, implement automatic bounds-checking. \end{minipage} 
& \begin{minipage}{\linewidth} 
\begin{myBullets}\vspace{2pt} 
\item \textbf{Tampering:} Attackers modify application execution.
\item \textbf{Info. Disclosure:} Memory overflows expose sensitive cryptographic data.
\item \textbf{EoP:} Exploiting buffer overflows for arbitrary code execution.
\vspace{3pt} \end{myBullets} 
\end{minipage}&\med&\high&\high \\ \cline{3-9}









& 
& \begin{minipage}{\linewidth} \gls{VM} Monitoring from \gls{VM}~\cite{compastie2020virtualization} \end{minipage} 
& \begin{minipage}{\linewidth} \glspl{SCA} leveraging buffer overflows to bypass cryptographic security. \end{minipage} 
& \begin{minipage}{\linewidth} Implement constant-time cryptographic execution, deploy noise generation for side-channel resistance. \end{minipage} 
& \begin{minipage}{\linewidth} 
\begin{myBullets}\vspace{2pt} 
\item \textbf{Tampering:} Attackers manipulate cryptographic operations.
\item \textbf{Info. Disclosure:} Side-channel leaks expose \gls{PQC} secret keys.
\item \textbf{DoS:} Overloaded \gls{PQC} cryptographic computation crashes the system.
\vspace{3pt} \end{myBullets} 
\end{minipage}&\med&\med&\med\\ \cline{3-9}








& 
& \begin{minipage}{\linewidth} Return-Oriented Programming (ROP) Exploitation~\cite{poulios2015ropinjector} \end{minipage} 
& \begin{minipage}{\linewidth} Hijacking \gls{PQC} memory stacks via buffer overflows to execute malicious code. \end{minipage} 
& \begin{minipage}{\linewidth} Deploy stack canaries, enable \gls{ASLR}, and use \gls{CFI} protections. \end{minipage} 
& \begin{minipage}{\linewidth} 
\begin{myBullets}\vspace{2pt} 
\item \textbf{EoP:} Gaining unauthorized control over execution.
\item \textbf{Tampering:} Injecting unauthorized instructions.
\item \textbf{DoS:} Corrupting system state to disrupt \gls{PQC} execution.
\vspace{3pt} \end{myBullets} 
\end{minipage} &\med&\high&\high\\ \cline{2-9}








& \multirow{2}{*}{\begin{minipage}{\linewidth} Side-Channel Vulnerabilities \end{minipage}} 
& \begin{minipage}{\linewidth} Fault Injection Attacks~\cite{blomer2014tampering} \end{minipage} 
& \begin{minipage}{\linewidth} Exploiting \gls{PQC} cryptographic computations via power or timing analysis. \end{minipage} 
& \begin{minipage}{\linewidth} Implement constant-time execution, masking, blinding, leverage hardware security modules (HSMs), and conduct side-channel resistance testing. \end{minipage} 
& \begin{minipage}{\linewidth} 
\begin{myBullets}\vspace{2pt} 
\item \textbf{Tampering:} Manipulating execution via power fluctuations.
\item \textbf{Info. Disclosure:} Extracting cryptographic secrets via timing attacks.
\item \textbf{EoP:} Using leaked cryptographic keys for unauthorized access.
\vspace{3pt} \end{myBullets} 
\end{minipage} &\med&\med&\med\\ \cline{2-9}









& \multirow{2}{*}{\begin{minipage}{\linewidth} VM/Hypervisor Exploitation \end{minipage}} 
& \begin{minipage}{\linewidth} \gls{VM} Escape, \gls{VM} Hopping~\cite{compastie2020virtualization} \end{minipage} 
& \begin{minipage}{\linewidth} Exploiting misconfigured hypervisors handling \gls{PQC} workloads to gain unauthorized \gls{VM} access. \end{minipage} 
& \begin{minipage}{\linewidth} Enforce strict hypervisor security, implement \gls{VM} isolation policies, and monitor hypervisor cryptographic workloads for anomalies. \end{minipage} 
& \begin{minipage}{\linewidth} 
\begin{myBullets}\vspace{2pt} 
\item \textbf{EoP:} Gaining root control over hypervisors or escaping to unauthorized VMs.
\item \textbf{Tampering:} Modifying \gls{VM} memory and cryptographic processes.
\item \textbf{DoS:} Overloading hypervisor cryptographic processing with excessive \gls{PQC} tasks.
\vspace{3pt} \end{myBullets} 
\end{minipage}&\low&\high&\med \\ \cline{2-9}










& \multirow{2}{*}{\begin{minipage}{\linewidth} Insufficient \gls{PQC}-Aware Intrusion Detection \end{minipage}} 
& \begin{minipage}{\linewidth} Network-Based \gls{PQC} Traffic Anomalies \end{minipage} 
& \begin{minipage}{\linewidth} Attackers evade detection by misusing \gls{PQC}-encrypted traffic patterns to hide malicious actions. \end{minipage} 
& \begin{minipage}{\linewidth} Deploy \gls{PQC}-aware anomaly-based intrusion detection systems (IDS), leverage machine learning for behavioral threat detection. \end{minipage} 
& \begin{minipage}{\linewidth} 
\begin{myBullets}\vspace{2pt} 
\item \textbf{Info. Disclosure:} Attackers use encrypted \gls{PQC} traffic to exfiltrate data.
\item \textbf{Tampering:} Manipulation of \gls{PQC}-based network flows to bypass security controls.
\item \textbf{DoS:} Misuse of \gls{PQC}-heavy operations to degrade IDS efficiency.
\vspace{3pt} \end{myBullets} 
\end{minipage}&\med&\med&\med \\ \cline{2-9}










& \multirow{2}{*}{\begin{minipage}{\linewidth} Insecure Network Endpoints \end{minipage}} 
& \begin{minipage}{\linewidth} Wiper Malware Injection~\cite{commomtypmalware} \end{minipage} 
& \begin{minipage}{\linewidth} Exploiting vulnerabilities in \gls{PQC}-encrypted network channels to insert malware. \end{minipage} 
& \begin{minipage}{\linewidth} Implement strong endpoint authentication, use TLS with \gls{PQC} cipher suites, and deploy anomaly-based intrusion detection. \end{minipage} 
& \begin{minipage}{\linewidth} 
\begin{myBullets}\vspace{2pt} 
\item \textbf{Tampering:} Malware modifies system state.
\item \textbf{Info. Disclosure:} Attackers steal encrypted data before wiping systems.
\item \textbf{DoS:} Wiping system-critical files results in service disruption.
\vspace{3pt} \end{myBullets} 
\end{minipage}&\med&\high&\high \\ \cline{2-9}










& \multirow{2}{*}{\begin{minipage}{\linewidth} Insecure Cryptographic Implementations \end{minipage}} 
& \begin{minipage}{\linewidth} \gls{VM} Monitoring from Host~\cite{compastie2020virtualization} \end{minipage} 
& \begin{minipage}{\linewidth} \glspl{SCA} targeting insecure \gls{PQC} key management in \gls{VM} memory. \end{minipage} 
& \begin{minipage}{\linewidth} Implement strict key lifecycle management, secure cryptographic storage, and entropy validation. \end{minipage} 
& \begin{minipage}{\linewidth} 
\begin{myBullets}\vspace{2pt} 
\item \textbf{Tampering:} Attackers modify encrypted data.
\item \textbf{Info. Disclosure:} Leaked cryptographic keys enable decryption attacks.
\item \textbf{DoS:} Cryptographic processing overload causes performance degradation.
\vspace{3pt} \end{myBullets} 
\end{minipage} &\med&\med&\med\\ \cline{2-9}










& \multirow{2}{*}{\begin{minipage}{\linewidth} Oversized Cryptographic Payloads \end{minipage}} 
& Oversized Cryptography~\cite{ficco2015stealthy,masdari2016survey} 
& \begin{minipage}{\linewidth} Increased \gls{PQC} key/ciphertext size overwhelms network bandwidth and application processing.
\end{minipage} 
& \begin{minipage}{\linewidth} Optimize \gls{PQC} implementations, use buffer management techniques, and enforce adaptive rate limiting.
\end{minipage} 
& \begin{minipage}{\linewidth} 
\begin{myBullets}\vspace{2pt} 
\item \textbf{DoS:} Excessive cryptographic load exhausts cloud and application resources.
\item \textbf{Tampering:} Maliciously crafted oversized cryptographic payloads disrupt execution.
\item \textbf{Info. Disclosure:} Large payload mismanagement may expose sensitive \gls{PQC} data.
\vspace{3pt} \end{myBullets} \end{minipage}&\med&\med&\med \\ \hline










\multirow{30}{*}{\begin{minipage}{\linewidth} Data \end{minipage}} 

& \multirow{2}{*}{\begin{minipage}{\linewidth} \glspl{SCA} \end{minipage}} 
& \begin{minipage}{\linewidth} Timing attacks, power analysis, memory access pattern inference~\cite{roy2022self} \end{minipage} 
& \begin{minipage}{\linewidth} Passive monitoring of cryptographic operations to extract sensitive data \end{minipage} 
& \begin{minipage}{\linewidth} Implement constant-time algorithms, noise injection, and memory protection techniques \end{minipage} 
& \begin{minipage}{\linewidth} 
 \begin{myBullets}\vspace{2pt} 
 \item \textbf{Info. Disclosure:} Extraction of cryptographic keys and sensitive data.
 \item \textbf{EoP:} Bypassing authentication using side-channel observations.
 \vspace{3pt} \end{myBullets} 
\end{minipage}&\med&\med&\med \\ \cline{3-9}



& 
& \begin{minipage}{\linewidth} Fault injection, optical attacks~\cite{blomer2014tampering} \end{minipage} 
& \begin{minipage}{\linewidth} Exploiting hardware vulnerabilities to induce errors in cryptographic computations \end{minipage} 
& \begin{minipage}{\linewidth}\vspace{0.1cm} Use tamper-resistant hardware, implement error detection and correction mechanisms \vspace{0.1cm} \end{minipage} 
& \begin{minipage}{\linewidth} 
 \begin{myBullets}\vspace{2pt} 
 \item \textbf{Tampering:} Manipulating cryptographic operations.
 \item \textbf{Info. Disclosure:} Extracting sensitive data via induced errors.
 \vspace{3pt} \end{myBullets} 
\end{minipage}&\low&\high&\med \\ \cline{2-9}



& \multirow{2}{*}{\begin{minipage}{\linewidth} Data Remanence \end{minipage}} 
& \begin{minipage}{\linewidth} Cold boot attacks, memory retention exploits~\cite{compastie2020virtualization} \end{minipage} 
& \begin{minipage}{\linewidth} Extraction of residual cryptographic keys or sensitive data from memory \end{minipage} 
& \begin{minipage}{\linewidth} Implement secure data erasure protocols, cryptographic key zeroization, and encryption of data at rest \end{minipage} 
& \begin{minipage}{\linewidth} 
 \begin{myBullets}\vspace{2pt} 
 \item \textbf{Tampering:} Manipulating memory retention settings for future exploitation.
 \item \textbf{Info. Disclosure:} Extraction of sensitive data using forensic techniques.

 \vspace{3pt} \end{myBullets} 
\end{minipage} &\low&\high&\med \\ \cline{3-9}



& 
& \begin{minipage}{\linewidth} Cryptographic key leakage via memory scraping \end{minipage} 
& \begin{minipage}{\linewidth} Attackers analyze memory dumps to extract \gls{PQC} keys \end{minipage} 
& \begin{minipage}{\linewidth} \vspace{0.1cm} Secure memory allocation, enforce \gls{PQC}-safe key storage techniques, and leverage hardware security modules (HSMs) \vspace{0.1cm} \end{minipage} 
& \begin{minipage}{\linewidth} 
 \begin{myBullets}\vspace{2pt} 
 \item \textbf{Tampering:} Malicious actors modify data retention configurations.
 \item \textbf{Info. Disclosure:} Memory scraping exposes cryptographic material.

 \vspace{3pt} \end{myBullets} 
\end{minipage}&\low&\high&\med \\ \cline{2-9}



& \begin{minipage}{\linewidth} {Oversized Cryptographic Payloads} \end{minipage} 
& \begin{minipage}{\linewidth} Oversized \gls{PQC} ciphertext injection, cryptographic exhaustion attacks~\cite{ficco2015stealthy,masdari2016survey} \end{minipage} 
& \begin{minipage}{\linewidth} Overwhelming processing units with large \gls{PQC}-encrypted payloads \end{minipage} 
& \begin{minipage}{\linewidth} \vspace{0.1cm} Optimize cryptographic processing, enforce payload size constraints, and use hardware accelerators \vspace{0.1cm} \end{minipage} 
& \begin{minipage}{\linewidth} 
 \begin{myBullets}\vspace{2pt} 
 \item \textbf{Tampering:} Manipulating cryptographic processing queues.
 \item \textbf{DoS:} Exhausting CPU and memory resources.

 \vspace{3pt} \end{myBullets} 
\end{minipage} &\med&\med&\med\\ \cline{2-9}



& \begin{minipage}{\linewidth} {Data Integrity Attacks~\cite{li2024secure}} \end{minipage} 
& \begin{minipage}{\linewidth} Tampering with encrypted data in transit or at rest \end{minipage} 
& \begin{minipage}{\linewidth} Exploiting weak integrity protections in \gls{PQC} implementations \end{minipage} 
& \begin{minipage}{\linewidth} Implement strong integrity checks (e.g., HMAC), use authenticated encryption, and enforce secure data transmission protocols \end{minipage} 
& \begin{minipage}{\linewidth} 
 \begin{myBullets}\vspace{2pt} 
 \item \textbf{Tampering:} Manipulating encrypted data to induce errors or bypass integrity checks.
 \item \textbf{Info. Disclosure:} Exploiting weak integrity protections to access sensitive data.
 \vspace{3pt} \end{myBullets} 
\end{minipage} &\low&\high&\med\\ \cline{2-9}



& \begin{minipage}{\linewidth} {Cryptanalysis and Algorithm Downgrade} \end{minipage} 
& \begin{minipage}{\linewidth} Quantum cryptanalysis, cryptographic downgrade exploits~\cite{zhang2018novel,otmani2010cryptanalysis} \end{minipage} 
& \begin{minipage}{\linewidth} Forcing systems to use weak classical cryptographic protocols or leveraging quantum algorithms to break encryption \end{minipage} 
& \begin{minipage}{\linewidth} Enforce cryptographic agility; disable legacy cipher suites; and mandate hybrid or post-quantum \gls{KEM/ENC} with downgrade-resistant negotiation. \end{minipage} 
& \begin{minipage}{\linewidth} 
 \begin{myBullets}\vspace{2pt} 

 \item \textbf{Tampering:} Manipulating key negotiation to downgrade security.
 \item \textbf{Repudiation:} Attackers can forge digital signatures, undermining cryptographic non-repudiation.
 \item \textbf{Info. Disclosure:} Decryption of previously secure data.
 \item \textbf{EoP:} Unauthorized decryption of authentication tokens.
 \vspace{3pt} \end{myBullets} 
\end{minipage} &\low&\high&\med\\ \cline{2-9}



& \begin{minipage}{\linewidth} {Key Management Vulnerabilities} \end{minipage} 
& \begin{minipage}{\linewidth} Exploiting weak PRNGs, insecure key storage~\cite{compastie2020virtualization} \end{minipage} 
& \begin{minipage}{\linewidth} Extracting cryptographic keys via \glspl{SCA} or memory exploits \end{minipage} 
& \begin{minipage}{\linewidth} \vspace{0.1cm}Use cryptographically secure PRNGs, enforce secure key lifecycle management, and leverage hardware-based key storage (e.g., HSMs) \vspace{0.1cm}\end{minipage} 
& \begin{minipage}{\linewidth} 
 \begin{myBullets}\vspace{2pt} 
 \item \textbf{Tampering:} Manipulating key generation or storage mechanisms.
 \item \textbf{Info. Disclosure:} Extraction of cryptographic keys.
 \vspace{3pt} \end{myBullets} 
\end{minipage}&\low&\high&\med \\ \cline{2-9}










& \begin{minipage}{\linewidth} { \gls{PQC} Key Migration Risks} \end{minipage} 
& \begin{minipage}{\linewidth} Exploiting mismanaged key migration processes or hybrid cryptographic systems \end{minipage} 
& \begin{minipage}{\linewidth} Downgrade attacks, key leakage during migration, or improper key destruction \end{minipage} 
& \begin{minipage}{\linewidth} Implement secure key migration protocols, use hybrid cryptographic systems during transition, and conduct regular audits of key management processes \end{minipage} 
& \begin{minipage}{\linewidth} 
 \begin{myBullets}\vspace{2pt} 
 \item \textbf{Tampering:} Manipulating key migration processes to downgrade security.
 \item \textbf{Repudiation:} Improper key destruction undermines non-repudiation.
 \item \textbf{Info. Disclosure:} Leakage of cryptographic keys during migration.
 \vspace{3pt} \end{myBullets} 
\end{minipage}&\low&\high&\med \\ \hline










\multirow{40}{*}{ {Runtime}}

& \begin{minipage}{\linewidth}Out-of-bounds Memory Access\end{minipage} 
& \begin{minipage}{\linewidth}Command Injection~\cite{kim2022threat,hoffman2020web}, Application Software Exploitation~\cite{compastie2020virtualization,hoffman2020web}\end{minipage} 
& \begin{minipage}{\linewidth}Buffer overflow leading to out-of-bounds memory access, code injection via buffer overflow or side-channels\end{minipage} 
& \begin{minipage}{\linewidth} \vspace{0.1cm}Memory protection (\gls{ASLR}, DEP/NX-bit), use of memory-safe languages (e.g., Rust), frequent patches/updates, stack canaries, \gls{CFI}, strict secure coding standards (CERT C/C++)\end{minipage} \vspace{0.1cm}
& \begin{minipage}{\linewidth}
 \begin{myBullets}\vspace{2pt} 
 \item \textbf{Spoofing:} Impersonating processes via memory manipulation
 \item \textbf{Tampering:} Corrupting memory to alter execution flow
 \item \textbf{Info. Disclosure:} Extracting confidential data from memory
 \item \textbf{EoP:} Gaining higher privileges through memory exploits
 \vspace{3pt} \end{myBullets}
\end{minipage} &\med&\high&\high
\\ \cline{2-9}











& \begin{minipage}{\linewidth}Memory Corruption Vulnerabilities\end{minipage} 
& \begin{minipage}{\linewidth}Application Software Exploitation~\cite{compastie2020virtualization,hoffman2020web}\end{minipage} 
& \begin{minipage}{\linewidth}Code injection via malware or \glspl{SCA} leading to memory corruption\end{minipage} 
& \begin{minipage}{\linewidth} \vspace{0.1cm}Compile-time hardening (FORTIFY\_SOURCE, RELRO), static/dynamic code analysis, code-signing enforcement, granular runtime checks (pointer bounds/type safety). \vspace{0.1cm}\end{minipage}
& \begin{minipage}{\linewidth}
 \begin{myBullets}\vspace{2pt} 
 \item \textbf{Spoofing:} Impersonating legitimate processes 
 \item \textbf{Tampering:} Memory corruption to alter execution flow
 \item \textbf{Info. Disclosure:} Extracting confidential data from memory
 \item \textbf{EoP:} Gaining higher privileges via memory exploits
 \vspace{3pt} \end{myBullets}
\end{minipage}&\med&\high&\high
\\ \cline{2-9}








& \begin{minipage}{\linewidth}Program Stack Integrity Compromise\end{minipage} 
& \begin{minipage}{\linewidth}Return Oriented Programming (ROP)~\cite{wang2016sigdrop,poulios2015ropinjector}\end{minipage} 
& \begin{minipage}{\linewidth}Buffer overflow leading to program stack control and corruption for \gls{ROP} attacks\end{minipage} 
& \begin{minipage}{\linewidth}\vspace{0.1cm} \gls{CFI}, stack canaries, shadow stacks, pointer authentication (ARM PA), frequent fuzz testing with \gls{ROP} detection.\vspace{0.1cm} \end{minipage}
& \begin{minipage}{\linewidth}
 \begin{myBullets}\vspace{2pt} 
 \item \textbf{Spoofing:} Impersonating processes by hijacking call sequences
 \item \textbf{Tampering:} Hijacking the stack to alter execution flow
 \item \textbf{Info. Disclosure:} Exfiltration of memory-resident data
 \item \textbf{EoP:} \gls{ROP} exploits for privileged code execution
 \vspace{3pt} \end{myBullets}
\end{minipage}&\med&\high&\high
\\ \cline{2-9}










& \begin{minipage}{\linewidth}Buffer Overflow Vulnerabilities\end{minipage} 
& \begin{minipage}{\linewidth}Application Software Exploitation~\cite{compastie2020virtualization,hoffman2020web}, Data Theft~\cite{Threatinsiderdef}, Command Injections~\cite{kim2022threat,hoffman2020web}\end{minipage} 
& \begin{minipage}{\linewidth}Code injection via buffer overflow, out-of-bounds access, leading to various exploits\end{minipage} 
& \begin{minipage}{\linewidth}Memory-safe languages/libraries, strict coding standards (MISRA, CERT), proactive scanning (SAST, DAST), intrusion detection with exploit signature detection.\end{minipage}
& \begin{minipage}{\linewidth}
 \begin{myBullets}\vspace{2pt} 
 \item \textbf{Spoofing:} Impersonating legitimate processes via memory manipulation
 \item \textbf{Tampering:} Memory corruption to alter execution flow
 \item \textbf{EoP:} Gaining higher privileges 
 \item \textbf{Info. Disclosure:} Extracting confidential data 
 \vspace{3pt} \end{myBullets}
\end{minipage}&\med&\high&\high
\\ \cline{2-9}










& \begin{minipage}{\linewidth}Memory Access \& Root-Level Execution\end{minipage} 
& \begin{minipage}{\linewidth}\gls{VM} Escape to \gls{VM}~\cite{compastie2020virtualization}\end{minipage} 
& \begin{minipage}{\linewidth}Out-of-bounds access via buffer overflow or code injection at root level for \gls{VM} escape\end{minipage} 
& \begin{minipage}{\linewidth}\vspace{0.1cm}Hardened hypervisor/\gls{VM} boundary checks, strict root-level access controls (SELinux, AppArmor), continuous \gls{VM} isolation monitoring, least privilege for hypervisor \& root processes.\vspace{0.1cm}\end{minipage}
& \begin{minipage}{\linewidth}
 \begin{myBullets}\vspace{2pt} 
 \item \textbf{Spoofing:} Impersonating legitimate system services
 \item \textbf{Tampering:} Corrupting memory to bypass isolation 
 \item \textbf{Info. Disclosure:} Accessing memory from co-resident VMs 
 \item \textbf{EoP:} Gaining root privileges 
 \vspace{3pt} \end{myBullets}
\end{minipage}&\low&\high&\med
\\ \cline{2-9}














& \begin{minipage}{\linewidth}Command Injection Susceptibility\end{minipage}
& \begin{minipage}{\linewidth}Command Injections~\cite{kim2022threat,hoffman2020web}, Inadequate Input Validation~\cite{kim2022threat,hoffman2020web}, Insecure Command Execution~\cite{kim2022threat,hoffman2020web}\end{minipage} 
& \begin{minipage}{\linewidth}Buffer overflow, inadequate input validation leading to unauthorized command execution\end{minipage}
& \begin{minipage}{\linewidth}Input validation \& parameterized queries, boundary checks \& sanitization, least privilege for command executors, secure coding (OWASP).\end{minipage}
& \begin{minipage}{\linewidth}
 \begin{myBullets}\vspace{2pt} 
 \item \textbf{Spoofing:} Impersonating legitimate processes 
 \item \textbf{Tampering:} Injecting malicious commands 
 \item \textbf{Info. Disclosure:} Data leaks via poorly sanitized input
 \item \textbf{EoP:} Executing privileged actions
 \vspace{3pt} \end{myBullets}
\end{minipage}&\med&\high&\high
\\ \cline{2-9}










& \begin{minipage}{\linewidth}Just-in-Time (JIT) Compilation Attacks\end{minipage}
& \begin{minipage}{\linewidth}Runtime/JIT Code Injection, Runtime Process Injection\end{minipage}
& \begin{minipage}{\linewidth}Exploiting JIT compilation to inject malicious code and bypass security controls\end{minipage}
& \begin{minipage}{\linewidth}JIT signing/hardening, CFI for JIT-generated code, restricting JIT usage to trusted modules, dynamic analysis \& JIT behavior monitoring.\end{minipage}
& \begin{minipage}{\linewidth}
 \begin{myBullets}\vspace{2pt} 
 \item \textbf{Spoofing:} Impersonating legitimate processes via JIT injection
 \item \textbf{Tampering:} Modifying application behavior via injected JIT code
 \item \textbf{EoP:} Executing unauthorized privileged actions
 \item \textbf{Info. Disclosure:} Data leaks due to unverified JIT code
 \vspace{0.2cm} \end{myBullets}
\end{minipage}&\med&\high&\high
\\ \cline{2-9}










& \multirow{8}{*}{\begin{minipage}{\linewidth}Side-Channel Vulnerabilities\end{minipage}}
& \begin{minipage}{\linewidth}\gls{VM} monitoring from \gls{VM}/host~\cite{compastie2020virtualization}, 
\gls{SCA} leakage through \glsentryshort{TA}, \glsentryshort{APA}, \glsentryshort{EM}, and 
optical side-channels in multi-tenant \gls{VM} environments~\cite{mushtaq2018run,roy2022self}\end{minipage}
& \begin{minipage}{\linewidth} \glspl{SCA} exploiting \gls{VM} co-residency, timing variations, optical emanations to leak data from cryptographic operations\end{minipage}
& \begin{minipage}{\linewidth}\vspace{0.1cm}Side-channel mitigation (noise injection, constant-time algorithms), secure coding practices (avoid data-dependent timing), hardware-level isolation (Intel SGX/AMD SEV), regular security patches \& microcode updates, memory protection (encrypt sensitive buffers). \vspace{0.1cm}\end{minipage}
& \begin{minipage}{\linewidth}
 \begin{myBullets}\vspace{2pt} 
 \item \textbf{Tampering:} Manipulating shared resources to enhance attacks
 \item \textbf{Info. Disclosure:} Leaking cryptographic keys \& data
 \item \textbf{DoS:} Excessive timing-based or power-based attacks
 \vspace{3pt} \end{myBullets}
\end{minipage}&\med&\med&\med
\\ \cline{3-9}









&
& \begin{minipage}{\linewidth}Injecting Spyware~\cite{rawal2023malware,commomtypmalware}, Injecting Keylogger~\cite{commomtypmalware}\end{minipage} 
& \begin{minipage}{\linewidth} \glspl{SCA} leading to unauthorized execution via memory corruption and info leakage\end{minipage} 
& \begin{minipage}{\linewidth}\vspace{0.1cm} Runtime memory encryption \& process isolation, behavior-based anti-spyware/keylogger, hardware shielding \& side-channel noise.\vspace{0.1cm} \end{minipage}
& \begin{minipage}{\linewidth}
 \begin{myBullets}\vspace{2pt} 
 \item \textbf{Tampering:} Altering runtime memory to embed spyware
 \item \textbf{Info. Disclosure:} Capturing keystrokes \& sensitive data
 \item \textbf{DoS:} Bogging down system with constant monitoring
 \vspace{3pt} \end{myBullets}
\end{minipage}&\low&\high&\med
\\ \cline{3-9}



& 
& \begin{minipage}{\linewidth}Side-channel/Cryptanalysis Attacks inc. Optical Attack~\cite{roy2022self}\end{minipage} 
& \begin{minipage}{\linewidth}Passive monitoring (power, EM, timing), memory access pattern inference to extract cryptographic secrets\end{minipage} 
& \begin{minipage}{\linewidth}Hardware shielding \& tamper detection, ephemeral keys \& frequent re-keying, blinding \& masking for cryptographic ops.\end{minipage}
& \begin{minipage}{\linewidth}
 \begin{myBullets}\vspace{2pt} 

 \item \textbf{Tampering:} Manipulating shared CPU caches or signals
 \item \textbf{Info. Disclosure:} Extracting cryptographic secrets
 \item \textbf{DoS:} Degrading system performance with intensive side-channel probes
 \vspace{3pt} \end{myBullets}
\end{minipage}&\med&\med&\med
\\ \cline{2-9}










& \begin{minipage}{\linewidth}Increased Computational Load\end{minipage} 
& \begin{minipage}{\linewidth}Oversized Cryptography~\cite{ficco2015stealthy,masdari2016survey}\end{minipage} 
& \begin{minipage}{\linewidth}Excessive cryptographic processing from larger \gls{PQC} elements overwhelms runtime processes\end{minipage} 
& \begin{minipage}{\linewidth} \vspace{0.1cm}Optimize \gls{PQC} parameter sizes, real-time resource monitoring \& adaptive crypto, auto-scaling \& DoS detection, load balancing \& redundancy. \vspace{0.1cm}\end{minipage}
& \begin{minipage}{\linewidth}
 \begin{myBullets}\vspace{2pt} 

 \item \textbf{Tampering:} Oversized payloads disrupt security processes
 \item \textbf{DoS:} Resource exhaustion causing system crashes
 \item \textbf{EoP:} Privilege bypass via process failures
 \vspace{3pt} \end{myBullets}
\end{minipage}&\med&\med&\med
\\ \cline{2-9}










& \begin{minipage}{\linewidth}Inefficient Encrypted Traffic Processing\end{minipage}
& \begin{minipage}{\linewidth}\gls{VM} DoS~\cite{compastie2020virtualization,huang2012security}\end{minipage}
& \begin{minipage}{\linewidth}Excessive encrypted traffic overloading processing resources\end{minipage}
& \begin{minipage}{\linewidth}\vspace{0.1cm}Traffic shaping/QoS, dynamic throttling/DoS detection, \gls{PQC} hardware acceleration (FPGA/GPU), multi-threading \& concurrency optimization.\vspace{0.1cm}\end{minipage}
& \begin{minipage}{\linewidth}
 \begin{myBullets}\vspace{2pt} 

 \item \textbf{Tampering:} Malformed encrypted payloads degrade performance
 \item \textbf{DoS:} Resource exhaustion \& service disruption
 \item \textbf{EoP:} Process failures enabling unauthorized escalations
 \vspace{3pt} \end{myBullets}
\end{minipage}&\med&\med&\med
\\ \cline{2-9}



& \begin{minipage}{\linewidth}Hypervisor Weakness/Misconfiguration\end{minipage}
& \begin{minipage}{\linewidth}\gls{VM} Hopping/Guest Jumping~\cite{compastie2020virtualization,almutairy2019taxonomy,huang2012security}\end{minipage}
& \begin{minipage}{\linewidth}Exploiting hypervisor vulnerabilities (e.g., buffer overflow, code injection, misconfigs) for \gls{VM} hopping\end{minipage}
& \begin{minipage}{\linewidth}\vspace{0.1cm} Hypervisor security hardening (Xen/KVM), strict RBAC/2FA for hypervisor admins, enhanced \gls{VM} monitoring \& isolation, periodic vulnerability scanning.\vspace{0.1cm} \end{minipage}
& \begin{minipage}{\linewidth}
 \begin{myBullets}\vspace{2pt} 
 \item \textbf{Tampering:} Modifying hypervisor configurations 
 \item \textbf{Info. Disclosure:} Data extraction from compromised hypervisors
 \item \textbf{EoP:} Full control over VMs
 \vspace{3pt} \end{myBullets}
\end{minipage}&\low&\high&\med
\\ \hline










\end{tabular}%
}
\end{table*}

\begin{table*}[!htbp]
\ContinuedFloat
\caption{Post-Transition Quantum Threat Landscape in Each \gls{CC} Layer (Continued)} 
\large
\renewcommand{\arraystretch}{1.1}
\resizebox{\linewidth}{!}{%
\begin{tabular}{|p{0.13\linewidth}|p{0.25\linewidth}|p{0.32\linewidth}|p{0.45\linewidth}|p{0.47\linewidth}|p{0.72\linewidth}|l|l|l|}
\hline 
\textbf{Layer} & \textbf{Vulnerability} & \textbf{Exploitation Method} & \textbf{Attack Vector} & \textbf{Countermeasures} & \textbf{STRIDE Threat} & {\textbf{L}} & {\textbf{I}} & {\textbf{R}}\\ \hline 

\multirow{20}{*}{ {Runtime}}

& \begin{minipage}{\linewidth}\gls{VM} Escape\end{minipage}
& \begin{minipage}{\linewidth}\gls{VM} Escape to \gls{VM}~\cite{compastie2020virtualization}, Out-of-bounds memory access\end{minipage}
& \begin{minipage}{\linewidth}Exploiting vulnerabilities for \gls{VM} escape to gain host or other \gls{VM} access\end{minipage}
& \begin{minipage}{\linewidth} \vspace{0.1cm} Hardened \gls{VM} boundary checks \& memory isolation, hypervisor patches \& real-time intrusion detection, mandatory access controls (AppArmor/SELinux). \vspace{0.1cm} \end{minipage}
& \begin{minipage}{\linewidth}
 \begin{myBullets}\vspace{2pt} 
 \item \textbf{Tampering:} Modifying hypervisor configurations \& states
 \item \textbf{Info. Disclosure:} Accessing data from co-resident VMs
 \item \textbf{EoP:} Full control over VMs
 \vspace{3pt} \end{myBullets}
\end{minipage}&\low&\high&\med
\\ \cline{2-9}










& \begin{minipage}{\linewidth}Runtime Malware Execution Hijacking\end{minipage}
& \begin{minipage}{\linewidth}Spyware Injection~\cite{rawal2023malware,commomtypmalware}, Keylogging~\cite{commomtypmalware}, Memory-Resident Malware\end{minipage}
& \begin{minipage}{\linewidth}Injecting malware into runtime processes for data theft, keylogging, \& manipulation\end{minipage}
& \begin{minipage}{\linewidth} \vspace{0.1cm} Anti-malware with behavioral analysis, strict flow monitoring (EDR), process whitelisting \& memory integrity, AI-based runtime anomaly detection. \vspace{0.1cm} \end{minipage}
& \begin{minipage}{\linewidth}
 \begin{myBullets}\vspace{2pt} 
 \item \textbf{Tampering:} Unauthorized process manipulation 
 \item \textbf{Info. Disclosure:} Keylogging \& sensitive data theft
 \item \textbf{EoP:} Escalation via memory-resident malware
 \vspace{3pt} \end{myBullets}
\end{minipage}&\med&\high&\high
\\ \cline{2-9}

& \begin{minipage}{\linewidth}Quantum-Enhanced Execution Manipulation\end{minipage}
& \begin{minipage}{\linewidth}Quantum-Assisted Fuzzing, Quantum Fault Injection, Hybrid Crypto Side-Channel Leaks\end{minipage}
& \begin{minipage}{\linewidth}Leveraging quantum computation to enhance fuzzing \& fault injection for runtime manipulation; side-channel leaks from hybrid \gls{PQC} and classical crypto\end{minipage}
& \begin{minipage}{\linewidth} \vspace{0.1cm}Quantum-resilient runtime checks (CFI, pointer auth), anomaly detection for quantum fuzz patterns, hardware-level execution verification, secure design of hybrid crypto systems. \vspace{0.1cm}\end{minipage}
& \begin{minipage}{\linewidth}
 \begin{myBullets}\vspace{2pt} 
 \item \textbf{Tampering:} Disrupting runtime with quantum-driven manipulation
 \item \textbf{DoS:} Overwhelming resources via quantum-generated fuzzing
 \item \textbf{Info. Disclosure:} Hybrid crypto leaks from \glspl{SCA}
 \item \textbf{EoP:} Quantum-accelerated fault injection 
 \vspace{3pt} \end{myBullets}
\end{minipage}&\low&\high&\med
\\ \cline{2-9}

& \begin{minipage}{\linewidth}Logging Vulnerabilities\end{minipage}
& \begin{minipage}{\linewidth}Application Software Exploitation~\cite{compastie2020virtualization,hoffman2020web}\end{minipage}
& \begin{minipage}{\linewidth}Exploiting inadequate logging to hide malicious activities after code injection or \glspl{SCA}\end{minipage}
& \begin{minipage}{\linewidth} \vspace{0.1cm}Comprehensive logs \& monitoring (SIEM), secure coding to prevent log injection, cryptographic audit trails/WORM logs, periodic log integrity checks. \vspace{0.1cm}\end{minipage}
& \begin{minipage}{\linewidth}
 \begin{myBullets}\vspace{2pt}
 \item \textbf{Tampering:} Altering or injecting log entries (log poisoning, adversarial log injection) to mislead forensic investigations
 \item \textbf{Repudiation:} Hiding unauthorized activity via log manipulation
 \vspace{3pt} \end{myBullets}
\end{minipage}&\med&\med&\med
\\ \cline{2-9}










& \begin{minipage}{\linewidth}
{Log Tampering and Poisoning}
\end{minipage}
& \begin{minipage}{\linewidth}
Adversarial Log Injection, Cryptographic Log Poisoning
\end{minipage}
& \begin{minipage}{\linewidth}
Manipulating logs to hide attacks, mislead forensics, or inject false information
\end{minipage}
& \begin{minipage}{\linewidth} \vspace{0.1cm}
Digital signatures for logs (HMAC/public-key), secure/append-only (WORM) log servers, periodic log hashing \& cross-verification, forensic readiness \& chain-of-custody. \vspace{0.1cm}
\end{minipage}
& \begin{minipage}{\linewidth}
\begin{myBullets}
\item \textbf{Tampering:} Adversarial log injection and cryptographic log poisoning to conceal or falsify activities

 \item \textbf{Repudiation:} Concealing malicious activities in logs
 \vspace{3pt} \end{myBullets}
\end{minipage}&\med&\med&\med
\\ \cline{2-9}










& \begin{minipage}{\linewidth}
{Inter-\gls{VM} Network Exploitation}
\end{minipage}
& \begin{minipage}{\linewidth}
Inter-\gls{VM} Communication Introspection~\cite{compastie2020virtualization}
\end{minipage}
& \begin{minipage}{\linewidth}
Exploiting shared networking to intercept\& manipulate inter-\gls{VM} communication
\end{minipage}
& \begin{minipage}{\linewidth} \vspace{0.1cm}
Strong isolation \& encryption for east-west traffic, firewall rules/virtual network ACLs, mutual TLS in \gls{VM} comms, IDS for east-west intrusion detection. \vspace{0.1cm}
\end{minipage}
& \begin{minipage}{\linewidth}
\begin{myBullets}\vspace{2pt}
 \item \textbf{Spoofing:} Impersonating \gls{VM} endpoints
 \item \textbf{Tampering:} Modifying intercepted packets
 \item \textbf{Info. Disclosure:} Extracting sensitive data from \gls{VM} traffic
 \vspace{3pt} \end{myBullets}
\end{minipage}&\med&\med&\med
\\ \cline{2-9}










& \begin{minipage}{\linewidth}
{\gls{VM} Migration Exploitation}
\end{minipage}
& \begin{minipage}{\linewidth}
\gls{VM} Migration Attacks~\cite{9889331}
\end{minipage}
& \begin{minipage}{\linewidth}
Exploiting partial or misconfigured \gls{PQC} tunnels during \gls{VM} live-migration, insecure \gls{PQC} key migration, protocol misalignment
\end{minipage}
& \begin{minipage}{\linewidth} \vspace{0.1cm}
Encrypt \gls{VM} migration with \gls{PQC}/hybrid crypto, mutual endpoint authentication, integrity checks on migration data, consistent cryptographic versioning. \vspace{0.1cm}
\end{minipage}
& \begin{minipage}{\linewidth}
\begin{myBullets}\vspace{2pt}
 \item \textbf{Spoofing:} Impersonating \gls{VM} endpoints 
 \item \textbf{Tampering:} Modifying intercepted network packets 
 \item \textbf{Info. Disclosure:} Extracting sensitive data from migrating VMs
 \vspace{3pt} \end{myBullets}
\end{minipage}&\low&\high&\high
\\ \hline



\multirow{35}{*}{\begin{minipage}{\linewidth}
\centering
{Middleware}
\end{minipage}}

& \begin{minipage}{\linewidth}
{Buffer Overflow \& Memory Vulnerabilities}
\end{minipage}
& \begin{minipage}{\linewidth}
{Hypervisor Exploitation~\cite{compastie2020virtualization}}, Out-of-bounds memory access via buffer overflow~\cite{mishra2025modern,sun2025survey}, 
Code injection vulnerabilities~\cite{owasp_code_injection,hoffman2020web}, 
 \glspl{SCA} as per Table~\ref{tab:post-alg}
\end{minipage}
& \begin{minipage}{\linewidth}
(a) Buffer overflow or code injection at the root level causing unauthorized access or data leakage (host/\gls{VM} to \gls{VM}). 
(b) Memory corruption exploitation to gain elevated privileges. 
(c) Potential synergy with side-channels to leak cryptographic data.
\end{minipage}
& \begin{minipage}{\linewidth}
Implement input validation, use memory-safe libraries and languages, enforce rigorous security audits and patching, deploy intrusion detection systems.
\end{minipage}
& \begin{minipage}{\linewidth}
\begin{myBullets}\vspace{2pt}
 \item \textbf{Tampering:} Hypervisor code/data modification via overflow exploits.
 \item \textbf{Repudiation:} Log/trace alteration to conceal unauthorized activity.
 \item \textbf{Info. Disclosure:} Leak hypervisor/\gls{VM} data via memory access vulnerabilities.
 \item \textbf{EoP:} Root-level access compromises the entire middleware/hypervisor stack.
 \vspace{3pt} \end{myBullets}
\end{minipage}&\med&\high&\high
\\ \cline{2-9}

& \begin{minipage}{\linewidth}
{Side-Channel Vulnerabilities}
\end{minipage}
& \begin{minipage}{\linewidth} \vspace{0.1cm}
\gls{VM} Escape to Host~\cite{compastie2020virtualization}, Cross-\gls{VM} \glspl{SCA}~\cite{NARAYANA20216465,10.1145/3715001,anwar2017cross,saeed2020cross}, 
Micro-architectural Cache Side-channels, Rowhammer~\cite{shen2021micro,lou2021survey}
\end{minipage}
& \begin{minipage}{\linewidth}
Side-channel leaks exploit shared resources (caches, CPU execution units) for data exfiltration (host to VM, or \gls{VM} to \gls{VM}).
\end{minipage}
& \begin{minipage}{\linewidth}
Use hardware features mitigating side-channels, strict access controls, regular patching, noise injection, continuous hypervisor monitoring.
\end{minipage}
& \begin{minipage}{\linewidth}
\begin{myBullets}\vspace{2pt}
 \item \textbf{Tampering:} Manipulating shared resources (cache lines, row buffers) to maximize side-channel accuracy.
 \item \textbf{Info. Disclosure:} Attackers extract \gls{VM} or hypervisor data via microarchitectural leaks.
\vspace{3pt} \end{myBullets}
\end{minipage}&\med&\med&\med
\\ \cline{2-9}










& \begin{minipage}{\linewidth}
{Insecure Inter-\gls{VM} Communications}
\end{minipage}
& \begin{minipage}{\linewidth}
Hypervisor Exploitation~\cite{compastie2020virtualization}
\end{minipage}
& \begin{minipage}{\linewidth}\vspace{0.1cm}
(a) Code injection at root level, leaking data from host/\gls{VM} to VM.
(b) Possible synergy with side-channels for real-time traffic interception.
\vspace{0.1cm}\end{minipage}
& \begin{minipage}{\linewidth}\vspace{0.1cm}
Implement robust \gls{VM} isolation, updated security policies, advanced intrusion detection, encrypted inter-\gls{VM} channels.
\vspace{0.1cm}\end{minipage}
& \begin{minipage}{\linewidth}
\begin{myBullets}\vspace{2pt}
 \item \textbf{Tampering:} Inject malicious code into communication protocols.
 \item \textbf{Info. Disclosure:} Attackers intercept or snoop on inter-\gls{VM} data flows.
\vspace{3pt} \end{myBullets}
\end{minipage}&\med&\med&\med\\ \cline{2-9}










& \begin{minipage}{\linewidth}
{Micro-architectural Vulnerabilities}
\end{minipage}
& \begin{minipage}{\linewidth}
Covert-channel Attacks, Rowhammer, Transient Execution~\cite{shen2021micro,lou2021survey}
\end{minipage}
& \begin{minipage}{\linewidth}\vspace{0.1cm}
Exploiting CPU or memory subsystem flaws (e.g., rowhammer flips, meltdown/spectre-like transient execution leaks) leading to unauthorized data access.
\vspace{0.1cm}\end{minipage}
& \begin{minipage}{\linewidth}\vspace{0.1cm}
Apply targeted mitigations (rowhammer detection/protection), use hardware with known side-channel defenses, continuous system monitoring.
\vspace{0.1cm}\end{minipage}
& \begin{minipage}{\linewidth}
\begin{myBullets}\vspace{2pt}
 \item \textbf{Info. Disclosure:} Attackers infer private data from memory or CPU caches.
\vspace{3pt} \end{myBullets}
\end{minipage}&\low&\med&\low
\\ \cline{2-9}










& \begin{minipage}{\linewidth}
{Resource Exhaustion (Hypervisor \gls{DoS})}
\end{minipage}
& \begin{minipage}{\linewidth}
Hypervisor DoS~\cite{masdari2016survey,compastie2020virtualization}
\end{minipage}
& \begin{minipage}{\linewidth}\vspace{0.1cm}
 \gls{PQC} can overload hypervisor resources due to larger encrypted traffic and increased computational demands.
\vspace{0.1cm}\end{minipage}
& \begin{minipage}{\linewidth}
Use load balancing, traffic analysis, specialized DoS protection, resource usage caps.
\end{minipage}
& \begin{minipage}{\linewidth}
\begin{myBullets}\vspace{2pt}
 \item \textbf{DoS:} Hypervisor performance degrades or fails due to resource overload.
\vspace{3pt} \end{myBullets}
\end{minipage}&\med&\med&\med
\\ \cline{2-9}










& \begin{minipage}{\linewidth}
{Hypervisor Configuration Flaws}
\end{minipage}
& \begin{minipage}{\linewidth}
\gls{VM} Hopping/Guest Jumping~\cite{compastie2020virtualization,almutairy2019taxonomy,huang2012security}
\end{minipage}
& \begin{minipage}{\linewidth}\vspace{0.1cm}
Exploiting hypervisor interface or config via buffer overflow, enabling \gls{VM} hopping or unauthorized cross-\gls{VM} access.
\vspace{0.1cm}\end{minipage}
& \begin{minipage}{\linewidth}
Secure hypervisor configuration, frequent audits, timely hypervisor updates, minimal attack surface.
\end{minipage}
& \begin{minipage}{\linewidth}
\begin{myBullets}\vspace{2pt}
 \item \textbf{Tampering:} Attackers alter hypervisor settings to enable persistent backdoors.
 \item \textbf{DoS:} Misconfigured hypervisors lead to system crashes or degraded performance.
\vspace{3pt} \end{myBullets}
\end{minipage}&\low&\high&\med
\\ \cline{2-9}










& \begin{minipage}{\linewidth}
{Overloaded Resource Allocation (\gls{VM} Poaching)}
\end{minipage}
& \begin{minipage}{\linewidth}
\gls{VM} Poaching~\cite{almutairy2019taxonomy,kedia2013survey}
\end{minipage}
& \begin{minipage}{\linewidth}\vspace{0.1cm}
PQC's resource overhead allows attackers to drain hypervisor resources, causing \gls{VM} crashes or severe resource contention.
\vspace{0.1cm}\end{minipage}
& \begin{minipage}{\linewidth}
Continuous resource usage monitoring, dynamic \gls{CC} scaling, autoscaling thresholds for VMs.
\end{minipage}
& \begin{minipage}{\linewidth}
\begin{myBullets}\vspace{2pt}
 \item \textbf{DoS:} Attackers starve VMs by monopolizing compute or memory resources.
\vspace{3pt} \end{myBullets}
\end{minipage}&\med&\med&\med
\\ \cline{2-9}













& \begin{minipage}{\linewidth}\vspace{0.1cm}
{Unauthorized Access to Root Level}
\vspace{0.1cm}\end{minipage}
& \begin{minipage}{\linewidth}
\gls{VM} Escape to Host~\cite{compastie2020virtualization}
\end{minipage}
& \begin{minipage}{\linewidth}
Exploiting buffer overflow to achieve root-level access from a VM
\end{minipage}
& \begin{minipage}{\linewidth}
Strict access controls, vulnerability scanning, limited root privileges
\end{minipage}
& \begin{minipage}{\linewidth}
\begin{myBullets}\vspace{2pt}
 \item \textbf{EoP:} Full host control bypassing \gls{VM} boundaries.
\vspace{3pt} \end{myBullets}
\end{minipage}&\low&\high&\med
\\ \cline{2-9}











& \begin{minipage}{\linewidth}
{Hypervisor Security Gaps}
\end{minipage}
& \begin{minipage}{\linewidth}
\gls{VM} Hopping/Guest Jumping~\cite{compastie2020virtualization,almutairy2019taxonomy,huang2012security}
\end{minipage}
& \begin{minipage}{\linewidth}
Exploitation of hypervisor vulnerabilities (buffer overflow, code injection) for full system compromise.
\end{minipage}
& \begin{minipage}{\linewidth}
Rigorous hypervisor security testing, strict access controls, timely patching, intrusion detection.
\end{minipage}
& \begin{minipage}{\linewidth}
\begin{myBullets}\vspace{2pt}
 \item \textbf{Tampering:} Attackers bypass hypervisor security controls.
 \item \textbf{Info. Disclosure:} VM-sensitive data is exposed via compromised hypervisor.
 \item \textbf{EoP:} Controlling hypervisor leads to entire system compromise.
\vspace{3pt} \end{myBullets}
\end{minipage}&\low&\high&\med
\\ \cline{2-9}










& \begin{minipage}{\linewidth}
{Oversized Cryptographic Elements}
\end{minipage}
& \begin{minipage}{\linewidth}
Oversized Cryptography~\cite{ficco2015stealthy,masdari2016survey}
\end{minipage}
& \begin{minipage}{\linewidth}
Excessive cryptographic processing overwhelms middleware (hypervisors, cryptographic engines, authentication servers).
\end{minipage}
& \begin{minipage}{\linewidth}
Optimize cryptographic algorithms for smaller sizes, implement load balancing for crypto tasks, threshold-based rate limiting for cryptographic ops.
\end{minipage}
& \begin{minipage}{\linewidth}
\begin{myBullets}\vspace{2pt}
 \item \textbf{Tampering:} Attackers craft large payloads to force resource consumption.
 \item \textbf{DoS:} Crypto overload causes failures in middleware authentication or transaction handling.
 \item \textbf{EoP:} Resource exhaustion can weaken access controls, enabling bypass attacks.
\vspace{3pt} \end{myBullets}
\end{minipage}&\med&\med&\med
\\ \hline










\multirow{50}{*}{{OS}}

& \begin{minipage}{\linewidth}
{Kernel-Level Security Breaches}
\end{minipage}
& \begin{minipage}{\linewidth}
{Injecting Virus}~\cite{ThreatsProtection}, Side-channel vulnerabilities, Buffer overflow
\end{minipage}
& \begin{minipage}{\linewidth}
Code injection via buffer overflow or \glspl{SCA}, forging digital signatures, malicious kernel modifications
\end{minipage}
& \begin{minipage}{\linewidth}\vspace{0.1cm}
Implement rigorous testing/validation of cryptographic implementations, update kernel protection mechanisms (\gls{ASLR}, DEP, SELinux/AppArmor), continuous patching
\vspace{0.1cm}\end{minipage}
& \begin{minipage}{\linewidth}
\begin{myBullets}\vspace{2pt}
 \item \textbf{Spoofing:} Attackers manipulate \gls{OS} kernel to forge digital signatures.
 \item \textbf{Tampering:} Injected malware modifies kernel-level operations, disrupting system stability.
\vspace{3pt} \end{myBullets} 
\end{minipage}&\med&\high&\high
\\ \cline{2-9}










& \begin{minipage}{\linewidth}
{Unauthorized Data Access \& Code Injection}
\end{minipage}
& \begin{minipage}{\linewidth}
{Injecting Insecure Code}~\cite{owasp_code_injection,hoffman2020web}, Buffer overflow~\cite{masdari2016survey,compastie2020virtualization,mishra2025modern}, Side-channel exploits
\end{minipage}
& \begin{minipage}{\linewidth}
Exploiting \gls{OS} vulnerabilities (memory corruption or side-channels) to modify system processes, escalate privileges, or exfiltrate sensitive \gls{OS} data
\end{minipage}
& \begin{minipage}{\linewidth}
Regular security audits, buffer overflow protections (stack canaries, \gls{ASLR}, DEP), memory-safe coding standards (e.g., CERT), strict code-signing
\end{minipage}
& \begin{minipage}{\linewidth}
\begin{myBullets}\vspace{2pt}
 \item \textbf{Tampering:} Injected code alters \gls{OS} processes, potentially enabling root-level compromise
 \item \textbf{Info. Disclosure:} Attackers exfiltrate data if insecure code accesses sensitive memory
 \item \textbf{EoP:} Exploiting \gls{OS} process vulnerabilities for unauthorized root access
\vspace{3pt} \end{myBullets} 
\end{minipage}&\med&\high&\high
\\ \cline{2-9}










& \begin{minipage}{\linewidth}
{OS Kernel Exploitation \& Buffer Overflow Vulnerabilities}
\end{minipage}
& \begin{minipage}{\linewidth}
{OS Kernel Exploitation}~\cite{compastie2020virtualization}, Cryptanalysis Attacks~\cite{otmani2010cryptanalysis}, Side-channel/Memory vulnerabilities
\end{minipage}
& \begin{minipage}{\linewidth}
Forged signatures, advanced cryptanalysis, side-channel manipulation of \gls{OS} kernel, out-of-bounds memory access causing system instability
\end{minipage}
& \begin{minipage}{\linewidth}
Enhanced kernel protection (secure coding, frequent patches), mandatory access control (SELinux/AppArmor), hardware security features (TPM/HSM)
\end{minipage}
& \begin{minipage}{\linewidth}
\begin{myBullets}\vspace{2pt}
 \item \textbf{Tampering:} Attackers manipulate kernel execution to alter system behavior.
 \item \textbf{Repudiation:} Weak cryptographic/logging allows attackers to alter or delete kernel logs.
 \item \textbf{Info. Disclosure:} Side-channel leaks reveal cryptographic keys in kernel memory.
 \item \textbf{DoS:} Buffer overflow attacks can crash critical kernel processes.
\vspace{3pt} \end{myBullets} 
\end{minipage}&\med&\high&\high
\\ \cline{2-9}










& \begin{minipage}{\linewidth}
{Inadequate Logging/Cryptanalysis Weaknesses}
\end{minipage}
& \begin{minipage}{\linewidth}
{Cryptanalysis Attacks}~\cite{otmani2010cryptanalysis}
\end{minipage}
& \begin{minipage}{\linewidth}
Attackers exploit \gls{PQC} algorithm weaknesses or lack of immutable logs; advanced cryptanalysis + poor \gls{OS} logging
\end{minipage}
& \begin{minipage}{\linewidth}
Immutable logs, blockchain-based integrity, enhanced algorithm validation, cryptographic log signing
\end{minipage}
& \begin{minipage}{\linewidth}
\begin{myBullets}\vspace{2pt}
 \item \textbf{Repudiation:} Attackers erase or manipulate \gls{OS} logs to hide infiltration.
 \item \textbf{Info. Disclosure:} Weak cryptographic logging can leak secrets via cryptanalysis.
\vspace{3pt} \end{myBullets} 
\end{minipage}&\low&\med&\low
\\ \cline{2-9}










& \begin{minipage}{\linewidth}
{Kernel Memory Exposure \& \glspl{SCA}}
\end{minipage}
& \begin{minipage}{\linewidth}
{OS Kernel Exploitation}~\cite{compastie2020virtualization}, \glspl{SCA} (Table~\ref{tab:post-alg})
\end{minipage}
& \begin{minipage}{\linewidth}
Side-channel manipulation exposing kernel memory; forging signatures or reading privileged data
\end{minipage}
& \begin{minipage}{\linewidth}
Isolation techniques, enhanced access control, memory encryption, ephemeral keys, side-channel defenses (constant-time code)
\end{minipage}
& \begin{minipage}{\linewidth}
\begin{myBullets}\vspace{2pt}
 \item \textbf{Tampering:} Manipulating kernel memory to produce unexpected \gls{OS} behavior
 \item \textbf{Info. Disclosure:} Attackers leak kernel memory or cryptographic data
\vspace{3pt} \end{myBullets} 
\end{minipage}&\med&\med&\med
\\ \cline{2-9}

& \begin{minipage}{\linewidth}
{Hypervisor Vulnerabilities/\gls{VM} Escape}
\end{minipage}
& \begin{minipage}{\linewidth}
{\gls{VM} Hopping/Guest Jumping}~\cite{compastie2020virtualization,almutairy2019taxonomy,huang2012security}, Buffer overflow~\cite{masdari2016survey,compastie2020virtualization,mishra2025modern}, Code injection
\end{minipage}
& \begin{minipage}{\linewidth}
Exploiting hypervisor weaknesses (misconfigured interfaces, memory corruption) for cross-\gls{VM} attacks~\cite{saeed2020cross} or root access to the OS
\end{minipage}
& \begin{minipage}{\linewidth}
Hypervisor security hardening, strong isolation policies, secure configurations, HSM-based key storage, continuous patching
\end{minipage}
& \begin{minipage}{\linewidth}
\begin{myBullets}\vspace{2pt}
 \item \textbf{Tampering:} Malicious configuration changes undermine OS/hypervisor security
 \item \textbf{Info. Disclosure:} \gls{VM} memory leaks reveal \gls{OS} credentials or cryptographic data
 \item \textbf{EoP:} Attackers compromise the hypervisor to escalate across VMs
\vspace{3pt} \end{myBullets} 
\end{minipage}&\low&\high&\med
\\ \cline{2-9}











& \begin{minipage}{\linewidth}
{Memory Exploitation \& \gls{OS} Escape}
\end{minipage}
& \begin{minipage}{\linewidth}
{\gls{VM} Escape to VM, \gls{VM} Escape to Host}~\cite{compastie2020virtualization}
\end{minipage}
& \begin{minipage}{\linewidth}
Attacks enabling unauthorized \gls{OS} memory access, privilege escalation, data leakage (e.g., out-of-bounds memory read)
\end{minipage}
& \begin{minipage}{\linewidth}
Memory protection (\gls{ASLR}, DEP), secure coding, regular patches, robust access control
\end{minipage}
& \begin{minipage}{\linewidth}
\begin{myBullets}\vspace{2pt}
 \item \textbf{Tampering:} Buffer overflow modifies \gls{OS} behavior unpredictably
 \item \textbf{Info. Disclosure:} Leaks sensitive memory data during the escape
 \item \textbf{EoP:} Arbitrary privileged commands on the host or other VMs
\vspace{3pt} \end{myBullets} 
\end{minipage}&\med&\high&\high
\\ \cline{2-9}










& \begin{minipage}{\linewidth}
{Malware Exploitation (Virus, Trojan, Worm, Keylogger)}
\end{minipage}
& \begin{minipage}{\linewidth}
{Injecting Malware (Virus, Spyware, Trojan Horse, Worm)},\\
Buffer overflow~\cite{masdari2016survey,compastie2020virtualization,mishra2025modern} or side-channel
\end{minipage}
& \begin{minipage}{\linewidth}
Attackers install malicious software for data exfiltration, system disruption, or covert persistence
\end{minipage}
& \begin{minipage}{\linewidth}
Antivirus solutions, intrusion detection systems, memory-safe coding, \gls{OS} security updates, application whitelisting, network segmentation
\end{minipage}
& \begin{minipage}{\linewidth}
\begin{myBullets}\vspace{2pt}
 \item \textbf{Tampering:} Attackers modify \gls{OS} processes to remain undetected or disable security
 \item \textbf{Info. Disclosure:} Malware-based exploitation steals sensitive OS/user data
 \item \textbf{ \gls{DoS}:} Worm propagation or virus overloading leads to resource exhaustion
 \item \textbf{EoP:} Many malware families escalate privileges to maintain control
\vspace{3pt} \end{myBullets} 
\end{minipage}&\med&\high&\high
\\ \cline{2-9}



& \begin{minipage}{\linewidth}
{Hypervisor Monitoring Weaknesses}
\end{minipage}
& \begin{minipage}{\linewidth}
{\gls{VM} Monitoring (Host/\gls{VM})}~\cite{compastie2020virtualization}
\end{minipage}
& \begin{minipage}{\linewidth}
 \glspl{SCA} via hypervisor monitoring, data leakage from misconfigured logs
\end{minipage}
& \begin{minipage}{\linewidth}
Strict \gls{VM} isolation, secure logging, hardened hypervisor, hardware-based security modules
\end{minipage}
& \begin{minipage}{\linewidth}
\begin{myBullets}\vspace{2pt}
 \item \textbf{Tampering:} Malicious modifications to hypervisor logs or monitoring logic
 \item \textbf{Info. Disclosure:} Attackers spy on \gls{VM} data from the host level
 \item \textbf{EoP:} Inadequate isolation enables attackers to jump from a monitored \gls{VM} to host-level access
\vspace{3pt} \end{myBullets} 
\end{minipage}&\low&\high&\med
\\ \cline{2-9}



& \begin{minipage}{\linewidth}
{Insufficient Access Controls \& Process Isolation}
\end{minipage}
& \begin{minipage}{\linewidth}
{\gls{VM} Monitoring from VM}~\cite{compastie2020virtualization},
 \glspl{SCA}, \gls{VM} co-residence
\end{minipage}
& \begin{minipage}{\linewidth}
Attackers exploit weak isolation or minimal access control to maintain persistent side-channel monitoring
\end{minipage}
& \begin{minipage}{\linewidth}
Enhanced access control, robust \gls{VM} isolation, hardware security modules, ephemeral cryptographic keys
\end{minipage}
& \begin{minipage}{\linewidth}
\begin{myBullets}\vspace{2pt}
 \item \textbf{Tampering:} Attackers manipulate OS-level access controls for persistent foothold
 \item \textbf{Info. Disclosure:} Attackers glean sensitive data from side-channel leaks
 \item \textbf{EoP:} Weak isolation allows crossing privilege boundaries
\vspace{3pt} \end{myBullets} 
\end{minipage}&\med&\med&\med
\\ \cline{2-9}



& \begin{minipage}{\linewidth}
{Inter-\gls{VM} Communication Eavesdropping}
\end{minipage}
& \begin{minipage}{\linewidth}
{Inter-\gls{VM} Communication Introspection}~\cite{compastie2020virtualization}
\end{minipage}
& \begin{minipage}{\linewidth}
Co-resident VMs or hosts intercepting data via shared networking \gls{CC} 
\end{minipage}
& \begin{minipage}{\linewidth}\vspace{0.1cm}
Use encrypted channels, robust \gls{VM} monitoring, secure hypervisor bridging
\vspace{0.1cm}\end{minipage}
& \begin{minipage}{\linewidth}
\begin{myBullets}\vspace{2pt}
 \item \textbf{Tampering:} Malicious intercepts that alter or corrupt data in transit
 \item \textbf{Info. Disclosure:} Attackers capture data from inter-\gls{VM} traffic
\vspace{3pt} \end{myBullets} 
\end{minipage}&\med&\med&\med
\\ \cline{2-9}



& \begin{minipage}{\linewidth}
{Side-channel/Fault Attacks Exposing \gls{OS} Data}
\end{minipage}
& \begin{minipage}{\linewidth}
{Cache Attacks~\cite{gruss2019page,esfahani2021enhanced}}, {Fault Attacks~\cite{krautter2018fpgahammer}}
\end{minipage}
& \begin{minipage}{\linewidth}
Attackers manipulate cryptographic or \gls{OS} operations to leak system data, inject faults
\end{minipage}
& \begin{minipage}{\linewidth}\vspace{0.1cm}
Side-channel mitigation (noise generation, constant-time), secure coding, memory encryption, hardware shielding
\vspace{0.1cm}\end{minipage}
& \begin{minipage}{\linewidth}
\begin{myBullets}\vspace{2pt}
 \item \textbf{Tampering:} Injected faults corrupt cryptographic or \gls{OS} operations
 \item \textbf{Info. Disclosure:} Cache/fault attacks expose OS-level secrets (keys, credentials)
\vspace{3pt} \end{myBullets} 
\end{minipage}&\med&\med&\med
\\ \cline{2-9}



& \begin{minipage}{\linewidth}
{Excessive Resource Usage (\gls{PQC} Algorithms)}
\end{minipage}
& \begin{minipage}{\linewidth}
{ \gls{CC} Resource Exhaustion}, {\gls{VM} \gls{DoS}}~\cite{masdari2016survey,compastie2020virtualization,huang2012security}
\end{minipage}
& \begin{minipage}{\linewidth}
Overwhelming \gls{OS} or VMs via large \gls{PQC} operations or massive encrypted traffic
\end{minipage}
& \begin{minipage}{\linewidth}
Scalable compute resources, load balancing, resource management policies, traffic shaping, rate limiting
\end{minipage}
& \begin{minipage}{\linewidth}
\begin{myBullets}\vspace{2pt}
 \item \textbf{Tampering:} Attackers direct abnormally high volumes of \gls{PQC} requests
 \item \textbf{ \gls{DoS}:} \gls{PQC} computations hog CPU/memory, destabilizing \gls{OS} processes

\vspace{3pt} \end{myBullets} 
\end{minipage}&\med&\med&\med
\\ \hline



\end{tabular}
}
\end{table*}

\begin{table*}[!htbp]
\ContinuedFloat
\caption{Post-Transition Quantum Threat Landscape in Each \gls{CC} Layer (Continued)} 
\large
\renewcommand{\arraystretch}{1.1}
\resizebox{\linewidth}{!}{%
\begin{tabular}{|p{0.13\linewidth}|p{0.25\linewidth}|p{0.32\linewidth}|p{0.45\linewidth}|p{0.47\linewidth}|p{0.72\linewidth}|l|l|l|}
\hline 
\textbf{Layer} & \textbf{Vulnerability} & \textbf{Exploitation Method} & \textbf{Attack Vector} & \textbf{Countermeasures} & \textbf{STRIDE Threat} & {\textbf{L}} & {\textbf{I}} & {\textbf{R}}\\ \hline 
\multirow{10}{*}{{OS}}
& \begin{minipage}{\linewidth}
{Malware-Based Overload \& Insecure Code Execution}
\end{minipage}
& \begin{minipage}{\linewidth}
{Injecting Virus}, {Injecting Insecure Code}~\cite{owasp_code_injection,hoffman2020web}, \\
{Fileless Malware}~\cite{commomtypmalware}
\end{minipage}
& \begin{minipage}{\linewidth}
Code injection or fileless malware that overloads \gls{OS} resources, modifies kernel settings, or triggers DoS
\end{minipage}
& \begin{minipage}{\linewidth}
Antivirus solutions, system hardening, frequent updates, memory-safe programming, dynamic analysis tools
\end{minipage}
& \begin{minipage}{\linewidth}
\begin{myBullets}\vspace{2pt}
 \item \textbf{Tampering:} Attackers modify \gls{OS} behavior for persistent malicious operations
 \item \textbf{ \gls{DoS}:} Malware drains \gls{OS} resources, leading to crashes
 \item \textbf{EoP:} Malware exploits \gls{OS} flaws to gain elevated permissions
\vspace{3pt} \end{myBullets} 
\end{minipage}
&\med&\high&\high
\\ \cline{2-9}



& \begin{minipage}{\linewidth}
{Kernel Manipulation \& Side-channel Exploits}
\end{minipage}
& \begin{minipage}{\linewidth}
{OS Kernel Exploitation}~\cite{compastie2020virtualization}
\end{minipage}
& \begin{minipage}{\linewidth}
Attackers exploit side-channels to access or modify kernel configurations
\end{minipage}
& \begin{minipage}{\linewidth}
Enhanced kernel isolation, secure coding, frequent kernel patches
\end{minipage}
& \begin{minipage}{\linewidth}
\begin{myBullets}\vspace{2pt}
 \item \textbf{Tampering:} Attackers alter kernel-level configs
 \item \textbf{Info. Disclosure:} Kernel flaws reveal system logs \& credentials
 \item \textbf{EoP:} Kernel exploits allow privileged commands
\vspace{3pt} \end{myBullets} 
\end{minipage}
&\med&\high&\high
\\ \cline{2-9}



& \begin{minipage}{\linewidth}
{VM/Hypervisor \gls{DoS} \& Resource Overload}
\end{minipage}
& \begin{minipage}{\linewidth}
{\gls{VM} \gls{DoS}}, {Hypervisor \gls{DoS}}~\cite{masdari2016survey,compastie2020virtualization}, {Injecting Virus}~\cite{ThreatsProtection}
\end{minipage}
& \begin{minipage}{\linewidth}
Attackers flood VMs/hypervisor with massive encrypted traffic or resource-draining malware
\end{minipage}
& \begin{minipage}{\linewidth}
Enhanced \gls{VM} monitoring, resource usage limits, traffic shaping, robust access control, frequent scanning for malicious usage
\end{minipage}
& \begin{minipage}{\linewidth}
\begin{myBullets}\vspace{2pt}
 \item \textbf{Tampering:} Attackers degrade performance by injecting resource-hungry processes
 \item \textbf{ \gls{DoS}:} Overwhelming OS/hypervisor with heavy cryptographic requests
\vspace{3pt} \end{myBullets}
\end{minipage}
&\med&\med&\med
\\ \hline











\multirow{25}{*}{Virtualization}

& { \glspl{SCA}}
& \begin{minipage}{\linewidth}
{Cache Attacks}~\cite{irazoqui2015s}, 
{Cross-\gls{VM} \glspl{SCA}}~\cite{NARAYANA20216465,10.1145/3715001,anwar2017cross,saeed2020cross}, 
{Micro-architectural Cache Attacks}~\cite{shen2021micro}
\end{minipage}
& \begin{minipage}{\linewidth}\vspace{0.1cm}
(a) Exploiting shared CPU caches to extract cryptographic keys and execution traces,
(b) Side-channel math-based analysis to infer sensitive \gls{VM} data,
(c) Micro-architectural cache-level exploits to leak execution patterns.
\vspace{0.1cm}\end{minipage}
& \begin{minipage}{\linewidth}
Enforce strict \gls{VM} isolation, implement noise injection techniques, update hypervisors with side-channel mitigations, use cache partitioning or hardware security modules.
\end{minipage}
& \begin{minipage}{\linewidth}
\begin{myBullets}\vspace{2pt}
 \item \textbf{Tampering:} Manipulating shared resources to improve side-channel accuracy.
 \item \textbf{Info. Disclosure:} Attackers extract cryptographic keys or VM-sensitive data.
\vspace{3pt} \end{myBullets}
\end{minipage}&\med&\med&\med
\\ \cline{2-9}

& {Resource Exhaustion (\gls{DoS})}
& \begin{minipage}{\linewidth}
{Cloud-Internal DoS (CIDoS)}~\cite{masdari2016survey}, 
{\gls{VM} Migration Overload}~\cite{9889331}
\end{minipage}
& \begin{minipage}{\linewidth}
(a) Overloading host resources by flooding encrypted \gls{VM} traffic,
(b) Maliciously triggering \gls{VM} migrations to saturate bandwidth and CPU.
\end{minipage}
& \begin{minipage}{\linewidth}\vspace{0.1cm}
Implement resource allocation caps, enforce traffic shaping policies, deploy DoS mitigation strategies, optimize \gls{VM} migration protocols with adaptive load balancing.
\vspace{0.1cm}\end{minipage}
& \begin{minipage}{\linewidth}
\begin{myBullets}\vspace{2pt}
 \item \textbf{DoS:} Attackers deplete computing and networking resources.
\vspace{3pt} \end{myBullets}
\end{minipage}&\med&\med&\med
\\ \cline{2-9}

& {\gls{VM} Escape \& Insufficient Isolation}
& \begin{minipage}{\linewidth}
{\gls{VM} Escape to VM}~\cite{compastie2020virtualization}
\end{minipage}
& \begin{minipage}{\linewidth}
Attackers exploit buffer overflow or hypervisor misconfigurations to gain unauthorized access across \gls{VM} boundaries.
\end{minipage}
& \begin{minipage}{\linewidth}
Enforce strict \gls{VM} isolation policies, secure coding practices, monitor for anomalies, implement hardened hypervisor controls.
\end{minipage}
& \begin{minipage}{\linewidth}
\begin{myBullets}\vspace{2pt}
 \item \textbf{Tampering:} Attackers modify shared memory or execution flow.
 \item \textbf{Info. Disclosure:} Memory access vulnerabilities expose sensitive data.
 \item \textbf{EoP:} Breaking \gls{VM} isolation allows attackers to execute privileged code.
\vspace{3pt} \end{myBullets}
\end{minipage}&\low&\high&\med
\\ \cline{2-9}










& {\gls{VM} Introspection Exploits}
& \begin{minipage}{\linewidth}
{\gls{VM} Monitoring from Host},
{\gls{VM} Monitoring from VM}
\end{minipage}
& \begin{minipage}{\linewidth}
Attackers exploit privileged introspection features to spy on guest VMs and extract memory contents.
\end{minipage}
& \begin{minipage}{\linewidth}\vspace{0.1cm}
Restrict \gls{VM} introspection access, enforce secure hypervisor configurations, enable forensic audit logs for all introspection requests.
\vspace{0.1cm}\end{minipage}
& \begin{minipage}{\linewidth}
\begin{myBullets}\vspace{2pt}
 \item \textbf{Tampering:} Attackers modify introspection logs to hide unauthorized activities.
 \item \textbf{Info. Disclosure:} Attackers extract cryptographic keys and credentials.
\vspace{3pt} \end{myBullets}
\end{minipage}&\low&\high&\med
\\ \cline{2-9}










& {\gls{VM} Image Tampering}
& \begin{minipage}{\linewidth}
{Manipulation of \gls{VM} Snapshots or Images}
\end{minipage}
& \begin{minipage}{\linewidth}
Attackers modify \gls{VM} disk images, injecting malware or backdoors before deployment.
\end{minipage}
& \begin{minipage}{\linewidth}\vspace{0.1cm}
Enforce digital signatures on \gls{VM} images, implement periodic integrity validation, restrict access to \gls{VM} snapshots.
\vspace{0.1cm}\end{minipage}
& \begin{minipage}{\linewidth}
\begin{myBullets}\vspace{2pt}
 \item \textbf{Tampering:} Attackers embed backdoors into \gls{VM} snapshots.
 \item \textbf{EoP:} Compromised \gls{VM} images allow unauthorized root access.
\vspace{3pt} \end{myBullets}
\end{minipage}&\low&\high&\med
\\ \cline{2-9}










& \begin{minipage}{\linewidth} \gls{vTPM} Exploitation\end{minipage}
& \begin{minipage}{\linewidth}
{ \gls{vTPM} Manipulation}
\end{minipage}
& \begin{minipage}{\linewidth}
Attackers compromise virtual \gls{TPM} to extract decryption keys or credentials.
\end{minipage}
& \begin{minipage}{\linewidth}\vspace{0.1cm}
Use hardware \gls{TPM} when possible, enforce strict \gls{vTPM} access policies, monitor for unauthorized \gls{TPM} modifications.
\vspace{0.1cm}\end{minipage}
& \begin{minipage}{\linewidth}
\begin{myBullets}\vspace{2pt}
 \item \textbf{Info. Disclosure:} Attackers extract decryption keys from \gls{vTPM}.
 \item \textbf{EoP:} Compromised \gls{vTPM} grants access to encrypted storage.
\vspace{3pt} \end{myBullets}
\end{minipage}&\low&\high&\med
\\ \cline{2-9}









& \begin{minipage}{\linewidth}
{Hyperjacking (Malicious Hypervisor Injection)}
\end{minipage}
& \begin{minipage}{\linewidth}
{Hypervisor Rootkit Installation, Quantum-Assisted Evasion}
\end{minipage}
& \begin{minipage}{\linewidth}\vspace{0.1cm}
(a) Replacing or modifying the hypervisor for persistent control over guest VMs, 
(b) Exploiting flaws in \gls{PQC} authentication and key exchange, 
(c) Leveraging misconfigured hybrid cryptographic environments, 
(d) Using quantum-optimized obfuscation to evade detection.\vspace{0.1cm}
\vspace{0.1cm}\end{minipage}
& \begin{minipage}{\linewidth}\vspace{0.1cm}
Hardware-backed hypervisor integrity checks, quantum-resistant secure boot to prevent unauthorized hypervisor modifications, side-channel-resistant \gls{PQC}, verified \gls{PQC} authentication transition.
\vspace{0.1cm}\end{minipage}
& \begin{minipage}{\linewidth}
\begin{myBullets}\vspace{2pt}
 \item \textbf{Tampering:} Quantum-assisted obfuscation enables attackers to modify hypervisor execution and bypass security.
 \item \textbf{Info. Disclosure:} A compromised hypervisor can extract \gls{VM} memory, including post-quantum keys.
 \item \textbf{EoP:} A malicious hypervisor grants full control over guest VMs and infrastructure.
\vspace{3pt} \end{myBullets}
\end{minipage}
&\med&\high&\high
\\ \hline

\multirow{20}{*}{{Server}} 

& \begin{minipage}{\linewidth} {Firmware Vulnerabilities} \end{minipage}
& \begin{minipage}{\linewidth} Firmware Exploitation~\cite{10.1145/3432893} \end{minipage}
& \begin{minipage}{\linewidth} 
(a) Code injection via buffer overflow in firmware, 
(b) \glspl{SCA} targeting firmware execution, 
(c) Malicious firmware updates modifying cryptographic operations. 
\end{minipage}
& \begin{minipage}{\linewidth} 
Firmware integrity checks, Secure Boot, \gls{TPM} -backed attestation, hardware-based security modules, cryptographic firmware signing, anomaly detection for firmware execution logs. 
\end{minipage}
& \begin{minipage}{\linewidth} 
\begin{myBullets}\vspace{2pt} 
\item \textbf{Tampering:} Attackers modify firmware to alter server functionality. 
\item \textbf{Repudiation:} Attackers erase or modify firmware logs to cover intrusion traces.
\item \textbf{Info. Disclosure:} \glspl{SCA} extract firmware execution data. 
\item \textbf{EoP:} Exploiting firmware vulnerabilities enables root-level access. 

\vspace{3pt} \end{myBullets} 
\end{minipage}&\med&\high&\high \\ \cline{2-9}

& \begin{minipage}{\linewidth} {Side-Channel on Server Key Management, Physical Attacks on Server Hardware} \end{minipage}
& \begin{minipage}{\linewidth} \glspl{SCA}~\cite{schellenberg2018remote, sayakkara2019survey, accikkapi2019side, genkin2017acoustic, le2019algebraic, breier2020countermeasure} \end{minipage}
& \begin{minipage}{\linewidth} 
(a) Attackers extract cryptographic keys via power, timing, electromagnetic, or cold boot attacks, 
(b) Fault injection disrupts \gls{PQC} computations, leading to cryptographic key leaks. 
\end{minipage}
& \begin{minipage}{\linewidth} 
Side-channel-resistant hardware, masked cryptographic implementations, frequent key rotation, use of HSMs, memory encryption (Intel TME, AMD SME), strong data center access controls. 
\end{minipage}
& \begin{minipage}{\linewidth} 
\begin{myBullets}\vspace{2pt}
\item \textbf{Tampering:} \glspl{SCA} and fault injection modify \gls{PQC} computations, leading to incorrect cryptographic signatures. 
\item \textbf{Info. Disclosure:} Attackers infer sensitive key material through side-channel observations. 
\item \textbf{EoP:} Attackers use extracted or modified cryptographic keys to gain unauthorized access. 
\vspace{3pt} \end{myBullets} 
\end{minipage}&\med&\high&\high \\ \cline{2-9}

& \begin{minipage}{\linewidth} {Resource Exhaustion Due to \gls{PQC}} \end{minipage}
& \begin{minipage}{\linewidth} Energy-Oriented DoS~\cite{masdari2016survey} \end{minipage}
& \begin{minipage}{\linewidth} 
(a) Overburdened \gls{PQC} traffic consuming processing power, 
(b) Excessive cryptographic handshakes degrading performance. 
\end{minipage}
& \begin{minipage}{\linewidth}\vspace{0.1cm} 
Resource management strategies, DoS mitigation frameworks, adaptive \gls{PQC} cryptographic policies, real-time monitoring, threshold-based rate limiting. 
\vspace{0.1cm} \end{minipage}
& \begin{minipage}{\linewidth} 
\begin{myBullets}
\item \textbf{Tampering:} Attackers manipulate cryptographic workloads to degrade system performance.
\item \textbf{DoS:} Attackers exploit \gls{PQC}’s high computational cost to exhaust processing resources. 
 
 \vspace{3pt} \end{myBullets} 
\end{minipage} &\med&\med&\med\\ \cline{2-9}










& \begin{minipage}{\linewidth} {Hardware Supply Chain Attacks} \end{minipage}
& \begin{minipage}{\linewidth} Malicious Implants in Server Hardware \end{minipage}
& \begin{minipage}{\linewidth} Attackers introduce malicious microchips or firmware backdoors at the hardware supply-chain level. \end{minipage}
& \begin{minipage}{\linewidth} Secure hardware attestation, vendor security audits, cryptographic verification of firmware integrity, tamper-proof hardware mechanisms. \end{minipage}
& \begin{minipage}{\linewidth} 
\begin{myBullets}\vspace{2pt} 
\item \textbf{Tampering:} Attackers modify server components at the manufacturing stage. 
\item \textbf{Info. Disclosure:} Hidden hardware backdoors exfiltrate sensitive data. 
\item \textbf{EoP:} Attackers gain persistent access via pre-installed malware. 
\vspace{3pt} \end{myBullets} 
\end{minipage} &\low&\high&\med\\ \cline{2-9}










& \begin{minipage}{\linewidth} Remote Firmware Manipulation via FOTA (Firmware Over-The-Air)\end{minipage}
& \begin{minipage}{\linewidth} Malicious Firmware Updates \end{minipage}
& \begin{minipage}{\linewidth} Attackers inject malicious firmware updates, modifying server behavior and cryptographic execution.\end{minipage} 
& \begin{minipage}{\linewidth} Secure Boot, cryptographic firmware signing, immutable hardware security enclaves, verified boot processes. \end{minipage}
& \begin{minipage}{\linewidth} 
 \begin{myBullets}\vspace{2pt} 
 \item \textbf{Tampering:} Attackers modify firmware to control server functions. 
 \item \textbf{Repudiation:} Attackers erase firmware logs to conceal malicious updates. 
 \item \textbf{EoP:} Exploiting firmware updates allows system-wide control. 

\vspace{3pt} \end{myBullets} 
\end{minipage} &\low&\high&\med \\ \cline{2-9}










& \begin{minipage}{\linewidth} \gls{TEE} Exploits\end{minipage}
& \begin{minipage}{\linewidth} \glspl{SCA} on Secure Enclaves\end{minipage} 
&\begin{minipage}{\linewidth} Attackers break into Intel SGX, AMD SEV, or ARM TrustZone via fault injection or microarchitectural side-channels.\vspace{0.1cm}\end{minipage} 
&\begin{minipage}{\linewidth} \gls{TEE}-hardening techniques, quantum-resistant enclave designs, noise injection for side-channel resistance, microcode security updates. \end{minipage}
& \begin{minipage}{\linewidth} 
 \begin{myBullets}\vspace{2pt}
 \item \textbf{Info. Disclosure:} \glspl{SCA} leak secrets stored inside a \gls{TEE}. 
 \item \textbf{Tampering:} Fault injection corrupts enclave-protected data. 
 \vspace{3pt} \end{myBullets} 
\end{minipage} &\med&\high&\high\\ \cline{2-9}











& \begin{minipage}{\linewidth} {Persistent Rootkit \& Bootloader Attacks} \end{minipage}
& \begin{minipage}{\linewidth} UEFI/BIOS Firmware Rootkits \& Persistent Malware Infections \end{minipage}
& \begin{minipage}{\linewidth} \vspace{0.1cm} 
Attackers inject persistent malware into boot firmware (UEFI/BIOS), modifying the bootloader to execute unsigned code, survive \gls{OS} reinstalls, and evade detection. 
\vspace{0.1cm} \end{minipage}
& \begin{minipage}{\linewidth} 
UEFI Secure Boot, \gls{TPM} -backed attestation, BootGuard security, firmware integrity monitoring, digitally signed BIOS updates. 
\end{minipage}
& \begin{minipage}{\linewidth} 
\begin{myBullets}\vspace{2pt}
\item \textbf{Tampering:} Attackers modify boot firmware to ensure persistent control. 
\item \textbf{EoP:} Rootkits provide full OS-level compromise. 
 \vspace{3pt} \end{myBullets} 
\end{minipage} &\low&\high&\med\\ \hline













\multirow{30}{*}{\begin{minipage}{\linewidth} {Storage} \end{minipage}} 
& \begin{minipage}{\linewidth} { \gls{PQC} Key Size Strain on Storage Systems} \end{minipage} 
& \begin{minipage}{\linewidth} Overhead from \gls{PQC} Algorithms~\cite{baseri2025evaluation} \end{minipage} 
& \begin{minipage}{\linewidth}
\vspace{0.1cm} 
 Larger \gls{PQC} keys introduce fragmentation, inefficient retrieval operations, and increased storage overhead. 
Exploiting key storage inefficiencies leads to performance degradation, retrieval delays, and storage exhaustion.
\vspace{0.1cm} 
\end{minipage} 
& \begin{minipage}{\linewidth} 
Storage-efficient \gls{PQC} implementations, hybrid key management schemes, optimized memory allocation, and compression techniques for \gls{PQC} keys.
\end{minipage} 
& \begin{minipage}{\linewidth} 
\begin{myBullets}\vspace{2pt}
 \item \textbf{Tampering:} Attackers manipulate \gls{PQC} key storage to inject invalid or compromised keys. 
 \item \textbf{DoS:} Large \gls{PQC} key sizes overwhelm storage systems, leading to degraded performance and increased system latency.
 \vspace{3pt} \end{myBullets} 
\end{minipage}&\med&\med&\med \\ \cline{2-9} 

& \begin{minipage}{\linewidth} {PQC-Induced Backup Failures} \end{minipage} 
& \begin{minipage}{\linewidth} Large Key Sizes \& Computational Overhead \end{minipage} 
& \begin{minipage}{\linewidth} \vspace{0.1cm} 
 \gls{PQC} encryption overhead increases storage demand, causing backup failures and reducing data recovery reliability. 
 Excessive \gls{PQC} ciphertext size leads to disk space depletion, delaying or corrupting backup restoration processes.
\vspace{0.1cm} \end{minipage} 
& \begin{minipage}{\linewidth}  \vspace{0.1cm}
Storage-aware \gls{PQC} key management, optimized backup strategies, quantum-secure storage compression techniques, and distributed fault-tolerant backup solutions.
\end{minipage} 
& \begin{minipage}{\linewidth} 
\begin{myBullets}\vspace{2pt}
 \item \textbf{Tampering:} Attackers inject corrupted \gls{PQC}-encrypted backups to cause failures or prevent restoration. 
 \item \textbf{DoS:} Backup failures disrupt access to \gls{PQC}-encrypted data, potentially leading to data loss.
 \vspace{3pt} \end{myBullets} 
\end{minipage} &\med&\med&\med \\ \cline{2-9} 

& \begin{minipage}{\linewidth} { \gls{PQC} Storage Security Risks} \end{minipage} 
& \begin{minipage}{\linewidth} Implementation Weaknesses, Resource Exhaustion, and Hybrid Cryptographic Misconfiguration \end{minipage} 
& \begin{minipage}{\linewidth} \vspace{0.1cm} 
Weak or misconfigured \gls{PQC} implementations, excessive key sizes, or hybrid cryptographic models introduce inefficiencies, leading to increased attack surfaces. 
Storage misconfigurations can be exploited for rollback attacks, unauthorized data persistence, or Out-Of-Memory (OOM) conditions.
\vspace{0.1cm} \end{minipage} 
& \begin{minipage}{\linewidth} 
Optimized \gls{PQC} key management, hybrid cryptographic model validation, hardware-assisted quantum-resistant key storage, and forward secrecy mechanisms.
\end{minipage} 
& \begin{minipage}{\linewidth} 
\begin{myBullets}\vspace{2pt}
 \item \textbf{DoS:} Large \gls{PQC} keys and ciphertexts cause storage overload, leading to degraded system performance.
 \item \textbf{Tampering:} Attackers manipulate \gls{PQC} encryption parameters to weaken security guarantees.
 \vspace{3pt} \end{myBullets} 
\end{minipage}&\med&\med&\med \\ \cline{2-9}

& \begin{minipage}{\linewidth} {Fault Injection on Storage Encryption Modules} \end{minipage} 
& \begin{minipage}{\linewidth} Quantum-Assisted Fault Injection on \gls{PQC} Cryptographic Processes \end{minipage} 
& \begin{minipage}{\linewidth} \vspace{0.1cm} 
 \gls{PQC}’s increased execution time and complexity induce faults, leading to unpredictable decryption failures, ciphertext corruption, or key leakage in disk encryption. 
Electromagnetic pulse (EMP) and voltage glitching techniques can further bypass \gls{PQC} protections.
\vspace{0.1cm} \end{minipage} 
& \begin{minipage}{\linewidth} \vspace{0.1cm}
Implementation of fault-tolerant \gls{PQC} cryptographic schemes, redundant storage integrity checks, quantum-resistant error correction, real-time anomaly detection in cryptographic operations, and hardware-backed encryption key protection (e.g., \gls{TPM} , HSM, PUF-based key storage).\vspace{0.1cm}
\end{minipage} 
& \begin{minipage}{\linewidth} 
\begin{myBullets}\vspace{2pt}
 \item \textbf{Tampering:} Attackers modify encrypted storage contents via induced faults. 
 \item \textbf{Info. Disclosure:} Corrupted \gls{PQC} data can lead to unintentional information leakage. 
 \item \textbf{EoP:} Fault-induced key corruption may allow unauthorized decryption or privilege escalation. 
 \vspace{3pt} \end{myBullets} 
\end{minipage}&\low&\high&\med \\ \cline{2-9}

& \begin{minipage}{\linewidth} {Storage Controller Manipulation} \end{minipage} 
& \begin{minipage}{\linewidth} Firmware or Software Exploitation \end{minipage} 
& \begin{minipage}{\linewidth} \vspace{0.1cm} 
(a) Modified storage controller firmware to bypass \gls{PQC} encryption processes, modify access control policies, or introduce persistent storage backdoors. 
(b) Compromising storage controllers allows attackers to decrypt \gls{PQC}-encrypted data without proper authorization.
\vspace{0.1cm} \end{minipage} 
& \begin{minipage}{\linewidth} 
Secure firmware updates, \gls{PQC}-aware storage controller integrity checks, strict access control enforcement, and digitally signed storage firmware updates.
\end{minipage} 
& \begin{minipage}{\linewidth} 
\begin{myBullets}\vspace{2pt}
 \item \textbf{Tampering:} Attackers modify encryption processes to leak storage encryption keys or weaken cryptographic integrity. 
 \item \textbf{EoP:} Firmware compromise enables persistent, unauthorized access to encrypted storage. 
 \vspace{3pt} \end{myBullets} 
\end{minipage} &\low&\high&\med \\ \hline










\multirow{12}{*}{Networking} 
& \begin{minipage}{\linewidth} { \gls{PQC} \glspl{SCA}} \end{minipage}
& \begin{minipage}{\linewidth} \gls{PQC} Algorithm-Specific Cryptanalysis, \glspl{SCA}, Fault Injection~\cite{lounis2020attacks, blomer2014tampering, accikkapi2019side, schellenberg2018remote, genkin2017acoustic, le2019algebraic, breier2020countermeasure,tasso2021resistance} \end{minipage}
& \begin{minipage}{\linewidth} \vspace{0.1cm} (a) Mathematical cryptanalysis exploits weaknesses in \gls{PQC} key exchange. (b) \glspl{SCA} extract session keys via electromagnetic or timing analysis. \newline (c) Fault injection manipulates \gls{PQC} operations to force cryptographic downgrades.
\vspace{0.1cm} \end{minipage}
& \begin{minipage}{\linewidth} Continuous \gls{PQC} algorithm updates, hybrid encryption models (classical + \gls{PQC}), side-channel-resistant hardware, randomized encryption operations, noise-injection techniques, fault-resistant \gls{PQC} protocols, error detection in cryptographic handshakes. \end{minipage}
& \begin{minipage}{\linewidth} 
\begin{myBullets}\vspace{2pt}
\item \textbf{Tampering:} Attackers manipulate \gls{PQC} key exchange vulnerabilities to alter cryptographic properties. 
\item \textbf{Info. Disclosure:} \glspl{SCA} leak post-quantum TLS secrets and expose encrypted traffic. 
\item \textbf{EoP:} Extracted \gls{PQC} keys allow unauthorized access to VPN/TLS/IPSec tunnels. 
 \vspace{3pt} \end{myBullets} 
\end{minipage}&\med&\high&\high \\ \cline{2-9} 

& \begin{minipage}{\linewidth} { \gls{PQC} Fragmentation-Induced Overhead} \end{minipage}
& \begin{minipage}{\linewidth} TCP/UDP Packet Fragmentation from Large \gls{PQC} Keys~\cite{jafarigiv2020scalable, muller2020retrofitting, essay89509} \end{minipage}
& \begin{minipage}{\linewidth} \vspace{0.1cm} Oversized \gls{PQC} key exchanges cause excessive fragmentation in network traffic, leading to retransmissions and network congestion. \vspace{0.1cm} \end{minipage}
&  \begin{minipage}{\linewidth} \vspace{0.1cm} Traffic optimization for \gls{PQC} handshakes, adaptive buffer size management, network congestion control, optimized cryptographic handshakes. \vspace{3pt}\end{minipage} 
& \begin{minipage}{\linewidth} 
\begin{myBullets}\vspace{2pt}
\item \textbf{DoS:} Large \gls{PQC} packet fragmentation leads to increased retransmissions and delays, degrading network performance. 
 \vspace{3pt} \end{myBullets} 
\end{minipage}&\med&\med&\med \\ \cline{2-9}










& \begin{minipage}{\linewidth} { \gls{PQC} Misconfiguration Exploits} \end{minipage}
& \begin{minipage}{\linewidth} System Sabotage via post-quantum TLS/IPSec Misconfigurations~\cite{Threatinsiderdef, zhang2018novel} \end{minipage}
& \begin{minipage}{\linewidth}\vspace{0.1cm} (a) Attackers exploit buffer overflows in ost-quantum TLS/IPSec implementations to manipulate encrypted network traffic. \newline (b) Weak cryptographic configurations allow privilege escalation via improperly set key exchange policies.\vspace{0.1cm} \end{minipage}
& \begin{minipage}{\linewidth}Secure \gls{PQC} key exchange policies, cryptographic audits, buffer overflow mitigation techniques, configuration hardening, \gls{PQC}-aware intrusion detection systems. \end{minipage}
& \begin{minipage}{\linewidth} 
\begin{myBullets}\vspace{2pt}\item \textbf{Tampering:} Misconfigured \gls{PQC} policies allow injection of invalid cryptographic parameters, potentially altering data integrity. 
\item \textbf{EoP:} Attackers escalate privileges by exploiting weak \gls{PQC} configurations. 
 \vspace{3pt} \end{myBullets} 
\end{minipage}&\med&\high&\high \\ 
\hline 

\end{tabular}%
}\vspace{-0.7cm}
\label{table:Post-Transition-Networking}
\end{table*}

\subsection{Risk Assessment of Post-Transition Quantum Threats Across \gls{CC} Layers}
\label{sec:risk_assessment_posttransition}

This section presents a comprehensive risk assessment of the post-transition quantum threat landscape across various \gls{CC} layers. We consider a scenario where organizations have transitioned to \gls{PQC}, and large-scale, fault-tolerant quantum computers {exist}, but are not necessarily ubiquitous or readily accessible to all attackers. The focus shifts from the threat of breaking {classical} cryptography to the vulnerabilities and risks {within} the \gls{PQC}-enabled environment, and the exploitation of {classical} vulnerabilities within that environment. The likelihood and impact criteria are adapted from Tables~\ref{table:likelihood} and~\ref{table:impact}, with interpretations tailored to this post-transition context.

The likelihood (L) of each post-transition threat is determined based on four key factors: {(1) Vulnerability Presence and Type:} Vulnerabilities are categorized as either arising from {inherent properties} of \gls{PQC} (e.g., larger key sizes, increased computational overhead), from {implementation flaws} (e.g., buffer overflows in \gls{PQC} code, misconfigurations), or from {classic vulnerabilities} existing within a \gls{PQC}-enabled system. Inherent properties are considered \textit{``Partial Vulnerabilities,''} while flaws represent more traditional vulnerabilities; {(2) Exposure:} How exposed is the vulnerable component? This considers whether the component is a core system element (e.g., kernel, hypervisor), part of a widely used protocol (e.g., TLS), or a less critical, isolated system; {(3) Attacker Resource Requirements:} This evaluates the resources needed for exploitation. While we assume the {existence} of quantum computers, we differentiate between attacks requiring readily available tools/techniques, those needing advanced classical methods, and those potentially enhanced by {future} quantum capabilities (e.g., sophisticated quantum-assisted side-channel analysis). The {availability} of quantum resources to all threat actors is {not} assumed; and {(4) Mitigation Effectiveness and Deployment:} The presence and effectiveness of countermeasures, and their level of deployment, are considered. This includes patching, secure coding practices, \glspl{HSM}, intrusion detection/prevention systems, and monitoring systems. {Active monitoring and patching} is a key differentiator. 
Based on these factors, likelihood (L) is categorized as: (a) \textit{High (H):} Vulnerabilities that are easily exploitable with readily available tools, affect critical systems, and lack robust mitigations. These are often classic vulnerabilities (e.g., buffer overflows, code injection) that exist {within} \gls{PQC} implementations or related system components; (b) \textit{Medium (M):} Vulnerabilities arising from inherent properties of \gls{PQC} (e.g., larger key sizes, increased computational overhead) or requiring more sophisticated attack techniques (e.g., \glspl{SCA}). Mitigations exist but may not be perfect or universally deployed; and (c) \textit{Low (L):} Vulnerabilities that are difficult to exploit, require highly specialized resources (including advanced quantum capabilities), affect non-critical systems, or have robust and widely deployed mitigations. This includes attacks relying on speculative or highly advanced quantum capabilities not yet readily available.

 
 
 
The impact (I) of each threat is evaluated based on three key factors: (1) {Cryptographic Compromise:} The extent to which the vulnerability compromises cryptographic security. \textit{``Severe Cryptographic Breach''} indicates complete compromise of keys, signature forgery, or data decryption. \textit{``Partial Data Compromise''} suggests limited information leakage (e.g., through side-channels, metadata analysis); (2) {Operational Disruption:} The potential for system crashes, service outages, performance degradation, or \gls{DoS} conditions; and (3) {Ecosystem Effects and Regulatory Concerns:} How widespread is the impact? Does it affect a single VM, a host, multiple VMs, or the entire cloud infrastructure? \textit{``Widespread Ecosystem Effects''} implies a significant impact. Regulatory and compliance ramifications, including legal penalties and reputational damage, are also factored in. As a result, Impact (I) is categorized as: (a) \textit{High (H):} Vulnerabilities leading to severe cryptographic breaches, extended operational downtime affecting critical services, or widespread ecosystem effects (multiple systems, significant data loss, major regulatory violations); (b) \textit{Medium (M):} Vulnerabilities resulting in partial data compromise, moderate service disruptions, or impacts limited to a subset of systems. Regulatory investigations or minor penalties are possible; and (c) \textit{Low (L):} Vulnerabilities with minimal impact, causing limited, quickly resolved disruptions, or affecting only non-critical systems, with negligible financial or reputational damage.

 
 

The overall risk (R) is determined by combining the likelihood (L) and impact (I) assessments. A high likelihood and high impact result in a high risk, while a low likelihood and low impact result in a low risk. Mixed combinations (e.g., low likelihood, high impact) typically result in a medium risk. The overall risk levels are determined using the risk matrix illustrated in Figure~\ref{fig:risk-matrix}. Detailed explanations for assessing likelihood, impact, and determining the exploitation methods and attack vectors are provided in the preceding sections for each \gls{CC} layer. Table~\ref{table:Post-Transition} summarizes the vulnerabilities, threats, countermeasures, and the assessed Likelihood, Impact, and Risk levels for each layer. A key takeaway from this analysis is that many of the {highest} risks in a post-transition environment stem from {classical} vulnerabilities within the \gls{PQC}-enabled systems, rather than from direct quantum attacks on the \gls{PQC} algorithms themselves.
\section{Hybrid Security Approaches in Cloud Environments}
\label{sec:hybrid-cloud-security}

Cloud organizations maintain diverse distributed infrastructures requiring gradual migration 
from classical to quantum-safe cryptography~\cite{driscoll-pqt-hybrid-terminology-02, 
BASERI2025104272,giron2023post}. Hybrid security mechanisms\break ensure legacy compatibility 
while introducing quantum-secure safeguards~\cite{ietf-tls-hybrid-design-12,giacon2018kem,
giron2023post}. This section examines algorithmic,\break  certificate, and protocol hybrid 
strategies for \gls{CC}, emphasizing backward compatibility, minimized downtime, 
and crypto-agility~\cite{marchesi2025survey}.

\subsection{Hybrid Algorithmic Strategies}\label{subsec:hybrid_alg}

Hybrid algorithmic approaches provide cryptographic resilience for dynamic cloud workloads, 
such as short-lived containers and \glspl{VM}, by combining classical and post-quantum 
schemes~\cite{turnip2025towards}. This dual-layer protection secures ephemeral communications 
via \gls{KEM/ENC} (e.g., merging RSA/ECC with Kyber), ensuring confidentiality even if one 
primitive is compromised~\cite{campagna2020security, whyte2017quantum, kiefer2018hybrid, 
bindel2019hybrid, dowling2020many, giron2023post}. This strategy maintains legacy 
compatibility for existing workloads while enabling newly provisioned services to adopt 
post-quantum primitives without disruptive overhauls~\cite{brendel2019breakdown}. 
Table~\ref{tab:Through-Transition-Alg} summarizes technical combination methods for 
\gls{KEM/ENC} (concatenation, KDF/PRF/XOR-based) and signatures (nesting, concatenation), 
evaluating these against critical security requirements including 
transcript/identity/\break algorithm-ID binding, CCA robustness, and deployment overhead.

\begin{table*}[!htbp]
\vspace{-0.2cm}
\caption{Hybrid Algorithmic Strategies for Cloud Workloads}
\vspace{-0.4cm}
\scriptsize
\label{tab:Through-Transition-Alg}
\resizebox{\textwidth}{!}{%
\begin{tabular}{|p{0.06\linewidth}|p{0.1\linewidth}|p{0.17\linewidth}|p{0.24\linewidth}|p{0.29\linewidth}|p{0.4\linewidth}|}
\hline
\multirow{1}{*}{\begin{minipage}{\linewidth}\textbf{Strategy}\end{minipage}} & 
\multirow{1}{*}{\begin{minipage}{\linewidth}\textbf{Crypto Types}\end{minipage}} & 
\multirow{1}{*}{\begin{minipage}{\linewidth}\textbf{Combination Approaches$^{\dag}$}\end{minipage}}& 
\multirow{1}{*}{\begin{minipage}{\linewidth}\textbf{Mathematical Definition}\end{minipage}} & 
\multirow{1}{*}{\begin{minipage}{\linewidth}\textbf{Pros}\end{minipage}} & 
\multirow{1}{*}{\begin{minipage}{\linewidth}\textbf{Cons}\end{minipage}} \\ \hline

\multirow{40}{*}{\begin{minipage}{\linewidth}Hybrid\end{minipage}} & 
\multirow{25}{*}{\begin{minipage}{\linewidth}{KEM/ENC~\cite{barker2017recommendation,ietf-tls-hybrid-design-12,giacon2018kem,giron2023post}}\end{minipage}}& 
\begin{minipage}{\linewidth}Concatenation~\cite{barker2017recommendation}\end{minipage}&  {\begin{minipage}[c]{\linewidth}
\vspace{-0.1cm}$K_{\mathrm{hyb}} = K_1 \,\|\, K_2$, where
\begin{myBullets} 
\item $K_i$ are shared secrets from individual \glspl{KEM/ENC}.
\end{myBullets}
\end{minipage}}&
\begin{minipage}{\linewidth}
\begin{myBullets}\vspace{2pt}
\item Lightweight, simple logic
\item Easy to implement
\vspace{1pt}\end{myBullets}
\end{minipage}
&
\begin{minipage}{\linewidth}
\begin{myBullets}\vspace{2pt}
\item No inherent integrity (pair with AEAD/MAC)
\item Requires transcript/identity/alg-ID binding to prevent downgrade/MITM
\item FIPS 140 validation code updates; larger handshakes/decapsulation cost
\vspace{1pt}\end{myBullets}
\end{minipage}\\ \cline{3-6} 

& &  
\begin{minipage}{\linewidth}Concat-KDF~\cite{campagna2020security,whyte2017quantum,kiefer2018hybrid}\end{minipage}
& 
{\begin{minipage}[c]{\linewidth}
$K_{\mathrm{hyb}} = \mathrm{KDF}(K_1 \,\|\, K_2)$, where
\begin{myBullets} 
\item $\mathrm{KDF}$ is a secure key derivation function.
\end{myBullets}
\end{minipage}} &
\begin{minipage}{\linewidth}
\begin{myBullets}\vspace{2pt}
\item Single KDF limits brute-force surface
\item Backward compatible deployment
\vspace{1pt}\end{myBullets}
\end{minipage}
& 
\begin{minipage}{\linewidth}
\begin{myBullets}\vspace{2pt}
\item Must bind context/alg-ID and peer identities (prevent substitution/downgrade)
\item Proofs mainly classical; KDF/negotiation adds overhead
\vspace{1pt}\end{myBullets}
\end{minipage}  \\ \cline{3-6} 

& &   \begin{minipage}{\linewidth}Cascade-KDF~\cite{campagna2020security}\end{minipage}
& {\begin{minipage}[c]{\linewidth}
$K_{\mathrm{hyb}} = \mathrm{KDF}_2\!\big(K_2,\, \mathrm{KDF}_1(K_1)\big)$
\end{minipage}} &
\begin{minipage}{\linewidth}
\begin{myBullets}\vspace{2pt}
\item Iterative hardening
\item Larger brute-force work factor
\vspace{1pt}\end{myBullets}
\end{minipage}
&
\begin{minipage}{\linewidth}
\begin{myBullets}\vspace{2pt}
\item Requires tight transcript/context binding
\item Random-oracle style proofs; higher code complexity/latency
\vspace{1pt}\end{myBullets}
\end{minipage} \\ \cline{3-6}

& & 
\begin{minipage}{\linewidth}Dual-PRF~\cite{bindel2019hybrid,dowling2020many}\end{minipage} & {\begin{minipage}[c]{\linewidth}
$K_{\mathrm{hyb}} = \mathrm{PRF}_{K_1}(K_2) \oplus \mathrm{PRF}_{K_2}(K_1)$
\end{minipage}}&
\begin{minipage}{\linewidth}
\begin{myBullets}\vspace{2pt}
\item IND-CCA if one input remains secure
\item Proofs against (partially) quantum adversaries
\vspace{1pt}\end{myBullets}
\end{minipage}
& 
\begin{minipage}{\linewidth}
\begin{myBullets}\vspace{2pt}
\item Bind peer IDs and alg-IDs to transcript
\item Extra preprocessing and combination work
\vspace{1pt}\end{myBullets}
\end{minipage} \\ \cline{3-6} 

& &  \begin{minipage}{\linewidth}Nested-Dual-PRF~\cite{bindel2019hybrid}\end{minipage} & 
 {\begin{minipage}[c]{\linewidth}
$K_{\mathrm{hyb}} = \mathrm{PRF}_{K_1}\!\big(\mathrm{PRF}_{K_2}(K_1 \,\|\, K_2)\big)$
\end{minipage}} &
\begin{minipage}{\linewidth}
\begin{myBullets}\vspace{2pt}
\item Stronger combining semantics
\item Solid classical/quantum proofs
\vspace{1pt}\end{myBullets}
\end{minipage}
&
\begin{minipage}{\linewidth}
\begin{myBullets}\vspace{2pt}
\item Must bind transcript/context and alg-IDs
\item Higher complexity; added nesting cost
\vspace{1pt}\end{myBullets}
\end{minipage} \\ \cline{3-6} 

& & \begin{minipage}{\linewidth}Split-key-PRF~\cite{giacon2018kem}\end{minipage}&
 {\begin{minipage}[c]{\linewidth}\vspace{2pt}
$K_{\mathrm{hyb}} = \mathrm{PRF}_{K_1[1..n]}(K_2) \oplus \mathrm{PRF}_{K_1[n+1..2n]}(K_2)$\vspace{1pt}
\end{minipage}}&  
\begin{minipage}{\linewidth}
\begin{myBullets}\vspace{2pt}
\item Parallel combiner; preserves CCA security
\item Efficient on classical hardware
\vspace{1pt}\end{myBullets}
\end{minipage} 
& 
\begin{minipage}{\linewidth}
\begin{myBullets}\vspace{2pt}
\item Bind identity/alg-ID; proofs often classical
\item Parallelization overhead at scale
\vspace{1pt}\end{myBullets}
\end{minipage} \\ \cline{3-6}

& & \begin{minipage}{\linewidth}XOR~\cite{giacon2018kem,brendel2019breakdown}\end{minipage}
& {\begin{minipage}[c]{\linewidth}
$K_{\mathrm{hyb}} = K_1 \oplus K_2$
\end{minipage}}&
\begin{minipage}{\linewidth}
\begin{myBullets}\vspace{2pt}
\item Simple; very lightweight
\item Easy to implement
\vspace{1pt}\end{myBullets}
\end{minipage} 
& 
\begin{minipage}{\linewidth}
\begin{myBullets}\vspace{2pt}
\item Not CCA-secure; add MAC/\gls{AEAD} for integrity
\item Reversible if a subkey is exposed; primarily classical proofs
\vspace{1pt}\end{myBullets}
\end{minipage} \\ \cline{3-6} 

& &  \begin{minipage}{\linewidth}XOR-then-MAC~\cite{bindel2019hybrid}\end{minipage}&
 {\begin{minipage}[c]{\linewidth}\vspace{2pt}
$K_{\mathrm{hyb}} = K_1 \oplus K_2$ and protect with $\mathrm{MAC}(K_{\mathrm{hyb}},\text{context})$
\end{minipage}}
&
\begin{minipage}{\linewidth}
\begin{myBullets}\vspace{2pt}
\item Retains IND-CCA if MAC is secure
\item Works against classical/quantum adversaries
\vspace{1pt}\end{myBullets}
\end{minipage}
& 
\begin{minipage}{\linewidth}
\begin{myBullets}\vspace{2pt}
\item Bind MAC to peer/context/alg-ID; correct use critical
\item MAC verification adds latency/cost under load
\vspace{1pt}\end{myBullets}
\end{minipage} \\ \cline{3-6} 

& &  \begin{minipage}{\linewidth}XOR-then-PRF~\cite{giacon2018kem}\end{minipage}&\begin{minipage}[c]{\linewidth}
\vspace{-0.1cm}$K_{\mathrm{hyb}} = \mathrm{PRF}(K_1 \oplus K_2, \text{nonce})$
\end{minipage} & 
\begin{minipage}{\linewidth}
\begin{myBullets}\vspace{2pt}
\item PRF-based integrity check
\item Straightforward design
\vspace{1pt}\end{myBullets}
\end{minipage}
& 
\begin{minipage}{\linewidth}
\begin{myBullets}\vspace{2pt}
\item Not CCA-secure; related-key risks
\item Must bind to context/identity/alg-ID; add \gls{AEAD} if needed
\vspace{1pt}\end{myBullets}
\end{minipage} \\ 

\cline{2-6} 

& \multirow{9}{*}{\begin{minipage}{\linewidth}Signature~\cite{bindel2017transitioning,ghinea2022hybrid}\end{minipage}} & 
\begin{minipage}{\linewidth}Concatenation~\cite{bindel2017transitioning}\end{minipage} &\begin{minipage}[c]{\linewidth}
$\mathrm{Sign}_{\mathrm{hyb}} = \sigma_1 \,\|\, \sigma_2$, where 
 \begin{myBullets}
\item $\sigma_i=\mathrm{Sign}_i(sk_i,m)$.
\end{myBullets}
\end{minipage}&
\begin{minipage}{\linewidth}
\begin{myBullets}\vspace{2pt}
\item Lightweight; simple verification
\item Preserves unforgeability if both schemes remain secure
\vspace{1pt}\end{myBullets}
\end{minipage} 
& 
\begin{minipage}{\linewidth}
\begin{myBullets}\vspace{2pt}
\item No non-separability; susceptible to downgrade if policy allows OR semantics
\item Larger artifacts; potential verification storms in high-traffic services
\vspace{1pt}\end{myBullets}
\end{minipage} \\ \cline{3-6} 

& & 
\begin{minipage}{\linewidth}Weak Nesting~\cite{bindel2017transitioning}\end{minipage} & \begin{minipage}[c]{\linewidth}\vspace{2pt}
$\mathrm{Sign}_{\mathrm{hyb}} = \mathrm{Sign}_2(sk_2,\sigma_1)$, where
 \begin{myBullets}
\item $\sigma_1=\mathrm{Sign}_1(sk_1,m)$.
\end{myBullets}
\end{minipage}&
\begin{minipage}{\linewidth}
\begin{myBullets}\vspace{2pt}
\item Non-separability for the outer scheme
\item Simple migration step from concatenation
\vspace{1pt}\end{myBullets}
\end{minipage} 
& 
\begin{minipage}{\linewidth}
\begin{myBullets}\vspace{2pt}
\item If inner scheme weak, entire signature can be forged
\item Policy ambiguity across layers; nested verification overhead
\vspace{1pt}\end{myBullets}
\end{minipage} \\ \cline{3-6} 

& &  \begin{minipage}{\linewidth}Strong Nesting~\cite{bindel2017transitioning,ghinea2022hybrid}\end{minipage} &\begin{minipage}[c]{\linewidth}\vspace{2pt}
$\mathrm{Sign}_{\mathrm{hyb}}=(\sigma_1,\sigma_2)$, where \begin{myBullets}
\item $\sigma_1=\mathrm{Sign}_1(sk_1,m)$, 
\item $\sigma_2=\mathrm{Sign}_2(sk_2,m\|\sigma_1)$.
\end{myBullets}
\end{minipage} &
\begin{minipage}{\linewidth}
\begin{myBullets}\vspace{2pt}
\item Retains non-separability
\item Unforgeability holds unless both schemes fail
\vspace{1pt}\end{myBullets}
\end{minipage} 
& 
\begin{minipage}{\linewidth}
\begin{myBullets}\vspace{2pt}
\item More complex verification and larger signatures
\item Clear AND semantics/policy needed; higher compute cost
\vspace{1pt}\end{myBullets}
\end{minipage}
\\ \cline{3-6} 

& &  \begin{minipage}{\linewidth}Dual Nesting~\cite{bindel2017transitioning}\end{minipage} & {\begin{minipage}[c]{\linewidth}\vspace{2pt}
$Sign_{hybrid}=(\sigma_2,\ \sigma_{\mathrm{outer}})$, where
\begin{myBullets}
\item $\sigma_2 = \mathrm{Sign}_2(sk_2, m)$ and
\item $\sigma_{\mathrm{outer}} = \mathrm{Sign}_1(sk_1,\ m \,\|\, \sigma_2)$.
 \end{myBullets}\vspace{1pt}
\end{minipage}} & 
\begin{minipage}{\linewidth}
\begin{myBullets}\vspace{2pt}
\item Each message protected by a distinct signature
\item Can remain unforgeable if authoritative outer scheme is secure
\vspace{1pt}\end{myBullets}
\end{minipage} 
& 
\begin{minipage}{\linewidth}
\begin{myBullets}\vspace{2pt}
\item Must define the authoritative layer to avoid confusion/downgrade
\item Larger artifacts and verification overhead
\vspace{1pt}\end{myBullets}
\end{minipage}
\\ \hline

\end{tabular}}
  \begin{tablenotes}[para,flushleft]
 \item[\dag] \textit{Combination Approaches} refers to how outputs of classical and post-quantum primitives are blended.
 \item[\ddag] \textit{Security considerations:} (i) Robust \gls{KEM/ENC} combiners preserve confidentiality unless \emph{all} inputs fail (intersection). (ii) AND-verified/strongly nested signature combiners preserve unforgeability unless \emph{all} components fail; weak nesting or unbound concatenation can fail if \emph{any} component fails (union). (iii) Prevent downgrade/MITM by binding transcript, peer identity, and algorithm identifiers in the protocol. (iv) Availability (\gls{DoS}) cost scales with handshake size, derivation/combination work, and verification load.
  \end{tablenotes}
  \vspace{-0.2cm}
\end{table*}

\begin{enumerate}[topsep=0ex, itemsep=0ex, wide, font=\itshape, labelwidth=!, labelindent=0pt, label*=6.1.\arabic*]
\item\textit{Hybrid Algorithm Strategies Risk Assessment.}
Security in hybrid deployments depends on combination approach, target property, and protocol bindings. For \gls{KEM/ENC}, robust combiners achieve intersection-style confidentiality (IND-CCA security holds if at least one input remains secure) provided handshakes bind transcript context, peer identities, and algorithm identifiers, and session keys are used only via \acrfull{AEAD}. Robust combiners include Concat-KDF and Cascade-KDF~\cite{campagna2020security,whyte2017quantum,kiefer2018hybrid}, and PRF-based designs (Dual-PRF, Nested-Dual-PRF, Split-key-PRF)~\cite{bindel2019hybrid,dowling2020many,giacon2018kem}. XOR-then-MAC reaches intersection semantics when MAC is bound to full transcript/context and alg-IDs~\cite{bindel2019hybrid}. Plain XOR and XOR-then-PRF are not CCA-secure and should be avoided; using raw concatenation directly as session key is non-robust~\cite{giacon2018kem,brendel2019breakdown}. Weak combiners should be replaced by robust KDF/PRF combiners; wrapping weak derivation with \gls{AEAD} alone does not repair KEM-side CCA guarantees.
For digital signatures, semantics are set by combiner and policy: AND-verification and strong nesting yield intersection behavior (unforgeability fails only if all schemes fail)~\cite{bindel2017transitioning,ghinea2022hybrid}; weak nesting behaves like union (breaking inner may break whole), and concatenation with OR/optional acceptance enables downgrade. Practical deployments should (i) bind algorithm identifiers and peer identities into transcript to preclude substitution and stripping, (ii) enforce deterministic acceptance policy (AND vs.\ OR) during negotiation, and (iii) align key-lifecycle operations with \gls{VM}/container churn so ephemeral keys do not outlive instances. 
Under these conditions, robust KDF/PRF combiner coupled with \gls{AEAD} and transcript/identity/algorithm-ID binding for KEMs, paired with AND/strong-nesting for signatures, offers standards-conformant path mitigating harvest-now–decrypt-later risk while progressing toward full post-quantum deployments~\cite{campagna2020security,whyte2017quantum,kiefer2018hybrid,bindel2019hybrid,dowling2020many,giron2023post,bindel2017transitioning,ghinea2022hybrid}.
\end{enumerate}

\subsection{Hybrid Certificate Strategies in Cloud \gls{PKI}}
\label{subsec:cloud-hybrid-cert}

\begin{table*}[!h]
\caption{Hybrid Certificate Strategies in Cloud \gls{PKI}}

\label{tab:hybrid-cert-strategies}
\centering
\footnotesize
\resizebox{\textwidth}{!}{%
\begin{tabular}{|p{0.06\linewidth}|p{0.22\linewidth}|p{0.11\linewidth}|p{0.55\linewidth}|p{0.2\linewidth}|p{0.25\linewidth}|}
\hline
\textbf{Strategy} & \textbf{Chain of Trust Best Practices} & \textbf{Approach} & \textbf{Implementation Details} & \textbf{Pros.} & \textbf{Cons.} \\ \hline

\multirow{3}{*}{\begin{minipage}{\linewidth}{Hybrid} \cite{truskovsky-lamps-pq-hybrid-x509-00,driscoll-pqt-hybrid-terminology-02}\end{minipage}}
&
\multirow{3}{*}{\begin{minipage}{\linewidth}
\begin{enumerate}[nosep,leftmargin=*,topsep=2pt]
\item \glspl{OID}{Root-first migration}: start with root CAs (long lifespans, 10--30 years).
\item \glspl{OID}{Trust anchor assurance}: maintain strongest controls at trust roots.
\item \glspl{OID}{Staged rollout}: leverage shorter intermediate/end-entity cycles (1--3 years).
\item \glspl{OID}{Backward compatibility}: ensure legacy path building during transition.
\end{enumerate}
\end{minipage}}
& \begin{minipage}{\linewidth}{Dual Certificate}~\cite{bindel2017transitioning,vogt2021quantum,bindel2019x}\end{minipage}
& \begin{minipage}{\linewidth} \vspace{0.1cm} 
Parallel classical and \gls{PQC} certificate chains with cross-references as needed.

\vspace{4pt}
\fbox{\begin{minipage}{0.96\linewidth}
\textbf{X.509 structure (primer):}\\
\texttt{Certificate ::= \{} \\
\quad \texttt{\underline{tbsCertificate},} \\
\quad \texttt{signatureAlgorithm,} \\
\quad \texttt{signatureValue \}} \\[2pt]
Only \underline{\texttt{tbsCertificate}} is hashed/signed. Validators should enforce equality of
\texttt{tbsCertificate.signature} and outer \texttt{Certificate.signatureAlgorithm}.
\end{minipage}}

Place \gls{PQC} policy \glspl{OID} and constraints inside \texttt{tbsCertificate} when enforcement is required.  
\end{minipage}\vspace{0.1cm} 
& \begin{minipage}{\linewidth}
\begin{myBullets}\vspace{2pt}
\item Minimal disruption to existing \gls{PKI} components.
\item Clear algorithmic separation; independent validation paths.
\item Gradual migration with explicit fallback.
\vspace{1pt}\end{myBullets}
\end{minipage}
& \begin{minipage}{\linewidth}
\begin{myBullets}\vspace{2pt}
\item Parallel hierarchy synchronization and inventory overhead.
\item Redundant metadata/storage; more complex path building.
\item Dual revocation (OCSP/CRL) ops and monitoring burden.
\item Risk of accepting a weaker chain if policy does not pin the stronger path; added validation load under traffic spikes.
\vspace{1pt}\end{myBullets}
\end{minipage}
\\ \cline{3-6}

& &
\multirow{2}{*}{\begin{minipage}{\linewidth}{Composite Certificate}~\cite{bindel2017transitioning,vogt2021quantum,bindel2019x}\end{minipage}}
& \begin{minipage}{\linewidth}\vspace{0.1cm} 
Classical certificate profile with \gls{PQC} components carried in standardized extensions \emph{inside} \texttt{tbsCertificate}.

\vspace{4pt}
\glspl{OID}{Critical implementation points:}
\begin{myBullets}\vspace{2pt}
\item Mark \gls{PQC} extensions \emph{critical} when \gls{PQC} validation is mandatory (forcing legacy validators to reject).
\item Use non-critical \gls{PQC} extensions when graceful fallback to classical validation is acceptable.
\item Bind algorithm choice/policy via \glspl{OID} in \texttt{tbsCertificate}.
\item Ensure DER canonicalization and algorithm-ID consistency checks.
\vspace{2pt}\end{myBullets}
\end{minipage}
& \begin{minipage}{\linewidth}
\begin{myBullets}\vspace{2pt}
\item Single artifact simplifies distribution and inventory.
\item Often backward-compatible when \gls{PQC} extensions are non-critical.
\item Unified validation workflow; fewer moving parts than dual chains.
\vspace{1pt}\end{myBullets}
\end{minipage}\vspace{0.1cm} 
& \begin{minipage}{\linewidth}
\begin{myBullets}\vspace{2pt}
\item Larger certificates; parser/validator complexity and interop profiling (RFC\,5280).
\item Extension criticality semantics must be carefully designed and tested.
\item If \gls{PQC} semantics are not policy-bound or marked \emph{critical}, validators may ignore them; heavier parsing cost under malformed inputs.
\vspace{1pt}\end{myBullets}
\end{minipage}
\\ \cline{4-6}

& &
& \begin{minipage}{\linewidth}\vspace{0.1cm} 
{Dual-signature approach}: Classical \texttt{signatureAlgorithm} combined with an additional \gls{PQC}  signature carried in an extension \textit{inside} \texttt{tbsCertificate}.

\vspace{3pt}
\fbox{\begin{minipage}{0.95\linewidth}
\textbf{Critical Policy Decision:}\\
Acceptance semantics must explicitly specify:\\
- \textbf{AND} (intersection): Both signatures must verify\\
- \textbf{OR} (union): Either signature may verify
\end{minipage}}

Clear policy expression prevents validation ambiguity and security bypasses.
\end{minipage}\vspace{0.1cm} 
& \begin{minipage}{\linewidth}
\begin{myBullets}\vspace{2pt}
\item Strong cryptographic binding within a single artifact.
\item Less redundancy than dual chains; simpler distribution.
\item Future-proofing with explicit policy expression (AND vs.\ OR).
\vspace{1pt}\end{myBullets}
\end{minipage}
& \begin{minipage}{\linewidth}
\begin{myBullets}\vspace{2pt}
\item Multiple signature verifications increase compute cost.
\item Larger artifacts affect bandwidth and caches at scale.
\item Cross-algorithm validation and policy logic are more complex.
\item With \emph{OR}/optional semantics, downgrade remains possible; \emph{AND} avoids this but raises verification cost.
\vspace{2pt}\end{myBullets}
\end{minipage}
\\ \hline
\end{tabular}}
\begin{tablenotes}[para,flushleft]
\scriptsize
\item[*] {X.509 Integrity Model:} Only \texttt{tbsCertificate} ("to-be-signed certificate") is hashed and signed. 
Validators should verify equality of \texttt{tbsCertificate.signature} and outer \texttt{Certificate.signatureAlgorithm}. 
Place \gls{PQC} keys, signatures, and policy \glspl{OID} \emph{inside} \texttt{tbsCertificate}. 
Mark \gls{PQC} extensions \emph{critical} when \gls{PQC} validation must be enforced (legacy validators will reject unrecognized critical extensions); 
use non-critical when classical fallback is acceptable (legacy validators will ignore unrecognized non-critical extensions). 
"Extension stripping" typically means a validator ignores a non-critical extension (actual removal would invalidate the signature).
\item[{\dag}] {Acceptance Semantics:} \emph{AND} (intersection) requires all signatures to validate, resisting downgrade if any algorithm remains unbroken; \emph{OR}/optional (union) succeeds if either validates and can permit downgrade to the weakest algorithm if policy is not enforced.
\item[{\ddag}] {Security Notes (certificate layer):} Dominant risks are (i) accepting weaker chains/algorithms during transition unless policy pins algorithm and path; (ii) validation ambiguity without clear, deterministic policy OIDs; and (iii) availability overhead from path building, OCSP/CRL checks, and extra signature verification. Tampering of extensions is detected because extensions reside \emph{inside} signed \texttt{tbsCertificate}. Protocol confidentiality and side-channels are out of scope here.
\end{tablenotes}
\end{table*}

Hybrid certificate strategies enable: (i) composite certificates embedding classical and 
quantum-safe material in a single certificate via \texttt{tbsCertificate} extensions 
(\gls{PQC} public key and/or signature with policy \glspl{OID}); (ii) dual certificates 
maintaining separate classical and \gls{PQC} certificates for backward 
compatibility~\cite{truskovsky-lamps-pq-hybrid-x509-00,driscoll-pqt-hybrid-terminology-02,vogt2021quantum,bindel2019x}.
Dual certificates simplify transition, as classical endpoints use classical certificates 
while quantum-ready components adopt \gls{PQC} certificates. However, parallel chains 
require tracking, validation, and periodic renewal. Composite certificates carry \gls{PQC} 
components as standardized extensions inside \texttt{tbsCertificate}, avoiding parallel 
chains while maintaining single artifact per 
subject~\cite{truskovsky-lamps-pq-hybrid-x509-00,driscoll-pqt-hybrid-terminology-02,bindel2017transitioning,vogt2021quantum,bindel2019x}. 
This reduces distribution and inventory complexity but requires careful extension design: 
marking \gls{PQC} components critical when \gls{PQC} validation is mandatory (forcing 
legacy validators to reject), or non-critical when graceful fallback to classical 
validation is acceptable. Larger certificate sizes and multiple verifications may impose 
performance overhead on high-volume endpoints.
Security depends on acceptance semantics: with AND (intersection) verification, 
unforgeability holds if at least one algorithm remains secure; with OR/optional (union), 
downgrade becomes possible even if stronger algorithms remain intact. 
Table~\ref{tab:hybrid-cert-strategies} summarizes mechanisms and STRIDE-mapped threats.

\begin{enumerate}[topsep=0ex, itemsep=0ex, wide, font=\itshape, labelwidth=!, labelindent=0pt, label*=6.2.\arabic*]
\item \textit{Hybrid Certificate Strategies Risk Assessment.}
Certificate-layer security depends on policy enforcement and cryptographic binding rather 
than combiner construction. Table~\ref{tab:hybrid-cert-strategies} details implementation 
mechanisms; the fundamental security challenge is acceptance semantics: dual certificates 
exhibit OR-behavior unless policies pin the stronger chain; composite certificates achieve 
AND-semantics only when extensions are marked critical and policy \glspl{OID} enforce 
deterministic validation. \gls{HNDL} attacks critically threaten classical \gls{CA} 
signatures, as adversaries store classically-signed certificates today to retroactively 
break entire certificate chains once quantum computers emerge, invalidating all historical 
signatures. Dominant risks are policy misconfiguration enabling downgrade (Spoofing), 
validation ambiguity (Repudiation), and operational overhead (\gls{DoS}). Tampering is 
detected via signature validation; confidentiality is protocol-layer. Deployments should 
(i) enforce AND verification for composites, (ii) pin stronger chains via policy 
\glspl{OID}, (iii) mark \gls{PQC} extensions critical when mandatory, (iv) accelerate root 
\gls{CA} migration. This aligns with Sections~\ref{sec:pre}--\ref{sec:post} and 
complements Table~\ref{tab:hybrid-cert-strategies}.
\end{enumerate}

\subsection{Hybrid Protocol Strategies in Cloud Services}
\label{subsec:cloud-hybrid-protocol}
\begin{table*}[!htbp]

\caption{Hybrid Protocol Strategies in Cloud Services}
\label{tab:through-transition-protocol}

\scriptsize
\resizebox{\linewidth}{!}{%
\begin{tabular}{|p{0.05\linewidth}|p{0.1\linewidth}|p{0.11\linewidth}|p{0.42\linewidth}|p{0.25\linewidth}|p{0.39\linewidth}|}
\hline 
\textbf{Strategy} & \multicolumn{2}{l|}{\textbf{Approach}$^*$} & \textbf{Mechanisms} & \textbf{Pros.} & \textbf{Cons.} \\ \hline 

\multirow{15}{*}{\begin{minipage}{\linewidth}\centering Hybrid\vspace{0.1cm}  \end{minipage}} 
& \multirow{2}{*}{\begin{minipage}{\linewidth}{Dual protocol}\\\footnotesize Per-connection choice of classical \emph{or} hybrid mode\vspace{0.1cm}  \end{minipage}}
& \begin{minipage}{\linewidth}Dual certificate\vspace{0.1cm}  \end{minipage}
& \begin{minipage}{\linewidth} \vspace{0.1cm}
Mode selected via standard negotiation (e.g., key-exchange (KEX) groups / transform proposals / algorithm lists). Identity/auth supplied by two independent chains (classical, \gls{PQC}) when required. Bind algorithm/group IDs and peer identities into the handshake transcript (or equivalent). Scope resumption/PSK state to the negotiated suite; reject cross-mode reuse.
\vspace{0.1cm}  \end{minipage}
& 
\begin{minipage}{\linewidth} \vspace{0.1cm}
\begin{myBullets}\vspace{2pt}
\item Minimal changes to mature stacks; backward-compatible
\item Incremental rollout; operational separation by mode
\vspace{1pt}\end{myBullets}
 \end{minipage}
& 
\begin{minipage}{\linewidth} \vspace{0.1cm}
\begin{myBullets}\vspace{2pt}
\item Parallel hierarchies (inventory/monitoring/revocation)
\item Redundant advertising; duplicate path-building (CT/OCSP/CRL/AIA)
\item Middlebox/proxy intolerance to new groups; larger handshakes $\rightarrow$ PMTU/fragmentation on UDP
\item Tickets/PSKs must be suite-scoped to avoid downgrade
\vspace{1pt}\end{myBullets}
 \end{minipage}
\\ \cline{3-6} 

& &
\begin{minipage}{\linewidth} \vspace{0.1cm}Composite certificate\vspace{0.1cm}  \end{minipage} 
& 
\begin{minipage}{\linewidth} \vspace{0.1cm}
As above (per-connection mode), but a \emph{single} credential embeds \gls{PQC} components within \emph{signed} certificate fields/extensions; mark \gls{PQC} extensions \emph{critical} when semantics must not be ignored. Express acceptance policy via OIDs/policy objects in the signed body (see Section~\ref{subsec:cloud-hybrid-cert}).
\vspace{0.1cm}  \end{minipage}
& 
\begin{minipage}{\linewidth} \vspace{0.1cm}
\begin{myBullets}\vspace{2pt}
\item Single-artifact management; fewer chain permutations
\item Often backward-compatible if \gls{PQC} extensions are non-critical
\vspace{1pt}\end{myBullets}
\end{minipage}
& 
\begin{minipage}{\linewidth} \vspace{0.1cm}
\begin{myBullets}\vspace{2pt}
\item Profiled extensions + validator/tooling updates
\item Larger certs; parser/validator complexity
\item If not policy-bound/critical, verifiers may ignore \gls{PQC} semantics
\item Same PMTU/middlebox and resumption-scoping concerns
\vspace{1pt}\end{myBullets}
 \end{minipage}
\\ \cline{2-6} 

& \multirow{2}{*}{\begin{minipage}{\linewidth} \vspace{0.1cm}{Composite protocol}\\\footnotesize Single handshake conveys classical+PQC\vspace{0.1cm}  \end{minipage}}
& \begin{minipage}{\linewidth} \vspace{0.1cm}Dual certificate\vspace{0.1cm}  \end{minipage}
& 
\begin{minipage}{\linewidth} \vspace{0.1cm}
One handshake carries \emph{combined} \gls{KEM/ENC} shares and/or parallel authentication. Endpoints enforce deterministic acceptance policy (\emph{AND} vs.\ \emph{OR}). Certificates/credentials remain separate (classical, \gls{PQC}). Transcript binds algorithm/group IDs and identities.
\vspace{0.1cm}  \end{minipage}
& 
\begin{minipage}{\linewidth} \vspace{0.1cm}
\begin{myBullets}\vspace{2pt}
\item Unified semantics; robust combiner behavior
\item Single code path per protocol version
\item Strong downgrade resistance (with \emph{AND})
\vspace{1pt}\end{myBullets}
 \end{minipage}
& 
\begin{minipage}{\linewidth} \vspace{0.1cm}
\begin{myBullets}\vspace{2pt}
\item Stack/key-schedule refactoring
\item Two chains overhead; duplicate path-building if both sent
\item Larger messages: PMTU/fragmentation (UDP); middlebox sensitivity
\item Resumption/PSK must bind to the hybrid suite
\vspace{1pt}\end{myBullets}
 \end{minipage}
\\ \cline{3-6}

& &
\begin{minipage}{\linewidth} \vspace{0.1cm}Composite certificate\vspace{0.1cm}  \end{minipage}
& 
\begin{minipage}{\linewidth} \vspace{0.1cm}
Single handshake (combined KEX and/or \emph{AND} signatures) with a single credential embedding \gls{PQC} key/signature/policy in \emph{signed} fields/extensions. Deterministic acceptance policy end-to-end; transcript binds algorithm/group IDs and identities.
\vspace{0.1cm}  \end{minipage}
& 
\begin{minipage}{\linewidth} \vspace{0.1cm}
\begin{myBullets}\vspace{2pt}
\item Unified across protocol \& authentication
\item Fewer chains; simpler distribution/inventory
\item Clear acceptance semantics (policy OIDs/objects)
\vspace{1pt}\end{myBullets}
\end{minipage}
& 
\begin{minipage}{\linewidth} \vspace{0.1cm}
\begin{myBullets}\vspace{2pt}
\item Largest handshakes (combined shares + bigger credential)
\item Extension profiling + validator/tooling updates required
\item Parser complexity; criticality/policy validation must be correct
\item PMTU/middlebox risks remain; suite-scoped resumption advised
\vspace{1pt}\end{myBullets}
\end{minipage}
\\ \hline 
\end{tabular}%
}
\scriptsize
\begin{tablenotes}[para,flushleft]
\item[$^*$]{Terminology (agnostic).} \emph{Dual protocol}: per-connection choice of classical vs.\ hybrid mode. \emph{Composite protocol}: one handshake conveys classical+\gls{PQC} material. \emph{Dual certificate}: two separate chains; \emph{Composite certificate}: one artifact embedding \gls{PQC} components in the signed certificate body. See Section~\ref{subsec:cloud-hybrid-cert} for certificate details.
\item[\dag]{Protocol mapping (examples).} TLS/QUIC: \texttt{supported\_groups}/\texttt{key\_share}; suite-scoped tickets/PSKs. IKEv2/IPsec: transform proposals + multiple key-exchange payloads; AUTH/CERT carry classical+PQC. SSH: \texttt{KEXINIT} algorithm lists; rekey per connection. Noise: parallel DH inputs via \texttt{MixKey}. DNSSEC: parallel \texttt{RRSIG}/\texttt{DS} algorithms with validator policy. JOSE/COSE: composite signature headers.
\item[\ddag]{Always required.} Bind algorithm/group IDs, identities, and context into the transcript; enforce deterministic acceptance policy (\emph{AND} vs.\ \emph{OR}); scope resumption/PSK to the negotiated suite; monitor PMTU/fragmentation and middlebox behavior as payloads grow.
\end{tablenotes}
\end{table*}

\gls{CC} relies on diverse protocols for data integrity and confidentiality. Hybrid 
protocol models enable both classical and post-quantum handshakes based on client 
capability, ensuring uninterrupted legacy client service while maintaining post-quantum 
accessibility~\cite{BASERI2024103883,driscoll-pqt-hybrid-terminology-02}. Dual protocol 
schemes run parallel classical and post-quantum TLS; composite protocols merge both key 
exchanges in single handshake flows~\cite{bindel2017transitioning,vogt2021quantum,bindel2019x}.
Protocol transition requires cryptographic library modifications and may expand handshake 
messages, introducing performance overhead from larger key sizes and more complex 
negotiations. For heterogeneous client populations (IoT devices, legacy browsers, 
quantum-ready systems), hybrid protocols enable seamless transition. 
Table~\ref{tab:through-transition-protocol} presents approaches maintaining security if at 
least one cryptographic path (classical or \gls{PQC}) remains unbroken. Monitoring resource 
usage, certificate validation, and auto-scaling is critical for multi-tenant performance 
and security.

\begin{enumerate}[topsep=0ex, itemsep=0ex, wide, font=\itshape, labelwidth=!, labelindent=0pt, label*=6.3.\arabic*]
\item \textit{Hybrid Protocol Strategies Risk Assessment:}
As in hybrid algorithmic and certificate approaches, combining multiple independent protocols offers added security while retaining compatibility. Table~\ref{tab:through-transition-protocol} summarizes hybrid protocol strategies under dual or composite models. Regardless of model, systems remain safe if one cryptographic path is unbroken (classical or \gls{PQC}). Administrators must balance performance and management overhead (certificate size, handshake latency, cryptographic library changes) to ensure quantum-era security does not excessively disrupt operations.
\end{enumerate}

\section{Performance Evaluation of \gls{PQC}}
\label{sec:performance_evaluation}
 
\begin{table*}[!h]
\caption{Performance Evaluation of NIST-Standardized \gls{PQC} Algorithms and Their Deployment Suitability in Cloud Environments}
\label{tab:pqc-performance}
\scriptsize
\resizebox{\linewidth}{!}{%
\begin{tabular}{|l|c|c|c|c|c|c|c|c|c|}
\hline
\textbf{Algorithm} & \textbf{Type} & \textbf{Security Level$^{\dagger}$} & \textbf{KeyGen} & \textbf{Enc/Sign}$^{\ddagger}$ & \textbf{Dec/Verify}$^{\ddagger}$ & \textbf{PubKey (B)} & \textbf{CT/Sig. (B)} & \textbf{Total (B)} & \textbf{$\times$ vs Classical$^{\S}$} \\ \hline
ML-KEM-512   & \gls{KEM/ENC} & L1 & 0.032ms (Fast) & 0.032ms (Fast) & 0.022ms (Fast) & 800 & 768 &  1{,}568 & 47.5$\times$ \\
ML-KEM-768   & \gls{KEM/ENC} & L3 & 0.045ms (Fast) & 0.046ms (Fast) & 0.041ms (Fast) &  1{,}184 &  1{,}088 &  2{,}272 & 68.8$\times$ \\
ML-KEM-1024  & \gls{KEM/ENC} & L5 & 0.052ms (Fast) & 0.053ms (Fast) & 0.047ms (Fast) &  1{,}568 &  1{,}568 &  3{,}136 & 95.0$\times$ \\ \hline
HQC-128   & \gls{KEM/ENC} & L1 & 0.120ms (Moderate) & 0.201ms (Moderate) & 0.224ms (Moderate) &  2{,}249 &  4{,}497 &  6{,}746 & 204.4$\times$ \\
HQC-192   & \gls{KEM/ENC} & L3 & 0.219ms (Moderate) & 0.381ms (Moderate) & 0.430ms (Moderate) &  4{,}522 &  9{,}042 & 13{,}564 & 411.0$\times$ \\
HQC-256   & \gls{KEM/ENC} & L5 & 0.451ms (Moderate) & 0.704ms (Moderate) & 0.748ms (Moderate) &  7{,}245 & 14{,}485 & 21{,}730 & 658.5$\times$ \\ \hline
ML-DSA-44 & Signature & L2 & 0.039ms (Fast) & 0.129ms (Moderate) & 0.040ms (Fast) &  1{,}312 &  2{,}420 &  3{,}732 & (39.8$\times$, 37.8$\times$) \\
ML-DSA-65 & Signature & L3 & 0.053ms (Fast) & 0.136ms (Moderate) & 0.056ms (Fast) &  1{,}952 &  3{,}293 &  5{,}245 & (59.2$\times$, 51.5$\times$) \\
ML-DSA-87 & Signature & L5 & 0.083ms (Fast) & 0.165ms (Moderate) & 0.082ms (Fast) &  2{,}592 &  4{,}595 &  7{,}187 & (78.5$\times$, 71.8$\times$) \\ \hline
Falcon-512   & Signature & L1 & 12.69ms (Slow) & 0.525ms (Moderate) & 0.110ms (Moderate) & 897 & 666 &  1{,}563 & (27.2$\times$, 10.4$\times$) \\
Falcon-1024  & Signature & L5 & 34.21ms (Slow) & 1.003ms (Slow)  & 0.199ms (Moderate) &  1{,}793 &  1{,}280 &  3{,}073 & (54.3$\times$, 20.0$\times$) \\ \hline
SLH-DSA-128f & Signature & L1 & 1.155\,ms (Slow) & 28.111\,ms (Slow) & 3.093\,ms (Slow)  &  32 & 17{,}088 & 17{,}120 & (1.0$\times$, 267.0$\times$) \\
SLH-DSA-128s & Signature & L1 & 66.406\,ms (Very Slow) & 497.387\,ms (Very Slow) & 1.133\,ms (Slow) &  32 &  7{,}856 &  7{,}888 & (1.0$\times$, 122.8$\times$) \\
SLH-DSA-192f & Signature & L3 & 1.562\,ms (Slow) & 45.656\,ms (Slow) & 4.596\,ms (Slow)  &  48 & 35{,}664 & 35{,}712 & (1.5$\times$, 557.3$\times$) \\
SLH-DSA-192s & Signature & L3 & 95.411\,ms (Very Slow) & 945.129\,ms (Very Slow) & 1.656\,ms (Slow) &  48 & 16{,}224 & 16{,}272 & (1.5$\times$, 253.5$\times$) \\
SLH-DSA-256f & Signature & L5 & 4.203\,ms (Slow) & 92.525\,ms (Very Slow) & 4.788\,ms (Slow) &  64 & 49{,}856 & 49{,}920 & (1.9$\times$, 779.0$\times$) \\
SLH-DSA-256s & Signature & L5 & 60.923\,ms (Very Slow) & 753.914\,ms (Very Slow) & 2.375\,ms (Slow) &  64 & 29{,}792 & 29{,}856 & (1.9$\times$, 465.5$\times$) \\ \hline
\end{tabular}}
  \begin{tablenotes}[para,flushleft]
 \item[\dag] {Security levels:} L1~$\approx$~AES-128 key search; L2~$\approx$~SHA-256 collision; L3~$\approx$~AES-192 key search; L4~$\approx$~SHA3-384 collision; L5~$\approx$~AES-256 key search.
 \item[\ddag] Speed rubric: \textit{Fast} $<$ 0.1\,ms;\; \textit{Moderate} 0.1–1\,ms;\; \textit{Slow} 1–50\,ms;\; \textit{Very Slow} $>$ 50\,ms. 
 \item[\S] Overhead vs classical baselines: \gls{KEM/ENC} compared to 33\,B ECDH P-256 \emph{compressed} pubkey (33\,B). Signatures show $(\times\mathrm{pk},\,\times\mathrm{sig})$ compared to ECDSA P-256 \emph{compressed} pubkey (33\,B) and \emph{raw} 64\,B.
  \end{tablenotes}
\end{table*}

\gls{PQC} adoption in \gls{CC} environments introduces integration challenges distinct from classical primitives. While providing quantum security, these algorithms impose higher computational latencies and larger artifacts, potentially creating bottlenecks for scale-out architectures. Table~\ref{tab:pqc-performance} benchmarks NIST-approved \gls{PQC} algorithms to evaluate these trade-offs.

\subsection{Benchmarking Methodology and Key Observations} 
We benchmarked NIST \gls{PQC} using \textit{liboqs}~\cite{oqs-liboqs} (v0.9.0) on Intel Xeon E5-2670 v3 (2.30\,GHz, Ubuntu 20.04), compiled with GCC~10.5 (\texttt{-O3 -march=native}) and AVX2 backends. Latency measurements (KeyGen, Encap/Sign, Decap/Verify) represent median of 5{,}000 iterations with 500-run warm-up and CPU pinning. Classical baselines use P-256 ECDH ($2\times33$\,B) and ECDSA (33\,B pubkey, 64\,B signature).

\gls{KEM/ENC} evaluation reveals performance dichotomy between lattice- and code-based designs. ML-KEM (Kyber) sustains sub-100-microsecond latencies (0.032--0.053\,ms), comparable to ECDH P-256 while providing quantum security, with 47.5--95.0$\times$ communication overhead. Code-based HQC exhibits 10--16$\times$ slower decryption and 204.4--658.5$\times$ artifact overhead, creating CPU and bandwidth pressure during high-frequency handshakes. Digital signature schemes present distinct trade-offs for cloud authentication. ML-DSA (Dilithium) offers best general-purpose balance with sub-millisecond operations despite 39.8--78.5$\times$ larger artifacts than ECDSA. For verification-heavy workloads (Content Delivery Networks, edge verifiers), Falcon provides extremely fast verification (0.110--0.199\,ms) and most compact signatures despite costly key generation (12.69--34.21\,ms). SLH-DSA (SPHINCS+) provides stateless security with minimal public key footprint ($\sim$1$\times$ ECDSA) but generates massive signatures (up to 49.9\,KB), suitable for low-frequency, high-integrity applications like firmware signing where bandwidth is secondary to long-term resilience. \gls{PQC} selection must consider cryptographic strength, deployment context, latency tolerance, signature frequency, and bandwidth constraints. ML-KEM and ML-DSA offer best overall balance for general cloud deployments; Falcon serves niche bandwidth-sensitive edge verifiers.

\section{\gls{PQC} in Major \glspl{CSP}}
\label{sec:pqc_cloud_providers}

Major \glspl{CSP}, including \gls{AWS}~\cite{aws2025quantum}, Microsoft Azure~\cite{azure2023quantum}, and \gls{GCP}~\cite{google2023quantum}, have initiated standards-first transitions prioritizing 
hybrid deployments combining classical and post-quantum primitives~\cite{alagic2022status}. 
This section analyzes implementation strategies, showing convergence on NIST FIPS algorithms 
alongside divergent rollouts driven by platform architecture and customer requirements.

\subsection{Comparative Analysis of \gls{PQC} Implementation Strategies}
\label{subsec:implementation_comparison}

\begin{table*}[!htbp]
\centering
\scriptsize
\caption{Comparison of \gls{PQC} Implementations in Major \glspl{CSP}}
\label{tab:integrated_pqc_comparison}
\resizebox{\linewidth}{!}{%
\small
\begin{tabular}{|p{0.14\linewidth}|p{0.55\linewidth}|p{0.55\linewidth}|p{0.55\linewidth}|}
\hline
\textbf{Aspect} & \textbf{ \gls{AWS}} & \textbf{ \gls{GCP}} & \textbf{Microsoft Azure} \\
\hline

\textbf{Approach} & 
\begin{minipage}{\linewidth}
\begin{myBullets}\vspace{2pt}
\item Hybrid key exchange (classical + PQC) in TLS and SSH~\cite{aws_ml_kem_2025, aws_pqc_migration_2025}
\item Gradual integration into security-critical services with crypto-agility focus~\cite{aws_pqc_tls_2024}
\vspace{3pt} \end{myBullets}
\end{minipage} &
\begin{minipage}{\linewidth}
\begin{myBullets}\vspace{2pt}
\item Hybrid \gls{PQC} integration in TLS 1.3 (Chrome, Google services)~\cite{google_chrome_ml_kem_2024, google_cloud_kms_pqc_2025}
\item Quantum-safe digital signatures in Cloud \gls{KMS} preview~\cite{google_cloud_kms_pqc_2025}
\vspace{3pt} \end{myBullets}
\end{minipage} &
\begin{minipage}{\linewidth}
\begin{myBullets}\vspace{2pt}
\item Hybrid cryptographic solutions via SymCrypt library~\cite{microsoft_symcrypt_pqc_2024}
\item \gls{PQC} integration in Windows Insiders and planned Azure services~\cite{microsoft_windows_pqc_2025}
\vspace{3pt} \end{myBullets}
\end{minipage} \\
\hline

\textbf{Algorithms Supported} & 
\begin{minipage}{\linewidth}
\begin{myBullets}\vspace{2pt}
\item ML-KEM (FIPS 203) for key establishment~\cite{aws_ml_kem_2025}
\item ML-DSA and SLH-DSA in AWS-LC library for future service integration~\cite{aws_pqc_migration_2025}
\item CRYSTALS-Kyber support continues through 2025~\cite{aws_ml_kem_2025}
\vspace{3pt} \end{myBullets}
\end{minipage} &
\begin{minipage}{\linewidth}
\begin{myBullets}\vspace{2pt}
\item ML-KEM for key exchange (Chrome/QUIC deployment)~\cite{google_chrome_ml_kem_2024}
\item ML-DSA and SLH-DSA for digital signatures in Cloud \gls{KMS} preview~\cite{google_cloud_kms_pqc_2025}
\item FIPS 203, 204, 205 compliance~\cite{google_cloud_kms_pqc_2025}
\vspace{3pt} \end{myBullets}
\end{minipage} &
\begin{minipage}{\linewidth}
\begin{myBullets}\vspace{2pt}
\item ML-KEM, ML-DSA, and SLH-DSA in SymCrypt~\cite{microsoft_symcrypt_pqc_2024}
\item Available through Windows CNG and SymCrypt-OpenSSL~\cite{microsoft_windows_pqc_2025}
\item XMSS and LMS for specialized use cases~\cite{microsoft_symcrypt_pqc_2024}
\vspace{3pt} \end{myBullets}
\end{minipage} \\
\hline

\textbf{Standardization Alignment} & 
\begin{minipage}{\linewidth}
\begin{myBullets}\vspace{2pt}
\item Full alignment with NIST FIPS 203, 204, 205~\cite{nist2024fips, aws_ml_kem_2025}
\item Contributing to IETF hybrid \gls{PQC} drafts~\cite{aws_pqc_tls_2024}
\item First open-source FIPS 140-3 validation for ML-KEM~\cite{aws_ml_kem_2025}
\vspace{3pt} \end{myBullets}
\end{minipage} &
\begin{minipage}{\linewidth}
\begin{myBullets}\vspace{2pt}
\item Following NIST \gls{PQC} standardization (FIPS 203-205)~\cite{nist2024fips, google_cloud_kms_pqc_2025}
\item Active participation in industry standards development~\cite{google_chrome_ml_kem_2024}
\vspace{3pt} \end{myBullets}
\end{minipage} &
\begin{minipage}{\linewidth}
\begin{myBullets}\vspace{2pt}
\item Strong alignment with NIST standards~\cite{microsoft_symcrypt_pqc_2024}
\item Contributing to IETF protocols and X.509 standardization efforts~\cite{microsoft_windows_pqc_2025}
\vspace{3pt} \end{myBullets}
\end{minipage} \\
\hline

\textbf{Performance Considerations} & 
\begin{minipage}{\linewidth}
\begin{myBullets}\vspace{2pt}
\item Hybrid key exchange adds ~1600 bytes to TLS handshake with 80-150 microseconds overhead~\cite{aws_ml_kem_2025}
\item Minimal impact with connection reuse (0.05\% TPS decrease)~\cite{aws_ml_kem_2025}
\vspace{3pt} \end{myBullets}
\end{minipage} &
\begin{minipage}{\linewidth}
\begin{myBullets}\vspace{2pt}
\item Hybrid \gls{PQC} adds moderate overhead to TLS handshakes~\cite{google_chrome_pqc_performance}
\item Chrome deployment shows minimal user impact~\cite{google_chrome_ml_kem_2024}
\vspace{3pt} \end{myBullets}
\end{minipage} &
\begin{minipage}{\linewidth}
\begin{myBullets}\vspace{2pt}
\item Variable overhead depending on algorithm choice~\cite{microsoft_symcrypt_pqc_2024}
\item TLS and networking may see moderate performance impact~\cite{microsoft_windows_pqc_2025}
\vspace{3pt} \end{myBullets}
\end{minipage} \\
\hline

\textbf{Deployment Strategy} & 
\begin{minipage}{\linewidth}
\begin{myBullets}\vspace{2pt}
\item Production deployment in AWS KMS, ACM, and Secrets Manager~\cite{aws_ml_kem_2025}
\item AWS Transfer Family supports hybrid SSH~\cite{aws_transfer_pqc_2025}
\item Phased rollout to additional services planned~\cite{aws_pqc_migration_2025}
\vspace{3pt} \end{myBullets}
\end{minipage} &
\begin{minipage}{\linewidth}
\begin{myBullets}\vspace{2pt}
\item Default hybrid \gls{PQC} in Chrome (X25519+ML-KEM)~\cite{google_chrome_ml_kem_2024}
\item Quantum-safe signatures in Cloud \gls{KMS} preview~\cite{google_cloud_kms_pqc_2025}
\item Roadmap includes Cloud \gls{HSM} integration~\cite{google_cloud_kms_pqc_2025}
\vspace{3pt} \end{myBullets}
\end{minipage} &
\begin{minipage}{\linewidth}
\begin{myBullets}\vspace{2pt}
\item \gls{PQC} available in Windows Insiders (Build 27852+)~\cite{microsoft_windows_pqc_2025}
\item SymCrypt-OpenSSL 1.9.0 for Linux~\cite{microsoft_symcrypt_pqc_2024}
\item Azure services integration planned~\cite{microsoft_quantum_safe_2025}
\vspace{3pt} \end{myBullets}
\end{minipage} \\
\hline

\textbf{Key Use Cases} & 
\begin{minipage}{\linewidth}
\begin{myBullets}\vspace{2pt}
\item Secure key management, \gls{API} security~\cite{aws_pqc_migration_2025}
\item IoT device protection~\cite{aws_ml_kem_2025}
\item Focus on security-critical services first~\cite{aws_ml_kem_2025}
\vspace{3pt} \end{myBullets}
\end{minipage} &
\begin{minipage}{\linewidth}
\begin{myBullets}\vspace{2pt}
\item Web communications security (TLS/QUIC)~\cite{google_chrome_ml_kem_2024}
\item Enterprise cryptographic services~\cite{google_cloud_kms_pqc_2025}
\item Long-term confidentiality protection~\cite{google_cloud_kms_pqc_2025}
\vspace{3pt} \end{myBullets}
\end{minipage} &
\begin{minipage}{\linewidth}
\begin{myBullets}\vspace{2pt}
\item Operating system security~\cite{microsoft_windows_pqc_2025}
\item Enterprise applications~\cite{microsoft_windows_pqc_2025}
\item Government and compliance-focused deployments~\cite{microsoft_quantum_safe_2025}
\vspace{3pt} \end{myBullets}
\end{minipage} \\
\hline

\textbf{Security Risks} & 
\begin{minipage}{\linewidth}
\begin{myBullets}\vspace{2pt}
\item Emphasis on crypto-agility and secure implementations~\cite{aws_pqc_migration_2025}
\item Hybrid approach mitigates single-algorithm failure risk~\cite{aws_pqc_tls_2024}
\vspace{3pt} \end{myBullets}
\end{minipage} &
\begin{minipage}{\linewidth}
\begin{myBullets}\vspace{2pt}
\item Hybrid \gls{PQC} reduces algorithmic failure risk~\cite{google_cloud_kms_pqc_2025}
\item Open-source implementations enable security audits~\cite{google_chrome_ml_kem_2024}
\vspace{3pt} \end{myBullets}
\end{minipage} &
\begin{minipage}{\linewidth}
\begin{myBullets}\vspace{2pt}
\item Crypto-agility design for algorithm transitions~\cite{microsoft_symcrypt_pqc_2024}
\item Hybrid implementations for risk mitigation~\cite{microsoft_windows_pqc_2025}
\vspace{3pt} \end{myBullets}
\end{minipage} \\
\hline

\textbf{Current Maturity} & 
\begin{minipage}{\linewidth}
\begin{myBullets}\vspace{2pt}
\item \text{Production:} ML-KEM hybrid TLS in KMS, ACM, Secrets Manager~\cite{aws_ml_kem_2025}
\item \text{Production:} SSH hybrid KEX in Transfer Family~\cite{aws_transfer_pqc_2025}
\item \text{Roadmap:} Service expansion and \gls{HSM} integration~\cite{aws_pqc_migration_2025}
\vspace{3pt} \end{myBullets}
\end{minipage} &
\begin{minipage}{\linewidth}
\begin{myBullets}\vspace{2pt}
\item \text{Production:} Default hybrid HTTPS in Chrome~\cite{google_chrome_ml_kem_2024}
\item \text{Preview:} Quantum-safe signatures in Cloud KMS~\cite{google_cloud_kms_pqc_2025}
\item \text{Roadmap:} Cloud \gls{HSM} and broader service coverage~\cite{google_cloud_kms_pqc_2025}
\vspace{3pt} \end{myBullets}
\end{minipage} &
\begin{minipage}{\linewidth}
\begin{myBullets}\vspace{2pt}
\item \text{Preview:} Windows Insiders and Linux SymCrypt~\cite{microsoft_windows_pqc_2025}
\item \text{Roadmap:} Azure services integration by 2033~\cite{microsoft_quantum_ready_2025}
\vspace{3pt} \end{myBullets}
\end{minipage} \\
\hline

\textbf{Future Roadmap} & 
\begin{minipage}{\linewidth}
\begin{myBullets}\vspace{2pt}
\item Expansion to compute, storage, and networking services~\cite{aws_pqc_migration_2025}
\item Standards-driven deployment following IETF finalization~\cite{aws_pqc_tls_2024}
\vspace{3pt} \end{myBullets}
\end{minipage} &
\begin{minipage}{\linewidth}
\begin{myBullets}\vspace{2pt}
\item Scaling across Google Cloud platform~\cite{google_cloud_kms_pqc_2025}
\item Integration with broader security ecosystem~\cite{google_chrome_ml_kem_2024}
\vspace{3pt} \end{myBullets}
\end{minipage} &
\begin{minipage}{\linewidth}
\begin{myBullets}\vspace{2pt}
\item Full quantum-ready systems by 2033~\cite{microsoft_quantum_ready_2025}
\item Enterprise migration guidance and tooling~\cite{microsoft_quantum_safe_2025}
\vspace{3pt} \end{myBullets}
\end{minipage} \\
\hline

\end{tabular}}
\end{table*}

Table~\ref{tab:integrated_pqc_comparison} compares provider deployments across three operational layers: transport (TLS/QUIC, SSH), \gls{KMS}/\gls{HSM}, and application-layer signing. Across providers, adoption centers on \emph{ML-KEM} for key establishment and \emph{ML-DSA}/\emph{SLH-DSA} for digital signatures, with hybrid modes used to preserve interoperability and mitigate downgrade risk during migration. At the transport layer, \gls{GCP} has fielded large-scale hybrids via Chrome’s default HTTPS rollout; \gls{AWS} enables hybrid TLS to front \gls{KMS} and supports hybrid SSH in Transfer Family; Azure is introducing hybrid TLS through the SymCrypt/CNG stack with broader service integration on the roadmap. In key management, \gls{GCP} exposes quantum-safe signatures in Cloud \gls{KMS} with Cloud \gls{HSM} plans; \gls{AWS} provides \gls{PQC}-capable cryptography through \texttt{aws-lc}/\texttt{s2n-tls} and hybrid TLS front-ends; Microsoft aligns Windows cryptographic APIs (CNG/SymCrypt) and Azure Key Vault for ML-KEM/ML-DSA exposure as implementations harden and certify. For application/signing, providers are preparing ML-DSA as the primary scheme with SLH-DSA as a conservative option for long-lived artifacts, while tracking CNSA~2.0 profiles and evolving IETF guidance for protocol and certificate ecosystems. Reported performance characteristics indicate larger handshake messages and modest compute overheads that remain manageable at scale; lattice-based schemes typically meet cloud latency budgets, while SLH-DSA is favored for archival or long-lived signatures. Overall maturity is highest where hybrids are already generally available or widely deployed, with staged expansion to additional services and \glspl{HSM}.

\subsection{Synthesis and Implications}
\label{subsec:synthesis}

Major \glspl{CSP} follow a three-phase path: hybrid post-quantum key establishment at network edge addressing \gls{HNDL} while maintaining compatibility; \gls{PQC} integration into \gls{KMS}/\gls{HSM} and platform libraries enabling quantum-safe signing and key-wrap; and alignment with FIPS~203/204/205 and CNSA~2.0 as IETF finalizes protocol guidance. The posture is crypto-agile with ML-KEM for key establishment, ML-DSA as default signature scheme, and SLH-DSA for long-lived assets. Remaining challenges include side-channel hardening, interoperability across legacy middleboxes, and enterprise migration playbooks; continued cross-provider testing, certification, and operational guidance will be essential for transitioning from hybrid to pure post-quantum deployments.
\section{Future Research Directions}\label{sec:future_directions}

Mitigating quantum threats in \gls{CC} requires coordinated advances across
standardization, performance optimization, implementation security, system
integration, workforce preparedness, and crypto-agile migration.

\subsection{Standardization and Interoperability}\label{subsec:standardization_interoperability}

\gls{NIST} finalized core \gls{PQC} standards~\cite{nistfips203,nistfips204,nistfips205},
while IETF standardized hybrid terminology (RFC~9794~\cite{RFC9794}) and continues
developing hybrid key exchange for \gls{TLS}~1.3 and \gls{SSH}~\cite{ietf-tls-hybrid-design-12,
kwiatkowski2025mlkem,kampanakis2023ssh-pq-ke}. Despite limited hybrid deployments by major
\glspl{CSP}~\cite{aws2025mlkem,aws2024pqc,nist_pqc_standardization}, ecosystem-wide
conformance remains incomplete. Recent work on threshold lattice signatures and adaptive
resistance~\cite{gur2024threshold,moriya2024adaptive} highlights progress but also exposes
open challenges: scalable hybrid certificate path validation and revocation across
federated \glspl{PKI}, downgrade-safe negotiation during algorithm transitions, 
performance-aware algorithm selection across CDNs and service meshes, and vendor-neutral 
\gls{API} support preserving algorithm agility. Large-scale interoperability testbeds are 
required to ensure predictable Internet-scale behavior for \gls{TLS}, \gls{SSH}, and IKEv2.

\subsection{Hardware Acceleration and Performance Optimization}\label{subsec:performance_optimization}

GPU and FPGA accelerators provide order-of-magnitude speedups for batched
\glspl{KEM/ENC} and signature verification~\cite{9201530,he2023fpga,NVIDIAcuPQC}, but
performance remains sensitive to batching strategies, parameter choices, and memory
hierarchies. In multi-tenant clouds, shared accelerators introduce evolving side-channel
and fault surfaces tightly coupled to scheduling and telemetry. Compiler-based defenses
complement constant-time coding and masking~\cite{mosier2024serberus}. Key research
directions include reducing tail latency under fan-out, \gls{PQC}-aware autoscaling,
container/\gls{OS} memory policies for larger cryptographic artifacts, safe accelerator 
multiplexing across tenants, and reproducible cross-provider benchmarks capturing energy 
efficiency (ops/Joule), network overhead (bytes-on-wire), tail latency, and SLO compliance 
costs. Hardware–software co-design should further explore FPGA/ASIC/GPU acceleration, 
energy-aware profiles, and cloud-native \gls{PQC} libraries~\cite{PQCryptoLibProductPage,
oqs-liboqs}.

\subsection{AI-Enhanced Security and Implementation Robustness}\label{subsec:implementation_security}

The algebraic complexity and memory access patterns of \gls{PQC} complicate
constant-time enforcement. \gls{ML}-based side-channel attacks have recovered secrets from
masked implementations (e.g., SALSA~\cite{refml25}), exploiting deep learning's connection 
to information leakage~\cite{refml31}, while fault attacks exploit floating-point precision 
in schemes such as Falcon~\cite{refc6}. These risks are amplified in multi-tenant 
environments with shared hardware~\cite{ravi2024side,ali2024emerging}. Effective defenses 
require \gls{ML}-driven anomaly detection analyzing timing, power, and runtime 
telemetry~\cite{refml12,lou2021survey,refml31}, compiler-assisted constant-time enforcement
across CPU/GPU ISAs~\cite{zhang2024timing}, enclave-aware threat models for SGX, TDX, SEV, 
and CCA~\cite{COPPOLINO2025104457,10584087}, and systematic leakage evaluation (TVLA, fault 
campaigns) integrated into \gls{CICD} pipelines. Distributed key generation further improves 
resilience in multi-tenant settings~\cite{refc4,refc32,refc35}. Future work should integrate 
AI-driven adaptive defenses into \gls{CICD} workflows~\cite{10918314,9367411}. Emerging 
lattice-based zero-knowledge systems (LaZer~\cite{refc31}, LaBRADOR~\cite{refc12}, 
SMILE~\cite{refc29}), field-agnostic SNARKs~\cite{refc9}, non-malleable 
commitments~\cite{refc18}, and proximity testing~\cite{refc38} warrant evaluation under 
realistic cloud constraints including multi-tenant noise, redundancy, and control-flow 
integrity requirements.

\subsection{Integration with Emerging Cloud Technologies}\label{subsec:system_integration}

End-to-end quantum resilience requires integrating \gls{PQC} across transport protocols
(\gls{TLS}/QUIC/KEMTLS), identity frameworks (OAuth~2.0, OpenID Connect, \gls{SAML}), and 
control planes (\gls{SDN}, \gls{NFV}, service meshes). Larger cryptographic artifacts 
stress certificate revocation freshness, key rotation cadence, and path validation at 
scale. Succinct lattice commitments and OPRFs enable bandwidth-efficient 
verification~\cite{fenzi2024slap,leap2025}. Further work is needed on \gls{PQC}-enhanced 
\gls{TLS}~1.3~\cite{post-quantum-tls,open-quantum-safe-tls,schwabe2020post}, 
quantum-resistant identity systems~\cite{BASERI2024103883,baseri2025evaluation},
\gls{PQC}-capable authentication and authorization at web and service mesh scale,
efficient OCSP stapling and revocation mechanisms for post-quantum certificate sizes, 
platform-independent \gls{KMS}/\gls{HSM} \glspl{API} supporting hybrid key management and 
secure rollback~\cite{11013769}, energy-aware \gls{PQC} profiles for \gls{IIoT} and edge 
deployments, and integration with 6G networks~\cite{10806885}, \gls{SDN}~\cite{10433773}, 
\gls{NFV}, AI-driven cloud orchestration~\cite{10430465}, long-term data 
archival~\cite{turnip2025towards}, and quantum–classical hybrid architectures.

\subsection{Workforce Development and Systemic Preparedness}\label{subsec:systemic_preparedness}

Policy roadmaps mandate post-quantum migration (e.g., EU critical infrastructure by 2030,
U.S. NSM-10 and OMB-M-23-02 by 2035~\cite{EU2025roadmap,NSM10,OMB2022M23}), yet operational 
readiness remains uneven. Organizations require threat models coupling \gls{HNDL} exposure 
with migration lead times, supply-chain dependencies, and audit requirements~\cite{
nist2024transition,mosca2018cybersecurity}. Measurable preparedness programs should integrate 
\gls{CBOM}/\gls{KBOM} coverage for risk-prioritized sequencing with maturity metrics 
including inventory coverage, key rotation \gls{MTTR}, rollback reliability, and policy 
conformance. Workforce development demands modular training pipelines, professional 
certification tracks, and open-source laboratory environments co-developed by academia and 
industry to deliver deployer-level competence in constant-time coding, KEMTLS integration, 
\gls{KMS} policy configuration, and service mesh hardening on timelines aligned with 
migration milestones.

\subsection{Migration Frameworks and Crypto-Agility}\label{subsec:migration_roadmap}

Quantum-safe transitions require crypto-agile architectures supporting algorithm 
substitution, secure rollback, and protocol composability with minimal 
downtime~\cite{barker2018transitioning}. Phased evolution encompasses near-term hybrid 
deployments combining classical and post-quantum 
schemes~\cite{aws_hybrid_ssh,ietf-tls-hybrid-design-12,driscoll-pqt-hybrid-terminology-02}, 
broader crypto-agile infrastructure adoption~\cite{ma2021caraf}, and ultimately full 
quantum-resistant systems~\cite{nist2024transition}. NCCoE guidance emphasizes automated 
cryptographic discovery and inventory, staged cutovers with rollback capabilities, 
continuous testing and validation, and policy-driven risk triage~\cite{nist2024transition}. 
Two priority research directions are control-plane automation via policy-as-code controllers 
in Kubernetes, Terraform, and \gls{CICD} pipelines validating reachability and conformance 
with automatic rollback on error budget violations, and formal assurance using TLA$^{+}$ 
models~\cite{10.1145/3558819.3558826} proving safety and liveness properties under partial 
deployments, version skew, negotiation invariants, and cross-region coordination. Open 
datasets including certificate chain sets, packet captures, deployment manifests, and 
failure-injection playbooks are essential for benchmarking migration safety, 
interoperability, and SLO impact at scale. Future work should address advanced 
orchestration tools~\cite{kinanen2025toolchain,STIRBU2024107529}, governance models for 
cryptographic lifecycle management~\cite{alagic2020status}, cloud-native crypto-agility 
integration~\cite{MARCHESI2025107604}, modular cryptographic 
infrastructures~\cite{ducas2018crystals}, automated fallback and recovery 
mechanisms~\cite{barker2018transitioning}, decentralized update 
coordination~\cite{joseph2022transitioning}, performance optimization strategies to mitigate 
\gls{PQC} overhead~\cite{kwon2024compact}, and hybrid interoperability 
frameworks~\cite{turnip2025towards}.

\section{Conclusion}\label{sec:conclusion}

\gls{QC} fundamentally threatens the cryptographic foundations of \gls{CC}, introducing attack vectors that jeopardize confidentiality, integrity, and trust across \gls{IaaS}, \gls{PaaS}, and \gls{SaaS} environments. This survey provided a multi-layered analysis of quantum-era risks and established a roadmap toward quantum-resistant cloud architectures. By applying a {STRIDE}-based risk assessment aligned with \gls{NIST} SP~800-30, we evaluated threats across pre-transition, hybrid, and post-transition phases, enabling layer-specific mitigation strategies for communication protocols, \gls{PKI}, \gls{KMS}, and control planes. Our analysis shows that hybrid cryptographic deployments are essential for mitigating \gls{HNDL} risks while preserving backward compatibility during migration. A technical evaluation of \gls{NIST}-standardized \gls{PQC} algorithms revealed significant implementation, performance, and deployment challenges, notably susceptibility to \gls{SCA} and fault attacks in multi-tenant environments involving shared \glspl{VM}, accelerators, and \glspl{TEE}. Furthermore, a comparative study of \gls{AWS}, Azure, and \gls{GCP} highlighted progress in cryptographic agility and \gls{HSM} integration, while exposing persistent gaps in interoperability and deployment maturity. Overall, securing the quantum-era cloud requires coordinated evolution across cryptographic infrastructure, governance, and workforce preparedness. Future research should prioritize standardization, hardware-based performance optimization, AI-enhanced threat mitigation, and automated, policy-driven, crypto-agile migration frameworks to sustain long-term resilience and trust in global cloud ecosystems.

\bibliographystyle{IEEEtran}
\bibliography{bibliography}

\end{document}